\documentclass{article}
\usepackage{amssymb,amsmath,amsthm,amsfonts,eurosym,geometry,ulem,graphicx,caption,color,xcolor,setspace,sectsty,comment,footmisc,caption,natbib,pdflscape,subfigure,array,hyperref,booktabs,float,multirow,wasysym,soul,tabularx,placeins,bm,rotating,adjustbox,afterpage}

\newtheorem{proposition}{Proposition}[section]
\usepackage{bbm}
\usepackage[title]{appendix} 

\hypersetup{
    colorlinks,
    linkcolor={red!50!black},
    citecolor={blue!50!black},
    urlcolor={blue!80!black}
}

\sethlcolor{red}

\DeclareMathOperator{\Var}{Var}
\DeclareMathOperator{\Cov}{Cov}
\DeclareMathOperator{\Corr}{Corr}
\begin{document}

\begin{titlepage}
\title{The Information Content of News Embeddings\footnote{I am grateful to Torben Andersen and Viktor Todorov for their valuable comments and suggestions. The use of AI tools, including ChatGPT, Codex and Claude Code, for coding assistance and minor language editing is acknowledged. Replication code is available at \url{https://github.com/rootchunhe/news-embeddings}. All errors are my own.}}
\author{Chun He\thanks{Kellogg School of Management, Northwestern University;  e-mail:  chun.he@kellogg.northwestern.edu.}}
\date{\today}
\maketitle
\begin{abstract}
Does the predictive success of news embeddings reflect a better measure of news surprises or other sources of return predictability in text? We decompose news embeddings by content into event-category, firm-identity, and idiosyncratic components, and by timing into historical levels and current deviations. At short horizons, event categories account for most of the predictive power of embedding forecasts, and simple forecasts based on historical average returns after news in each event category outperform them. At longer horizons, firm information already present in earlier news coverage accounts for a substantial share of return predictability, even after residualization on firm characteristics. The dominance of event categories persists, and historical levels still contribute in the first months, when using a language model pretrained entirely before our sample period. For Llama-2, even embeddings constructed from news more than three years old predict returns at long horizons. However, its predictive advantage over RoBERTa is concentrated in firm identity and the historical level, and its advantage in these components largely disappears by its pretraining cutoff. A calibration shows that pretraining memory, a source of look-ahead bias, can account for this pattern. Describing news more fully is not the same as measuring news surprises, and persistent return predictability from news text does not by itself establish slow incorporation of news into prices.

\vspace{0.5em}

\bigskip
\end{abstract}
\setcounter{page}{0}
\thispagestyle{empty}
\end{titlepage}
\pagebreak

\doublespacing
\newpage

\section{Introduction}

Advances in large language models (LLMs) have opened new possibilities for extracting economic information from text. A longstanding challenge in finance is to measure the information conveyed by news and understand how prices respond to it. \citet{chen2022expected} (henceforth CKX) represent news articles using high-dimensional embeddings extracted from LLMs and find that embedding-based forecasts predict returns over the next few trading days. \citet{didisheim2026inefficient} (henceforth DKPT) seek to isolate ``news shocks'' by removing the part of embeddings explained by firm characteristics and find return predictability up to 18 months ahead. Both studies interpret their findings as evidence that news is gradually incorporated into prices. But does this predictive success reflect a better measure of news surprises, or other sources of return predictability embedded in text?

For short-horizon return prediction, traditional event studies provide a natural benchmark. \citet{ball1968empirical} first note evidence of post-announcement return drift based on the sign of unexpected earnings, while \citet{bernard1989post} show that returns continue to drift in the direction of earnings surprises over subsequent months. Although conventional post-earnings-announcement drift has weakened considerably over time \citep{martineau2022rest}, text-based measures from earnings conference calls continue to predict post-announcement returns \citep{meursault2023pead}. These findings highlight how both simple classifications of news, such as the sign of earnings surprises, and richer textual information can predict returns. For broader news coverage, event categories provide a similar classification. Embedding forecasts may draw on the average returns associated with these categories as well as on information beyond them. However, extracting weak return signals from high-dimensional embeddings is challenging, particularly when the signal is small relative to return noise \citep{shen2025can}. The relevant question is how much embedding-based forecasts add beyond the predictability of the assigned event categories.

At longer horizons, distinguishing delayed price adjustment from persistent differences in expected returns becomes more difficult. This reflects the joint-hypothesis problem of market efficiency \citep{fama1970efficient}, which is particularly consequential for long-horizon return tests \citep{fama1998market}. This concern is relevant for news embeddings because news articles repeatedly describe firms' businesses, prospects, and other persistent attributes. Embeddings may capture firm states associated with expected returns, whether through risk premia \citep{fama1993common,fama2015five} or persistent mispricing \citep{stambaugh2017mispricing,daniel2020short}. If embeddings capture persistent firm states, they may predict returns long after the news appears, even if the information revealed by the news was immediately incorporated into prices. Because observed firm characteristics need not span these latent states, residualizing embeddings against them does not necessarily eliminate this source of return predictability.

We provide a simple framework in which return predictability can arise from price adjustment following news events or persistent differences in expected returns. To examine how these sources of predictability are reflected in news embeddings, we introduce two complementary decompositions. A content decomposition separates embedding variation into components associated with event categories, firm identity, and the remaining idiosyncratic variation. The decomposition reflects the semantic structure of financial news and connects embedding-based forecasts to traditional event studies, allowing us to assess how much short-horizon predictability comes from event categories and how much lies beyond them. A timing decomposition separates a firm's historical embedding level from deviations in recent coverage, allowing us to assess whether long-horizon predictability comes from the current news or from firm information already present in earlier news. We use observable measures of news surprises and firm fundamentals to examine the economic information represented in embeddings, and then attribute return predictability across horizons to the embedding components. 

Our decompositions trace much of short-horizon predictability to event categories and long-horizon predictability to components constructed from earlier firm coverage. Using Dow Jones headlines encoded with RoBERTa and Llama-2, we reproduce the main predictive patterns documented by CKX and DKPT and examine the information underlying forecasts constructed using their methods.

A simple forecast based on historical average returns by event category, such as analyst downgrades, performs better than the embedding forecasts. Even a cruder rule based on broad event groups and simple directional indicators extracted from headlines outperforms them. Previous research documents distinct market reactions across different types of corporate news \citep{neuhierl2013market, kwon2026extreme}. We find that event categories account for most of the covariance between embedding forecasts and next-day returns. This predictability decays rapidly after the news, consistent with the average return drift associated with each event category.

We next examine what embeddings add beyond event categories. In the full sample of news headlines, embedding forecasts remain statistically significant after controlling for category averages, but their incremental out-of-sample gains are modest. These findings suggest that embeddings capture information beyond event labels but may recover category-level return patterns less effectively. The comparison also varies across event groups. For revenue-related news, where category averages have relatively weak predictive power, embedding forecasts rank returns substantially better.

We also examine the economic information represented in news embeddings. For earnings announcements, embeddings explain variation in standardized unexpected earnings (SUE) beyond event categories, but return-trained forecasts retain little of this information, even though observed earnings surprises strongly predict subsequent returns. Controlled headline rewrites further show that forecasts respond much more strongly to the direction of earnings news than to reported earnings figures. Turning to firm fundamentals, we find that news-derived firm representations explain valuations and analysts' long-term growth forecasts beyond conventional firm characteristics. Firms that news embeddings place close together also share operating and risk characteristics, for some characteristics more closely than peers defined by conventional industry classifications.

At long horizons, DKPT attribute the persistent predictability of their residualized-embedding strategy to the slow incorporation of news shocks into prices. However, the residualization need not completely remove persistent firm information associated with expected returns. We first examine firm identity, a component estimated from news published before the test year and held fixed throughout it. Firm identity accounts for a substantial share of the replicated strategy's returns even two years after the signal. Because the underlying news may have been published relatively recently, this predictability could still reflect delayed adjustment to that news. We examine whether firm identity estimated three years earlier still predicts returns. Using this lagged component in place of monthly embeddings, while otherwise following DKPT's strategy construction procedure, yields substantial strategy returns for Llama-2, including in the 36th month after the signal. For RoBERTa, the lagged component also predicts returns, primarily at shorter horizons. Explaining this predictability through delayed price adjustment would require a response to news published more than three years earlier. These findings suggest that the historical level, mainly through firm identity, contributes substantially to long-horizon return predictability, which cannot be attributed solely to the delayed incorporation of the current month's news.

We then extend the analysis to firm information that evolves over time by separating each embedding into a historical level, constructed from the firm's news coverage over the preceding 12 months, and a deviation from that level. The historical level accounts for most of the strategy's monthly returns at every horizon up to three years for Llama-2 and at the majority of these horizons for RoBERTa. Even in the 36th month, it accounts for a substantial share of returns for both models. In contrast, the contribution of deviations drops sharply within the first two months, while that of the historical level declines more gradually. Beyond the first two months, deviations predict returns mainly at a few horizons that are multiples of three months, a seasonal pattern absent from the historical level.

The historical level's predictability could still be interpreted as delayed price adjustment to news published in earlier months. To examine this possibility, we test whether deviations from those months contain predictive information beyond the current historical level. We find little incremental predictability apart from the same quarterly-spaced pattern, while the current level's predictive power remains largely unchanged across lags. In our framework, deviations can capture changes in a firm's persistent state not yet reflected in earlier coverage. As subsequent news reflects these changes, the updated historical level absorbs the return-relevant information previously contained in the deviations. Apart from this quarterly pattern, return predictability beyond the first months comes from information accumulated in earlier coverage, not from the current month's news.

To assess whether our findings are affected by look-ahead bias in pretrained LLMs, we repeat our analysis using ChronoGPT$_{1999}$ of \citet{he2026chronologically}, pretrained entirely before our sample begins. Event categories still account for most of the next-day predictive covariance, and the historical level continues to contribute substantially to the monthly strategy's returns, although it fades after the first few months.

Nevertheless, a remaining concern is look-ahead bias in LLMs with stronger representation capabilities. Persistent firm information and pretraining memory of subsequently realized outcomes can generate similar patterns of return predictability before an LLM's pretraining cutoff. This concern is particularly relevant for Llama-2, whose predictive advantage over RoBERTa is concentrated in the firm-identity and historical-level components of news embeddings. In our replication, CKX's short-horizon predictability is driven mainly by event categories, so their post-cutoff and anonymization checks may be insensitive to such memory. DKPT find nearly identical total performance using ChronoGPT trained with and without later text, but ChronoGPT's weaker representation capacity relative to Llama-2 limits what this comparison can reveal about larger LLMs.

We examine Llama-2 around its pretraining cutoff, focusing on separate embedding components because overall predictive performance may mask changes in the sources of predictability. Llama-2's historical-level advantage over RoBERTa largely disappears from 2021, before its cutoff, and does not recover after it, while its deviation component shows no comparable decline. RoBERTa's historical level shows no similar decline around its own cutoff. We quantify these changes using the ratio of post- to pre-cutoff rank information coefficients (ICs) for Llama-2 relative to RoBERTa, which equals one if both encoders' predictability changes by the same proportion. The observed relative decline is difficult to explain by Llama-2 measuring persistent firm states better than RoBERTa. A calibration allowing for pretraining memory can account for this pattern, suggesting that some of the additional predictability of larger LLMs may reflect look-ahead bias.

\paragraph{Related literature.} 
Our paper contributes to the literature on measuring news and its relation to stock returns. This literature represents news through earnings surprises as a single number measured against expectations \citep{bernard1989post}, the price move accompanying public news \citep{chan2003stock,tetlock2010does,savor2012stock}, and classifications of corporate events \citep{neuhierl2013market,boudoukh2019information,kwon2026extreme}. Text measures include counts of negative or dictionary words \citep{tetlock2008more}, learned or supervised sentiment \citep{heston2017news, ke2026enhancing}, and LLM-scored sentiment \citep{lopez2026can}. CKX and DKPT use high-dimensional embeddings of news text to forecast returns. Because embedding dimensions have no direct interpretation, a related question is what information drives embedding-based predictions. CKX cluster news articles into topics and find that analyst-related news yields the most profitable signals. \citet{chen2025financial} replace dense embeddings with sparse-autoencoder features labeled by plain-English concepts and find that sentiment features matter most for next-day return forecasts; DKPT use these features to trace their residualized-embedding strategy to news themes such as corporate guidance. \citet{didisheim2026factor} construct factors from sparse-autoencoder features of news embeddings and find that earnings, analyst actions, and guidance account for much of their power to span the traditional factor zoo. Our content decomposition separates event-category, firm-identity, and idiosyncratic components of news embeddings and traces their contributions to return forecasts. Using historical event-category returns as a benchmark, we assess how much short-horizon predictability comes from event categories and what economic information embeddings contain beyond them.

Our paper also contributes to the literature on measuring firms' economic attributes. Examples include product-market similarity from filing descriptions \citep{hoberg2010product, hoberg2016text}, changes in disclosure text \citep{cohen2020lazy}, asset representations from investor holdings \citep{gabaix2025asset}, and business networks from business descriptions \citep{breitung2025global}. Relatedly, \citet{sarkar2026economic} constructs news-based representations of firms' perceived business models and studies how these representations and their changes relate to valuations and stock returns. We examine the economic information in firm-level representations through valuations, analysts' growth forecasts, and similarities in firms' operating and risk characteristics. We then examine the contributions of firm identity and historical levels to long-horizon return predictability, distinguishing information already reflected in earlier firm coverage from changes associated with recent news.

A concern for LLM-based return prediction is that models may encode information from text published after the news \citep{sarkar2024lookahead,lopez2025memorization}, and memorization increases with model size \citep{didisheim2025ai}. Responses include chronological training \citep{he2026chronologically,kelly2026scaling}, identity removal \citep{glasserman2024assessing,engelberg2025entity}, which can also remove economically relevant information \citep{wu2025anonymization}, recall diagnostics \citep{gao2025detecting}, and CKX's and DKPT's aggregate performance checks discussed above. DKPT argue that look-ahead bias is likely minor in their embedding-based return regressions because return prediction and token prediction are separate objectives, and attribute their larger model's advantage over ChronoGPT to scale. However, \citet{levy2026caution} finds that LLM representations of financial statements predict earnings changes primarily when the models were pretrained on text published after those statements, consistent with memorization. Pretraining memory may confound the gains attributed to model scale. We examine this possibility using component-specific predictability around pretraining cutoffs and a calibrated reduced-form framework.

\section{Event Drift and Short-Horizon Predictability}
\label{sec:event_drift}

News articles report recognizable events, such as analyst upgrades and downgrades, earnings increases and declines, and insider purchases and sales. These categories distinguish both the nature of an event and, where applicable, the direction of the reported change. We use the term event drift to describe the average post-news return path associated with an event category. This definition allows subsequent returns to continue or reverse the initial price movement. We first document these paths and then examine how forecasts based on historical category returns compare with forecasts estimated from news embeddings.

Our sample consists of Dow Jones news obtained through RavenPack and matched to CRSP U.S. common stocks. News histories begin in 2000, and out-of-sample evaluation covers 2004--2025. Following CKX, we retain single-company news with nonmissing subsequent open-to-open returns and remove duplicate headlines. We define single-company news as headlines with exactly one company assigned RavenPack's maximum relevance score of 100; other companies may appear with lower scores. Deduplication follows CKX's rule: we remove a headline if its binary bag-of-words representation has cosine similarity of at least 0.8 with an earlier headline about the same firm within the preceding five trading days, including earlier headlines on the same day.

We impose an additional size filter because Dow Jones RavenPack covers substantially more small firms than CKX's Reuters sample. We require market capitalization at or above the NYSE fifth percentile, using both capitalization and the NYSE breakpoint from the preceding trading day. This restriction brings the monthly number of firms with news closer to the 2,593 reported by CKX. For both news flashes and full articles, our embeddings use only the headline.\footnote{Our RavenPack data contain only headlines. CKX compare embeddings of the headline and the body text of the same articles and find modestly higher equal-weighted Sharpe ratios for body text, with headlines performing better in some value-weighted specifications (their Table~16).}

RavenPack assigns each story to an event category in a predefined taxonomy by applying fixed linguistic templates to the news text. For example, \texttt{analyst-\allowbreak{}ratings-\allowbreak{}change-\allowbreak{}positive} and \texttt{analyst-\allowbreak{}ratings-\allowbreak{}change-\allowbreak{}negative} separate upgrades from downgrades within the \texttt{analyst-ratings} group.\footnote{Categories and groups come from RavenPack's \texttt{CATEGORY} and \texttt{GROUP} fields. When a story reports multiple events for the focal company, we select the event with the highest \texttt{EVENT\_RELEVANCE}, breaking ties by its order in the story.} Although the assignment rules are fixed, the taxonomy itself was designed by the vendor. Most categories record an event that the story states, which leaves little room for hindsight.\footnote{For example, categories separate insider sales from purchases, earnings increases from decreases, analyst upgrades from downgrades, and dividend increases from dividend announcements.} To address the possibility that some finer distinctions reflect hindsight about which events matter for prices, we also use RavenPack's 38 broad event groups as an arguably ex ante classification. These groups name only the subject of the news, such as earnings or analyst ratings, without distinguishing outcomes. We split each group into positive, negative, or neutral direction by a fixed list of headline words.\footnote{The direction rule uses only ten word pairs: up/down, increase/decrease, above/below, upgrade/downgrade, positive/negative, buy/sell, bullish/bearish, raise/cut, gain/loss, and overbought/oversold, including regular inflections. A headline is negative if it contains any negative word, positive if it contains a positive word and no negative word, and neutral otherwise. The rule uses headline text alone and is not fitted to returns.}

We encode each headline with two pretrained language models, RoBERTa-large \citep{liu2019roberta} and the 13-billion-parameter base version of Llama-2 \citep{touvron2023llama}, whose pretraining data extend to July 2019 and September 2022, respectively. As in CKX, we keep the model weights fixed, use mean-pooled last-layer hidden states over non-padding tokens as headline embeddings, and estimate a ridge regression of next-day open-to-open returns on these embeddings. For each test year $T$, the regression is estimated on years $\max(T-8,2000)$ through $T-3$, and the following two years are used for validation. Because CKX do not specify the validation criterion, we choose the ridge penalty from their grid to maximize the rank correlation between forecasts and returns in the validation years. The fitted coefficients are held fixed during the test year. Forecasts are averaged across headlines about the same firm assigned to the same trading day.\footnote{We convert RavenPack's \texttt{TIMESTAMP\_UTC}, which records receipt by its servers, to New York time. News received before 09:00 on a trading day is assigned to that day's open; later news is assigned to the next trading day's open. Weekend and holiday news is assigned to the next eligible trading open. The prediction target is the return from that open to the following trading day's open. Training returns are winsorized at $\pm50\%$; forecast evaluation uses unmodified realized returns.}

\begin{figure}[htbp]
\centering
\caption{Cumulative Returns after Analyst Rating News}
\includegraphics[width=1\textwidth]{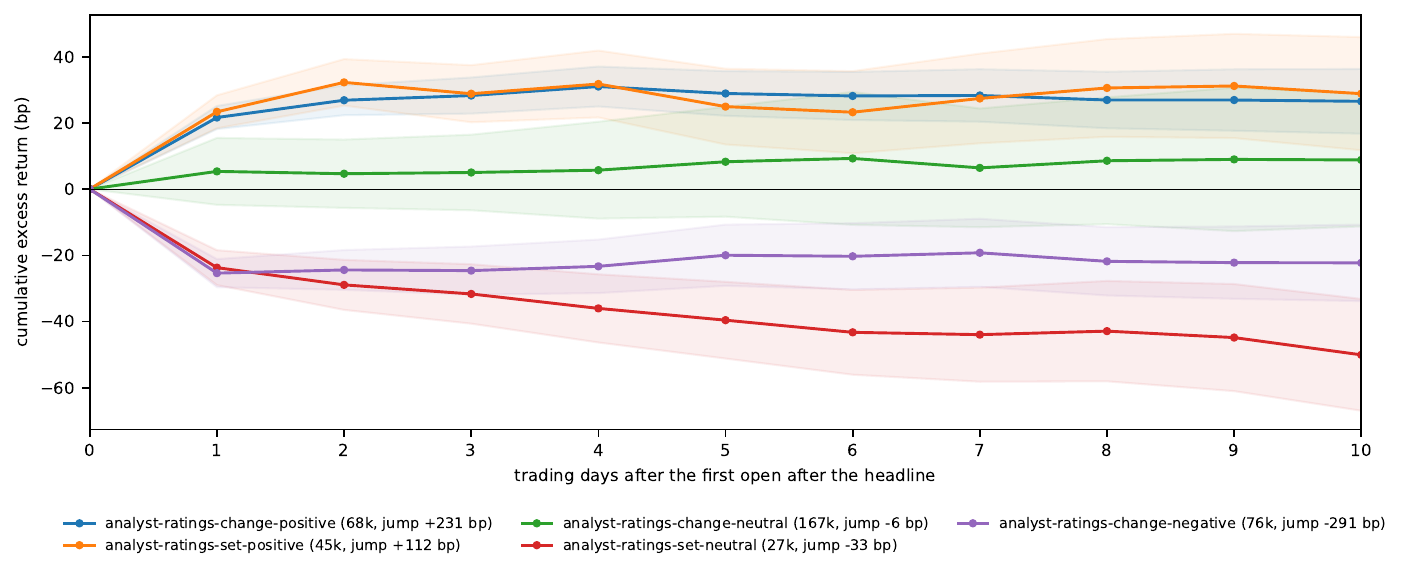}
\label{fig:fig1_drift}
\begin{minipage}{\linewidth}\footnotesize
This figure plots cumulative returns after news in each analyst-rating category, 2004--2025. The horizontal axis counts trading days after the first eligible open. Curves sum subsequent open-to-open returns, in basis points, in excess of the mean return of all headlines on the same trading day, averaged equally across trading days. The sample includes headlines with an observed initial return and a complete ten-day return path, and each plotted category has at least 5,000 such headlines. Each legend entry gives the category's number of headlines, in thousands, and its jump, the initial excess return to the eligible open from the preceding close or, for news published during the trading session (09:30--16:00), from that session's open. Curves start at zero at the eligible open and exclude the jump. Shading shows 95\% Newey--West confidence intervals with $\max(5,h)$ lags at day $h$.
\end{minipage}
\end{figure}

Figure~\ref{fig:fig1_drift} illustrates event drift using the analyst-ratings group as an example, plotting cumulative returns over ten trading days from the first eligible open after the news.\footnote{For overnight news, the initial movement is measured from the preceding close to the eligible open. For news released during a trading session, it starts at that session's open and includes returns before the news timestamp. The plotted paths begin at the eligible open and exclude this initial movement.} Table~\ref{tab:tabA_drift} reports, for categories with at least 20,000 headlines in the displayed event groups, the price movement to that open and the subsequent returns. Several categories exhibit continuation. Positive analyst-rating changes are associated with an initial increase of 231 basis points, followed by a further 22 basis points on the first day and 27 basis points over ten days. Negative rating changes are associated with an initial decline of 291 basis points, followed by declines of 25 and 22 basis points over the corresponding windows. Insider purchases and sales also separate subsequent returns: their ten-day cumulative returns are approximately 31 and $-16$ basis points, respectively.

The paths differ substantially across event categories, including within the same broad event group. News classified as \texttt{earnings-negative} is followed by a further decline of 40 basis points on the first day and 50 basis points over ten days. By contrast, \texttt{earnings-up} has a positive initial movement of 93 basis points followed by a ten-day decline of 20 basis points. Similarly, \texttt{revenue-up} has a positive initial movement followed by negative subsequent returns. Categories describing positive earnings levels exhibit little subsequent drift. These differences show why the direction of the reported event alone need not determine the direction of subsequent returns. Estimating a separate historical return for each category accommodates both continuation and reversal.

We construct the category forecast from the average next-day return following each category over up to eight preceding years. To reduce estimation noise for infrequent events, category means are shrunk toward their broader event-group means.\footnote{For category $c$ in group $G(c)$, the forecast is
\[
\widehat r_{c,T}
=
\frac{n_{c,T}}{n_{c,T}+200}\bar r_{c,T}
+
\frac{200}{n_{c,T}+200}\bar r_{G(c),T},
\]
where $n_{c,T}$ is the number of historical headlines in the category. An unseen category receives its group mean, or the pooled mean if the group is also unseen.} Each new headline receives its category's forecast, and forecasts are averaged within firm-days using the same aggregation as the embedding forecasts. We evaluate all signals using daily rank IC, the mean cross-sectional Spearman correlation with subsequent returns, and equal- and value-weighted portfolios that buy the highest forecast decile and sell the lowest.\footnote{CKX's baseline portfolios use quintiles; deciles appear in their robustness checks.}

\begin{table}[htbp]
\centering
\caption{Return Forecasts from Event Categories and Headline Embeddings}
\label{tab:tab1}
\begin{minipage}{\linewidth}\footnotesize
This table reports out-of-sample forecasts of next open-to-open returns from news classifications and headline embeddings over 2004--2025. Category forecasts are mean returns over the one or eight years before test year $T$, shrunk toward their event group's mean with weight $n/(n+200)$ on the category mean, where $n$ is the category's headline count. Embedding forecasts use ridge regressions on RoBERTa-large and Llama-2 (13B) embeddings, trained on $T-8$ through $T-3$ and validated on $T-2$ and $T-1$. Coarser forecasts are eight-year means by RavenPack event group or RavenPack type, headline direction from a fixed word rule, or group by direction. LMMD is the Loughran--McDonald positive-minus-negative word count divided by the headline's word count. ESS is RavenPack's event sentiment score. Rank IC is the mean daily Spearman correlation with returns, in percent, with Newey--West $t$-statistics (5 lags). SR is the annualized Sharpe ratio of equal-weighted and value-weighted decile long-short portfolios, with value weights from previous-day capitalization. Net SR deducts 10 basis points per 100\% turnover, or 20 below the NYSE 20th capitalization percentile, when $\gamma$ of the gap between drifted holdings and the target portfolio is traded each day.
\end{minipage}
\medskip

{\footnotesize
\begin{adjustbox}{max width=\linewidth, max totalheight=0.75\textheight}
\begin{tabular}{lcccccccccc}
\toprule
\multicolumn{2}{c}{} & \multicolumn{2}{c}{Rank IC (\%)} & \multicolumn{3}{c}{Equal-weighted} & \multicolumn{3}{c}{Value-weighted} &  \\
\cmidrule(lr){3-4} \cmidrule(lr){5-7} \cmidrule(lr){8-10}
\multicolumn{2}{c}{} & \multicolumn{2}{c}{} &  & \multicolumn{2}{c}{Net SR} &  & \multicolumn{2}{c}{Net SR} & Names \\
\cmidrule(lr){6-7} \cmidrule(lr){9-10}
Signal & \# Labels & Mean & $t$ & SR & $\gamma=1$ & $\gamma=0.5$ & SR & $\gamma=1$ & $\gamma=0.5$ & per leg \\
\midrule
Category mean (8-yr) & 873 & 3.33 & 37.55 & 6.35 & \phantom{$-$}2.50 & \phantom{$-$}2.79 & 2.30 & $-$0.72 & $-$0.07 & 49 \\
Category mean (1-yr) & 873 & 2.94 & 33.36 & 6.16 & \phantom{$-$}1.92 & \phantom{$-$}2.29 & 2.18 & $-$0.84 & $-$0.34 & 49 \\
RoBERTa forecast &  & 2.41 & 28.18 & 4.92 & \phantom{$-$}0.79 & \phantom{$-$}1.04 & 1.21 & $-$1.58 & $-$1.04 & 49 \\
Llama-2 forecast &  & 2.67 & 29.83 & 4.97 & \phantom{$-$}1.24 & \phantom{$-$}1.48 & 1.90 & $-$0.70 & $-$0.20 & 49 \\
RavenPack ESS &  & 2.63 & 29.34 & 5.02 & \phantom{$-$}0.75 & \phantom{$-$}1.10 & 1.72 & $-$1.66 & $-$1.25 & 49 \\
\midrule
\multicolumn{11}{l}{\textit{Coarser classifications and dictionary tone}} \\
Event-group mean & \phantom{0}38 & 0.88 & 11.37 & 2.16 & $-$2.33 & $-$1.44 & 0.23 & $-$2.81 & $-$2.06 & 49 \\
Direction mean & \phantom{00}3 & 2.24 & 24.79 & 4.56 & $-$0.04 & \phantom{$-$}0.46 & 1.60 & $-$1.68 & $-$1.13 & 49 \\
RavenPack-type mean & 279 & 2.43 & 29.15 & 4.83 & \phantom{$-$}0.59 & \phantom{$-$}1.09 & 0.99 & $-$2.07 & $-$1.51 & 49 \\
Group $\times$ direction mean & 107 & 2.88 & 32.62 & 6.36 & \phantom{$-$}2.27 & \phantom{$-$}2.47 & 2.04 & $-$0.97 & $-$0.57 & 49 \\
LMMD sentiment &  & 1.08 & 12.45 & 2.11 & $-$1.42 & $-$0.86 & 0.40 & $-$2.63 & $-$1.84 & 46 \\
\bottomrule
\end{tabular}
\end{adjustbox}
}
\end{table}

Consistent with CKX, both embedding forecasts predict next-day returns. Table~\ref{tab:tab1} shows rank ICs of 2.41\% for RoBERTa and 2.67\% for Llama-2, with $t$-statistics near 30. RavenPack's event sentiment score (ESS), which varies little within event categories (Table~\ref{tab:tabA_mz}), has a similar rank IC of 2.63\%. The category forecast performs better, with a rank IC of 3.33\%. Its annualized equal-weighted Sharpe ratio is 6.35, compared with 4.92 for RoBERTa and 4.97 for Llama-2. These differences in rank IC and equal-weighted Sharpe ratios are statistically significant under Diebold--Mariano and Jobson--Korkie tests. Value-weighted Sharpe ratios are lower for all forecasts, consistent with stronger predictability among smaller stocks. The category forecast still has the highest value-weighted Sharpe ratio, although its advantage over Llama-2 is not statistically significant (Table~\ref{tab:tabA_subsamples}). Under CKX's trading costs with daily rebalancing, equal-weighted net Sharpe ratios fall to 2.50, 0.79, and 1.24, respectively. Slower rebalancing reduces this decline: with $\gamma=0.5$, the net Sharpe ratios are 2.79, 1.04, and 1.48 (Table~\ref{tab:tabA_costs} reports other values of $\gamma$). Value-weighted net Sharpe ratios are generally negative for all three forecasts.

The comparison is robust to alternative estimation windows and several sample restrictions. A category forecast using only the preceding year achieves a rank IC of 2.94\% and an equal-weighted Sharpe ratio of 6.16. Restricting category histories to years through $T-3$, the end of the embedding forecasts' training window, yields corresponding values of 3.25\% and 6.34 (Table~\ref{tab:tabA_windows}). The category forecast also has higher rank IC in all three size groups and after excluding insider-trading headlines or news describing price movements (Table~\ref{tab:tabA_subsamples}). Its advantage is particularly pronounced for headline-only news flashes.\footnote{RavenPack's \texttt{NEWS\_TYPE} field records the source format. News flashes, which consist of a headline only, account for 39.4\% of headlines and full articles for 23.8\%. The remainder consists mainly of press releases (16.1\%), tabular material (11.5\%), and SEC Form 144 and 13F filings (5.2\% and 3.3\%).} Among full articles, all still represented by their headlines, the rank-IC advantage is smaller and the embedding forecasts achieve higher equal-weighted Sharpe ratios (Table~\ref{tab:tabA_subsamples}).\footnote{In the Table~\ref{tab:tabA_subsamples} sample, earnings account for 27\% of headline-only news flashes, analyst ratings for 18\%, revenues for 17\%, and price targets for 11\%, while insider trading accounts for 68\% of full articles, stock prices for 11\%, and earnings for 10\%.} The predictive performance of both embedding and classification-based signals generally declines over the sample period (Figure~\ref{fig:annual_performance}).

The detailed taxonomy is not necessary to obtain strong predictive performance. We combine RavenPack's 38 broad event groups with a positive, negative, or neutral indicator extracted directly from each headline, producing 107 observed group-direction combinations. This coarser forecast achieves a rank IC of 2.88\% and an equal-weighted Sharpe ratio of 6.36. Broad event groups alone achieve a rank IC of only 0.88\%, while direction alone achieves 2.24\%. Combining the event group with reported direction captures substantially more predictive information than either alone.

The forecast-sorted portfolios also reveal how quickly these returns accrue. In Table~\ref{tab:tabA_drift}, the category portfolio earns 37 basis points on the first day and 50 basis points cumulatively over ten days. The corresponding figures are 27 and 38 basis points for RoBERTa, and 30 and 52 basis points for Llama-2. Thus, a large share of the ten-day return accrues on the first day for all three signals. The category forecast has the strongest first-day spread, while Llama-2 accumulates more additional return over the following days. This speed differs by firm size: among micro stocks, the category portfolio keeps drifting for about 20 trading days, whereas among big stocks the drift ends within two days (Figure~\ref{fig:figA_drift_size}).

\begin{table}[htbp]
\centering
\caption{Return Forecasts Combining Categories and Embeddings}
\label{tab:tab2}
\begin{minipage}{\linewidth}\footnotesize
This table reports the contribution of embeddings beyond event-category forecasts of next open-to-open returns over 2004--2025. Ridge estimation for test year $T$ uses $T-8$ through $T-3$ and validation $T-2$ and $T-1$. Panel A reports daily Fama--MacBeth regressions of returns on signals standardized within day; coefficients are in basis points per standard deviation, with Newey--West $t$-statistics (5 lags) in parentheses. The group $\times$ direction mean is the eight-year mean by RavenPack event group and headline direction. Panel B adds to the eight-year category mean a ridge forecast of the return minus its category's training-window mean, estimated on headline embeddings minus their category's training-window mean embedding. No combination weight is estimated; the ridge penalty maximizes the validation rank IC of the combined forecast. Rank IC is the mean daily Spearman correlation with next open-to-open returns, in percent. DM $t$ and JK $z$ test rank-IC and Sharpe-ratio differences from the category benchmark. OOS $R^2$, in basis points, is measured against a zero forecast; its $t$-statistic tests the mean daily difference in summed squared errors from the category benchmark. SR is the annualized Sharpe ratio of equal-weighted (EW) and value-weighted (VW) decile long-short portfolios.
\end{minipage}
\medskip

{\footnotesize
\textit{Panel A: Fama--MacBeth regressions, standardized signals (bp per standard deviation)}\\[2pt]
\begin{adjustbox}{max width=\linewidth, max totalheight=0.33\textheight}
\begin{tabular}{lcccccc}
\toprule
 & \multicolumn{3}{c}{RoBERTa} & \multicolumn{3}{c}{Llama-2} \\
\cmidrule(lr){2-4} \cmidrule(lr){5-7}
 & (1) & (2) & (3) & (4) & (5) & (6) \\
\midrule
Embedding forecast & \phantom{(0}7.56\phantom{)} & \phantom{(0}3.02\phantom{)} & \phantom{(0}4.23\phantom{)} & \phantom{(0}8.29\phantom{)} & \phantom{(0}3.93\phantom{)} & \phantom{(0}5.12\phantom{)} \\
 & (24.35) & \phantom{0}(9.24) & (12.71) & (23.59) & (10.35) & (13.28) \\
Category mean (8-yr) &  & \phantom{(0}9.61\phantom{)} &  &  & \phantom{(0}9.13\phantom{)} &  \\
 &  & (27.71) &  &  & (25.90) &  \\
Group $\times$ direction mean &  &  & \phantom{(0}7.95\phantom{)} &  &  & \phantom{(0}7.65\phantom{)} \\
 &  &  & (25.25) &  &  & (23.68) \\
\bottomrule
\end{tabular}
\end{adjustbox}
\par\medskip
\textit{Panel B: Out of sample: the category mean plus a ridge forecast of its residual}\\[2pt]
\begin{adjustbox}{max width=\linewidth, max totalheight=0.22\textheight}
\begin{tabular}{lcccccccc}
\toprule
 & \multicolumn{2}{c}{Rank IC (\%)} & \multicolumn{2}{c}{EW SR} & \multicolumn{2}{c}{VW SR} & \multicolumn{2}{c}{OOS $R^2$} \\
\cmidrule(lr){2-3} \cmidrule(lr){4-5} \cmidrule(lr){6-7} \cmidrule(lr){8-9}
Forecast & Mean & DM $t$ & SR & JK $z$ & SR & JK $z$ & bp & $t$ \\
\midrule
Category mean (8-yr) & 3.33 &  & 6.35 &  & 2.30 &  & 8.75 &  \\
+ residual ridge, RoBERTa & 3.39 & 2.20 & 6.92 & 4.44 & 2.47 & 1.26 & 9.04 & \phantom{$-$}0.78 \\
+ residual ridge, Llama-2 & 3.54 & 4.43 & 6.83 & 3.05 & 2.45 & 0.87 & 7.93 & $-$1.24 \\
\bottomrule
\end{tabular}
\end{adjustbox}
}
\end{table}

Embedding forecasts nevertheless contribute predictive information conditional on the category forecast. In the daily Fama--MacBeth regressions of Table~\ref{tab:tab2}, the coefficient on the standardized RoBERTa forecast falls from 7.6 to 3.0 basis points when the category forecast is included; the Llama-2 coefficient falls from 8.3 to 3.9 basis points. Both remain statistically significant. We also construct out-of-sample forecasts by adding to the category mean a ridge forecast of its residual, estimated on embeddings demeaned within category. This raises rank IC from 3.33\% to 3.39\% for RoBERTa and 3.54\% for Llama-2, with statistically significant increases for both. Equal-weighted Sharpe ratios rise to 6.92 and 6.83, respectively, although improvements in value-weighted Sharpe ratios are not statistically significant and neither combined forecast significantly improves mean squared prediction error. The two signals are not redundant: embedding forecasts have correlations of only around 0.4 with category forecasts (Table~\ref{tab:tabA_mz}), and neither forecast-sorted portfolio spans the other (Table~\ref{tab:tabA_spanning}). The category portfolio's alpha relative to the embedding portfolio exceeds the reverse alpha. Historical category returns provide a strong standalone forecast, while embeddings supply additional predictive information, mainly in equal-weighted portfolios.

\section{A Framework for News Predictability}
\label{sec:framework}

Return predictability from news can arise from delayed price adjustment following a news event or from persistent differences in expected returns across firms. We model these sources and compare forecasts based on event categories with forecasts based on news embeddings. We examine how the sources of predictability shape its persistence.

\subsection{Event Drift}
\label{sec:fw_drift}

We begin with the behavioral model of post-earnings-announcement drift in \citet{banerjee2020discussion} (henceforth BBE). Consider a stock in zero net supply that pays a terminal dividend $D$ at $t=2$, equal to $s$ plus independent noise with mean zero. A public signal $s$, representing the surprise relative to the market's prior expectation, arrives at $t=1$, but each investor observes an imperfect interpretation $s_i=s+\varepsilon_i$, where interpretation errors are independent across investors and independent of the signal. Investors agree to disagree and do not learn from prices or from one another's interpretations of the news. With $s\sim\mathcal N(0,\sigma_s^2)$ and $\varepsilon_i\sim\mathcal N(0,\sigma_\varepsilon^2)$, each investor's expectation is
\[
\mathbb E[D\mid s_i]=\kappa s_i,
\qquad
\kappa=\frac{\sigma_s^2}{\sigma_s^2+\sigma_\varepsilon^2}<1.
\]
The price averages these expectations. Because investors underweight the public signal, the price initially underreacts and subsequently drifts toward the terminal value:
\[
P_1=\kappa s,
\qquad
\mathbb E[P_2-P_1\mid s]=(1-\kappa)s.
\]

Although BBE focus on earnings announcements, the same mechanism can apply to other news events. Let $K$ denote the event category and write the signal as
\[
s=m_K+\nu,
\qquad
m_K=\mathbb E[s\mid K],
\qquad
\mathbb E[\nu\mid K]=0.
\]
Here $m_K$ represents the average signal associated with an event category, while $\nu$ represents information specific to the individual news item. Conditioning on the event category gives
\[
\mathbb E[P_2-P_1\mid K]=(1-\kappa)m_K.
\]
If $m_K$ varies across categories, event types can predict differences in subsequent returns without requiring a separate measure of the surprise in each news item. This provides a simple economic interpretation of the category-based forecasts in Section~\ref{sec:event_drift}.

To allow price adjustment to unfold over multiple periods, suppose investors gradually refine their interpretations of the original signal. We summarize the resulting return adjustment by
\[
\mathbb E[r_h\mid K,\nu]=\delta_h(m_K+\nu),
\]
where $r_h$ is the return in period $h$ after the news and $\delta_h$ measures the response to the original signal at that horizon. Repeated independent interpretations in the BBE setting generate a positive response that diminishes over time. More generally, we leave the return response in reduced form and assume that the effect of an individual event eventually fades, with $\delta_h\to0$. The behavioral model illustrates continuation, while the empirical event drifts documented in Section~\ref{sec:event_drift} also exhibit reversals. In a reduced form, we can accommodate reversals by deviating from BBE's $\kappa<1$, so that the price overreacts at the announcement, and by allowing $\delta_h$ to be negative afterward.

\subsection{Short-Horizon Predictability}
\label{sec:fw_short}

The researcher does not directly observe the underlying signal $s$. Traditional event studies condition on observable event categories:
\[
\mathbb E[r_h\mid K]=\delta_hm_K.
\]
An embedding-based forecast instead conditions on a high-dimensional representation $E$ of the news text:
\[
\mathbb E[r_h\mid E]
=
\delta_h\left(
\mathbb E[m_K\mid E]+\mathbb E[\nu\mid E]
\right).
\]

The embedding may capture information specific to an individual news item beyond its event category. At the same time, a high-dimensional representation need not preserve all return-relevant distinctions between event categories. The additional information conveyed by embeddings may come at the expense of accurately recovering category-level return patterns.

To illustrate this tradeoff, define the population predictive $R^2$ associated with an information set $W$ as
\[
\mathcal R^2(W)
=
\frac{\operatorname{Var}(\mathbb E[r_h\mid W])}
{\operatorname{Var}(r_h)}.
\]
For simplicity, suppose the embedding-based measures of category information and story-specific detail are uncorrelated:
\[
\operatorname{Cov}\left(
\mathbb E[m_K\mid E],
\mathbb E[\nu\mid E]
\right)=0.
\]
Then the difference in predictive power can be expressed as
\[
\mathcal R^2(E)-\mathcal R^2(K)
=\frac{\delta_h^2}{\operatorname{Var}(r_h)}
\Bigg[
\underbrace{
\operatorname{Var}\left(\mathbb E[\nu\mid E]\right)
}_{\text{Enhanced details}}
-\underbrace{
\left(
\operatorname{Var}(m_K)
-
\operatorname{Var}\left(\mathbb E[m_K\mid E]\right)
\right)
}_{\text{Blurred types}}
\Bigg].
\]
The first term captures additional predictive information about story-specific details. The second captures category-level predictive information that is not fully recovered from the embedding. Both quantities are nonnegative. Embeddings may therefore contain more information than event labels while still providing weaker return predictions when category-level information is imperfectly represented.

This distinction becomes more important in finite samples. CKX estimate return forecasts using ridge regression on high-dimensional embeddings. Even when the representation contains additional predictive information, the corresponding signals may be weak and difficult to estimate reliably \citep{shen2025can}. The decomposition above concerns population information content. Actual forecasting performance also depends on how effectively the information can be extracted from embeddings.

\subsection{Long-Horizon Predictability}
\label{sec:fw_long}

At longer horizons, return predictability need not arise from delayed price adjustment to the current news. Firms also differ in persistent economic attributes associated with expected returns, independently of when news about these attributes appears.

Let $z_t$ denote a latent firm state on trading day $t$, which may be multidimensional. For simplicity, consider a scalar state consisting of a permanent component and a slowly evolving component:
\[
z_t=\bar z+x_t,
\qquad
x_{t+1}=\rho x_t+u_{t+1},
\qquad 0<\rho<1.
\]
Suppose the state contributes $\pi z_t$ to expected returns.\footnote{\citet{keloharju2021long} find that cross-sectional differences in expected returns largely disappear within five years and that the persistent components of firm characteristics do not predict returns, whereas \citet{baba2024persistent} find that the compensation for persistent and transitory components varies across characteristics and is large for the persistent components of size and book-to-market. For simplicity, both components of the state carry the same loading $\pi$.}

Combining this source of predictability with the event drift described above, the expected return in period $h$ following a news item can be written as\footnote{We assume that event categories and story-specific details are independent across news items and of the firm state, so the event-drift and firm-state terms add and $\mathbb E[r_h\mid K]=\delta_h m_K$ continues to hold.}
\[
\mathbb E[r_h\mid E]
={}
\underbrace{
\delta_h\left(
\mathbb E[m_K\mid E]+\mathbb E[\nu\mid E]
\right)
}_{\text{Event drift}}
+
\underbrace{
\pi\left(
\mathbb E[\bar z\mid E]
+
\rho^{h-1}\mathbb E[x_t\mid E]
\right)
}_{\text{Firm state}}.
\]

The two components have different implications for the persistence of return predictability. The event-drift component reflects the response to an individual event and diminishes as the event ages. The firm-state component reflects persistent firm attributes and can remain predictive long after the news appears, even if the information revealed by that news was immediately incorporated into prices. The innovations $u_t$ may themselves be revealed by news, as the newly arriving information that DKPT's news shocks aim to capture. When $\rho$ is close to one, a single innovation is a small part of the state, which mostly reflects innovations revealed by earlier coverage. The firm-state component may reflect risk premia or persistent mispricing, and we do not distinguish between them.

This distinction is particularly relevant to DKPT, who residualize embeddings against observed firm characteristics to isolate news shocks. Residualization removes the part of the embedding that characteristics predict, but not the part that the firm's earlier news predicts beyond these characteristics. Observed characteristics need not span latent firm states and may measure these states with noise. Residualized embeddings may therefore retain persistent firm information, making it difficult to distinguish the event-drift and firm-state components of return predictability.

\section{News Measurement with Embeddings}
\label{sec:measure}

The preceding framework distinguishes the news signal, which drives event drift, from the persistent firm state. Both may be represented in the same news embedding. Connecting these economic sources to return forecasts requires understanding what embeddings measure and which information contributes to predictability. We first decompose embeddings by content and timing, then use observable benchmarks to measure how much information embeddings contain about news surprises and firm fundamentals. Finally, we derive additive attributions of return predictability across forecasting horizons.

\subsection{Decomposing News Embeddings}
\label{sec:decomp}

A news item names the firm, reports an event, and provides details about the firm's circumstances, often without revealing what investors previously expected. We suppress the firm index throughout this subsection. We index news items by $t$, the trading day of the news. Consider a population linear projection of the embedding onto event-category indicators, the story-specific news signal, and the latent firm state:
\begin{equation}\label{eq:measure}
E_t=
\underbrace{a_K}_{\text{Event type}}
+
\underbrace{q\nu}_{\text{Story detail}}
+
\underbrace{gz_t}_{\text{Firm state}}
+
\underbrace{\zeta_t}_{\text{Language noise}},
\end{equation}
where $a_K$ captures the representation of the event category, $q\nu$ captures information associated with the individual news signal through the loading $q$, $gz_t$ captures the firm's persistent state through the loading $g$, and $\zeta_t$ represents variation assumed to be unrelated to future returns. Because the embedding also loads on the event category, the firm state, and variation unrelated to returns, a more informative representation of news is not necessarily a better measure of the surprise relative to market expectations.

\paragraph{Content decomposition.}
We first separate embeddings into components associated with event categories, firm identity, and the remaining variation. We use a two-way projection onto event-category and firm indicators to obtain
\begin{equation}\label{eq:content}
E_t-\bar E
=
\underbrace{C_t}_{\text{Event category}}
+
\underbrace{F}_{\text{Firm identity}}
+
\underbrace{I_t}_{\text{Idiosyncratic}},
\end{equation}
where $\bar E$ is the mean embedding across all headlines of all firms in the embedding forecasts' training window. 

In terms of the conceptual measurement equation, the components roughly correspond to
\[
C_t\approx a_K,
\qquad
F\approx g\bar z,
\qquad
I_t\approx q\nu+gx_t+\zeta_t,
\]
up to centering and projection normalization. These approximations treat the evolving state as varying around zero over the years used to estimate firm identity. When $\rho$ is close to one, firm identity also absorbs much of the evolving state. The category component captures information systematically associated with recognizable event types. Firm identity captures persistent attributes repeatedly described in news about the same company. The idiosyncratic component contains more detailed information, including news-specific information, changes in firm states, and language noise. 

\paragraph{Timing decomposition.}
The content decomposition separates types of information, but it does not distinguish information newly conveyed by the current news from information already available in earlier coverage. In particular, the idiosyncratic component may contain persistent firm information beyond the fixed firm-identity component.

We separate each embedding into a historical level constructed from the firm's preceding news coverage and a deviation from that level:
\begin{equation}\label{eq:timing}
E_t-\bar E
=
\underbrace{E_{L,t}}_{\text{Historical level}}
+
\underbrace{E_{D,t}}_{\text{Current deviation}}.
\end{equation}
The historical level is constructed from news published before the current news. Applying the same split to the content components gives
\begin{equation}\label{eq:full}
E_t-\bar E
=
\underbrace{
C_{L,t}+F+I_{L,t}
}_{\text{Historical level}}
+
\underbrace{
C_{D,t}+I_{D,t}
}_{\text{Current deviation}}.
\end{equation}

The category level $C_{L,t}$ captures the firm's historical mix of event types, while $C_{D,t}$ captures the current event category relative to that mix. By construction, firm identity $F$ belongs entirely to the historical level because it is estimated from earlier news and remains fixed during the test year, unlike the category and idiosyncratic components, which vary across news items. 

More importantly, the timing decomposition helps distinguish persistent firm information from details specific to the current news. In the measurement framework, the idiosyncratic level and the idiosyncratic deviation correspond to
\[
I_{L,t}\approx gx_t,
\qquad
I_{D,t}\approx q\nu+\zeta_t.
\]
These approximations hold when the state changes little between the earlier news and the current news.
Story details and language noise largely average out in the historical level. The idiosyncratic level then captures persistent descriptions of the firm's evolving state, and the idiosyncratic deviation captures story details and language noise. These decompositions separate sources of embedding variation without requiring observed firm characteristics to span the latent state.

When the state does change, earlier coverage does not yet reflect the change. An innovation $u_t$ revealed by the current news therefore first enters the idiosyncratic deviation. As later coverage reflects the updated state, it moves into the idiosyncratic level and, once the projection window includes that coverage, into firm identity (the parameterized framework in Appendix~\ref{sec:appD1} gives the exact state terms of each component). Because its effect on expected returns decays only at rate $\rho^{h-1}$, the idiosyncratic deviation can predict distant returns. A single innovation is a small part of the state when $\rho$ is close to one, so this predictability is small relative to that of the historical level. If prices instead absorbed the current news slowly over months, as DKPT argue for news shocks, the deviation, the only component that contains the current news, would keep predicting returns over these months. The timing decomposition thus tells us whether persistent predictability comes from the current news or from information in earlier coverage.

For each test year $T$, we estimate the two-way projection once by least squares on the headline embeddings of all firms in years $\max(T-6,2000)$ through $T-1$, the projection window. The category and firm effects stay fixed within the test year. We use the firm effect as firm identity $F$ only for firms with at least ten headlines in the window, spread over at least two calendar years, and set $F=0$ for all other firms.\footnote{Estimating firm identity from only recent headlines would make $F$ absorb the idiosyncratic level and mechanically reduce $I_L$. Setting it to zero keeps the recency of $F$ comparable across firms, and we retain these firms to preserve the CKX/DKPT sample. They account for 6\% of firm-days and 9\% of firm-months. We recenter the estimated $F$ to average zero over the headlines of the firms with estimated $F$, so that $F=0$ assigns the remaining firms the average firm effect. Because the category effects are not re-estimated after recentering, the subtracted mean moves into the idiosyncratic component of the firms with estimated $F$.} The category component $C_t$ of a news item is the effect of its event category $K_t$ and changes only with the category.\footnote{$C_t$ also contains the difference between the mean embedding of the projection window and $\bar E$, the mean embedding of the ridge forecast's training window. This constant is kept out of $C_L$ and does not affect cross-sectional comparisons.} The idiosyncratic component $I_t=E_t-\bar E-C_t-F$ is the remainder, so for firms with $F=0$ it also holds their persistent information.

We index calendar months by $\tau$. Let $X$ be a variable defined on news, such as $E$, $C$, or $I$, with value $X_t$ for the firm's news on trading day $t$. As in DKPT, its firm-month value $X_\tau$ is the equally weighted mean of $X_t$ over the firm's news published in month $\tau$.\footnote{For the daily forecasts, news published after 09:00 on the last trading day of a month belongs to the next month, following CKX's timing.} The level operator gives $(\mathcal L X)_\tau$, the equally weighted mean of the firm-month values over the firm's news months among months $\tau-12$ through $\tau-1$, and zero when there are none. The historical level and deviation of firm-month $\tau$ are
\[
X_{L,\tau}=(\mathcal L X)_\tau,
\qquad
X_{D,\tau}=X_\tau-(\mathcal L X)_\tau,
\]
and news on trading day $t$ takes the historical level of the month $\tau(t)$ to which it belongs:
\[
X_{L,t}=X_{L,\tau(t)},
\qquad
X_{D,t}=X_t-X_{L,\tau(t)}.
\]
Applying $\mathcal L$ to $C$ and $I$, with $F$ entirely in the level, gives the full decomposition in equation~\eqref{eq:full}. The historical level is fixed before the month begins, while the deviation measures current news relative to that level.

\subsection{Measuring News Surprises}

An event category indicates the type of news reported, but does not measure its surprise relative to market expectations. In our framework, the news signal $s$ is not directly observable. Earnings announcements provide a setting in which this signal can be proxied by SUE. We examine how much information about earnings surprises is captured by headline embeddings beyond event categories.\footnote{In related work, \citet{koijen2026assessing} find that prompt- and context-optimized LLMs substantially improve the explanation of earnings-announcement returns beyond earnings surprises.} Applying the timing decomposition in equation~\eqref{eq:timing}, we measure the historical level $E_L$ from the firm’s news in the 12 months preceding the headline’s month. We then examine how much $E$, $E_L$, and $E_D$ add to event categories $K$ in explaining SUE, using the CKX embedding-based return forecast $\hat y$ as a benchmark.

\begin{table}[htbp]
\centering
\caption{Explaining Earnings Surprises and Predicting Next-Day Returns}
\label{tab:tab3}
\begin{minipage}{\linewidth}\footnotesize
This table reports how well earnings headlines explain earnings surprises and predict next-day returns, 2008--2025. Headlines arrive between a timed IBES announcement and 09:00 New York time on its assigned trading day. SUE is actual earnings minus consensus, divided by the previous day's price, winsorized at the 1st and 99th percentiles and expressed as a percentile rank among the announcements of the preceding 12 months. $K$ denotes category dummies, $\hat y$ the CKX forecast, and $E$ the headline embedding. $E_L$ is the firm's historical level, the mean of its monthly average embeddings over its news months in the preceding 12 months, and $E_D=E-E_L$; in $K+E_D+E_L$ they have separate coefficients and ridge penalties. All specifications use the same headlines; coefficients use all earlier years, and ridge penalties are chosen on the preceding year. Panel A reports out-of-sample $R^2$ relative to the training mean and Spearman correlations, averaged over test years. Panel B's $R^2$ is relative to a zero forecast, and its rank IC is the mean daily Spearman correlation with next open-to-open returns, in percent.
\end{minipage}
\medskip

{\footnotesize
\textit{Panel A: Explaining the SUE rank}\\[2pt]
\begin{adjustbox}{max width=\linewidth, max totalheight=0.24\textheight}
\begin{tabular}{llcccc}
\toprule
\multicolumn{2}{c}{} & \multicolumn{2}{c}{RoBERTa} & \multicolumn{2}{c}{Llama-2} \\
\cmidrule(lr){3-4} \cmidrule(lr){5-6}
Specification & Method & OOS $R^2$ (\%) & Spearman & OOS $R^2$ (\%) & Spearman \\
\midrule
$K$ & OLS & 1.87 & 0.12 & 1.87 & 0.12 \\
$K+\hat y$ & OLS & 1.87 & 0.12 & 1.91 & 0.13 \\
$K+E$ & Ridge & 2.71 & 0.16 & 5.07 & 0.22 \\
$K+E_D$ & Ridge & 2.64 & 0.16 & 4.59 & 0.22 \\
$K+E_L$ & Ridge & 2.86 & 0.18 & 4.71 & 0.22 \\
$K+E_D+E_L$ & Ridge & 3.40 & 0.19 & 5.84 & 0.25 \\
\bottomrule
\end{tabular}
\end{adjustbox}
\par\medskip
\textit{Panel B: Predicting the next-day return}\\[2pt]
\begin{adjustbox}{max width=\linewidth, max totalheight=0.31\textheight}
\begin{tabular}{llcccc}
\toprule
\multicolumn{2}{c}{} & \multicolumn{2}{c}{RoBERTa} & \multicolumn{2}{c}{Llama-2} \\
\cmidrule(lr){3-4} \cmidrule(lr){5-6}
Specification & Method & OOS $R^2$ (\%) & Rank IC (\%) & OOS $R^2$ (\%) & Rank IC (\%) \\
\midrule
$K$ & OLS & 0.07 & 2.49 & 0.07 & 2.49 \\
$K+\hat y$ & OLS & 0.08 & 2.97 & 0.10 & 3.49 \\
$K+E$ & Ridge & 0.06 & 2.74 & 0.09 & 3.33 \\
$K+E_D+E_L$ & Ridge & 0.06 & 2.99 & 0.10 & 3.97 \\
\midrule
$K+\mathrm{SUE}$ & OLS & 0.73 & 8.69 & 0.73 & 8.69 \\
$K+\mathrm{SUE}+\hat y$ & OLS & 0.73 & 8.69 & 0.77 & 8.99 \\
$K+\mathrm{SUE}+E$ & Ridge & 0.71 & 8.81 & 0.77 & 9.24 \\
$K+\mathrm{SUE}+E_D+E_L$ & Ridge & 0.72 & 8.87 & 0.78 & 9.24 \\
\bottomrule
\end{tabular}
\end{adjustbox}
}
\end{table}

Panel A of Table~\ref{tab:tab3} examines the variation in SUE explained by event categories and headline embeddings. Event categories alone achieve an out-of-sample $R^2$ of 1.87\%. Adding RoBERTa and Llama-2 embeddings increases $R^2$ to 2.71\% and 5.07\%, respectively, indicating that embeddings contain information about earnings surprises beyond event categories. We further distinguish current deviations $E_D$ from historical levels $E_L$. Adding $E_D$ to event categories yields $R^2$ of 2.64\% and 4.59\%, while adding $E_L$ yields 2.86\% and 4.71\%. Allowing the two components to enter jointly with separate coefficients and penalties raises $R^2$ to 3.40\% and 5.84\%, exceeding both individual specifications and the full-embedding specification. These results suggest that current deviations and historical levels contain complementary information about earnings surprises. The historical-level results are particularly informative: the historical level, constructed from news preceding the announcement, contains information about the subsequent earnings surprise relative to analysts' expectations. In contrast, the scalar forecast $\hat y$ explains little additional SUE variation beyond event categories, with $R^2$ of 1.87\% and 1.91\%.

Panel B examines the return predictability associated with these measures. Controlling for event categories, observed SUE raises the next open-to-open return rank IC from 2.49\% to 8.69\%, substantially more than either headline embeddings or ridge forecasts. The ridge forecast $\hat y$, trained on all news, raises the rank IC to 2.97\% and 3.49\%, about as much as embeddings whose coefficients are re-estimated on earnings news alone. The return $R^2$ similarly increases from 0.07\% with categories alone to 0.73\% with SUE, compared with 0.06\% and 0.09\% when adding full headline embeddings. Allowing current deviations and historical levels to enter separately improves the rank IC to 2.99\% for RoBERTa and 3.97\% for Llama-2, but these gains remain small relative to the predictability of observed SUE. Adding the two components to categories and observed SUE produces only modest further improvements.

Together, the two panels reveal a gap between representation and extraction: embeddings contain information about earnings surprises beyond event categories, but return-trained ridge forecasts retain little of this information, even though observed SUE strongly predicts subsequent returns.

We complement the SUE-based analysis with controlled rewrites of earnings headlines to examine how embeddings and return forecasts respond to changes in headline content. These headlines are short and formulaic, typically consisting of the firm name, the fiscal period, and the reported per-share figure, as in ``Microsoft 2Q EPS 14c.'' This structure allows us to change one detail while holding the rest of the headline fixed. We use 541,008 earnings headlines that begin with a firm name and report a numerical figure. We multiply or divide reported figures by ten, reverse unambiguous directional expressions, such as ``EPS'' to ``Loss/Shr'' or ``Beats'' to ``Misses'', replace firm names with those of similarly sized firms in the same six-digit GICS industry, or remove firm names and tickers. Each rewritten headline is re-embedded with the same frozen model and scored using the replicated CKX ridge coefficients of its test year, without re-estimation (Table~\ref{tab:tabA_rewrites}).

Reversing earnings-news direction changes forecasts by 1.77 standard deviations for RoBERTa and 2.03 for Llama-2 in absolute value, almost always in the implied direction. To assess whether this response reflects event-category information, we construct category centroids by averaging embeddings within each event category in the training window. The leading 32 principal components (PCs) of variation across these centroids account for 74.4\% and 74.5\% of the squared length of the average embedding change, far above the random-direction benchmarks of 3.1\% and 0.6\%. In contrast, multiplying or dividing reported figures by ten changes forecasts by only around 0.3 and 0.15 standard deviations for RoBERTa and Llama-2, respectively. Swapping or removing firm names produces larger changes, of around 0.6 standard deviations for RoBERTa and 0.75 for Llama-2. For Llama-2, roughly 40\%--47\% of the squared length of the average embedding change from name rewrites lies in the subspace spanned by the leading 32 PCs of firm effects. Overall, return forecasts respond most strongly to earnings-news direction, which event categories already capture, and also to firm identity, but much less to reported numerical figures.

\subsection{Measuring Firm Fundamentals}

In the measurement equation~\eqref{eq:measure}, embeddings also contain the firm's persistent state through $gz_t$, beyond the signal of an individual news event. We assess this component in two ways. First, we examine whether news-derived firm representations explain cross-sectional variation in equity valuations and analysts' long-term growth forecasts, as in \citet{sarkar2026economic}. Second, in the spirit of \citet{hoberg2010product, hoberg2016text}, we examine whether peers selected using headline embeddings, particularly firm identity and news-category mix, share similar operating and risk characteristics beyond what conventional industry classifications capture.

\begin{table}[htbp]
\centering
\caption{Valuations and Growth Forecasts Explained by Firm Embeddings}
\label{tab:tab4}
\begin{minipage}{\linewidth}\footnotesize
This table reports $R^2$, in percent, on held-out firms from cross-sectional ridge regressions of year-end valuations and growth forecasts on firm-level news embeddings and firm characteristics, 2004--2025. Outcomes are log price-to-book and log price-to-earnings, the latter for positive earnings, and IBES median long-term growth forecasts (LTG). News inputs use the headlines of up to six calendar years through the year end. The two-way category--firm projection of their embeddings gives firm identity $F$; $C^{\mathrm{mix}}$ is the firm's mean category component and $E^{\mathrm{avg}}$ its average embedding. Controls are Fama--French 49 industry indicators, 60-month CAPM beta, log book equity, asset growth, gross profits to assets, and dividends to assets. Each embedding input, the characteristics, and the industry indicators have their own ridge penalty, chosen by leave-one-out error. Each of three fixed company folds is held out in turn. Panel A fits on the other two folds; Panel B, which starts in 2005, uses slopes fitted on those folds one year earlier and refits only the intercept. $R^2$ is measured against the held-out fold mean and averaged over folds and years.
\end{minipage}
\medskip

{\footnotesize
\textit{Panel A: Split sample}\\[2pt]
\begin{adjustbox}{max width=\linewidth, max totalheight=0.28\textheight}
\begin{tabular}{lcccccc}
\toprule
 & \multicolumn{3}{c}{RoBERTa} & \multicolumn{3}{c}{Llama-2} \\
\cmidrule(lr){2-4} \cmidrule(lr){5-7}
Inputs & $\log(P/B)$ & $\log(P/E)$ & LTG & $\log(P/B)$ & $\log(P/E)$ & LTG \\
\midrule
Controls & 25.25 & 13.69 & 16.10 & 25.25 & 13.69 & 16.10 \\
$E^{\mathrm{avg}}$ & 21.42 & 11.01 & 13.60 & 30.93 & 18.25 & 19.77 \\
$C^{\mathrm{mix}}$ & 23.59 & 12.05 & 16.84 & 23.92 & 12.30 & 16.99 \\
$F$ & 17.30 & \phantom{0}9.68 & 11.15 & 28.97 & 17.56 & 18.48 \\
$C^{\mathrm{mix}}+F$ & 26.94 & 13.44 & 17.45 & 32.86 & 18.70 & 20.45 \\
Controls + $E^{\mathrm{avg}}$ & 40.82 & 15.41 & 18.18 & 51.57 & 19.61 & 21.22 \\
Controls + $C^{\mathrm{mix}}+F$ & 46.77 & 16.89 & 19.97 & 54.43 & 20.03 & 21.10 \\
\bottomrule
\end{tabular}
\end{adjustbox}
\par\medskip
\textit{Panel B: Fitted on the previous window}\\[2pt]
\begin{adjustbox}{max width=\linewidth, max totalheight=0.28\textheight}
\begin{tabular}{lcccccc}
\toprule
 & \multicolumn{3}{c}{RoBERTa} & \multicolumn{3}{c}{Llama-2} \\
\cmidrule(lr){2-4} \cmidrule(lr){5-7}
Inputs & $\log(P/B)$ & $\log(P/E)$ & LTG & $\log(P/B)$ & $\log(P/E)$ & LTG \\
\midrule
Controls & 20.12 & \phantom{0}7.74 & 10.61 & 20.12 & \phantom{0}7.74 & 10.61 \\
$E^{\mathrm{avg}}$ & 18.95 & \phantom{0}7.64 & \phantom{0}8.53 & 27.15 & 12.42 & 12.22 \\
$C^{\mathrm{mix}}$ & 20.41 & \phantom{0}8.47 & 11.86 & 20.45 & \phantom{0}8.48 & 11.23 \\
$F$ & 15.40 & \phantom{0}6.72 & \phantom{0}7.34 & 25.49 & 11.96 & 11.62 \\
$C^{\mathrm{mix}}+F$ & 23.55 & \phantom{0}8.91 & 11.66 & 28.76 & 12.09 & 12.73 \\
Controls + $E^{\mathrm{avg}}$ & 34.89 & \phantom{0}9.36 & 11.61 & 45.61 & 12.36 & 12.89 \\
Controls + $C^{\mathrm{mix}}+F$ & 40.31 & 10.06 & 12.91 & 48.18 & 12.27 & 12.69 \\
\bottomrule
\end{tabular}
\end{adjustbox}
}
\end{table}

Table~\ref{tab:tab4} evaluates how much news-based firm representations explain valuations and growth forecasts, alone and beyond conventional firm characteristics from \citet{jensen2023there} (henceforth JKP) and industry indicators. At each year end, we use the two-way projection on the headlines of the preceding six years to summarize each firm by its identity $F$, its news-category mix $C^{\mathrm{mix}}$, and its average embedding $E^{\mathrm{avg}}$, the equally weighted means of the category components and of the embeddings of its headlines. We relate these inputs to log price-to-book, log price-to-earnings, and IBES long-term growth forecasts at the same year end. All $R^2$ are computed on held-out firms, with coefficients fitted on other firms at the same year end in Panel A and at the previous year end in Panel B.

News-derived firm representations explain substantial variation in equity valuations and growth forecasts beyond conventional characteristics. In Panel A, the controls alone explain 25.3\% of the variation in log price-to-book. Adding the Llama-2 average embedding raises $R^2$ to 51.6\%, while using its news-category mix and firm identity raises it to 54.4\%. For analysts' long-term growth forecasts, the corresponding values are 16.1\%, 21.2\%, and 21.1\%. The explanatory power of the average embedding is comparable to that reported by \citet{sarkar2026economic}, who pretrains a language model on historical text and fine-tunes it on full news articles so that articles about the same firm lie close together. The incremental explanatory power persists when slopes are fitted at the previous year end. In Panel B, adding the Llama-2 news-category mix and firm identity to the controls raises $R^2$ from 20.1\% to 48.2\% for log price-to-book, while the gain for long-term growth forecasts is smaller, from 10.6\% to 12.7\%.

The decomposition provides further insight into the information captured by news representations. The news-category mix explains similar variation in log price-to-book for RoBERTa and Llama-2 (23.6\% and 23.9\%), whereas firm identity explains substantially more for Llama-2 (29.0\%) than for RoBERTa (17.3\%).\footnote{For simplicity, the framework in Section~\ref{sec:framework} treats event categories as independent of the firm state, so $C^{\mathrm{mix}}$ would carry no information about valuations. Its explanatory power instead indicates that the types of events reported about a firm vary with its fundamentals, for example more rating downgrades for firms with weaker fundamentals.} The advantage of Llama-2 lies primarily in its firm-identity representation rather than its news-category mix. Combining the two components explains more than either alone in all but one comparison, suggesting that they contain complementary information about firm fundamentals.

\begin{table}[htbp]
\centering
\caption{Firm Networks and Industry Classifications}
\label{tab:tab5}
\begin{minipage}{\linewidth}\footnotesize
This table reports adjusted $R^2$, in percent, from pooled regressions of year-end firm characteristics on their mean over each firm's peers, with year fixed effects, 2004--2025. Industry peers share the firm's SIC-3, NAICS-4, or GICS-6 code. Embedding peers are the ten most similar firms under each signal, where similarity is cosine similarity minus the mean of the two firms' median cosine similarities with all other firms. Signals use the headlines of the up to six calendar years ending at each year end: the two-way category--firm projection gives firm identity $F$; $C^{\mathrm{mix}}$ is the firm's mean category component and $E^{\mathrm{avg}}$ its average embedding. Two-signal rows use two peer means: the ten most similar firms and the next ten for $E^{\mathrm{avg}}$, and each signal's ten for $F$, $C^{\mathrm{mix}}$. OI is operating income before depreciation; $\Delta$Sales is one-year sales growth; market $\beta$ is 60-month CAPM beta; asset $\beta$ is market $\beta$ times the ratio of market equity to market equity plus debt. For each characteristic, all rows use the same firm-years. Overlap is the percentage of a firm's peers in its own industry.
\end{minipage}
\medskip

{\footnotesize
\begin{adjustbox}{max width=\linewidth, max totalheight=0.75\textheight}
\begin{tabular}{lcccccccc}
\toprule
 & \multicolumn{5}{c}{Firm characteristics} & \multicolumn{3}{c}{Overlap} \\
\cmidrule(lr){2-6} \cmidrule(lr){7-9}
Signal & OI/Sales & OI/Assets & $\Delta$Sales & Market $\beta$ & Asset $\beta$ & SIC-3 & NAICS-4 & GICS-6 \\
\midrule
\multicolumn{9}{l}{\textit{Panel A: Industry classifications}} \\
SIC-3 & 10.63 & 33.94 & 6.71 & 26.39 & 32.18 &  & 78.3 & 62.3 \\
NAICS-4 & 10.50 & 33.83 & 6.91 & 25.71 & 32.15 & 76.3 &  & 59.2 \\
GICS-6 & 13.67 & 38.29 & 7.68 & 30.79 & 37.93 & 44.7 & 43.9 &  \\
\midrule
\multicolumn{9}{l}{\textit{Panel B: RoBERTa}} \\
$E^{\mathrm{avg}}$ & 10.65 & 41.23 & 4.64 & 22.65 & 28.08 & 24.6 & 26.7 & 26.8 \\
$F$ & \phantom{0}9.78 & 32.84 & 3.90 & 18.23 & 24.57 & 24.4 & 26.5 & 27.1 \\
$C^{\mathrm{mix}}$ & 12.82 & 45.17 & 4.00 & 19.22 & 24.28 & 13.9 & 14.7 & 13.8 \\
$E^{\mathrm{avg}}$, $E^{\mathrm{avg}}$ & 13.80 & 42.52 & 5.46 & 24.22 & 29.82 & 22.6 & 24.8 & 24.1 \\
$F$, $C^{\mathrm{mix}}$ & 15.31 & 49.07 & 5.25 & 25.87 & 31.61 & 18.9 & 20.4 & 20.2 \\
\midrule
\multicolumn{9}{l}{\textit{Panel C: Llama-2}} \\
$E^{\mathrm{avg}}$ & 13.55 & 50.63 & 5.44 & 34.17 & 39.02 & 34.7 & 36.3 & 39.5 \\
$F$ & 12.17 & 47.72 & 5.00 & 33.16 & 38.95 & 38.2 & 39.6 & 44.4 \\
$C^{\mathrm{mix}}$ & 13.37 & 44.73 & 3.83 & 18.04 & 22.77 & 13.4 & 14.1 & 13.3 \\
$E^{\mathrm{avg}}$, $E^{\mathrm{avg}}$ & 16.52 & 51.18 & 6.41 & 35.56 & 40.40 & 31.0 & 32.7 & 34.9 \\
$F$, $C^{\mathrm{mix}}$ & 16.53 & 54.13 & 6.03 & 35.45 & 40.62 & 25.5 & 26.6 & 28.4 \\
\bottomrule
\end{tabular}
\end{adjustbox}
}
\end{table}

Table~\ref{tab:tab5} turns from valuations to the firm's operating and risk characteristics. \citet{hoberg2016text} measure product similarity by the cosine similarity of firms' 10-K word vectors, net of each firm's median similarity with all other firms. We adapt this measure to embedding signals, subtracting the average of the two firms' median cosine similarities. We define a firm's peers under each signal as its ten most similar firms \citep{hoberg2010product}. We then ask how much of a characteristic's cross-sectional variation is explained by its mean across the firm's peers. Panel A reports the benchmarks from conventional industry classifications, among which GICS-6 performs best. Panels B and C show that firm identity and news-category mix select different sets of peers. For Llama-2, 44.4\% of firm-identity peers share the firm's GICS-6 industry, compared with only 13.3\% of news-category-mix peers. Firm identity may reflect recurring descriptions of firms' businesses and operating characteristics, whereas the news-category mix reflects the distribution of reported event types, which may be associated with firms' operating conditions and activities across industry boundaries.

These differences are reflected in the characteristics of the selected peers. Llama-2 firm-identity peers outperform GICS-6 in explaining operating income to assets and, to a lesser extent, market and asset betas, while RoBERTa firm-identity peers perform less well. News-category-mix peers, despite their limited industry overlap, explain more variation in operating income to assets than GICS-6 peers (45.2\% and 44.7\% versus 38.3\%), but are less informative about other operating characteristics and betas. As in Table~\ref{tab:tab4}, results for the news-category mix are similar across encoders, while firm identity is substantially more informative under Llama-2. These findings suggest that news-category mix captures a dimension of operating profitability not fully reflected in conventional industry classifications, although its advantage is limited to this characteristic. Combining firm-identity and news-category-mix peer means further improves explanatory power for operating income to assets relative to using two sets of average-embedding peers, from 42.5\% to 49.1\% for RoBERTa and from 51.2\% to 54.1\% for Llama-2.

\subsection{Attribution of Return Predictability}
\label{sec:attribution}

Evidence that embeddings capture news surprises and firm fundamentals does not establish which components contribute to return predictability. The content and timing decompositions allow us to examine which information in news embeddings accounts for return predictability and how its contribution varies across forecasting horizons. CKX's daily return forecasts and DKPT's monthly portfolio weights are both linear in embeddings. With the estimated coefficients held fixed, this linearity allows an exact additive attribution of return predictability to the components of the decompositions.

Let $\mathbf E_t$ be the $N_t\times d$ matrix of CKX's firm-day news embeddings of dimension $d$ for the $N_t$ firms with news on trading day $t$, and let $\mathbf r_{t+h}$ denote the corresponding firms' returns $h$ trading days later. For DKPT, $\mathbf E_\tau$ and $\mathbf r_{\tau+h}$ are the corresponding firm-month embeddings of month $\tau$ and returns $h$ months later. Either decomposition satisfies
\[
\mathbf E_t-\mathbf 1\bar E'
=\sum_{j\in\mathcal J}\mathbf E_t^{(j)},
\]
and likewise for $\mathbf E_\tau$, where $\mathcal J$ denotes the set of components.

For CKX, ridge regression produces forecasts
\[
\hat{\mathbf y}_t
=\bar y\mathbf 1+(\mathbf E_t-\mathbf 1\bar E')\beta_T,
\]
where $\beta_T$ denotes the ridge coefficients, fixed within test year $T$, and $\bar y$ the intercept. Long-short portfolios sort stocks into quantiles of the forecast (quintiles in CKX, deciles here), so their returns are not linear in it, and we focus on the forecast's covariance with returns. Let $\mathsf H_t=\mathbf I_{N_t}-\mathbf 1\mathbf 1'/N_t$ denote the cross-sectional demeaning matrix. The within-day forecast-return covariance is proportional to
\[
(\mathsf H_t\hat{\mathbf y}_t)'\mathbf r_{t+h}
=\sum_{j\in\mathcal J}
(\mathsf H_t\mathbf E_t^{(j)}\beta_T)'\mathbf r_{t+h}.
\]
Applying the same normalization and aggregation across days yields an exact additive decomposition of predictive covariance.

For DKPT, let $\mathsf M_\tau$ denote the cross-sectional residual-maker for firm characteristics in month $\tau$, including an intercept, and let $\mathsf Q_\tau$ be the diagonal matrix that recursively standardizes each embedding dimension by its standard deviation. DKPT estimate $b_\tau$ by maximum Sharpe ratio regression (MSRR). Each column of $\mathsf M_{\tau'}\mathbf E_{\tau'}\mathsf Q_{\tau'}$ gives the stock weights of one portfolio, so the portfolio returns in month $\tau'+1$ are $\mathsf Q_{\tau'}\mathbf E_{\tau'}'\mathsf M_{\tau'}\mathbf r_{\tau'+1}$, and
\[
b_\tau=\arg\min_b\sum_{\tau'<\tau}\bigl(1-b'\mathsf Q_{\tau'}\mathbf E_{\tau'}'\mathsf M_{\tau'}\mathbf r_{\tau'+1}\bigr)^2+\lambda b'b,
\]
where the ridge penalty $\lambda$ is chosen by leave-one-out cross-validation on an expanding window. Without the penalty, $b_\tau$ gives the maximum Sharpe ratio weights of these portfolios. Because $\mathsf M_\tau\mathbf 1=0$, the centering of embeddings is removed by residualization. Before a common volatility scaling, the strategy assigns stock weights
\[
\mathbf w_\tau
=\mathsf M_\tau\mathbf E_\tau\mathsf Q_\tau b_\tau
=\sum_{j\in\mathcal J}\mathsf M_\tau\mathbf E_\tau^{(j)}\mathsf Q_\tau b_\tau.
\]
Each component is residualized on the current month's characteristics, so firm identity and the historical level contribute only through information not spanned by these characteristics. The strategy return satisfies
\[
R_{\tau+h}^{\mathrm{DKPT}}
=\mathbf w_\tau'\mathbf r_{\tau+h}
=\sum_{j\in\mathcal J}
(\mathsf M_\tau\mathbf E_\tau^{(j)}\mathsf Q_\tau b_\tau)'\mathbf r_{\tau+h}.
\]

These attributions allow us to examine whether short-horizon predictability is primarily associated with current news events, particularly event-category and idiosyncratic deviations, and whether long-horizon predictability instead reflects information accumulated in historical news coverage.

\section{Decomposing Short-Horizon Predictability}
\label{sec:short}
News embeddings represent event types, story-specific details, and firm information. We use the content and timing decompositions developed in Section~\ref{sec:measure} to examine their contributions to short-horizon return predictability across event groups and trading-day horizons.

\subsection{Decomposing Next-Day Predictability}

We begin by decomposing the short-horizon return forecasts of CKX. Table~\ref{tab:tab6} attributes their predictive covariance with next open-to-open returns to event-category, firm-identity, and idiosyncratic components, holding the original ridge coefficients fixed.

\begin{table}[htbp]
\centering
\caption{Decomposition of the Embedding Forecast}
\label{tab:tab6}
\begin{minipage}{\linewidth}\footnotesize
This table decomposes embedding forecasts of next open-to-open returns over 2004--2025. A two-way least-squares projection of the headline embeddings of the up to six years before each test year on category and firm indicators gives category effects ($C$), firm effects ($F$), and the idiosyncratic component ($I$). $F$ is the firm effect for firms with at least ten headlines in the projection window, spread over at least two of its calendar years, and zero for other firms. Applying the forecast's ridge coefficients gives $\hat y-\bar y=\hat y^C+\hat y^F+\hat y^I$, where $\bar y$ is the ridge intercept. Covariance shares divide each component's covariance with returns by the forecast's; variance shares divide its variance by the forecast's. Moments pool firm-days after removing day means. Covariance shares sum to 100; variance shares need not. Standalone columns use each component as a signal. Rank IC is the mean daily Spearman correlation with next open-to-open returns, in percent. SR is the annualized Sharpe ratio of equal-weighted (EW) and value-weighted (VW) decile long-short portfolios.
\end{minipage}
\medskip

{\footnotesize
\begin{adjustbox}{max width=\linewidth, max totalheight=0.75\textheight}
\begin{tabular}{llccccc}
\toprule
\multicolumn{2}{c}{} & \multicolumn{2}{c}{Share (\%)} & \multicolumn{3}{c}{Standalone performance} \\
\cmidrule(lr){3-4} \cmidrule(lr){5-7}
Model & Decomposition & $\mathrm{Cov}(\hat y^j,r)/\mathrm{Cov}(\hat y,r)$ & $\mathrm{Var}(\hat y^j)/\mathrm{Var}(\hat y)$ & Rank IC (\%) & EW SR & VW SR \\
\midrule
RoBERTa & Event category $\hat y^C$ & \phantom{0}80.0 & \phantom{0}37.3 & 2.90 & 5.94 & \phantom{$-$}1.94 \\
 & Firm identity $\hat y^F$ & \phantom{00}6.3 & \phantom{0}21.4 & 0.47 & 0.51 & $-$0.33 \\
 & Idiosyncratic $\hat y^I$ & \phantom{0}13.6 & \phantom{0}46.7 & 0.58 & 1.29 & \phantom{$-$}0.36 \\
 & Total forecast $\hat y$ & 100.0 & 100.0 & 2.41 & 4.92 & \phantom{$-$}1.21 \\
\midrule
Llama-2 & Event category $\hat y^C$ & \phantom{0}65.5 & \phantom{0}39.6 & 2.75 & 5.66 & \phantom{$-$}1.79 \\
 & Firm identity $\hat y^F$ & \phantom{0}15.7 & \phantom{0}32.7 & 1.09 & 0.93 & \phantom{$-$}0.88 \\
 & Idiosyncratic $\hat y^I$ & \phantom{0}18.8 & \phantom{0}39.2 & 0.47 & 1.42 & \phantom{$-$}0.84 \\
 & Total forecast $\hat y$ & 100.0 & 100.0 & 2.67 & 4.97 & \phantom{$-$}1.90 \\
\bottomrule
\end{tabular}
\end{adjustbox}
}
\end{table}

Event categories account for most of the predictive covariance: 80.0\% for RoBERTa and 65.5\% for Llama-2. Thus, despite the high dimensionality of news embeddings, most of the return predictability extracted by CKX can be traced to a single categorical label indicating the type of news event.

The dominance of event categories is particularly striking relative to their contribution to forecast variation. Category effects explain only 37.3\% of RoBERTa's forecast variance and 39.6\% of Llama-2's, compared with their much larger shares of predictive covariance. Conversely, idiosyncratic information accounts for 46.7\% and 39.2\% of forecast variance but contributes only 13.6\% and 18.8\% of predictive covariance.

The standalone performance of the components reinforces this conclusion. The category component achieves rank ICs of 2.90\% and 2.75\% for RoBERTa and Llama-2, respectively, exceeding the full forecasts' 2.41\% and 2.67\%. In contrast, the idiosyncratic components achieve rank ICs of only 0.58\% and 0.47\%. 

We further examine whether this finding depends on the particular decomposition or the original forecasting coefficients. Table~\ref{tab:tabA_centroid} replaces the category component with category centroids, computed by averaging training-sample embeddings within each event category across all firms. These centroids also account for most of the predictive covariance. Table~\ref{tab:tabA_probe} further examines forecasts trained on category templates, which replace each headline with a fixed sentence consisting of ``Company'' followed by its RavenPack category name (e.g., ``Company analyst ratings change positive''). These forecasts achieve rank ICs of 3.19\% and 3.21\%, exceeding those trained on full headline embeddings, even though the templates contain neither firm identities nor story-specific details. They are close to, but still below, the rank IC of the category mean (3.33\% in Table~\ref{tab:tab1}), consistent with the difficulty of extracting weak category-level signals from high-dimensional embeddings in finite samples, as discussed in Section~\ref{sec:framework}.

To examine how predictive information is distributed across the embedding spectrum, we project the category, firm, and idiosyncratic components of headline embeddings, estimated here on the embedding forecasts' training window, onto the PCs of the full embeddings. Table~\ref{tab:tabA_spectrum} and Figure~\ref{fig:figA_spectrum} report their contributions to embedding variance and predictive covariance across PCs. The highest-variance PCs are dominated by the category component but contribute little predictive covariance. Most of the category component's predictive covariance instead comes from lower-variance PCs, where it accounts for a relatively small share of variance but a large share of predictive covariance. A finer decomposition into the group-by-direction combinations of Section~\ref{sec:event_drift} shows that much of this predictability comes from the headline direction rather than from the event group. Ridge keeps about 70\% of PCs 33--128 and about a third of PCs 129--512, and it shrinks PCs beyond 512, which the idiosyncratic component dominates, by about 90\% or more. Thus, ridge attenuates some of the PCs carrying substantial category-related predictability, even as it largely preserves the highest-variance PCs.

The timing decomposition further shows where this predictability originates. Table~\ref{tab:tabA_decomposition} shows that the current category deviation accounts for 77.9\% of RoBERTa's predictive covariance and 63.6\% of Llama-2's, while category levels contribute only 2.2\% and 1.9\%. Category levels summarize the types of events a firm reported over the preceding year, whose drift has largely run its course, while the current deviation reflects the event just reported. Idiosyncratic deviations account for another 12.5\% and 17.1\%, respectively. Including firm identity and idiosyncratic levels, historical levels jointly contribute 9.6\% and 19.3\%.

The concentration of predictability in current category deviations is consistent with event drift: the type of event reported today carries information about subsequent returns beyond the firm's historical news coverage.

The decomposition also varies across event groups (Table~\ref{tab:tabA_groups}). For analyst ratings, insider trading, and credit ratings, category means have higher rank ICs than both embedding forecasts, and the category component accounts for most of each forecast's predictive covariance. Revenue news shows a different pattern. The category mean has a rank IC of 1.32\%, compared with 2.32\% for RoBERTa and 3.58\% for Llama-2. For RoBERTa, the idiosyncratic and firm identity components account for 46.7\% and 35.3\% of predictive covariance, compared with 18.1\% for the category component. For Llama-2, the corresponding shares are 48.3\% and 43.7\%, compared with 7.9\% for the category component. Most revenue headlines fall in the \texttt{revenue-up}, \texttt{revenue-down}, and \texttt{revenues} categories, whose next open-to-open returns are similar, at $-15$, $-14$, and $-13$ basis points (Table~\ref{tab:tabA_drift}), so category means give these headlines similar forecasts. Firm identity and the historical level, both constructed from earlier news, account for larger shares of predictive covariance than for earnings news. For revenue news, the level's share is 32.1\% for RoBERTa and 44.3\% for Llama-2, compared with 15.2\% and 26.8\% for earnings news.

\subsection{Predictability Across Trading-Day Horizons}

\begin{figure}[htbp]
\centering
\caption{Daily Forecast Covariance with Returns across Horizons}
\includegraphics[width=1\textwidth]{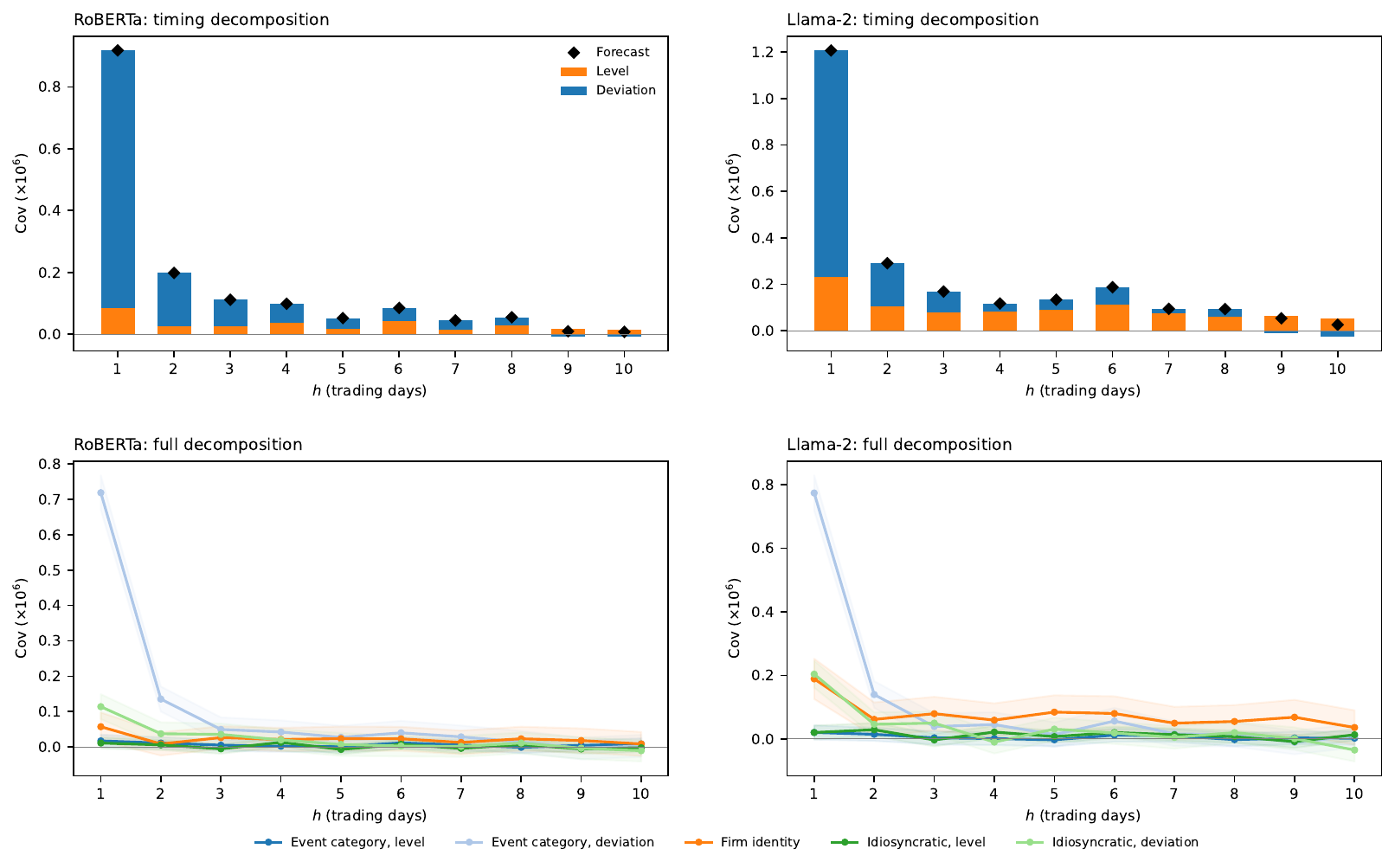}
\label{fig:fig2a}
\begin{minipage}{\linewidth}\footnotesize
This figure decomposes the covariance of the daily ridge forecast with returns across trading-day horizons, for signal years 2004--2025. Horizon $h$ is the $h$-th open-to-open return after the news. The historical level is the mean of the firm's monthly average forecast over its news months in the preceding 12 months, computed with the current coefficients; without such months it is the forecast from the firm effect $F$ of a two-way category--firm projection of the headline embeddings of the up to six years before the signal year, or zero when $F=0$. The deviation is the remainder. Upper panels separate level and deviation; lower panels show the category, firm, and idiosyncratic components of this projection, the first and last split into level and deviation, with firm identity entirely in the level. Diamonds show the total, to which the components sum. Covariances are cross-sectional, pooled across firm-days, and multiplied by $10^6$. Shading shows 95\% Newey--West confidence intervals with 5 lags.
\end{minipage}
\end{figure}

Event categories account for most next-day predictability. If this predictability mainly reflects event drift, the category deviation should fade within days, leaving the historical level relatively more important at longer horizons. Figure~\ref{fig:fig2a} plots the covariance of the daily forecast with returns over subsequent trading days, separating the historical level from the deviation, and Panel A of Table~\ref{tab:tab7} reports selected horizons.\footnote{The daily panels of Tables~\ref{tab:tab7} and~\ref{tab:tabA_horizons} keep firm-days with returns at every horizon, so their first-horizon shares differ slightly from Table~\ref{tab:tabA_decomposition}.}

Predictability declines within a few trading days. For RoBERTa, the forecast's covariance with returns, multiplied by $10^6$, falls from 0.92 on the first day to 0.20 on the second and is no longer significant on the fifth (0.05, $t=1.71$). For Llama-2, it falls from 1.21 to 0.29 and remains significant at 0.13 on the fifth day. The decline comes from the deviation, whose covariance falls from 0.83 to 0.03 for RoBERTa and from 0.98 to 0.04 for Llama-2 between the first and fifth days. The category deviation declines fastest (Table~\ref{tab:tabA_horizons}). Its covariance falls from 0.72 to 0.03 for RoBERTa and from 0.77 to 0.01 for Llama-2, consistent with the concentration of event drift in the first day after the news (Figure~\ref{fig:fig1_drift}). The idiosyncratic deviation is also insignificant by the fifth day for both models.

As the deviation fades, the historical level accounts for a growing share of the remaining predictability. After the first day, Llama-2's historical level contributes 0.08 to 0.10 on each day shown, and its share of the forecast's covariance reaches 66.8\% on the fifth. This persistence comes from firm identity, whose share rises from 15.7\% on the first day to 63.4\% on the fifth; the idiosyncratic level contributes 5.5\%. Firm identity is estimated from news published before the test year, so its predictability need not reflect a delayed reaction to the current news. RoBERTa's historical level contributes 0.09 on the first day and is insignificant afterward. The level's rank IC, which is less sensitive to extreme returns than the covariance, is significant on each of the first ten trading days for both models, while the deviation's is close to zero from the third day (Figure~\ref{fig:figA_persistence_a}).

\section{Decomposing Long-Horizon Predictability}
\label{sec:long}

In daily forecasts, the deviation from the historical level loses its predictability within days, while that of the historical level is more persistent. Because the historical level is constructed from earlier news, its predictability cannot come from the response to the current news. It may reflect a delayed response to earlier news or persistent firm information that predicts returns independently of news arrivals. We examine these possibilities at monthly horizons. We first relate daily ridge forecasts to monthly returns and examine forecasts estimated directly on monthly embeddings, without characteristic residualization. We then decompose DKPT's residualized-embedding strategy across return horizons and compare MSRR strategies based on different information components. Finally, we examine whether predictability persists when the historical level and firm identity are measured further in the past and whether past news deviations predict returns beyond the current historical level.

\subsection{Monthly Aggregation of News Embeddings}
\label{sec:monthagg}

Relating the daily forecast to monthly returns requires aggregating news that arrives on individual days. We average each firm's news embeddings within the month, as DKPT do. For a firm with $n$ news items arriving at $t_1,\ldots,t_n$, averaging the measurement equation~\eqref{eq:measure} gives
\[
\frac{1}{n}\sum_{k=1}^n E_{t_k}
=\frac{1}{n}\sum_{k=1}^n a_{K_k}
+q\,\frac{1}{n}\sum_{k=1}^n \nu_k
+g\,\frac{1}{n}\sum_{k=1}^n z_{t_k}
+\frac{1}{n}\sum_{k=1}^n \zeta_{t_k}.
\]
Averaging affects these terms differently. The category term reflects the mix of events during the month, while story-specific information and language noise may partly offset across news items. When $\rho$ is close to one, the firm state changes little within the month, so information about persistent firm states is largely preserved. The historical level is also unchanged because all news within a month shares the same level, while the deviations are averaged. Moreover, much of the event drift following news released during the month may have already run its course by the time the following month begins. Innovations to the firm state revealed during the month are not reflected in the historical level and remain in the averaged deviations. If prices absorbed these innovations slowly over subsequent months, as DKPT argue for news shocks, the deviations would remain predictive in later months. These differences suggest that persistent firm information may become relatively more important at monthly horizons as the event drift of individual news items fades.

\begin{table}[htbp]
\centering
\caption{Forecast Covariance with Returns across Horizons}
\label{tab:tab7}
\begin{minipage}{\linewidth}\footnotesize
This table reports forecast covariances with returns across horizons for signal years 2004--2025. Panel A uses the daily ridge forecast, Panel B its headline-weighted monthly average, and Panel C a ridge forecast of next-month JKP excess returns fitted to monthly average embeddings. For test year $T$, both ridge forecasts are trained on years $\max(T-8,2000)$ through $T-3$ and validated on $T-2$ and $T-1$, with Panel C's penalty selected by monthly rank IC. Each forecast uses the same coefficients at every horizon. Daily $h=0$ is the nontradable return to the eligible open and $h\ge1$ the $h$-th subsequent open-to-open return. Monthly panels assign news by trading month, and monthly $h$ counts months after the signal month. The historical level is the mean of the firm's monthly average forecast over its news months in the preceding 12 months, computed with the current coefficients; without such months it is the forecast from the firm effect of a two-way category--firm projection of the headline embeddings of the up to six years before the signal year, or zero when the firm has $F=0$. The deviation is the remainder. Cov is the cross-sectional covariance times $10^6$, pooled across firm-days or firm-months; Share is the percentage of the forecast's covariance, omitted when the latter has $|t|<2$. Newey--West $t$-statistics use 5 daily or 6 monthly lags.
\end{minipage}
\medskip

{\footnotesize
\textit{Panel A: Daily forecast}\\[2pt]
\begin{adjustbox}{max width=\linewidth, max totalheight=0.15\textheight}
\begin{tabular}{llcccccccc}
\toprule
\multicolumn{2}{c}{} & \multicolumn{2}{c}{Forecast} & \multicolumn{3}{c}{Level $\hat y_L$} & \multicolumn{3}{c}{Deviation $\hat y_D$} \\
\cmidrule(lr){3-4} \cmidrule(lr){5-7} \cmidrule(lr){8-10}
Model & $h$ (trading days) & Cov & $t$ & Cov & $t$ & Share (\%) & Cov & $t$ & Share (\%) \\
\midrule
RoBERTa & 0 & 2.73 & 40.01 & $-$0.14 & $-$5.99 & \phantom{0}$-$5.1 & 2.87 & 43.99 & 105.1 \\
 & 1 & 0.92 & 22.73 & \phantom{$-$}0.09 & \phantom{$-$}3.18 & \phantom{$-$0}9.3 & 0.83 & 26.87 & \phantom{0}90.7 \\
 & 2 & 0.20 & \phantom{0}6.12 & \phantom{$-$}0.03 & \phantom{$-$}1.11 & \phantom{$-$}12.9 & 0.17 & \phantom{0}7.10 & \phantom{0}87.1 \\
 & 3 & 0.11 & \phantom{0}3.60 & \phantom{$-$}0.03 & \phantom{$-$}1.20 & \phantom{$-$}23.6 & 0.09 & \phantom{0}3.70 & \phantom{0}76.4 \\
 & 5 & 0.05 & \phantom{0}1.71 & \phantom{$-$}0.02 & \phantom{$-$}0.77 &  & 0.03 & \phantom{0}1.52 &  \\
\midrule
Llama-2 & 0 & 2.72 & 33.88 & $-$0.32 & $-$8.86 & $-$11.9 & 3.04 & 44.00 & 111.9 \\
 & 1 & 1.21 & 21.03 & \phantom{$-$}0.23 & \phantom{$-$}5.56 & \phantom{$-$}19.0 & 0.98 & 25.76 & \phantom{0}81.0 \\
 & 2 & 0.29 & \phantom{0}6.69 & \phantom{$-$}0.10 & \phantom{$-$}2.90 & \phantom{$-$}36.1 & 0.19 & \phantom{0}6.54 & \phantom{0}63.9 \\
 & 3 & 0.17 & \phantom{0}4.08 & \phantom{$-$}0.08 & \phantom{$-$}2.24 & \phantom{$-$}47.2 & 0.09 & \phantom{0}3.24 & \phantom{0}52.8 \\
 & 5 & 0.13 & \phantom{0}3.25 & \phantom{$-$}0.09 & \phantom{$-$}2.50 & \phantom{$-$}66.8 & 0.04 & \phantom{0}1.61 & \phantom{0}33.2 \\
\bottomrule
\end{tabular}
\end{adjustbox}
\par\medskip
\textit{Panel B: Monthly average of the daily forecast}\\[2pt]
\begin{adjustbox}{max width=\linewidth, max totalheight=0.20\textheight}
\begin{tabular}{llcccccccc}
\toprule
\multicolumn{2}{c}{} & \multicolumn{2}{c}{Forecast} & \multicolumn{3}{c}{Level $\hat y_L$} & \multicolumn{3}{c}{Deviation $\hat y_D$} \\
\cmidrule(lr){3-4} \cmidrule(lr){5-7} \cmidrule(lr){8-10}
Model & $h$ (months) & Cov & $t$ & Cov & $t$ & Share (\%) & Cov & $t$ & Share (\%) \\
\midrule
RoBERTa & 1 & \phantom{$-$}1.31 & \phantom{$-$}2.45 & \phantom{$-$}0.69 & \phantom{$-$}1.42 & \phantom{0}52.6 & \phantom{$-$}0.62 & \phantom{$-$}3.46 & \phantom{$-$}47.4 \\
 & 2 & \phantom{$-$}1.16 & \phantom{$-$}1.90 & \phantom{$-$}0.63 & \phantom{$-$}1.27 &  & \phantom{$-$}0.52 & \phantom{$-$}2.44 &  \\
 & 4 & \phantom{$-$}0.10 & \phantom{$-$}0.20 & \phantom{$-$}0.39 & \phantom{$-$}0.79 &  & $-$0.29 & $-$1.58 &  \\
 & 6 & \phantom{$-$}0.52 & \phantom{$-$}0.90 & \phantom{$-$}0.25 & \phantom{$-$}0.49 &  & \phantom{$-$}0.27 & \phantom{$-$}1.42 &  \\
 & 12 & \phantom{$-$}0.32 & \phantom{$-$}0.54 & \phantom{$-$}0.19 & \phantom{$-$}0.35 &  & \phantom{$-$}0.14 & \phantom{$-$}0.74 &  \\
 & 18 & $-$0.21 & $-$0.34 & $-$0.05 & $-$0.09 &  & $-$0.16 & $-$0.66 &  \\
 & 24 & $-$0.01 & $-$0.02 & $-$0.12 & $-$0.21 &  & \phantom{$-$}0.11 & \phantom{$-$}0.47 &  \\
\midrule
Llama-2 & 1 & \phantom{$-$}2.82 & \phantom{$-$}3.76 & \phantom{$-$}2.20 & \phantom{$-$}3.12 & \phantom{0}78.2 & \phantom{$-$}0.61 & \phantom{$-$}2.92 & \phantom{$-$}21.8 \\
 & 2 & \phantom{$-$}2.73 & \phantom{$-$}3.26 & \phantom{$-$}2.20 & \phantom{$-$}3.19 & \phantom{0}80.5 & \phantom{$-$}0.53 & \phantom{$-$}2.07 & \phantom{$-$}19.5 \\
 & 4 & \phantom{$-$}1.92 & \phantom{$-$}2.71 & \phantom{$-$}2.17 & \phantom{$-$}3.25 & 113.1 & $-$0.25 & $-$1.04 & $-$13.1 \\
 & 6 & \phantom{$-$}2.11 & \phantom{$-$}2.55 & \phantom{$-$}1.60 & \phantom{$-$}2.24 & \phantom{0}76.2 & \phantom{$-$}0.50 & \phantom{$-$}2.02 & \phantom{$-$}23.8 \\
 & 12 & \phantom{$-$}2.17 & \phantom{$-$}2.40 & \phantom{$-$}1.73 & \phantom{$-$}2.17 & \phantom{0}79.6 & \phantom{$-$}0.44 & \phantom{$-$}1.81 & \phantom{$-$}20.4 \\
 & 18 & \phantom{$-$}1.05 & \phantom{$-$}1.18 & \phantom{$-$}1.02 & \phantom{$-$}1.28 &  & \phantom{$-$}0.03 & \phantom{$-$}0.11 &  \\
 & 24 & \phantom{$-$}1.54 & \phantom{$-$}1.88 & \phantom{$-$}0.89 & \phantom{$-$}1.05 &  & \phantom{$-$}0.65 & \phantom{$-$}2.66 &  \\
\bottomrule
\end{tabular}
\end{adjustbox}
\par\medskip
\textit{Panel C: Monthly-trained forecast}\\[2pt]
\begin{adjustbox}{max width=\linewidth, max totalheight=0.20\textheight}
\begin{tabular}{llcccccccc}
\toprule
\multicolumn{2}{c}{} & \multicolumn{2}{c}{Forecast} & \multicolumn{3}{c}{Level $\hat y_L$} & \multicolumn{3}{c}{Deviation $\hat y_D$} \\
\cmidrule(lr){3-4} \cmidrule(lr){5-7} \cmidrule(lr){8-10}
Model & $h$ (months) & Cov & $t$ & Cov & $t$ & Share (\%) & Cov & $t$ & Share (\%) \\
\midrule
RoBERTa & 1 & \phantom{$-$0}7.60 & \phantom{$-$}1.63 & \phantom{$-$0}1.70 & \phantom{$-$}0.43 &  & \phantom{$-$0}5.90 & \phantom{$-$}2.25 &  \\
 & 2 & \phantom{$-$0}5.58 & \phantom{$-$}1.33 & \phantom{$-$0}0.90 & \phantom{$-$}0.24 &  & \phantom{$-$0}4.69 & \phantom{$-$}2.13 &  \\
 & 4 & \phantom{$-$0}2.48 & \phantom{$-$}0.64 & \phantom{$-$0}1.40 & \phantom{$-$}0.42 &  & \phantom{$-$0}1.08 & \phantom{$-$}0.53 &  \\
 & 6 & \phantom{$-$0}2.23 & \phantom{$-$}0.56 & \phantom{0}$-$0.47 & $-$0.15 &  & \phantom{$-$0}2.70 & \phantom{$-$}1.31 &  \\
 & 12 & \phantom{$-$0}3.01 & \phantom{$-$}0.72 & \phantom{0}$-$0.09 & $-$0.03 &  & \phantom{$-$0}3.10 & \phantom{$-$}0.92 &  \\
 & 18 & \phantom{0}$-$3.47 & $-$0.83 & \phantom{0}$-$2.27 & $-$0.59 &  & \phantom{0}$-$1.20 & $-$0.55 &  \\
 & 24 & \phantom{0}$-$0.20 & $-$0.05 & \phantom{0}$-$0.82 & $-$0.23 &  & \phantom{$-$0}0.62 & \phantom{$-$}0.26 &  \\
\midrule
Llama-2 & 1 & \phantom{$-$}25.91 & \phantom{$-$}3.03 & \phantom{$-$}11.51 & \phantom{$-$}1.65 & 44.4 & \phantom{$-$}14.40 & \phantom{$-$}2.98 & 55.6 \\
 & 2 & \phantom{$-$}15.39 & \phantom{$-$}2.02 & \phantom{$-$}13.34 & \phantom{$-$}1.83 & 86.7 & \phantom{$-$0}2.05 & \phantom{$-$}0.51 & 13.3 \\
 & 4 & \phantom{$-$}12.80 & \phantom{$-$}1.57 & \phantom{$-$0}8.60 & \phantom{$-$}1.27 &  & \phantom{$-$0}4.20 & \phantom{$-$}1.11 &  \\
 & 6 & \phantom{$-$}11.24 & \phantom{$-$}1.47 & \phantom{$-$0}6.53 & \phantom{$-$}0.88 &  & \phantom{$-$0}4.71 & \phantom{$-$}0.93 &  \\
 & 12 & \phantom{$-$}18.54 & \phantom{$-$}2.19 & \phantom{$-$}10.70 & \phantom{$-$}1.38 & 57.7 & \phantom{$-$0}7.84 & \phantom{$-$}1.91 & 42.3 \\
 & 18 & \phantom{0}$-$0.72 & $-$0.09 & \phantom{0}$-$0.91 & $-$0.12 &  & \phantom{$-$0}0.19 & \phantom{$-$}0.04 &  \\
 & 24 & \phantom{$-$0}8.57 & \phantom{$-$}0.86 & \phantom{$-$0}4.36 & \phantom{$-$}0.38 &  & \phantom{$-$0}4.20 & \phantom{$-$}0.92 &  \\
\bottomrule
\end{tabular}
\end{adjustbox}
}
\end{table}

Table~\ref{tab:tab7} examines which components of embedding-based forecasts predict returns at monthly horizons. Panel A reports the daily ridge forecast as a benchmark. Panel B averages this forecast across each firm's news within the month and relates it to subsequent monthly returns, holding the ridge coefficients fixed. This allows us to examine which components of the daily forecast remain predictive at monthly horizons. Because coefficients trained on next-day returns may underweight information relevant to monthly returns, Panel C estimates the ridge forecast directly on monthly embeddings and returns.

The averaging argument suggests that the historical level should gain weight relative to the deviations once the daily forecast is averaged within the month. For Llama-2, the level and the deviation reverse roles. The level accounts for 19.0\% of predictive covariance at the next-day horizon but 78.2\% in the first month in Panel B, mostly through firm identity and the idiosyncratic level (Figure~\ref{fig:fig2b}). For RoBERTa, the level's share rises from 9.3\% to 52.6\%, although its covariance is not significant. In the first month, the deviation's predictability comes mainly from the category deviation for both models, while the idiosyncratic deviation contributes little (Table~\ref{tab:tabA_horizons}). The category deviation's covariance is several times larger among micro stocks than among small and big stocks (Figure~\ref{fig:figA_bridge_size}), consistent with the longer event drift among micro stocks (Figure~\ref{fig:figA_drift_size}).

If a month's news is incorporated into prices slowly, as DKPT argue, the deviations should keep predicting returns in the months after the news arrives, because the historical level contains no news from that month. In Panel B, neither the category nor the idiosyncratic deviation predicts returns steadily (Table~\ref{tab:tabA_horizons}). The category deviation predicts returns mainly in the first month or two, and the idiosyncratic deviation only at isolated horizons. The historical level's predictability, in contrast, persists. For Llama-2, the level retains about four-fifths of its first-month covariance in the twelfth month, and its rank IC stays near 3.7\% over 36 months. For both models, the level's rank IC exceeds the deviation's at every reported horizon (Figure~\ref{fig:figA_persistence_b}).

The deviations' limited predictability in Panel B may reflect the next-day training target rather than a lack of information about later returns. Panel C trains the forecast on monthly returns. These forecasts are less precise than the averaged daily forecasts, possibly because they estimate as many coefficients from far fewer firm-months. For Llama-2, the deviation's share of first-month predictive covariance rises from 21.8\% to 55.6\%, mainly through the idiosyncratic deviation. The deviation's covariance, however, falls sharply after the first month (Figure~\ref{fig:fig2c}), whereas the level's point estimate is nearly as large in the twelfth month as in the first, and its rank IC remains above the deviation's at every reported horizon (Figure~\ref{fig:figA_persistence_c}). For RoBERTa, the deviation predicts returns in the first two months, and the level contributes little, although the forecast as a whole is not significant. Training the forecast on monthly returns thus raises the deviations' share of first-month predictive covariance without making them more persistent.

Both the historical level and the deviations predict monthly returns in forecasts built from unresidualized embeddings, with their relative contributions depending on how the forecast is estimated. The predictability that persists beyond the first months, however, comes mainly from the historical level, most clearly for Llama-2.

\subsection{Persistent Firm Information after Characteristic Residualization}

DKPT treat the part of news that firm characteristics predict as ``old news'' already incorporated in prices, and removing it more than doubles the return predictive power of news embeddings. They attribute the persistent returns of the resulting residualized-embedding strategy to slow incorporation of news shocks, which they describe as ``the irregular, event-driven component of news.'' The historical level, however, is likewise constructed from news published before the month, because it averages the firm's news over the preceding 12 months, and the deviation measures the month's news relative to this level. If the persistent returns of DKPT's strategy came from slow incorporation of the month's news, they would come from the deviation rather than from the historical level, which is constructed before the month begins.

\afterpage{%
\begin{landscape}
\begin{table}[htbp]
\centering
\caption{Decomposition of the Residualized-Embedding Strategy}
\label{tab:tab8}
\begin{minipage}{\linewidth}\footnotesize
This table decomposes the returns of the residualized-embedding strategy for signal months 2004--2025. As in DKPT, firm-month embeddings are recursively standardized and residualized each month on 132 JKP characteristics, and managed portfolios are combined out of sample by maximum Sharpe ratio regression (MSRR). News is assigned by arrival month. Signal month $\tau$ fixes the firms traded, those with news in month $\tau$, and the month of the characteristics. The strategy earns their returns in month $\tau+h$, the $h$-th subsequent return month, whether or not they have news then. Every horizon uses the coefficients of signal month $\tau$, fitted in an expanding window from 2000 to the next-month portfolio returns of earlier signal months. A two-way projection of the headline embeddings of the up to six years before the signal year on category and firm indicators gives category effects ($C$), firm effects ($F$), and the idiosyncratic component ($I$). Historical levels average the category and idiosyncratic components' monthly means over the firm's news months in the preceding 12 months; firm identity lies entirely in the level, and deviations are the remainders. Each component is residualized like the full embedding and uses the same coefficients, so component returns sum to the total. Ret is the annualized mean in percent after scaling the strategy to 10\% annual volatility at each horizon, with the same scale for every component. Share is the percentage of the total mean return, omitted when the total has $|t|<2$; SR is annualized. Newey--West $t$-statistics use 6 lags.
\end{minipage}
\medskip

{\footnotesize
\textit{Panel A: Timing decomposition}\\[2pt]
\begin{adjustbox}{max width=\linewidth, max totalheight=0.31\textheight}
\begin{tabular}{llccccccccc}
\toprule
\multicolumn{2}{c}{} & \multicolumn{3}{c}{Embedding $E$} & \multicolumn{3}{c}{Level $E_L$} & \multicolumn{3}{c}{Deviation $E_D$} \\
\cmidrule(lr){3-5} \cmidrule(lr){6-8} \cmidrule(lr){9-11}
Model & $h$ (months) & SR & Ret & $t$ & Ret & $t$ & Share (\%) & Ret & $t$ & Share (\%) \\
\midrule
RoBERTa & 1 & 1.79 & 17.9 & \phantom{0}7.96 & \phantom{0}9.1 & \phantom{0}5.09 & \phantom{0}51.0 & \phantom{$-$}8.8 & \phantom{$-$}4.02 & \phantom{$-$}49.0 \\
 & 2 & 1.03 & 10.3 & \phantom{0}4.82 & \phantom{0}8.1 & \phantom{0}5.14 & \phantom{0}78.7 & \phantom{$-$}2.2 & \phantom{$-$}1.37 & \phantom{$-$}21.3 \\
 & 4 & 0.55 & \phantom{0}5.5 & \phantom{0}2.19 & \phantom{0}6.1 & \phantom{0}4.14 & 109.7 & $-$0.5 & $-$0.31 & \phantom{0}$-$9.7 \\
 & 6 & 0.69 & \phantom{0}6.9 & \phantom{0}2.82 & \phantom{0}5.6 & \phantom{0}3.88 & \phantom{0}80.6 & \phantom{$-$}1.3 & \phantom{$-$}0.76 & \phantom{$-$}19.4 \\
 & 12 & 0.97 & \phantom{0}9.7 & \phantom{0}4.28 & \phantom{0}4.1 & \phantom{0}2.14 & \phantom{0}42.7 & \phantom{$-$}5.5 & \phantom{$-$}3.87 & \phantom{$-$}57.3 \\
 & 18 & 0.45 & \phantom{0}4.5 & \phantom{0}2.38 & \phantom{0}3.6 & \phantom{0}2.12 & \phantom{0}80.0 & \phantom{$-$}0.9 & \phantom{$-$}0.54 & \phantom{$-$}20.0 \\
 & 24 & 1.23 & 12.3 & \phantom{0}5.69 & \phantom{0}5.4 & \phantom{0}2.54 & \phantom{0}43.9 & \phantom{$-$}6.9 & \phantom{$-$}3.45 & \phantom{$-$}56.1 \\
\midrule
Llama-2 & 1 & 3.62 & 36.2 & 18.56 & 27.6 & 11.36 & \phantom{0}76.3 & \phantom{$-$}8.6 & \phantom{$-$}4.77 & \phantom{$-$}23.7 \\
 & 2 & 2.70 & 27.0 & 11.04 & 23.3 & 11.59 & \phantom{0}86.1 & \phantom{$-$}3.8 & \phantom{$-$}2.74 & \phantom{$-$}13.9 \\
 & 4 & 2.43 & 24.3 & 10.25 & 23.0 & 11.21 & \phantom{0}94.5 & \phantom{$-$}1.3 & \phantom{$-$}0.77 & \phantom{$-$0}5.5 \\
 & 6 & 2.18 & 21.8 & \phantom{0}7.17 & 20.2 & 11.00 & \phantom{0}92.9 & \phantom{$-$}1.6 & \phantom{$-$}0.91 & \phantom{$-$0}7.1 \\
 & 12 & 2.24 & 22.4 & \phantom{0}9.93 & 19.1 & \phantom{0}9.39 & \phantom{0}85.5 & \phantom{$-$}3.3 & \phantom{$-$}2.28 & \phantom{$-$}14.5 \\
 & 18 & 1.71 & 17.1 & \phantom{0}7.49 & 14.4 & \phantom{0}7.04 & \phantom{0}84.3 & \phantom{$-$}2.7 & \phantom{$-$}2.36 & \phantom{$-$}15.7 \\
 & 24 & 1.99 & 19.9 & \phantom{0}9.23 & 14.0 & \phantom{0}7.78 & \phantom{0}70.2 & \phantom{$-$}5.9 & \phantom{$-$}4.18 & \phantom{$-$}29.8 \\
\bottomrule
\end{tabular}
\end{adjustbox}
\par\medskip
\textit{Panel B: Full decomposition}\\[2pt]
\begin{adjustbox}{max width=\linewidth, max totalheight=0.35\textheight}
\begin{tabular}{llccccccccccccccc}
\toprule
\multicolumn{2}{c}{} & \multicolumn{3}{c}{Event category, level $C_L$} & \multicolumn{3}{c}{Event category, deviation $C_D$} & \multicolumn{3}{c}{Firm identity $F$} & \multicolumn{3}{c}{Idiosyncratic, level $I_L$} & \multicolumn{3}{c}{Idiosyncratic, deviation $I_D$} \\
\cmidrule(lr){3-5} \cmidrule(lr){6-8} \cmidrule(lr){9-11} \cmidrule(lr){12-14} \cmidrule(lr){15-17}
Model & $h$ (months) & Ret & $t$ & Share (\%) & Ret & $t$ & Share (\%) & Ret & $t$ & Share (\%) & Ret & $t$ & Share (\%) & Ret & $t$ & Share (\%) \\
\midrule
RoBERTa & 1 & 2.7 & 4.05 & 15.1 & \phantom{$-$}5.8 & \phantom{$-$}3.54 & \phantom{$-$}32.7 & \phantom{0}5.1 & \phantom{0}4.42 & 28.6 & 1.3 & 1.71 & \phantom{0}7.3 & \phantom{$-$}2.9 & \phantom{$-$}2.19 & \phantom{$-$}16.4 \\
 & 2 & 2.0 & 3.12 & 19.4 & \phantom{$-$}2.1 & \phantom{$-$}1.49 & \phantom{$-$}20.1 & \phantom{0}4.5 & \phantom{0}3.53 & 43.2 & 1.7 & 2.42 & 16.1 & \phantom{$-$}0.1 & \phantom{$-$}0.12 & \phantom{$-$0}1.2 \\
 & 4 & 1.4 & 2.29 & 25.6 & \phantom{$-$}0.5 & \phantom{$-$}0.32 & \phantom{$-$0}8.2 & \phantom{0}3.2 & \phantom{0}2.73 & 58.3 & 1.4 & 2.26 & 25.7 & $-$1.0 & $-$1.00 & $-$17.9 \\
 & 6 & 1.9 & 3.29 & 27.5 & \phantom{$-$}0.9 & \phantom{$-$}0.64 & \phantom{$-$}12.6 & \phantom{0}2.7 & \phantom{0}2.20 & 39.2 & 1.0 & 1.10 & 13.9 & \phantom{$-$}0.5 & \phantom{$-$}0.43 & \phantom{$-$0}6.8 \\
 & 12 & 0.3 & 0.46 & \phantom{0}3.5 & \phantom{$-$}4.1 & \phantom{$-$}3.04 & \phantom{$-$}42.0 & \phantom{0}2.3 & \phantom{0}1.77 & 23.8 & 1.5 & 1.63 & 15.5 & \phantom{$-$}1.5 & \phantom{$-$}1.16 & \phantom{$-$}15.3 \\
 & 18 & 1.0 & 1.42 & 21.7 & $-$0.2 & $-$0.15 & \phantom{0}$-$4.5 & \phantom{0}2.4 & \phantom{0}1.66 & 52.3 & 0.3 & 0.31 & \phantom{0}5.9 & \phantom{$-$}1.1 & \phantom{$-$}0.75 & \phantom{$-$}24.5 \\
 & 24 & 0.3 & 0.29 & \phantom{0}2.4 & \phantom{$-$}5.2 & \phantom{$-$}2.79 & \phantom{$-$}42.1 & \phantom{0}4.0 & \phantom{0}2.63 & 32.6 & 1.1 & 1.31 & \phantom{0}8.8 & \phantom{$-$}1.7 & \phantom{$-$}1.32 & \phantom{$-$}14.1 \\
\midrule
Llama-2 & 1 & 1.6 & 2.34 & \phantom{0}4.3 & \phantom{$-$}5.8 & \phantom{$-$}4.03 & \phantom{$-$}16.1 & 18.8 & 10.10 & 52.1 & 7.2 & 5.26 & 19.9 & \phantom{$-$}2.8 & \phantom{$-$}2.11 & \phantom{$-$0}7.6 \\
 & 2 & 0.8 & 1.30 & \phantom{0}3.0 & \phantom{$-$}2.4 & \phantom{$-$}2.19 & \phantom{$-$0}8.9 & 15.8 & \phantom{0}9.41 & 58.5 & 6.7 & 5.75 & 24.7 & \phantom{$-$}1.4 & \phantom{$-$}1.34 & \phantom{$-$0}5.0 \\
 & 4 & 0.7 & 0.99 & \phantom{0}2.7 & \phantom{$-$}1.3 & \phantom{$-$}1.14 & \phantom{$-$0}5.5 & 15.4 & \phantom{0}9.41 & 63.3 & 6.9 & 5.38 & 28.6 & $-$0.0 & $-$0.01 & \phantom{0}$-$0.0 \\
 & 6 & 0.9 & 1.45 & \phantom{0}4.1 & \phantom{$-$}1.5 & \phantom{$-$}1.25 & \phantom{$-$0}6.9 & 13.0 & \phantom{0}8.17 & 59.9 & 6.3 & 6.12 & 28.9 & \phantom{$-$}0.1 & \phantom{$-$}0.06 & \phantom{$-$0}0.3 \\
 & 12 & 0.4 & 0.84 & \phantom{0}2.0 & \phantom{$-$}1.8 & \phantom{$-$}1.60 & \phantom{$-$0}7.9 & 12.6 & \phantom{0}6.63 & 56.1 & 6.1 & 5.44 & 27.3 & \phantom{$-$}1.5 & \phantom{$-$}1.25 & \phantom{$-$0}6.6 \\
 & 18 & 0.6 & 1.21 & \phantom{0}3.3 & \phantom{$-$}1.3 & \phantom{$-$}1.20 & \phantom{$-$0}7.4 & \phantom{0}9.7 & \phantom{0}5.37 & 56.9 & 4.1 & 3.79 & 24.0 & \phantom{$-$}1.4 & \phantom{$-$}1.48 & \phantom{$-$0}8.3 \\
 & 24 & 0.6 & 1.03 & \phantom{0}3.0 & \phantom{$-$}3.3 & \phantom{$-$}3.26 & \phantom{$-$}16.7 & 10.5 & \phantom{0}7.43 & 52.7 & 2.9 & 3.24 & 14.6 & \phantom{$-$}2.6 & \phantom{$-$}2.78 & \phantom{$-$}13.0 \\
\bottomrule
\end{tabular}
\end{adjustbox}
}
\end{table}

\end{landscape}
}

\begin{figure}[htbp]
\centering
\caption{MSRR Strategy Return Decomposition across Horizons}
\includegraphics[width=1\textwidth]{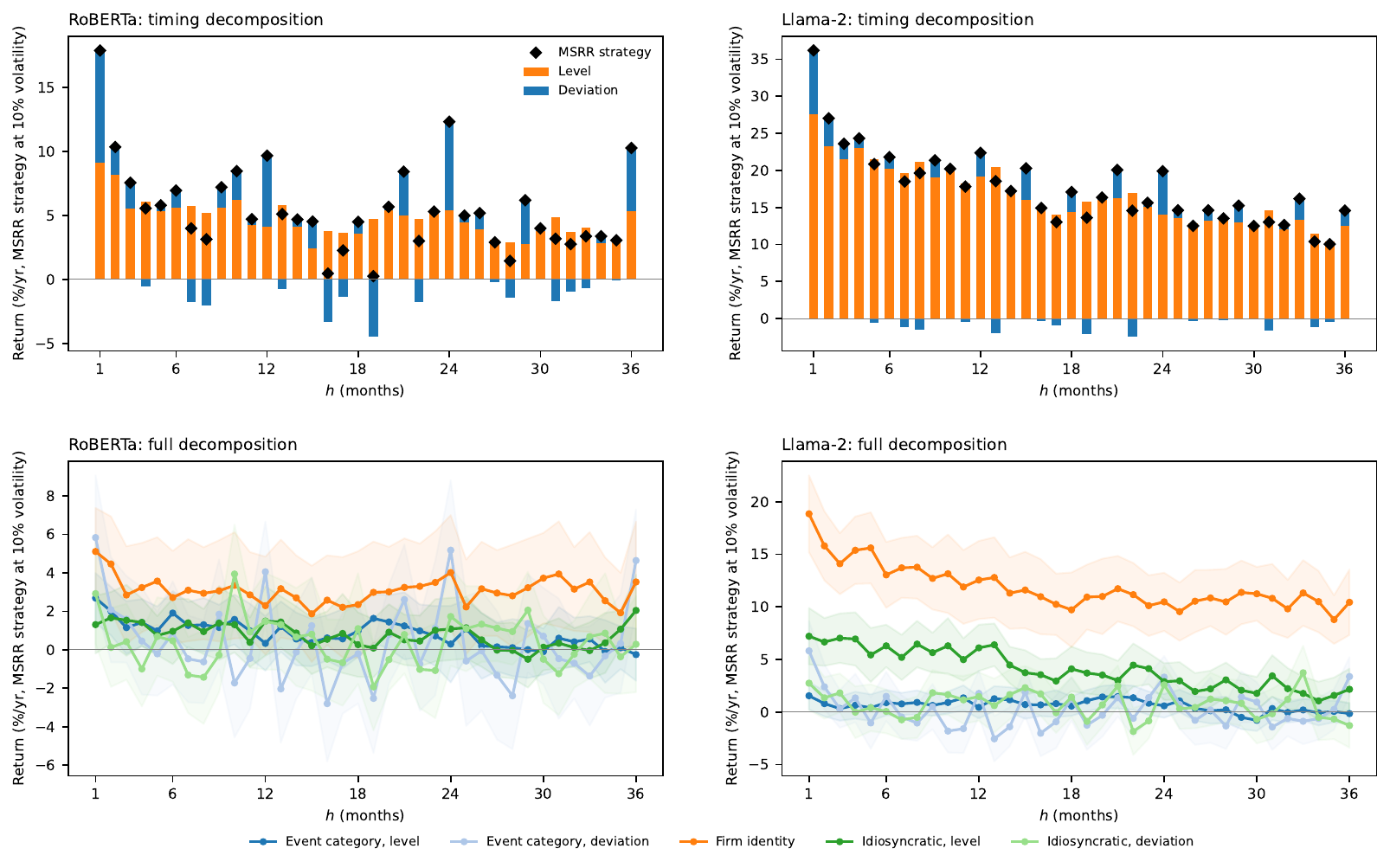}
\label{fig:fig3_attribution}
\begin{minipage}{\linewidth}\footnotesize
This figure decomposes monthly maximum Sharpe ratio regression (MSRR) strategy returns for signal years 2004--2025. Firm-month embeddings are recursively standardized, residualized each month on 132 JKP characteristics, and combined out of sample using MSRR. News is assigned by arrival month. Signal month $\tau$ fixes the firms traded, those with news in month $\tau$, and the month of the characteristics. The strategy earns their returns in month $\tau+h$, $h=1,\ldots,36$, whether or not they have news then. Every horizon uses the coefficients of signal month $\tau$, fitted in an expanding window from 2000 to the next-month portfolio returns of earlier signal months. Components come from the two-way category--firm projection of the headline embeddings of the up to six years before the signal year and are residualized like the embedding. Historical levels average the category and idiosyncratic components' monthly means over the firm's news months in the preceding 12 months; firm identity lies entirely in the level, and deviations are the remainders. Upper panels separate historical level and deviation; lower panels show the category, firm, and idiosyncratic components, the first and last split into level and deviation. Diamonds show the total, to which the components sum. Returns are annualized percentages after scaling the strategy to 10\% annual volatility at each horizon, with the same scale for every component. Shading shows 95\% Newey--West confidence intervals with 6 lags.
\end{minipage}
\end{figure}

Table~\ref{tab:tab8} and Figure~\ref{fig:fig3_attribution} decompose the strategy's returns across subsequent return months, holding the original MSRR coefficients fixed. Every component is residualized on JKP characteristics like the full embedding, so the components' contributions sum to the strategy's return. Despite the residualization, the historical level accounts for most of Llama-2's strategy return, 76.3\% in the first month and 70.2\% in month 24, and its contribution is significant at every horizon through 36 months. For RoBERTa, the level is significant at every reported horizon and accounts for more than half of mean strategy return at most of them. For both models, this contribution is more persistent than in the unresidualized forecast of Figure~\ref{fig:fig2b}.\footnote{Two differences may contribute. DKPT's weights are fitted to monthly rather than next-day returns, and residualizing on current characteristics removes the part of the level spanned by fast-decaying predictors such as past returns. Both may shift the level toward the more persistent part of the firm state.} The level's rank IC also exceeds the deviation's at every reported horizon for both models and changes little over 36 months for Llama-2 (Figure~\ref{fig:figA_persistence_strategy}).

The largest component of the level's contribution is firm identity, which is estimated from headlines published before the test year and held fixed during it, so it contains no news from the test year. For Llama-2, firm identity accounts for more than half of mean strategy return at every reported horizon through 24 months, and the rest of the level's contribution comes primarily from the idiosyncratic level. For RoBERTa, firm identity contributes positively at every reported horizon and significantly at most of them. The deviation's first-month contribution, like the category deviation's predictability in Figure~\ref{fig:figA_bridge_size}, comes mostly from micro stocks. Firm identity's contribution, in contrast, is significant among big stocks at nearly every horizon for both models and, for Llama-2, of similar size among micro, small, and big stocks (Figure~\ref{fig:figA_attribution_size}). Equal-weighted daily portfolios sorted on unresidualized firm identity or the historical level load positively on the profitability and value factors (Table~\ref{tab:tabA_factors}), so these components measure firm information related to profitability and value. Residualizing firm identity on current characteristics does not turn information from earlier years into news of the current month.

DKPT find that the predictability of news shocks roughly halves after one month but remains significant for at least 18 months. If this persistence came from slow incorporation of the current month's news, the deviation's contribution would decline gradually over later months. Instead, it falls from 8.6\% per year in the first month to 1.3\% in the fourth for Llama-2 and from 8.8\% to $-0.5\%$ for RoBERTa. Later, it is significant mainly at a few horizons that are multiples of three months, such as 12 and 24 months, where it reaches 5.9\% per year for Llama-2 and 6.9\% for RoBERTa at 24 months (Table~\ref{tab:tab8}). These later contributions come mostly from the category deviation for RoBERTa and about equally from both deviations for Llama-2, and show a three-month seasonality (Figure~\ref{fig:fig3_attribution}). The deviation's rank IC in Figure~\ref{fig:figA_persistence_strategy} shows this seasonality more clearly, and the historical level shows none.\footnote{From the third month on, the deviation's rank IC averages 0.29\% for Llama-2 and 0.26\% for RoBERTa at horizons that are multiples of three months and $-0.14\%$ and $-0.10\%$ one month later, while the level's rank IC barely changes. Over the same horizons, the category deviation's contribution to the strategy averages 1.3\% and 1.7\% per year at multiples of three months and $-0.9\%$ and $-1.2\%$ one month later.} Because the category deviation measures the month's mix of event categories relative to the firm's historical mix, this seasonality may be related to the quarterly cycle of earnings announcements. In the framework, an innovation to the firm state revealed by the current month's news enters the idiosyncratic deviation. After the first month, this component's contribution is insignificant at every reported horizon except month 24 for Llama-2 (Table~\ref{tab:tab8}, Panel B).

The fixed coefficients are estimated on next-month returns and may give the deviation too little weight at longer horizons. Table~\ref{tab:tabA_reoptimized} re-estimates the MSRR coefficients separately for each return horizon, using only returns realized before the return month, as DKPT do for 12-month-ahead returns, and decomposes each resulting strategy using its own coefficients. Re-estimation raises the deviation's contribution in the second month but does not make it more persistent, and the historical level continues to account for most of the return for Llama-2 and for much of it for RoBERTa in the first months.

Residualization on firm characteristics thus does not remove the historical level, which accounts for most of Llama-2's strategy returns beyond the first months and a substantial share of RoBERTa's, mainly through firm identity. In the framework of Appendix~\ref{sec:appendixD}, the state that characteristics do not span stays in the residual (Proposition~\ref{prop:D3}). Residualization can still raise the Sharpe ratio, as DKPT find, because it removes embedding variation unrelated to returns (Proposition~\ref{prop:D4}). The level contains no news from the signal month, so its contribution is not a delayed response to that month's news. It may still reflect slow incorporation of the news from the preceding 12 months from which the level is measured. At 36 months after the signal month, however, the level's contribution is still significant for both models. Slow incorporation would then require prices to take more than three years to absorb that news, beyond the 18-month horizon emphasized by DKPT. The next subsection uses embeddings constructed only from news published more than three years earlier.

\subsection{Long-Horizon Predictability from Distant Information}

\begin{figure}[!htb]
\centering
\caption{MSRR Strategy Sharpe Ratios by Input and Horizon}
\includegraphics[width=1\textwidth]{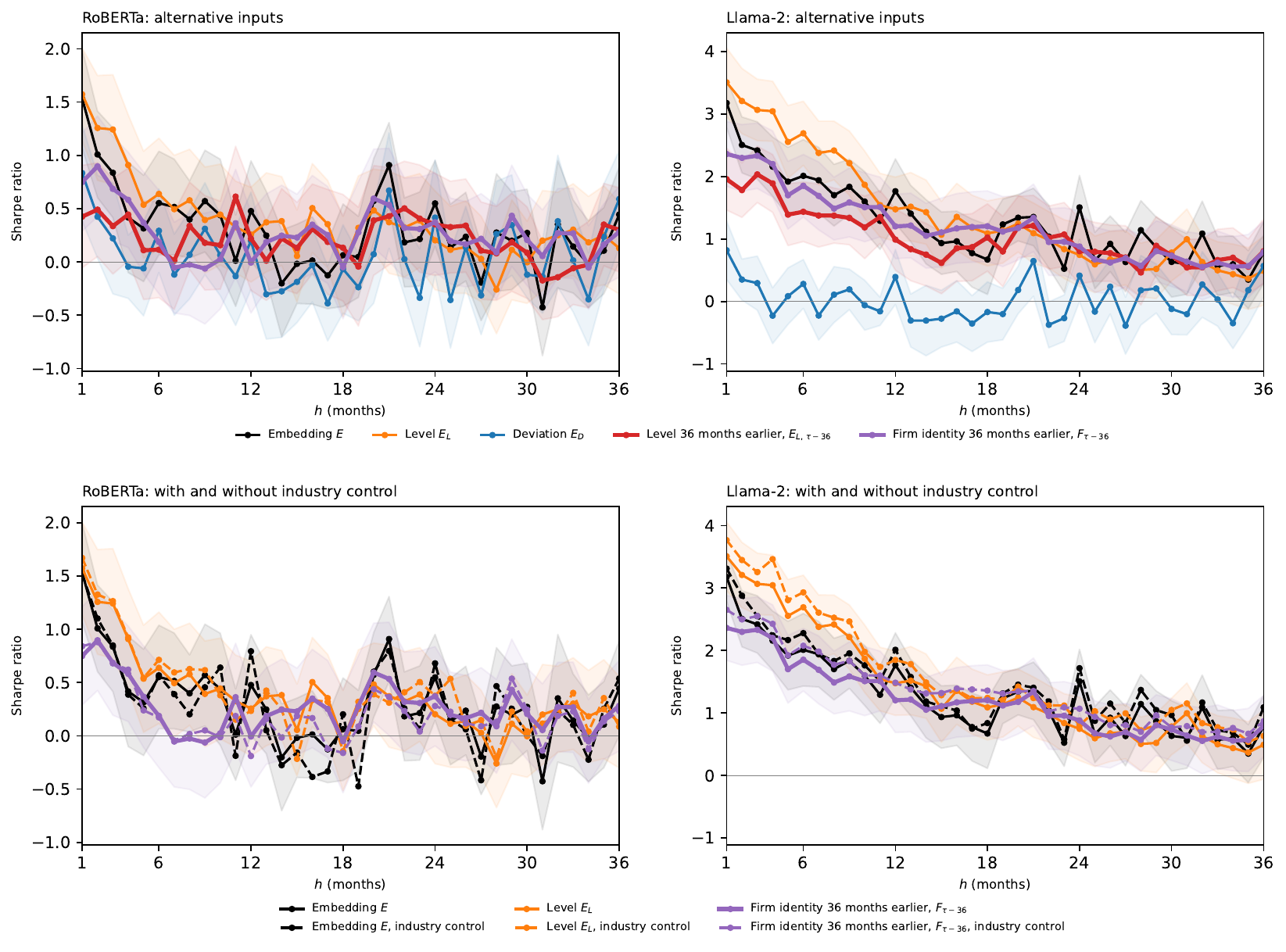}
\label{fig:fig4_inputs}
\begin{minipage}{\linewidth}\footnotesize
This figure reports annualized maximum Sharpe ratio regression (MSRR) Sharpe ratios by input and horizon $h$, for signal months $\tau$ from January 2005 through 2025. News is assigned by arrival month. Signal month $\tau$ fixes the firms traded, those with news in month $\tau$, and the month of the characteristics. The strategy earns their returns in month $\tau+h$, the $h$-th subsequent return month. The historical level $E_{L,\tau}$ is the mean of the firm's monthly average embedding over its news months from $\tau-12$ to $\tau-1$, or, when the firm has no such months, its firm effect from a two-way category--firm projection of the headline embeddings of the up to six years before the year of $\tau$. The deviation $E_{D,\tau}$ is the remainder of the month-$\tau$ embedding. The lagged level $E_{L,\tau-36}$ is the same mean over news months from $\tau-48$ to $\tau-37$. Lagged firm identity $F_{\tau-36}$ is the firm effect of the same projection fitted three years earlier, on the headlines of years $Y-9$ to $Y-4$ (from 2000 at the earliest), where $Y$ is the year of $\tau$; these headlines end at least 37 months before $\tau$. Upper panels report strategies on the embedding, $E_{L,\tau}$, $E_{D,\tau}$, $E_{L,\tau-36}$, and $F_{\tau-36}$. Lower panels report strategies on the embedding, $E_{L,\tau}$, and $F_{\tau-36}$, without (solid lines) or with (dashed lines) GICS-4 industry indicators added to the residualization. Each strategy trades only firms for which its input exists. Each input, including $E_{L,\tau-36}$ and $F_{\tau-36}$, is standardized with the coordinate means and standard deviations of all firm-month embeddings through month $\tau-1$, as in DKPT, and residualized on the month-$\tau$ values of 132 JKP characteristics. Managed portfolios are combined out of sample by MSRR. For each input and $h$, the coefficients for signal month $\tau$ are fitted in an expanding window to the horizon-$h$ portfolio returns of all earlier signal months, using only returns realized by month $\tau+h-1$. Signal months start in January 2004 for every input, and the first 12 initialize the fit. Shading divides 95\% Newey--West confidence intervals of mean returns by sample volatility and is shown for the solid lines.
\end{minipage}
\end{figure}

We measure the historical level and firm identity three years earlier, $E_{L,\tau-36}$ and $F_{\tau-36}$, where $\tau$ is the signal month, so that neither contains news from the three years before month $\tau$. Figure~\ref{fig:fig4_inputs} and Table~\ref{tab:tabA_constructions} report MSRR strategies built on each of these inputs and, as benchmarks, on the embedding, the current historical level, and the deviation. Each input is standardized and residualized on JKP characteristics as in DKPT, and for each horizon $h$ its MSRR coefficients are fitted to returns $h$ months after the signal month. Only the news inputs are lagged: residualization uses the signal month's characteristics, and the strategy trades the firms covered in the signal month.\footnote{All strategies in Figure~\ref{fig:fig4_inputs} are estimated from 2004, with the first 12 months used for initialization, whereas the decomposition in Table~\ref{tab:tab8} applies coefficients estimated from 2000 and held fixed across horizons, so their Sharpe ratios and persistence differ.} For both models, the current level alone earns a higher Sharpe ratio than the full embedding over the first six months and about as high a one on average afterwards, whereas the deviation earns little beyond the first months.

For Llama-2, strategies on the two lagged inputs, which use only news published more than three years earlier, earn Sharpe ratios of about two in the first months. This is more than twice the deviation's 0.8 and more than half of the current level's 3.5. They still earn about 0.8 in month 36, when the underlying news is more than six years old, and slow incorporation would require prices to be still absorbing it. For RoBERTa, lagged firm identity earns Sharpe ratios of 0.6 to 0.9 in the first four months, and both lagged inputs earn little afterwards.

In the framework of Section~\ref{sec:framework}, a level measured three years earlier retains the permanent component of the firm state but less of the slowly evolving component, whose contribution to expected returns declines with the horizon. Consistent with this, Llama-2's lagged inputs earn less than the current level during the first year. From the second year on, the strategies on the lagged inputs, the current level, and the full embedding earn similar Sharpe ratios on average, which suggests that predictability at these horizons comes from firm information that changes little over three years.

The lower panels of Figure~\ref{fig:fig4_inputs} add GICS-4 industry indicators to the residualization, as in DKPT's industry robustness test. Each strategy is then industry neutral: it has zero net weight in every GICS-4 industry each month, so returns common to all firms in an industry cancel. The pattern across horizons is unchanged. For Llama-2, the Sharpe ratios of the strategies on the embedding, the current level, and lagged firm identity rise slightly at nearly every horizon, as DKPT find for their residualized-embedding strategy; for RoBERTa, the changes are small and in both directions.

The historical level still contains the news of the preceding 12 months, which slow incorporation could explain. Table~\ref{tab:tab9} examines whether the news of earlier months predicts returns beyond the current historical level. It regresses next-month returns on the deviation of month $\tau-k$ and the historical level of month $\tau$, and Figure~\ref{fig:figA_old_news} plots the deviation's coefficient at every lag. For both models, the level's coefficient changes little with the lag and exceeds the lagged deviation's. Beyond the first months, the deviation's coefficient does not decline gradually with the lag. This holds both at lags within the 12 months that the current level averages and at longer lags. It is significant only at a few lags and roughly follows the three-month cycle of the deviation's rank IC in Figure~\ref{fig:figA_persistence_strategy}. The lagged deviation's coefficients are of similar size for the two models, whereas the level's coefficient is about three times as large for Llama-2 as for RoBERTa. The return-relevant information in earlier months' news is thus largely contained in the current level.

\begin{table}[htbp]
\centering
\caption{Return Regressions on Lagged Deviations and Current Levels}
\label{tab:tab9}
\begin{minipage}{\linewidth}\footnotesize
This table reports monthly Fama--MacBeth regressions of next-month returns on the deviation of month $\tau-k$ and the historical level of month $\tau$, for firms with news in both months, 2004--2025. The historical level is the mean of the firm's monthly average embedding over its news months in the preceding 12 months, or, without such months, the firm effect of a two-way category--firm projection of the headline embeddings of the up to six years before the signal year; the deviation is the remainder. Each component is standardized and residualized on JKP characteristics, converted into a score by its inner product with the residualized-embedding strategy's maximum Sharpe ratio regression (MSRR) coefficients of its own signal month, and standardized cross-sectionally. Coefficients are percentage returns per standard deviation of the score, with Newey--West $t$-statistics (6 lags) in parentheses. News is assigned by arrival month, and the return month is the $(k+1)$-th month after the deviation's month.
\end{minipage}
\medskip

{\footnotesize
\begin{adjustbox}{max width=\linewidth, max totalheight=0.75\textheight}
\begin{tabular}{llcccccc}
\toprule
Model &  & $k=1$ & $k=3$ & $k=5$ & $k=11$ & $k=17$ & $k=23$ \\
\midrule
RoBERTa & Deviation $E_{D,\,\tau-k}$ & \phantom{(0}0.031\phantom{)} & \phantom{(0}$-$0.011\phantom{)} & \phantom{(0}0.023\phantom{)} & \phantom{(0}0.050\phantom{)} & \phantom{(0}$-$0.010\phantom{)} & \phantom{(}0.046\phantom{)} \\
 &  & \phantom{0}(1.84)\phantom{0} & \phantom{0}($-$0.56)\phantom{0} & \phantom{0}(1.42)\phantom{0} & \phantom{0}(2.79)\phantom{0} & \phantom{0}($-$0.65)\phantom{0} & (2.48)\phantom{0} \\
 & Level $E_{L,\,\tau}$ & \phantom{(0}0.114\phantom{)} & \phantom{$-$(0}0.101\phantom{)} & \phantom{(0}0.103\phantom{)} & \phantom{(0}0.092\phantom{)} & \phantom{$-$(0}0.105\phantom{)} & \phantom{(}0.103\phantom{)} \\
 &  & \phantom{0}(5.71)\phantom{0} & \phantom{$-$0}(5.02)\phantom{0} & \phantom{0}(5.23)\phantom{0} & \phantom{0}(4.48)\phantom{0} & \phantom{$-$0}(5.06)\phantom{0} & (4.80)\phantom{0} \\
\midrule
Llama-2 & Deviation $E_{D,\,\tau-k}$ & \phantom{(0}0.039\phantom{)} & \phantom{$-$(0}0.012\phantom{)} & \phantom{(0}0.028\phantom{)} & \phantom{(0}0.005\phantom{)} & \phantom{$-$(0}0.023\phantom{)} & \phantom{(}0.053\phantom{)} \\
 &  & \phantom{0}(2.08)\phantom{0} & \phantom{$-$0}(0.67)\phantom{0} & \phantom{0}(1.47)\phantom{0} & \phantom{0}(0.27)\phantom{0} & \phantom{$-$0}(1.42)\phantom{0} & (3.03)\phantom{0} \\
 & Level $E_{L,\,\tau}$ & \phantom{(0}0.329\phantom{)} & \phantom{$-$(0}0.323\phantom{)} & \phantom{(0}0.316\phantom{)} & \phantom{(0}0.299\phantom{)} & \phantom{$-$(0}0.288\phantom{)} & \phantom{(}0.282\phantom{)} \\
 &  & (12.02)\phantom{0} & \phantom{$-$}(12.81)\phantom{0} & (11.78)\phantom{0} & (11.54)\phantom{0} & \phantom{$-$}(10.36)\phantom{0} & (8.93)\phantom{0} \\
\bottomrule
\end{tabular}
\end{adjustbox}
}
\end{table}

For Llama-2, inputs constructed only from news published more than three years earlier still predict returns 36 months after the signal month, which slow incorporation of news is unlikely to explain. For RoBERTa, such inputs predict returns mostly in the first months, mainly through lagged firm identity. Because this evidence rests mainly on Llama-2, whose pretraining data extend to September 2022, Section~\ref{sec:lookahead} examines whether it reflects information available when the news was published or information the model learned from later text during pretraining.

\section{Assessing Potential Look-Ahead Bias}
\label{sec:lookahead}

RoBERTa and Llama-2 were pretrained on text published after much of our evaluation sample. Their embeddings may reflect information unavailable when the news was published. The concern matters most for the historical level, which carries Llama-2's long-horizon predictability, including inputs built only from news published more than three years earlier (Section~\ref{sec:long}). This predictability may reflect better extraction of information available at the news date, or information about subsequent developments that the model learned from later text during pretraining. We examine these explanations by comparing encoders with different pretraining cutoffs, including ChronoGPT$_{1999}$, whose pretraining ends before our sample begins, by comparing predictability across the pretraining cutoffs, and by extending the framework of Section~\ref{sec:framework} to allow for look-ahead in the embeddings.

\subsection{Robustness to Pretraining Cutoffs}

An encoder pretrained only on text published before our sample cannot encode information about later developments. We repeat our analysis using ChronoGPT$_{1999}$ of \citet{he2026chronologically}, whose pretraining ends in December 1999, before our news sample begins.\footnote{We use the Hugging Face release \url{https://huggingface.co/manelalab/chrono-gpt-v1-19991231} (commit \texttt{45d6568}, July 2025), which has about 1.6 billion parameters and was pretrained over about ten epochs on a corpus of 7 billion tokens of text published before 2000 \citep{he2026chronologically}. Under greedy decoding, it completes ``The company with the largest stock market value in the world is'' with ``the United States Steel corporation. It's stock is worth \$1, ,000, ,000,'' and ``IBM is a company that makes'' with ``money. The company's earnings in 1860 were \$1,000, 900.'' RoBERTa names Apple and computers, respectively.} Table~\ref{tab:tab10} compares its results with RoBERTa and Llama-2, with detailed results reported in Appendix \ref{sec:appendixB}. ChronoGPT produces significant return predictability under both the CKX and DKPT procedures, although its overall performance is weaker. Its daily forecast achieves a rank IC of 1.90\%, compared with 2.41\% for RoBERTa and 2.67\% for Llama-2, and the equal-weighted long-short decile portfolio sorted on it has an annualized Sharpe ratio of 4.50, close to the 4.92 and 4.97 of the other two. Its first-month MSRR strategy has an annualized Sharpe ratio of 1.38, compared with 1.79 and 3.62, respectively.

\begin{table}[htbp]
\centering
\caption{Forecast, MSRR Strategy, and Valuation Results by Encoder}
\label{tab:tab10}
\begin{minipage}{\linewidth}\footnotesize
This table reports the main results for ChronoGPT$_{1999}$, RoBERTa-large, and Llama-2 (13B), whose pretraining ends in December 1999, July 2019, and September 2022. The coefficient is the Fama--MacBeth coefficient on the forecast controlling for the eight-year category mean, in basis points per standard deviation, with its Newey--West $t$-statistic (5 lags) in parentheses. Rank IC is the mean daily Spearman correlation with next open-to-open returns, in percent. SR is the annualized Sharpe ratio. Daily-forecast portfolios are equal-weighted (EW) or value-weighted (VW) decile long-short portfolios. The event-category, firm-identity, and idiosyncratic components come from a two-way category--firm projection of the headline embeddings of the up to six years before the test year. Covariance shares divide each component's covariance with returns by the forecast's. Maximum Sharpe ratio regression (MSRR) strategies trade firms with news in signal month $\tau$, residualize their input on month-$\tau$ JKP characteristics, and earn returns in month $\tau+h$. The DKPT strategy trades the embedding with the coefficients of signal month $\tau$, fitted in an expanding window from 2000 to next-month returns, at both $h=1$ and $h=12$. The historical level $E_L$ is the mean of the firm's monthly average embedding over its news months from $\tau-12$ to $\tau-1$, or the firm effect of the projection when the firm has no such months. The level 36 months earlier is the mean over news months from $\tau-48$ to $\tau-37$. Its strategy trades only firms with such months and has its own coefficients, fitted in an expanding window from 2004 to next-month returns, with the first 12 signal months initializing the fit. Level shares are percentages of the strategy's mean return. Entries are blank at horizons with fewer than 24 signal months, and level shares also when the strategy's mean return has $|t|<2$. Firm-identity $R^2$ is the $R^2$ of log price-to-book on held-out firms, with its increment over controls for Fama--French 49 industries, 60-month CAPM beta, log book equity, asset growth, gross profits to assets, and dividends to assets. The $F$-network rows report the adjusted $R^2$ of log price-to-book on its mean over the ten most similar firms under $F$, and its difference from SIC-3 on the same firm-years. Valuation and network windows cover the up to six years through each December, no earlier than 2000 in Panel A and 2023 in Panel B.
\end{minipage}
\medskip

{\footnotesize
\textit{Panel A: Signal years 2004--2025}\\[2pt]
\begin{adjustbox}{max width=\linewidth, max totalheight=0.28\textheight}
\begin{tabular}{llccc}
\toprule
 &  & ChronoGPT$_{1999}$ & RoBERTa & Llama-2 \\
\midrule
CKX daily forecast & Rank IC (\%) & \phantom{$-$(0}1.90\phantom{)} & \phantom{$-$(0}2.41\phantom{)} & \phantom{(0}2.67\phantom{)} \\
 & EW SR & \phantom{$-$(0}4.50\phantom{)} & \phantom{$-$(0}4.92\phantom{)} & \phantom{(0}4.97\phantom{)} \\
 & VW SR & \phantom{$-$(0}1.02\phantom{)} & \phantom{$-$(0}1.21\phantom{)} & \phantom{(0}1.90\phantom{)} \\
 & Coefficient with the category mean & \phantom{$-$(0}2.04\phantom{)} & \phantom{$-$(0}3.02\phantom{)} & \phantom{(0}3.93\phantom{)} \\
 &  & \phantom{$-$0}(7.00) & \phantom{$-$0}(9.24) & (10.35) \\
 & Covariance share, event category (\%) & \phantom{$-$(}82.2\phantom{)0} & \phantom{$-$(}80.0\phantom{)0} & \phantom{(}65.5\phantom{)0} \\
 & Covariance share, firm identity (\%) & \phantom{$-$(0}4.2\phantom{)0} & \phantom{$-$(0}6.3\phantom{)0} & \phantom{(}15.7\phantom{)0} \\
 & Covariance share, idiosyncratic (\%) & \phantom{$-$(}13.6\phantom{)0} & \phantom{$-$(}13.6\phantom{)0} & \phantom{(}18.8\phantom{)0} \\
\midrule
DKPT monthly MSRR strategy & SR, $h=1$ & \phantom{$-$(0}1.38\phantom{)} & \phantom{$-$(0}1.79\phantom{)} & \phantom{(0}3.62\phantom{)} \\
 & Share of the level $E_L$, $h=1$ (\%) & \phantom{$-$(}43.5\phantom{)0} & \phantom{$-$(}51.0\phantom{)0} & \phantom{(}76.3\phantom{)0} \\
 & SR, $h=12$ & \phantom{$-$(0}0.57\phantom{)} & \phantom{$-$(0}0.97\phantom{)} & \phantom{(0}2.24\phantom{)} \\
 & Share of the level $E_L$, $h=12$ (\%) & \phantom{$-$(}41.2\phantom{)0} & \phantom{$-$(}42.7\phantom{)0} & \phantom{(}85.5\phantom{)0} \\
 & SR, level 36 months earlier, $h=1$ & \phantom{$-$(0}0.21\phantom{)} & \phantom{$-$(0}0.43\phantom{)} & \phantom{(0}1.96\phantom{)} \\
\midrule
Firm identity $F$ & $R^2$ of $\log(P/B)$ (\%) & \phantom{$-$(}10.56\phantom{)} & \phantom{$-$(}17.30\phantom{)} & \phantom{(}28.97\phantom{)} \\
 & Increment over the controls (pp) & \phantom{$-$(0}2.17\phantom{)} & \phantom{$-$(0}9.03\phantom{)} & \phantom{(}22.93\phantom{)} \\
 & $F$ network: adjusted $R^2$ of $\log(P/B)$ (\%) & \phantom{$-$(0}8.37\phantom{)} & \phantom{$-$(}17.44\phantom{)} & \phantom{(}24.72\phantom{)} \\
 & $F$ network minus SIC-3 (pp) & \phantom{(}$-$11.46\phantom{)} & \phantom{(0}$-$2.39\phantom{)} & \phantom{(0}4.89\phantom{)} \\
\bottomrule
\end{tabular}
\end{adjustbox}
\par\medskip
\textit{Panel B: Signal years 2023--2025}\\[2pt]
\begin{adjustbox}{max width=\linewidth, max totalheight=0.28\textheight}
\begin{tabular}{llccc}
\toprule
 &  & ChronoGPT$_{1999}$ & RoBERTa & Llama-2 \\
\midrule
CKX daily forecast & Rank IC (\%) & \phantom{$-$(0}1.65\phantom{)} & \phantom{$-$(0}2.14\phantom{)} & \phantom{(0}2.30\phantom{)} \\
 & EW SR & \phantom{$-$(0}3.67\phantom{)} & \phantom{$-$(0}3.73\phantom{)} & \phantom{(0}4.15\phantom{)} \\
 & VW SR & \phantom{$-$(0}0.54\phantom{)} & \phantom{$-$(0}1.29\phantom{)} & \phantom{(0}1.27\phantom{)} \\
 & Coefficient with the category mean & \phantom{$-$(0}3.10\phantom{)} & \phantom{$-$(0}3.97\phantom{)} & \phantom{(0}4.34\phantom{)} \\
 &  & \phantom{$-$0}(3.40) & \phantom{$-$0}(3.74) & \phantom{0}(3.65) \\
 & Covariance share, event category (\%) & \phantom{$-$(}73.4\phantom{)0} & \phantom{$-$(}66.1\phantom{)0} & \phantom{(}61.7\phantom{)0} \\
 & Covariance share, firm identity (\%) & \phantom{(0}$-$2.7\phantom{)0} & \phantom{$-$(0}7.4\phantom{)0} & \phantom{(0}5.5\phantom{)0} \\
 & Covariance share, idiosyncratic (\%) & \phantom{$-$(}29.3\phantom{)0} & \phantom{$-$(}26.5\phantom{)0} & \phantom{(}32.8\phantom{)0} \\
\midrule
DKPT monthly MSRR strategy & SR, $h=1$ & \phantom{$-$(0}2.02\phantom{)} & \phantom{$-$(0}1.70\phantom{)} & \phantom{(0}2.57\phantom{)} \\
 & Share of the level $E_L$, $h=1$ (\%) & \phantom{$-$(}16.7\phantom{)0} & \phantom{$-$(}19.7\phantom{)0} & \phantom{(}17.8\phantom{)0} \\
 & SR, $h=12$ & \phantom{$-$(0}0.59\phantom{)} & \phantom{$-$(0}0.73\phantom{)} & \phantom{(0}0.10\phantom{)} \\
 & Share of the level $E_L$, $h=12$ (\%) &  &  &  \\
 & SR, level 36 months earlier, $h=1$ & \phantom{(0}$-$0.63\phantom{)} & \phantom{(0}$-$0.18\phantom{)} & \phantom{(0}0.04\phantom{)} \\
\midrule
Firm identity $F$ & $R^2$ of $\log(P/B)$ (\%) & \phantom{$-$(0}8.50\phantom{)} & \phantom{$-$(}15.48\phantom{)} & \phantom{(}24.41\phantom{)} \\
 & Increment over the controls (pp) & \phantom{$-$(0}1.36\phantom{)} & \phantom{$-$(}11.49\phantom{)} & \phantom{(}20.01\phantom{)} \\
 & $F$ network: adjusted $R^2$ of $\log(P/B)$ (\%) & \phantom{$-$(0}1.42\phantom{)} & \phantom{$-$(}10.57\phantom{)} & \phantom{(}15.13\phantom{)} \\
 & $F$ network minus SIC-3 (pp) & \phantom{(0}$-$9.08\phantom{)} & \phantom{$-$(0}0.07\phantom{)} & \phantom{(0}4.63\phantom{)} \\
\bottomrule
\end{tabular}
\end{adjustbox}
}
\end{table}

Despite these differences in overall performance, the sources of predictability exhibit similar patterns across encoders. Event categories account for 82.2\% of ChronoGPT's daily predictive covariance, compared with 80.0\% for RoBERTa and 65.5\% for Llama-2. Adding a within-category ridge forecast raises Sharpe ratios but does not significantly improve rank IC over the category benchmark (Table~\ref{tab:tabB_residual}). For ChronoGPT, the historical level also contributes to the characteristic-residualized monthly strategy, accounting for 43.5\% of first-month strategy returns, compared with 51.0\% and 76.3\% for the other two encoders. Its level contribution comes from both the category level and firm identity and fades after the sixth month, while the idiosyncratic level contributes little (Table~\ref{tab:tabB_attribution}). An encoder pretrained entirely before the evaluation sample thus shows both the dominance of event categories at short horizons and the contribution of the historical level after characteristic residualization.

The amount of firm-specific information captured by the encoders nevertheless differs substantially. Firm identity accounts for only 4.2\% of ChronoGPT's daily predictive covariance, compared with 6.3\% for RoBERTa and 15.7\% for Llama-2. Adding ChronoGPT's firm identity to conventional controls increases the cross-sectional $R^2$ for log price-to-book by only 2.2 percentage points, compared with 9.0 and 22.9 percentage points for the other encoders. In the firm-similarity analysis, peers selected by ChronoGPT's firm embeddings explain a firm's log price-to-book 11.5 percentage points worse than SIC-3 industries, compared with 2.4 points worse for RoBERTa and 4.9 points better for Llama-2 (Table~\ref{tab:tab10}). ChronoGPT captures relatively little persistent firm-specific information beyond event categories in its daily forecasts and firm representations.\footnote{An encoder need not have encountered a firm during pretraining to represent information about it. Subword tokenization allows previously unseen company names or tickers to be encoded without a learned association with the firm, while contextual hidden states can capture economic information from the surrounding news text. Moreover, firm identity in our decomposition is estimated by projecting embeddings across headlines linked to the same firm using RavenPack identifiers; the encoder itself need not recognize the firm. Repeated references to a firm's operations, products, and financial conditions can generate a persistent firm component without prior knowledge of the company.} The same gap appears for information measured three years before the signal month. The strategy on the historical level lagged 36 months, which uses only news published more than three years earlier, has a first-month Sharpe ratio of 0.21 for ChronoGPT and 0.43 for RoBERTa, neither significant, against 1.96 for Llama-2 (Table~\ref{tab:tab10}). The stronger firm information captured by Llama-2 may reflect its greater ability to extract information from news, but exposure to later pretraining data offers another possible explanation. Comparisons across encoders alone cannot distinguish these effects.

We restrict evaluation to periods after each encoder's pretraining cutoff, using 2020--2025 for RoBERTa and 2023--2025 for Llama-2. Appendix \ref{sec:appendixC} reports the corresponding results. The main short-horizon findings remain intact. Event categories account for 77.4\% and 61.7\% of daily predictive covariance for RoBERTa and Llama-2, respectively (Table~\ref{tab:tabC_post_split}). Embedding forecasts also retain significant incremental predictability beyond category averages, with improvements in rank IC for both encoders (Table~\ref{tab:tabC_post_residual}). Firm representations constructed exclusively from news published after pretraining remain economically informative. For Llama-2, firm identity explains 24.4\% of the cross-sectional variation in log price-to-book, compared with 29.0\% in the full sample (Tables~\ref{tab:tabC_post_representation} and \ref{tab:tab4}). Firms with similar embeddings also continue to exhibit similar operating and risk characteristics (Table~\ref{tab:tabC_post_networks}).

The return predictability associated with firm identity and the historical level is less stable. For Llama-2, firm identity's share of daily predictive covariance declines from 15.7\% to 5.5\% after the cutoff (Tables~\ref{tab:tab6} and \ref{tab:tabC_post_split}), while the historical level's contribution to first-month MSRR returns falls from 76.3\% to 17.8\%.

\subsection{Predictability across the Pretraining Cutoffs}
\label{sec:cutoff}

\begin{figure}[htbp]
\centering
\caption{Cumulative MSRR Strategy Returns by Component and Encoder}
\includegraphics[width=0.85\textwidth]{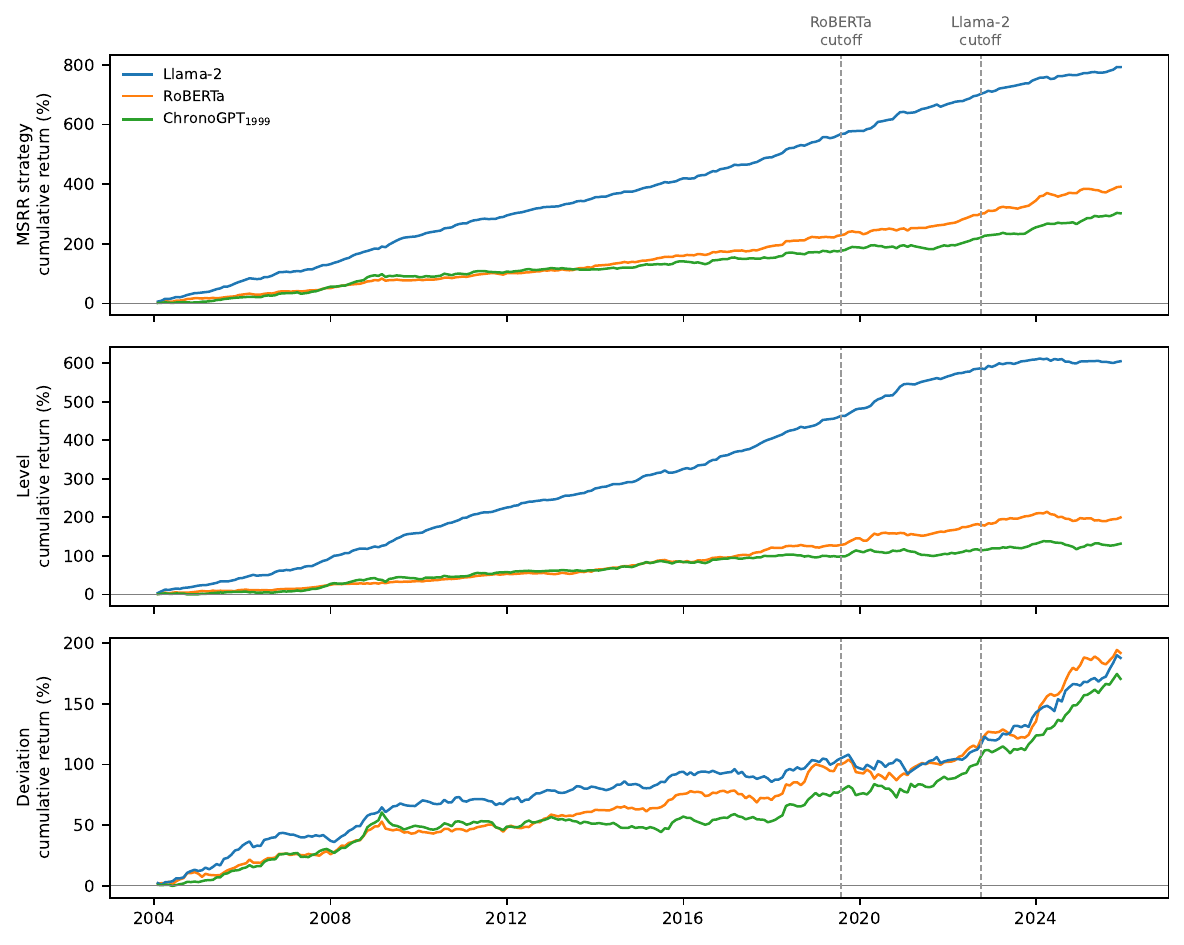}
\label{fig:fig5_cutoff_returns}
\begin{minipage}{\linewidth}\footnotesize
This figure plots cumulative returns of the monthly maximum Sharpe ratio regression (MSRR) strategy and of its level and deviation components for RoBERTa-large, Llama-2 (13B), and ChronoGPT$_{1999}$, for signal years 2004--2025. News is assigned by arrival month. Firm-month embeddings are residualized each month on JKP characteristics and combined out of sample using MSRR; returns are those of the month after the signal month. The historical level is the mean of the firm's monthly average embedding over its news months in the preceding 12 months; the deviation is the remainder. Each component is residualized like the embedding and uses the same coefficients, so component returns sum to the strategy return. Each encoder's strategy is scaled to 10\% annual volatility over the plotted months, with the same scale for its components. Cumulative returns sum monthly percentage returns, dated at the end of the signal month, over the signal months shared by the three encoders. Dashed lines mark the pretraining cutoffs of RoBERTa (July 2019) and Llama-2 (September 2022).
\end{minipage}
\end{figure}

Figure~\ref{fig:fig5_cutoff_returns} plots cumulative returns of the monthly MSRR strategies, split into the contributions of the historical level and the deviation. The cumulative returns of the total strategies show no obvious break at the pretraining cutoffs. Llama-2 remains the strongest performer after its September 2022 cutoff, with an annualized Sharpe ratio of 2.57 over 2023--2025, compared with 1.70 for RoBERTa and 2.02 for ChronoGPT (Table~\ref{tab:tab10}). The components, however, show a different pattern. Under the strategy's weights, Llama-2's historical level contributes strongly to cumulative returns before its cutoff but little afterwards, while the deviation accounts for most subsequent gains. Thus, aggregate strategy performance can conceal substantial changes in the sources of predictability. 

Portfolios formed on the daily forecasts show the same contrast: their total returns are similar across encoders, while their historical-level and deviation components differ (Figure~\ref{fig:figA_cutoff_returns_daily}). This distinction matters for look-ahead checks based on total strategy performance, such as DKPT's comparison of point-in-time and foresight ChronoGPT models.

The change is not unique to Llama-2. After 2022, all three encoders earn relatively little from their historical levels, whose share of first-month returns falls from 51.0\% to 19.7\% for RoBERTa and from 43.5\% to 16.7\% for ChronoGPT (Table~\ref{tab:tab10}), while their deviation returns increase, including for ChronoGPT, whose pretraining ends before the sample begins. A decline in Llama-2's historical-level returns around its cutoff could therefore reflect changes shared across encoders rather than look-ahead bias. Moreover, our sample provides only 35 post-cutoff signal months for Llama-2's monthly strategy. To distinguish its change from those of the other encoders, we turn to the daily forecasts and compare their rank ICs on common trading days.

\begin{figure}[htbp]
\centering
\caption{Rolling Rank IC of Daily Forecast Levels and Deviations}
\includegraphics[width=0.85\textwidth]{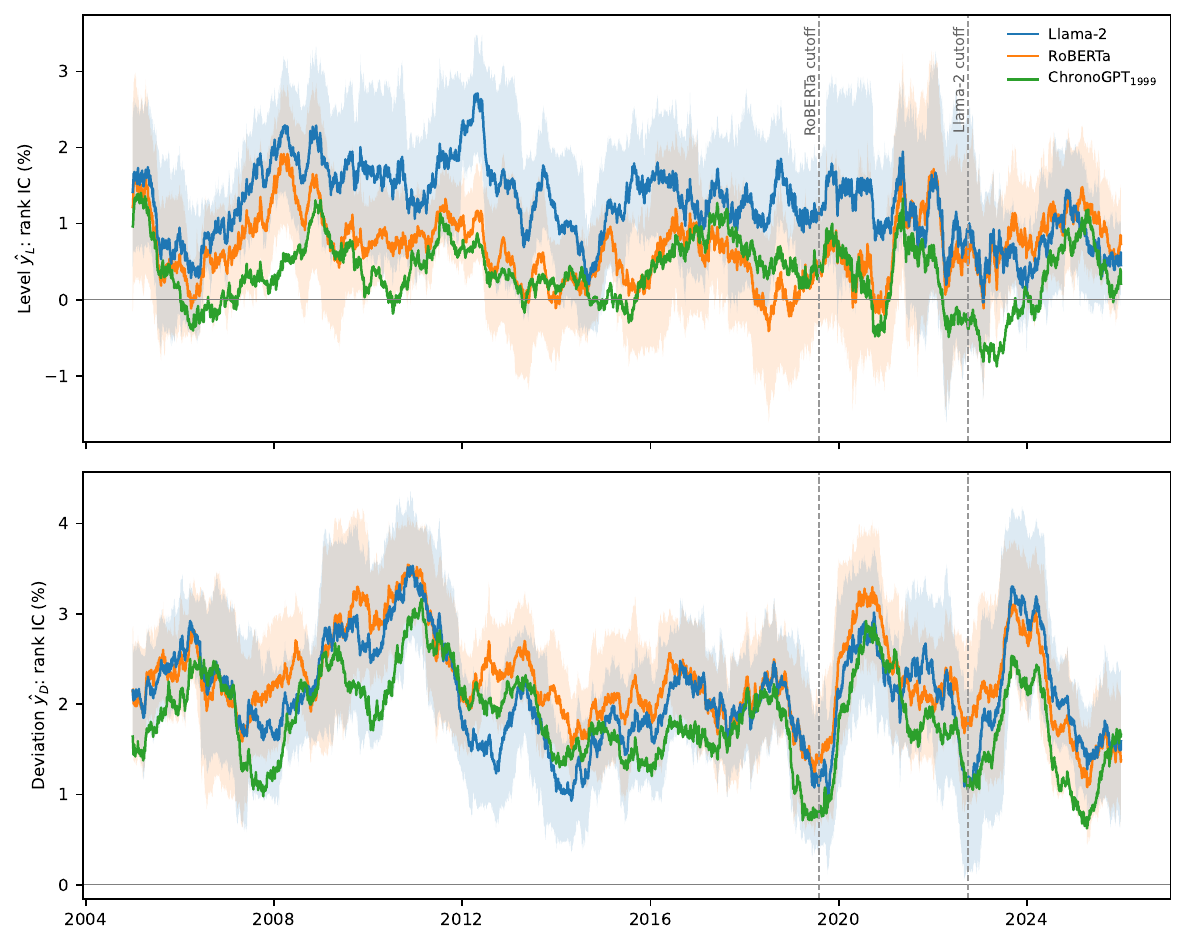}
\label{fig:fig6_cutoff}
\begin{minipage}{\linewidth}\footnotesize
This figure reports the rank IC of the level (upper panel) and deviation (lower panel) of the daily ridge forecast for RoBERTa-large, Llama-2 (13B), and ChronoGPT$_{1999}$, for signal years 2004--2025. Forecasts are averaged within firm-days with a return to the eligible open and ten subsequent open-to-open returns, on the trading days shared by the three encoders. Rank IC is the cross-sectional Spearman correlation with next open-to-open returns, in percent, averaged over the preceding 252 trading days. The historical level is the mean of the firm's monthly average forecast over its news months in the preceding 12 months, computed with the current coefficients; without such months it is the forecast from the firm effect of a two-way category--firm projection of the headline embeddings of the up to six years before the signal year, or zero when the firm has $F=0$. The deviation is the remainder. Shading shows 95\% Newey--West confidence intervals for RoBERTa and Llama-2 with 40 lags within each window. Dashed lines mark the pretraining cutoffs of RoBERTa (July 2019) and Llama-2 (September 2022).
\end{minipage}
\end{figure}

Figure~\ref{fig:fig6_cutoff} plots rolling rank ICs of the historical levels and deviations of the daily ridge forecasts. Llama-2's historical level substantially outperforms RoBERTa's and ChronoGPT's through 2020, but its advantage over RoBERTa largely disappears from 2021, before its September 2022 cutoff, and does not recover afterwards.\footnote{The monthly MSRR signals show a similar narrowing during 2021--2022 (Figure~\ref{fig:figA_cutoff_monthly}). A possible reason is that Llama-2's effective knowledge cutoff is earlier than its reported one. \citet{cheng2024dated} suggest that Llama-2, like many other models, has an effective knowledge cutoff for Wikipedia well before its reported date, and attribute such gaps partly to newer web crawls that contain much older text.} RoBERTa's historical-level IC shows no comparable decline around its own July 2019 cutoff or after Llama-2's cutoff. The deviations display a different pattern: their ICs fluctuate together and remain close across encoders. The monthly MSRR signals show a similar distinction between historical levels and deviations (Figure~\ref{fig:figA_cutoff_monthly}).

Table~\ref{tab:tab11} reports relative IC ratios, defined as Llama-2's ratio of post- to pre-cutoff mean rank IC divided by the corresponding ratio for RoBERTa. We use RoBERTa rather than ChronoGPT as the control because ChronoGPT's historical-level IC is small and insignificant after 2022, which makes the ratio imprecise.\footnote{With ChronoGPT as the control, the relative IC ratio is 0.48 for the historical level and 1.21 for the deviation. ChronoGPT's historical-level IC from January 2023 has a Newey--West $t$-statistic of 1.47, and the 90\% interval for the historical-level ratio is $[-1.24, 2.66]$.} We split the sample at January 2023, the start of the first full calendar year after Llama-2's pretraining cutoff. This split places 2021--2022 in the earlier period, which lowers Llama-2's earlier mean IC and thus the measured relative decline. Llama-2's historical-level IC falls from 1.34\% to 0.61\%, while RoBERTa's rises from 0.68\% to 0.85\%. The relative IC ratio for the historical level is 0.36, with a 90\% confidence interval of [0.21, 0.68]. In contrast, the ratio for the deviation is 1.23, with an interval of [1.03, 1.36]. Llama-2's loss of relative predictability is concentrated in the historical level rather than the deviation.

\begin{table}[htbp]
\centering
\caption{Rank ICs across the Llama-2 Pretraining Cutoff by Firm Size and News Intensity}
\label{tab:tab11}
\begin{minipage}{\linewidth}\footnotesize
This table reports mean daily cross-sectional Spearman correlations (rank ICs), in percent, of the historical level and the deviation of the daily ridge forecast with next open-to-open returns for RoBERTa-large and Llama-2 (13B), before and from January 2023, the first calendar year after Llama-2's September 2022 pretraining cutoff. Forecasts are averaged within firm-days with a return to the eligible open and ten subsequent open-to-open returns, on the trading days both encoders share. The historical level is the mean of the firm's monthly average forecast over its news months in the preceding 12 months, computed with the current coefficients; without such months it is the forecast from the firm effect of a two-way category--firm projection of the headline embeddings of the up to six years before the signal year, or zero when the firm has $F=0$. The deviation is the remainder. Micro, small, and big stocks are below the NYSE 20th capitalization percentile, between it and the NYSE median, and at or above the median. News intensity is the firm's number of sample headlines in the 12 calendar months before the trading day's month; terciles are formed each day. The relative IC ratio divides Llama-2's ratio of its later to its earlier mean IC by the same ratio for RoBERTa. Brackets report 90\% moving-block bootstrap intervals with 40-day blocks drawn within each period. A relative IC ratio is omitted when Llama-2's earlier mean IC or RoBERTa's later one has a Newey--West $|t|<2$.
\end{minipage}
\medskip

{\footnotesize
\begin{adjustbox}{max width=\linewidth, max totalheight=0.75\textheight}
\begin{tabular}{llccccccc}
\toprule
\multicolumn{3}{c}{} & \multicolumn{3}{c}{Size} & \multicolumn{3}{c}{News intensity} \\
\cmidrule(lr){4-6} \cmidrule(lr){7-9}
 &  & All & Micro & Small & Big & Low & Mid & High \\
\midrule
\multicolumn{9}{l}{\textit{Level $\hat y_L$}} \\
RoBERTa & 2004--2022 & 0.68 & 1.28 & 0.54 & 0.27 & 1.03 & 0.49 & \phantom{$-$}0.48 \\
 & 2023--2025 & 0.85 & 1.30 & 0.93 & 0.71 & 1.09 & 1.14 & \phantom{$-$}0.88 \\
\midrule
Llama-2 & 2004--2022 & 1.34 & 2.07 & 1.20 & 0.80 & 1.84 & 1.01 & \phantom{$-$}1.00 \\
 & 2023--2025 & 0.61 & 0.55 & 0.70 & 0.21 & 0.91 & 0.96 & \phantom{$-$}0.30 \\
\midrule
Relative IC ratio &  & 0.36 & 0.26 & 0.34 & 0.10 & 0.47 & 0.41 & \phantom{$-$}0.16 \\
 &  & [0.21, 0.68] & [$-$0.11, 0.57] & [0.06, 0.76] & [$-$0.08, 0.39] & [0.24, 0.81] & [0.18, 0.79] & [$-$0.09, 0.51] \\
\midrule
\multicolumn{9}{l}{\textit{Deviation $\hat y_D$}} \\
RoBERTa & 2004--2022 & 2.29 & 3.78 & 2.66 & 1.53 & 3.11 & 2.17 & \phantom{$-$}1.54 \\
 & 2023--2025 & 1.87 & 4.65 & 2.04 & 0.67 & 3.34 & 1.63 & $-$0.10 \\
\midrule
Llama-2 & 2004--2022 & 2.09 & 3.42 & 2.46 & 1.49 & 2.80 & 1.92 & \phantom{$-$}1.54 \\
 & 2023--2025 & 2.10 & 4.27 & 2.56 & 0.94 & 3.58 & 1.49 & \phantom{$-$}0.37 \\
\midrule
Relative IC ratio &  & 1.23 & 1.02 & 1.36 & 1.44 & 1.19 & 1.04 &  \\
 &  & [1.03, 1.36] & [0.85, 1.14] & [1.10, 1.82] & [0.81, 2.30] & [1.02, 1.32] & [0.76, 1.42] &  \\
\bottomrule
\end{tabular}
\end{adjustbox}
}
\end{table}

The historical level's relative IC ratio is below one in every firm-size and news-intensity group, with all 90\% confidence intervals lying below one. In contrast, the deviation's relative IC ratios are at least one in every group for which they are reported. The historical-level ratio is particularly low among firms with high recent news intensity (0.16), compared with 0.47 for firms with low news intensity. It is also lower among big stocks (0.10) than micro stocks (0.26), although the confidence intervals overlap in both comparisons. These patterns suggest stronger look-ahead bias among more widely covered firms, which may also appear more frequently in pretraining text.

The concentration of Llama-2's relative decline in the historical level, and not in the deviation, suggests pretraining memory of subsequently realized outcomes. If the two encoders' memory-free rank ICs change by the same proportion, persistent firm information alone would produce a relative IC ratio of one. The relative IC ratios alone, however, do not identify how much predictability comes from memory, because they depend on memory in the control encoder.

\subsection{Calibration}
\label{sec:calibration}

We extend the framework in Section~\ref{sec:framework} to examine whether pretraining memory of subsequently realized outcomes can account for the relative changes in predictability documented above. Appendix~\ref{sec:appendixD} provides the full model specification, derivations, and calibration details. In addition to the return response to published news and the premium associated with a persistent firm state, we allow returns to contain unexpected daily cash-flow-news shocks $\theta_{i,t}$, which are unpredictable from investors' information before their realization. An encoder trained on later text may nevertheless retain information about these outcomes. We represent its memory of firm $i$'s realized cash-flow news by
\[
\tilde\theta_i^\ast
=
\frac{1}{\sigma_\theta\sqrt{N_\theta}}
\sum_{t=t^\ast-21N_\theta+1}^{t^\ast}\theta_{i,t},
\]
where $t^\ast$ is the encoder's pretraining cutoff, $N_\theta$ is the memory-window length in months, and $\sigma_\theta^2$ is the variance of monthly cash-flow news, with 21 trading days per month. We set $N_\theta=360$ months, so the memory windows of RoBERTa and Llama-2 cover the sample through their respective cutoffs.\footnote{\citet{crane2025total} find that an LLM recalls GDP growth, inflation, and unemployment with similar errors for the 1990s, 2000s, and 2010s and with larger errors for earlier decades. \citet{lopez2025memorization} document LLMs' recall of realized financial data, including individual stock prices, before their pretraining cutoffs and find no evidence of recall afterward.} ChronoGPT's cutoff precedes the sample, so it remembers none of the sample-period shocks.

We augment the measurement equation~\eqref{eq:measure} with this memory term:
\[
E_t=a_K+q\nu+gz_t+\eta\tilde\theta_i^\ast+\zeta_t.
\]
The loading $\eta$ governs the strength of memory in the embedding. As specified in Appendix~\ref{sec:appD2}, the memory loading acts along a direction orthogonal to the news and state signals. The three encoders share the same economic environment, including event-drift dynamics, the persistent firm-state process, and the state premium, but differ in their loadings on the news signal, the firm state, and memory.

Because memory is constant across a firm's news items, it enters primarily firm identity $F$ and the historical level $E_L$ rather than deviations. Its predictive covariance with returns vanishes once the return dates pass the encoder's cutoff, whereas persistent firm information can continue to predict returns. The Pearson correlation between a forecast component and returns is their covariance divided by the cross-sectional standard deviations of the component, $\sigma$, and of returns, $\sigma_r$. Under joint normality and small correlations, as for daily returns, rank IC is approximately proportional to the Pearson correlation. Splitting the covariance into a part unrelated to memory, $\Gamma$, and a part due to memory, $M$, the rank IC over a period satisfies approximately
\[
\mathrm{IC}\propto\frac{\Gamma+M}{\sigma\,\sigma_r}.
\]
Let $\omega=\Gamma/(\Gamma+M)$ denote the memory-free share. After January 2023, both encoders are past their pretraining cutoffs, so $M=0$ and $\omega=1$. Before the split, memory raises the IC by the factor $1/\omega$. Let $\mathrm{IC}_{j,1}^o$ and $\mathrm{IC}_{j,2}^o$ be encoder $o$'s mean rank IC for component $j$ before and after January 2023. Under the assumption that memory-free rank ICs change by the same proportion for both encoders, this common change cancels from the relative IC ratio:
\[
\mathrm{RIC}_j
\equiv
\frac{\mathrm{IC}_{j,2}^{\mathrm{Llama2}}/\mathrm{IC}_{j,1}^{\mathrm{Llama2}}}
{\mathrm{IC}_{j,2}^{\mathrm{RoBERTa}}/\mathrm{IC}_{j,1}^{\mathrm{RoBERTa}}}
\approx
\frac{\omega_j^{\mathrm{Llama2}}}
{\omega_j^{\mathrm{RoBERTa}}},
\]
where $\omega_j^o$ is encoder $o$'s memory-free share for component $j$ before January 2023. Because the ratio also depends on RoBERTa's memory, we use a second comparison across January 2020, around RoBERTa's cutoff, with Llama-2 as the control. This comparison ends in September 2022, before Llama-2's cutoff, and assumes that Llama-2's memory-free share remains unchanged across the two periods.\footnote{Llama-2's historical-level IC appears to decline from 2021 (Figure~\ref{fig:fig6_cutoff}). Such a decline raises this ratio and understates RoBERTa's memory share and, through the comparison across January 2023, also Llama-2's.} This ratio \eqref{eq:D29} then approximates RoBERTa's own memory-free share before January 2020.

We calibrate the model in two steps. We first determine the event-drift parameters from announcement returns and their subsequent decay. We then jointly choose 16 parameters to match 49 moments: daily predictive covariances by component, monthly MSRR component returns across return horizons, and relative IC ratios around both pretraining cutoffs. The cutoff ratios primarily discipline memory, while the allocation and decay of predictability across components and horizons discipline the state premium, its persistence, and the encoders' measurement of firm information. The calibration also ties cash-flow-news variance to long-horizon return variation. Appendix~\ref{sec:appD3} reports the parameters and targeted moments (Tables~\ref{tab:D2}--\ref{tab:D4}). We use the calibration to assess the relative importance of these mechanisms rather than interpret the parameters as separately identified structural estimates.

In the calibrated model, the three encoders measure the news signal similarly well but differ substantially in how reliably they measure the firm state (Table~\ref{tab:D2}). The loadings $a$, $q$, $g$, and $\eta$ measure how strongly the embedding responds to the standardized category signal, the story detail, the firm state, and memory, relative to language noise. The category-signal loadings $a$ are 0.88, 0.95, and 1.03 for ChronoGPT, RoBERTa, and Llama-2, and the story-detail loadings $q$ are 0.14, 0.14, and 0.18. Daily moments determine the state loading $g$ only jointly with the state premium, so we fix Llama-2's $g$ at one and compare reliabilities $\phi(g)=g^2/(1+g^2)$. ChronoGPT's and RoBERTa's are about a quarter and a third of Llama-2's. Memory is also concentrated in Llama-2, whose memory reliability $\phi(\eta)$ is 0.83, compared with 0.01 for RoBERTa.

\begin{table}[htbp]
\centering
\caption{Relative IC Ratios across the Pretraining Cutoffs}
\label{tab:tab12}
\begin{minipage}{\linewidth}\footnotesize
This table reports, by component $j$ of the daily ridge forecast, data and model values of the relative IC ratios $\mathrm{RIC}_j$ in \eqref{eq:D28}, Llama-2 relative to RoBERTa across January 2023, and $\mathrm{RIC}_j^*$ in \eqref{eq:D29}, RoBERTa relative to Llama-2 across January 2020 with dates through September 2022, with mean daily rank ICs with next open-to-open returns on common trading days. Each ratio divides one encoder's ratio of its later to its earlier mean IC by the same ratio for the other encoder. The historical level is the mean of the firm's monthly average forecast over its news months in the preceding 12 months, computed with the current coefficients; without such months it is the forecast from the firm effect of a two-way category--firm projection of the headline embeddings of the up to six years before the signal year, or zero when the firm has $F=0$. The deviation is the remainder. The dispersion ratio replaces each IC with the component's mean cross-sectional standard deviation. Firm identity $F$, the idiosyncratic level $I_L$, and the category deviation $C_D$ are the forecast components from the projection's firm effect, the historical level of its idiosyncratic component, and the deviation of its category component. Memory share is the percentage of the pre-split predictive covariance of the encoder whose cutoff defines the split, Llama-2 in the upper block and RoBERTa in the lower, attributable to memory, $1-\omega_j^o$. Intervals come from a moving-block bootstrap with 40-trading-day blocks drawn separately before and after the split. Targeted marks the calibration targets.
\end{minipage}
\medskip

{\footnotesize
\begin{adjustbox}{max width=\linewidth}
\begin{tabular}{lcccccc}
\toprule
 & & \multicolumn{3}{c}{Data} & \multicolumn{2}{c}{Model} \\
\cmidrule(lr){3-5}\cmidrule(lr){6-7}
Component & Targeted & IC ratio & 90\% interval & Dispersion ratio & IC ratio & Memory share (\%) \\
\midrule
\multicolumn{7}{@{}l}{\textit{Llama-2 across January 2023, relative to RoBERTa}} \\
Firm identity $F$ & Yes & 0.32 & [0.13, 0.61] & 1.37 & 0.33 & 70 \\
Historical level $E_L$ & Yes & 0.36 & [0.21, 0.68] & 1.30 & 0.41 & 62 \\
Idiosyncratic level $I_L$ & No & 1.08 & [$-$3.19, 9.33] & 1.38 & 0.78 & 23 \\
Deviation $E_D$ & No & 1.23 & [1.03, 1.36] & 1.38 & 1.00 & 0 \\
Category deviation $C_D$ & No & 1.10 & [1.02, 1.16] & 1.27 & 1.00 & 0 \\
\midrule
\multicolumn{7}{@{}l}{\textit{RoBERTa across January 2020, relative to Llama-2}} \\
Firm identity $F$ & Yes & 1.18 & [$-$2.01, 5.08] & 0.95 & 0.89 & 11 \\
Historical level $E_L$ & Yes & 1.20 & [$-$0.27, 2.42] & 0.94 & 0.92 & 8 \\
Idiosyncratic level $I_L$ & No & 0.18 & [$-$3.96, 5.17] & 0.88 & 0.98 & 2 \\
Deviation $E_D$ & No & 0.99 & [0.85, 1.22] & 0.89 & 1.00 & 0 \\
Category deviation $C_D$ & No & 0.95 & [0.90, 1.05] & 0.91 & 1.00 & 0 \\
\bottomrule
\end{tabular}
\end{adjustbox}
}
\end{table}

The baseline calibration attributes 70\% of Llama-2's firm-identity predictive covariance and 62\% of its historical-level predictive covariance before January 2023 to memory (Table~\ref{tab:tab12}). The model-implied relative IC ratios are 0.33 for $F$ and 0.41 for $E_L$, compared with the targeted 0.32 and 0.36. RoBERTa's calibrated memory shares before its own cutoff are much smaller, at 11\% for $F$ and 8\% for $E_L$, although its cutoff comparisons have wide confidence intervals. Overall, the calibration attributes a much larger share of the predictability of firm identity and the historical level to memory for Llama-2 than for RoBERTa.

The calibration also speaks to the long-horizon strategies of Section~\ref{sec:long}. For Llama-2, memory accounts for 47\% of the model-implied first-month mean return of the MSRR strategy fitted to the historical level, averaged over its signal months, compared with 24\% for RoBERTa (Table~\ref{tab:D2}). This differs from the 62\% above, which measures memory's share of daily historical-level predictive covariance before January 2023. For the strategy fitted to the historical level 36 months earlier, the memory contribution rises to 66\% for Llama-2 and 40\% for RoBERTa. Lagging the historical level weakens its association with the evolving firm state but does not similarly reduce firm-level memory. Memory can therefore account for a substantial part of the predictability of Llama-2's historical level lagged 36 months, whose strategy earns an annualized first-month Sharpe ratio of 1.96 (Table~\ref{tab:tab10}). Persistent firm information also contributes: the baseline calibration implies a 36-month autocorrelation of the firm state of 0.67.

The deviation components provide a check on the assumption that memory-free rank ICs change by the same proportion for both encoders. In the model, the deviation and category deviation contain little or no memory, so under this assumption their relative IC ratios equal one. Their observed ratios are 1.23 and 1.10, respectively. If similar relative improvements in Llama-2's memory-free predictability occurred in $F$ and $E_L$, the calibrated memory shares would be conservative, understating rather than overstating the contribution of memory. The use of rank ICs removes differences in forecast scale, and the relative changes in component dispersions are similar across components, including those without memory (Table~\ref{tab:tab12}). As a further check, Figure~\ref{fig:D1} reports placebo comparisons at January splits from 2008 to 2016. The historical-level ratio across January 2023 is below every reported placebo ratio, and the firm-identity ratio is below all but one. These checks suggest that the decline across Llama-2's cutoff is unusual within the sample and that the calibrated memory shares are conservative if the deviations' departure from the common-change assumption extends to $F$ and $E_L$.

\begin{table}[htbp]
\centering
\caption{Model and Data Statistics outside the Targets}
\label{tab:tab13}
\begin{minipage}{\linewidth}\footnotesize
This table reports model and data values of statistics that the calibration does not target, for RoBERTa-large and Llama-2 (13B). The upper panel reports statistics that no targeted moment determines, and the lower panel reports statistics that the model's structure ties to targeted moments. Model values include memory, and the IC and Mincer--Zarnowitz rows approximate the forecast's variance along the memory direction by the mean memory component of its covariance with returns. The monthly-averaged daily forecast is the headline-weighted monthly average of a firm's daily ridge forecasts. $F+I_L$ is firm identity plus the idiosyncratic level. The month-1 $F+I_L$ and day-one rows are covariances with returns, multiplied by $10^6$. The lagged-level row divides the month-1 Sharpe ratio of an MSRR strategy on the historical level 36 months earlier, the mean embedding over news months $\tau-48$ to $\tau-37$ for signal month $\tau$, by that of the strategy on the current historical level. In the data, each strategy has its own coefficients fitted to next-month returns, trades the firms with news in month $\tau$ for which its input exists, and uses month-$\tau$ characteristics. Valuation is log price-to-book. Model SUE gains are population values and data gains out of sample, while model ICs are Pearson and data ICs rank correlations on firm-days with returns over the ten subsequent trading days. Norm.\ marks values matched by construction.
\end{minipage}
\medskip

{\footnotesize
\begin{adjustbox}{max width=\linewidth}
\begin{tabular}{lccccl}
\toprule
 & \multicolumn{2}{c}{Model} & \multicolumn{2}{c}{Data} & \\
\cmidrule(lr){2-3}\cmidrule(lr){4-5}
Statistic & RoBERTa & Llama-2 & RoBERTa & Llama-2 & Equation \\
\midrule
\multicolumn{6}{@{}l}{\textit{Not pinned by the targets}} \\
Monthly-averaged daily forecast, $F+I_L$, month 1 & 0.23 & 1.67 & 0.57 & 2.02 & \eqref{eq:D14}, \eqref{eq:D25} \\
Lagged historical level, month-1 Sharpe ratio over the current level's & 0.68 & 0.76 & 0.27 & 0.56 & \eqref{eq:D16}, \eqref{eq:D31} \\
SUE $R^2$ gain from the embedding (pp) & 1.8 & norm. & 0.8 & 3.2 & \eqref{eq:D7} \\
Firm-identity valuation $R^2$ from the state, Llama-2 over RoBERTa & \multicolumn{2}{c}{1.03} & \multicolumn{2}{c}{1.67} & \eqref{eq:D20} \\
\midrule
\multicolumn{6}{@{}l}{\textit{Model restrictions}} \\
Firm identity, day one & 0.007 & 0.066 & 0.057 & 0.189 & \eqref{eq:D32} \\
Monthly-averaged daily forecast, $F+I_L$, month 12 over month 1 & 0.89 & 0.95 & 0.33 & 0.82 & \eqref{eq:D14}, \eqref{eq:D25} \\
MSRR strategy, historical-level share, month 1 (\%) & 92.1 & 95.3 & 51.0 & 76.3 & \eqref{eq:D10}, \eqref{eq:D35}--\eqref{eq:D37} \\
Mincer--Zarnowitz slope & 0.86 & 0.89 & 0.70 & 0.66 & \eqref{eq:D9} \\
Embedding forecast IC (\%) & norm. & 2.73 & 2.41 & 2.66 & \eqref{eq:D8} \\
\bottomrule
\end{tabular}
\end{adjustbox}
}
\end{table}

The model fits the targeted moments well. All jointly targeted blocks have root mean squared $t$-statistics no greater than 1.14, with values of 0.23 and 0.26 for the two cutoff comparisons (Table~\ref{tab:D4}). The untargeted statistics in Table~\ref{tab:tab13} partly agree with the data, and the remaining differences show where the model's simplifying assumptions bind. First, the model overstates the historical level's share of monthly predictability. It assigns 92\% of RoBERTa's and 95\% of Llama-2's first-month MSRR returns to the historical level, compared with 51\% and 76\% in the data (Table~\ref{tab:tab8}). The higher model shares reflect its simplified treatment of event drift. It assumes a common, rapidly decaying response to the news signal, whereas drift among micro stocks persists for about 20 trading days. Second, its common firm-state process implies more persistence than RoBERTa's data show. For RoBERTa, the historical level lagged 36 months retains 27\% of the current level's Sharpe ratio in the data, compared with 68\% in the model. Third, memory in the model is lost only when return dates pass the cutoff, so it does not explain why Llama-2's advantage fades during 2021--2022. Finally, the model's state-only valuation equation implies a Llama-2-to-RoBERTa $R^2$ ratio of 1.03, compared with 1.67 in the data. This calculation excludes the possibility that embeddings contain memory of information already reflected in valuations, which may contribute to the difference.

The memory attribution is relatively stable across the alternative specifications in Table~\ref{tab:D5}. Llama-2's memory shares range from 67\% to 76\% for $F$ and from 61\% to 66\% for $E_L$, while RoBERTa's corresponding shares remain below half of Llama-2's in every specification. The implied autocorrelation of the firm state over 36 months also remains high, ranging from 0.58 to 0.71 against 0.67 in the baseline. Its decomposition into permanent and evolving components is less stable: the evolving state's half-life ranges from 11 to 67 months, and removing the permanent component leaves the fit nearly unchanged. The calibration thus accounts for key patterns of long-horizon predictability through two mechanisms: persistent firm information and Llama-2's memory of subsequently realized outcomes. The memory attribution remains conditional on the modeled mechanism and the assumptions underlying the cutoff comparisons.

\section{Conclusion}

We ask whether the predictive success of news embeddings reflects a better measure of news surprises or other sources of return predictability in text. At short horizons, most of their predictive power comes from event categories, simple labels of the news such as analyst downgrades. A forecast based on historical average returns after news in each event category predicts next-day returns better than the embedding forecasts, and event drift is concentrated in the first few days. Embeddings nevertheless contain information beyond event categories. They explain variation in earnings surprises beyond event labels, although return-trained forecasts retain little of this information. Their firm representations also explain valuations and analysts' growth forecasts beyond conventional characteristics.

At long horizons, news can predict returns either because prices absorb it slowly or because it reveals or describes persistent differences in expected returns, and the two are hard to separate from returns alone. The timing decomposition instead asks which information carries the predictability. We define the historical level as the average of the firm's monthly news embeddings over the preceding year, which separates information in earlier coverage from changes in the embedding during the current month. The deviation's predictive contribution falls sharply within the first two months, apart from a quarterly pattern, whereas the historical level remains predictive at longer horizons. After residualization on firm characteristics, the historical level, mainly firm identity, carries most of the return of DKPT's residualized-embedding strategy for Llama-2 and a substantial share for RoBERTa. Observed characteristics need not span the latent firm state that news repeatedly describes, so residualized embeddings can retain persistent firm information. Inputs built only from news more than three years old still predict returns, at long horizons for Llama-2 and in the first few months for RoBERTa. Much of the persistent predictability that DKPT attribute to slow incorporation of news shocks can thus be traced to the firm's earlier coverage rather than to the current month's news.

Pretraining memory complicates the interpretation of predictability from historical news. For ChronoGPT$_{1999}$, pretrained before our sample, event categories still dominate next-day predictability, and the historical level still contributes substantially to the residualized strategy's returns in the first months. Llama-2's advantage over RoBERTa, however, is concentrated in firm identity and the historical level, and its advantage in these components largely disappears by its pretraining cutoff, while its deviation shows no comparable decline. A calibration allowing for both persistent firm information and pretraining memory can account for this pattern, with memory accounting for more than half of Llama-2's pre-cutoff predictive covariance from firm identity and the historical level but a much smaller share of RoBERTa's. Persistent firm information and pretraining memory can generate similar patterns of return predictability before the cutoff, and aggregate performance can mask a decline in predictability concentrated in particular components. Look-ahead bias is a particular concern when the predictive advantage of LLMs with stronger representation capabilities is concentrated in these components. More generally, describing news more fully is not the same as measuring news surprises, and persistent return predictability from news text does not by itself establish slow incorporation of news into prices.

\clearpage
\setlength\bibsep{0pt}
\bibliographystyle{econ}
\bibliography{finllm}

\newpage
\begin{appendices}

\renewcommand{\theequation}{\thesection.\arabic{equation}}
\renewcommand{\thefigure}{\thesection.\arabic{figure}}
\renewcommand{\thetable}{\thesection.\arabic{table}}

\counterwithin{equation}{section}
\counterwithin{figure}{section}
\counterwithin{table}{section}

\section{Additional Tables and Figures}

\begin{table}[htbp]
\centering
\caption{Returns around News by Event Category}
\label{tab:tabA_drift}
\begin{minipage}{\linewidth}\footnotesize
This table reports returns around news for categories in the displayed event groups, 2004--2025. Jump is the return to the first eligible open, from the preceding close or, for news published during the trading session (09:30--16:00), from that session's open. Day 1 is the subsequent open-to-open return; Days 1--10 sums ten such returns. Returns are in basis points, in excess of the mean return of all headlines on the same trading day, and averaged equally across trading days. The sample includes headlines with an observed Jump and a complete ten-day return path. The Headlines column counts each category's headlines, and the table shows categories with at least 20,000 of them. Panel B reports decile long-short portfolios sorted on category and embedding forecasts. Newey--West $t$-statistics use 5 lags for Jump and Day 1 and 10 for Days 1--10.
\end{minipage}
\medskip

{\footnotesize
\textit{Panel A: Event categories}\\[2pt]
\begin{adjustbox}{max width=\linewidth, max totalheight=0.47\textheight}
\begin{tabular}{llccccccc}
\toprule
\multicolumn{3}{c}{} & \multicolumn{2}{c}{Jump} & \multicolumn{2}{c}{Day 1} & \multicolumn{2}{c}{Days 1--10} \\
\cmidrule(lr){4-5} \cmidrule(lr){6-7} \cmidrule(lr){8-9}
Group & Category & Headlines & bp & $t$ & bp & $t$ & bp & $t$ \\
\midrule
analyst-ratings & analyst-ratings-change-positive & \phantom{0}67,738 & \phantom{$-$}231.4 & \phantom{$-$}70.18 & \phantom{$-$}21.6 & \phantom{$-$}12.13 & \phantom{$-$}26.5 & \phantom{$-$}5.32 \\
 & analyst-ratings-change-neutral & 167,439 & \phantom{00}$-$5.8 & \phantom{0}$-$1.05 & \phantom{$-$0}5.4 & \phantom{$-$0}1.04 & \phantom{$-$0}8.8 & \phantom{$-$}0.86 \\
 & analyst-ratings-change-negative & \phantom{0}75,990 & $-$291.2 & $-$56.43 & $-$25.4 & $-$11.71 & $-$22.3 & $-$3.77 \\
 & analyst-ratings-set-positive & \phantom{0}44,689 & \phantom{$-$}111.9 & \phantom{$-$}44.37 & \phantom{$-$}23.4 & \phantom{$-$0}9.20 & \phantom{$-$}28.8 & \phantom{$-$}3.31 \\
 & analyst-ratings-set-neutral & \phantom{0}27,064 & \phantom{0}$-$32.6 & $-$15.47 & $-$23.7 & \phantom{0}$-$8.81 & $-$50.0 & $-$5.81 \\
\midrule
dividends & dividend-up & \phantom{0}31,539 & \phantom{$-$0}35.7 & \phantom{$-$}13.18 & \phantom{$-$}21.4 & \phantom{$-$0}7.35 & \phantom{$-$}27.6 & \phantom{$-$}3.35 \\
 & dividend & \phantom{0}76,873 & \phantom{$-$00}9.5 & \phantom{$-$0}6.31 & \phantom{$-$0}7.8 & \phantom{$-$0}5.67 & \phantom{$-$0}6.0 & \phantom{$-$}1.11 \\
\midrule
earnings & earnings-positive & 123,247 & \phantom{$-$0}36.7 & \phantom{$-$0}8.12 & \phantom{0}$-$1.5 & \phantom{0}$-$0.41 & \phantom{$-$0}3.3 & \phantom{$-$}0.44 \\
 & earnings-up & \phantom{0}80,584 & \phantom{$-$0}92.7 & \phantom{$-$}18.13 & $-$13.4 & \phantom{0}$-$3.40 & $-$19.5 & $-$2.38 \\
 & earnings & 252,165 & \phantom{0}$-$13.0 & \phantom{0}$-$5.91 & \phantom{$-$0}2.4 & \phantom{$-$0}1.37 & \phantom{$-$0}1.3 & \phantom{$-$}0.31 \\
 & earnings-down & \phantom{0}36,012 & $-$136.4 & $-$20.07 & \phantom{0}$-$5.5 & \phantom{0}$-$1.06 & \phantom{$-$0}2.1 & \phantom{$-$}0.19 \\
 & earnings-negative & \phantom{0}49,157 & $-$104.9 & $-$14.59 & $-$39.5 & \phantom{0}$-$6.22 & $-$49.7 & $-$3.45 \\
 & earnings-per-share-positive & 188,193 & \phantom{$-$0}27.8 & \phantom{$-$0}6.80 & \phantom{0}$-$1.3 & \phantom{0}$-$0.40 & \phantom{$-$0}1.3 & \phantom{$-$}0.20 \\
 & earnings-per-share-up & \phantom{0}35,665 & \phantom{$-$0}84.5 & \phantom{$-$}12.59 & $-$10.1 & \phantom{0}$-$1.84 & \phantom{$-$0}8.0 & \phantom{$-$}0.77 \\
 & earnings-per-share-negative & \phantom{0}49,728 & $-$121.9 & $-$15.89 & $-$30.3 & \phantom{0}$-$4.45 & $-$50.9 & $-$3.56 \\
 & earnings-per-share-guidance & \phantom{0}83,767 & \phantom{0}$-$41.2 & \phantom{0}$-$7.43 & \phantom{$-$0}6.5 & \phantom{$-$0}1.70 & \phantom{$-$}14.9 & \phantom{$-$}2.15 \\
\midrule
insider-trading & insider-buy & 328,262 & \phantom{$-$0}24.7 & \phantom{$-$}20.72 & \phantom{$-$}19.8 & \phantom{$-$}17.21 & \phantom{$-$}30.7 & \phantom{$-$}8.05 \\
 & insider-sell & 477,205 & \phantom{0}$-$14.4 & $-$13.66 & \phantom{0}$-$8.2 & \phantom{0}$-$8.51 & $-$15.8 & $-$3.45 \\
 & insider-gift & \phantom{0}41,463 & \phantom{00}$-$7.8 & \phantom{0}$-$5.04 & \phantom{0}$-$0.9 & \phantom{0}$-$0.52 & \phantom{$-$0}0.9 & \phantom{$-$}0.12 \\
 & insider-sell-registration & 290,177 & \phantom{0}$-$10.0 & \phantom{0}$-$7.78 & \phantom{0}$-$7.2 & \phantom{0}$-$6.61 & $-$15.9 & $-$3.12 \\
 & insider-surrender & 254,101 & \phantom{0}$-$12.0 & $-$10.05 & \phantom{0}$-$4.6 & \phantom{0}$-$3.09 & \phantom{0}$-$1.9 & $-$0.38 \\
\midrule
labor-issues & board-member-appointment & \phantom{0}32,496 & \phantom{00}$-$3.8 & \phantom{0}$-$1.68 & \phantom{$-$0}3.8 & \phantom{$-$0}1.48 & \phantom{0}$-$7.0 & $-$0.88 \\
 & executive-appointment & 104,700 & \phantom{0}$-$12.7 & \phantom{0}$-$7.09 & \phantom{0}$-$7.2 & \phantom{0}$-$4.47 & $-$22.0 & $-$4.89 \\
\midrule
price-targets & price-target-upgrade & 123,369 & \phantom{$-$0}97.9 & \phantom{$-$}21.29 & \phantom{$-$0}8.1 & \phantom{$-$0}2.81 & \phantom{$-$}20.3 & \phantom{$-$}2.12 \\
 & price-target-downgrade & \phantom{0}89,592 & $-$180.1 & $-$23.66 & $-$16.5 & \phantom{0}$-$3.86 & $-$28.2 & $-$2.39 \\
 & price-target-set & \phantom{0}35,022 & \phantom{$-$0}66.6 & \phantom{$-$}15.75 & \phantom{$-$0}3.8 & \phantom{$-$0}1.28 & \phantom{$-$0}7.0 & \phantom{$-$}0.65 \\
\midrule
revenues & revenue-up & \phantom{0}76,225 & \phantom{$-$0}39.0 & \phantom{$-$0}8.34 & $-$15.2 & \phantom{0}$-$4.08 & $-$27.0 & $-$3.50 \\
 & revenue-down & \phantom{0}20,020 & $-$101.8 & $-$15.62 & $-$13.6 & \phantom{0}$-$2.49 & $-$14.9 & $-$1.30 \\
 & revenue-guidance-up & \phantom{0}25,371 & \phantom{$-$0}89.8 & \phantom{$-$}13.02 & \phantom{$-$0}7.1 & \phantom{$-$0}1.49 & \phantom{0}$-$9.3 & $-$1.00 \\
 & revenue-guidance & \phantom{0}67,881 & \phantom{0}$-$59.6 & \phantom{0}$-$7.92 & \phantom{$-$0}3.3 & \phantom{$-$0}0.70 & \phantom{$-$0}2.4 & \phantom{$-$}0.30 \\
 & revenues & 170,535 & \phantom{0}$-$12.9 & \phantom{0}$-$3.07 & $-$12.9 & \phantom{0}$-$3.96 & $-$14.0 & $-$2.10 \\
\bottomrule
\end{tabular}
\end{adjustbox}
\par\medskip
\textit{Panel B: Decile long-short portfolios}\\[2pt]
\begin{adjustbox}{max width=\linewidth, max totalheight=0.08\textheight}
\begin{tabular}{lcccccc}
\toprule
 & \multicolumn{2}{c}{Jump} & \multicolumn{2}{c}{Day 1} & \multicolumn{2}{c}{Days 1--10} \\
\cmidrule(lr){2-3} \cmidrule(lr){4-5} \cmidrule(lr){6-7}
Signal & bp & $t$ & bp & $t$ & bp & $t$ \\
\midrule
Category mean (8-yr) & 107.4 & 40.95 & 37.0 & 27.43 & 49.7 & 10.19 \\
RoBERTa forecast & \phantom{0}76.9 & 45.65 & 27.0 & 21.71 & 37.8 & \phantom{0}6.48 \\
Llama-2 forecast & \phantom{0}68.4 & 34.69 & 30.1 & 21.55 & 52.4 & \phantom{0}7.98 \\
\bottomrule
\end{tabular}
\end{adjustbox}
}
\end{table}

\begin{table}[htbp]
\centering
\caption{Signal Performance by Subsample}
\label{tab:tabA_subsamples}
\begin{minipage}{\linewidth}\footnotesize
This table reports category-mean, embedding, and RavenPack event sentiment score (ESS) forecasts of next open-to-open returns across subsamples, 2004--2025. Micro, small, and big stocks are below the NYSE 20th capitalization percentile, between it and the NYSE median, and at or above the median, on the preceding trading day. Price-derived news comprises stock prices, technical analysis, order imbalances, indexes, and stock picks. Headline-only and full-article are source formats; all forecasts use headlines. Headlines (\%) is the subsample's percentage of all headlines. Panel B reports full-sample differences from the eight-year category mean; DM $t$ and JK $z$ test the rank-IC and Sharpe-ratio differences. Rank IC is the mean daily Spearman correlation with next open-to-open returns, in percent. SR is the annualized Sharpe ratio of equal-weighted (EW) and value-weighted (VW) decile long-short portfolios.
\end{minipage}
\medskip

{\footnotesize
\textit{Panel A: By subsample}\\[2pt]
\begin{adjustbox}{max width=\linewidth, max totalheight=0.45\textheight}
\begin{tabular}{lcccccc}
\toprule
\multicolumn{2}{c}{} & \multicolumn{2}{c}{Category mean} & \multicolumn{2}{c}{Forecast} & RavenPack \\
\cmidrule(lr){3-4} \cmidrule(lr){5-6}
Sample & Headlines (\%) & 8-yr & 1-yr & RoBERTa & Llama-2 & ESS \\
\midrule
\multicolumn{7}{l}{\textit{Rank IC (\%)}} \\
All firm-days & 100.0 & 3.33 & 2.94 & 2.41 & 2.67 & 2.63 \\
Micro & \phantom{0}17.6 & 5.47 & 4.56 & 4.31 & 4.57 & 4.70 \\
Small & \phantom{0}27.9 & 4.05 & 3.60 & 2.60 & 2.87 & 3.21 \\
Big & \phantom{0}54.5 & 2.17 & 1.95 & 1.44 & 1.62 & 1.55 \\
Without insider trading & \phantom{0}74.9 & 3.23 & 2.66 & 2.20 & 2.57 & 2.10 \\
Without price-derived news & \phantom{0}86.4 & 3.98 & 3.46 & 2.81 & 3.11 & 3.12 \\
Headline-only news & \phantom{0}39.4 & 4.73 & 4.11 & 3.24 & 3.73 & 3.75 \\
Full-article news & \phantom{0}23.8 & 3.45 & 3.30 & 3.01 & 3.39 & 3.16 \\
\midrule
\multicolumn{7}{l}{\textit{EW SR}} \\
All firm-days & 100.0 & 6.35 & 6.16 & 4.92 & 4.97 & 5.02 \\
Micro & \phantom{0}17.6 & 3.86 & 4.13 & 3.30 & 3.18 & 3.66 \\
Small & \phantom{0}27.9 & 4.14 & 3.80 & 3.09 & 3.02 & 3.40 \\
Big & \phantom{0}54.5 & 3.21 & 3.05 & 2.41 & 2.85 & 2.75 \\
Without insider trading & \phantom{0}74.9 & 5.47 & 5.16 & 3.73 & 3.95 & 4.19 \\
Without price-derived news & \phantom{0}86.4 & 6.61 & 6.20 & 5.06 & 4.97 & 5.01 \\
Headline-only news & \phantom{0}39.4 & 4.77 & 4.29 & 3.57 & 3.98 & 4.25 \\
Full-article news & \phantom{0}23.8 & 2.53 & 1.79 & 3.11 & 2.90 & 1.15 \\
\midrule
\multicolumn{7}{l}{\textit{VW SR}} \\
All firm-days & 100.0 & 2.30 & 2.18 & 1.21 & 1.90 & 1.72 \\
Micro & \phantom{0}17.6 & 3.80 & 4.20 & 3.33 & 3.04 & 3.68 \\
Small & \phantom{0}27.9 & 3.79 & 3.53 & 2.77 & 2.73 & 3.14 \\
Big & \phantom{0}54.5 & 2.20 & 2.03 & 0.98 & 1.96 & 1.59 \\
Without insider trading & \phantom{0}74.9 & 2.24 & 2.07 & 1.05 & 1.68 & 1.76 \\
Without price-derived news & \phantom{0}86.4 & 2.62 & 2.45 & 1.34 & 2.08 & 2.05 \\
Headline-only news & \phantom{0}39.4 & 2.97 & 2.36 & 1.41 & 1.74 & 2.25 \\
Full-article news & \phantom{0}23.8 & 1.30 & 0.97 & 1.25 & 1.02 & 0.41 \\
\bottomrule
\end{tabular}
\end{adjustbox}
\par\medskip
\textit{Panel B: Differences from the 8-year category mean, all firm-days}\\[2pt]
\begin{adjustbox}{max width=\linewidth, max totalheight=0.10\textheight}
\begin{tabular}{lcccccc}
\toprule
 & \multicolumn{2}{c}{Rank IC (\%)} & \multicolumn{2}{c}{EW SR} & \multicolumn{2}{c}{VW SR} \\
\cmidrule(lr){2-3} \cmidrule(lr){4-5} \cmidrule(lr){6-7}
Signal & $\Delta$ & DM $t$ & $\Delta$ & JK $z$ & $\Delta$ & JK $z$ \\
\midrule
Category mean (1-yr) & $-$0.39 & \phantom{0}$-$5.85 & $-$0.19 & $-$0.96 & $-$0.11 & $-$0.49 \\
RoBERTa forecast & $-$0.91 & $-$10.54 & $-$1.43 & $-$5.78 & $-$1.09 & $-$4.18 \\
Llama-2 forecast & $-$0.65 & \phantom{0}$-$7.33 & $-$1.38 & $-$5.66 & $-$0.40 & $-$1.47 \\
RavenPack ESS & $-$0.70 & \phantom{0}$-$9.12 & $-$1.34 & $-$5.34 & $-$0.58 & $-$2.23 \\
\bottomrule
\end{tabular}
\end{adjustbox}
}
\end{table}

\begin{table}[htbp]
\centering
\caption{Portfolio Sharpe Ratios before and after Trading Costs}
\label{tab:tabA_costs}
\begin{minipage}{\linewidth}\footnotesize
This table reports annualized Sharpe ratios before and after trading costs of daily equal- and value-weighted decile long-short portfolios held from open to open, 2004--2025. Signals are the one- and eight-year category means, the embedding ridge forecasts, RavenPack's event sentiment score (ESS), Loughran--McDonald sentiment (LMMD), and eight-year means by RavenPack event group, RavenPack type, headline direction from a fixed word rule, or group by direction. Rebalancing follows $w_t=(1-\gamma)\tilde w_{t-1}+\gamma w_t^*$, where $\tilde w_{t-1}$ is the drifted holding and $w_t^*$ the target. Costs per 100\% turnover are 10 basis points for stocks at or above the NYSE 20th capitalization percentile and 20 basis points below it. Turnover is half the sum of absolute changes in the long-short weights; holdings drift with each stock's return relative to the long-short return.
\end{minipage}
\medskip

{\footnotesize
\textit{Panel A: Gross Sharpe ratio}\\[2pt]
\begin{adjustbox}{max width=\linewidth, max totalheight=0.28\textheight}
\begin{tabular}{lcccccccccc}
\toprule
 & \multicolumn{5}{c}{Equal-weighted} & \multicolumn{5}{c}{Value-weighted} \\
\cmidrule(lr){2-6} \cmidrule(lr){7-11}
Signal & $\gamma=0.1$ & $\gamma=0.2$ & $\gamma=0.3$ & $\gamma=0.5$ & $\gamma=1$ & $\gamma=0.1$ & $\gamma=0.2$ & $\gamma=0.3$ & $\gamma=0.5$ & $\gamma=1$ \\
\midrule
Category mean (8-yr) & 2.44 & 3.74 & 4.65 & 5.74 & 6.35 & \phantom{$-$}0.73 & \phantom{$-$}1.38 & \phantom{$-$}1.80 & 2.25 & 2.30 \\
Category mean (1-yr) & 2.59 & 3.78 & 4.59 & 5.57 & 6.16 & \phantom{$-$}0.70 & \phantom{$-$}1.22 & \phantom{$-$}1.59 & 2.04 & 2.18 \\
RoBERTa forecast & 1.34 & 2.27 & 2.98 & 3.96 & 4.92 & \phantom{$-$}0.13 & \phantom{$-$}0.45 & \phantom{$-$}0.68 & 0.96 & 1.21 \\
Llama-2 forecast & 1.66 & 2.53 & 3.18 & 4.09 & 4.97 & \phantom{$-$}0.77 & \phantom{$-$}1.13 & \phantom{$-$}1.39 & 1.71 & 1.90 \\
RavenPack ESS & 1.78 & 2.78 & 3.48 & 4.38 & 5.02 & \phantom{$-$}0.13 & \phantom{$-$}0.50 & \phantom{$-$}0.85 & 1.34 & 1.72 \\
Event-group mean & 0.59 & 1.08 & 1.42 & 1.82 & 2.16 & $-$0.28 & $-$0.18 & $-$0.08 & 0.06 & 0.23 \\
Direction mean & 1.71 & 2.51 & 3.08 & 3.87 & 4.56 & \phantom{$-$}0.50 & \phantom{$-$}0.85 & \phantom{$-$}1.11 & 1.41 & 1.60 \\
RavenPack-type mean & 1.75 & 2.72 & 3.40 & 4.25 & 4.83 & $-$0.03 & \phantom{$-$}0.28 & \phantom{$-$}0.51 & 0.79 & 0.99 \\
Group $\times$ direction mean & 1.99 & 3.32 & 4.28 & 5.48 & 6.36 & \phantom{$-$}0.18 & \phantom{$-$}0.70 & \phantom{$-$}1.11 & 1.65 & 2.04 \\
LMMD sentiment & 0.39 & 0.89 & 1.23 & 1.69 & 2.11 & \phantom{$-$}0.16 & \phantom{$-$}0.29 & \phantom{$-$}0.36 & 0.40 & 0.40 \\
\bottomrule
\end{tabular}
\end{adjustbox}
\par\medskip
\textit{Panel B: Net Sharpe ratio}\\[2pt]
\begin{adjustbox}{max width=\linewidth, max totalheight=0.28\textheight}
\begin{tabular}{lcccccccccc}
\toprule
 & \multicolumn{5}{c}{Equal-weighted} & \multicolumn{5}{c}{Value-weighted} \\
\cmidrule(lr){2-6} \cmidrule(lr){7-11}
Signal & $\gamma=0.1$ & $\gamma=0.2$ & $\gamma=0.3$ & $\gamma=0.5$ & $\gamma=1$ & $\gamma=0.1$ & $\gamma=0.2$ & $\gamma=0.3$ & $\gamma=0.5$ & $\gamma=1$ \\
\midrule
Category mean (8-yr) & \phantom{$-$}1.49 & \phantom{$-$}2.09 & \phantom{$-$}2.46 & \phantom{$-$}2.79 & \phantom{$-$}2.50 & \phantom{$-$}0.01 & \phantom{$-$}0.10 & \phantom{$-$}0.10 & $-$0.07 & $-$0.72 \\
Category mean (1-yr) & \phantom{$-$}1.42 & \phantom{$-$}1.87 & \phantom{$-$}2.12 & \phantom{$-$}2.29 & \phantom{$-$}1.92 & $-$0.12 & $-$0.15 & $-$0.19 & $-$0.34 & $-$0.84 \\
RoBERTa forecast & \phantom{$-$}0.56 & \phantom{$-$}0.82 & \phantom{$-$}0.95 & \phantom{$-$}1.04 & \phantom{$-$}0.79 & $-$0.42 & $-$0.56 & $-$0.72 & $-$1.04 & $-$1.58 \\
Llama-2 forecast & \phantom{$-$}0.95 & \phantom{$-$}1.23 & \phantom{$-$}1.38 & \phantom{$-$}1.48 & \phantom{$-$}1.24 & \phantom{$-$}0.20 & \phantom{$-$}0.12 & \phantom{$-$}0.02 & $-$0.20 & $-$0.70 \\
RavenPack ESS & \phantom{$-$}0.69 & \phantom{$-$}0.94 & \phantom{$-$}1.05 & \phantom{$-$}1.10 & \phantom{$-$}0.75 & $-$0.75 & $-$0.94 & $-$1.04 & $-$1.25 & $-$1.66 \\
Event-group mean & $-$0.35 & $-$0.60 & $-$0.90 & $-$1.44 & $-$2.33 & $-$0.82 & $-$1.21 & $-$1.53 & $-$2.06 & $-$2.81 \\
Direction mean & \phantom{$-$}0.66 & \phantom{$-$}0.68 & \phantom{$-$}0.62 & \phantom{$-$}0.46 & $-$0.04 & $-$0.37 & $-$0.58 & $-$0.77 & $-$1.13 & $-$1.68 \\
RavenPack-type mean & \phantom{$-$}0.83 & \phantom{$-$}1.06 & \phantom{$-$}1.13 & \phantom{$-$}1.09 & \phantom{$-$}0.59 & $-$0.73 & $-$0.96 & $-$1.17 & $-$1.51 & $-$2.07 \\
Group $\times$ direction mean & \phantom{$-$}1.09 & \phantom{$-$}1.73 & \phantom{$-$}2.11 & \phantom{$-$}2.47 & \phantom{$-$}2.27 & $-$0.49 & $-$0.49 & $-$0.49 & $-$0.57 & $-$0.97 \\
LMMD sentiment & $-$0.37 & $-$0.47 & $-$0.60 & $-$0.86 & $-$1.42 & $-$0.52 & $-$0.90 & $-$1.25 & $-$1.84 & $-$2.63 \\
\bottomrule
\end{tabular}
\end{adjustbox}
}
\end{table}

\begin{table}[htbp]
\centering
\caption{Category Means over Alternative Estimation Windows}
\label{tab:tabA_windows}
\begin{minipage}{\linewidth}\footnotesize
This table reports category forecasts of next open-to-open returns from category means over the one, three, or eight years preceding test year $T$, an expanding window from 2000 through $T-1$, or an expanding window through $T-3$, on the same firm-days over 2004--2025. Category means shrink toward the mean return of their event group with weight $n/(n+200)$ on the category mean, where $n$ is the category's headline count in the window; a category absent from the window takes its group's mean. Rank IC is the mean daily Spearman correlation with next open-to-open returns, in percent. L--S is the mean daily decile long-short return in basis points. SR is the annualized Sharpe ratio of equal-weighted (EW) and value-weighted (VW) decile long-short portfolios.
\end{minipage}
\medskip

{\footnotesize
\begin{adjustbox}{max width=\linewidth, max totalheight=0.75\textheight}
\begin{tabular}{lcccccc}
\toprule
Window & Rank IC (\%) & $t$ & EW L--S (bp) & EW SR & VW L--S (bp) & VW SR \\
\midrule
Category mean (1-yr) & 2.94 & 33.36 & 32.9 & 6.16 & 13.7 & 2.18 \\
Category mean (3-yr) & 3.22 & 37.61 & 35.3 & 6.20 & 14.2 & 2.32 \\
Category mean (8-yr) & 3.33 & 37.55 & 37.2 & 6.35 & 14.4 & 2.30 \\
Category mean (expanding) & 3.34 & 36.81 & 38.8 & 6.76 & 16.5 & 2.73 \\
Category mean (through $T-3$) & 3.25 & 35.76 & 36.3 & 6.34 & 14.9 & 2.36 \\
\bottomrule
\end{tabular}
\end{adjustbox}
}
\end{table}

\begin{table}[htbp]
\centering
\caption{Mincer--Zarnowitz Slopes and Correlation with Category Benchmarks}
\label{tab:tabA_mz}
\begin{minipage}{\linewidth}\footnotesize
This table reports Mincer--Zarnowitz slopes and category-related variation in forecasts, 2004--2025. Slopes come from daily regressions of next open-to-open returns on forecasts in return units; $t(b=1)$ tests a unit slope. No slope is reported for RavenPack's event sentiment score, which is not a return forecast. Variance explained by category is the headline-level $R^2$ on category indicators, in percent. The final columns report firm-day correlations with the eight-year category mean and with the eight-year mean by RavenPack event group and headline direction.
\end{minipage}
\medskip

{\footnotesize
\begin{adjustbox}{max width=\linewidth, max totalheight=0.75\textheight}
\begin{tabular}{lccccc}
\toprule
 & \multicolumn{2}{c}{Mincer--Zarnowitz} & Var. explained & \multicolumn{2}{c}{Correlation with} \\
\cmidrule(lr){2-3} \cmidrule(lr){5-6}
Signal & Slope $b$ & $t(b=1)$ & by category (\%) & Category & Group $\times$ dir. \\
\midrule
RoBERTa forecast & 0.70 & $-$8.76 & 22.87 & 0.42 & 0.38 \\
Llama-2 forecast & 0.66 & $-$9.48 & 24.26 & 0.45 & 0.40 \\
Category mean (8-yr) & 0.78 & $-$8.49 &  &  & 0.73 \\
Group $\times$ direction mean & 0.76 & $-$9.65 & 50.86 & 0.73 &  \\
RavenPack ESS &  &  & 97.23 & 0.56 & 0.51 \\
\bottomrule
\end{tabular}
\end{adjustbox}
}
\end{table}

\begin{table}[htbp]
\centering
\caption{Spanning Regressions}
\label{tab:tabA_spanning}
\begin{minipage}{\linewidth}\footnotesize
This table regresses daily decile long-short returns on the indicated benchmark portfolio, alone or with the Fama--French five factors and the momentum, short-term reversal, and long-term reversal factors, 2004--2025. Category mean is the eight-year category mean. Test and benchmark portfolios use the same weighting, equal (EW) or value (VW). Benchmark-only regressions use open-to-open returns; factor regressions follow CKX in valuing both portfolios at the close to match factor timing while keeping the trading rule. $\beta$ is the benchmark loading, $\alpha$ is in basis points per day, and Appraisal is annualized alpha divided by residual standard deviation. Newey--West $t$-statistics use 5 lags.
\end{minipage}
\medskip

{\footnotesize
\begin{adjustbox}{max width=\linewidth, max totalheight=0.75\textheight}
\begin{tabular}{lllcccccccccc}
\toprule
\multicolumn{3}{c}{} & \multicolumn{5}{c}{Benchmark only} & \multicolumn{5}{c}{Benchmark + factors} \\
\cmidrule(lr){4-8} \cmidrule(lr){9-13}
Weighting & Test portfolio & Benchmark & $\beta$ & $\alpha$ (bp) & $t$ & $R^2$ & Appraisal & $\beta$ & $\alpha$ (bp) & $t$ & $R^2$ & Appraisal \\
\midrule
EW & RoBERTa & Category mean & 0.36 & 13.7 & 10.23 & 0.15 & 2.69 & 0.40 & 11.9 & 10.22 & 0.28 & 2.53 \\
\cmidrule(lr){2-13}
 & Llama-2 & Category mean & 0.43 & 14.9 & 10.27 & 0.16 & 2.64 & 0.45 & 13.6 & \phantom{0}9.42 & 0.27 & 2.53 \\
\cmidrule(lr){2-13}
 & Category mean & RoBERTa & 0.41 & 26.1 & 18.17 & 0.15 & 4.83 & 0.50 & 23.7 & 15.14 & 0.21 & 4.49 \\
 &  & Llama-2 & 0.38 & 25.5 & 17.37 & 0.16 & 4.75 & 0.43 & 23.9 & 15.20 & 0.20 & 4.52 \\
\midrule
VW & RoBERTa & Category mean & 0.28 & \phantom{0}4.0 & \phantom{0}2.92 & 0.07 & 0.63 & 0.25 & \phantom{0}4.3 & \phantom{0}3.21 & 0.09 & 0.69 \\
\cmidrule(lr){2-13}
 & Llama-2 & Category mean & 0.25 & 10.0 & \phantom{0}6.36 & 0.05 & 1.43 & 0.23 & 10.2 & \phantom{0}6.73 & 0.09 & 1.52 \\
\cmidrule(lr){2-13}
 & Category mean & RoBERTa & 0.25 & 12.4 & \phantom{0}9.35 & 0.07 & 2.05 & 0.22 & 12.8 & \phantom{0}9.64 & 0.06 & 2.17 \\
 &  & Llama-2 & 0.19 & 11.8 & \phantom{0}9.08 & 0.05 & 1.93 & 0.18 & 12.1 & \phantom{0}9.34 & 0.05 & 2.03 \\
\bottomrule
\end{tabular}
\end{adjustbox}
}
\end{table}

\begin{table}[htbp]
\centering
\caption{Forecast and Embedding Changes under Headline Rewrites}
\label{tab:tabA_rewrites}
\begin{minipage}{\linewidth}\footnotesize
This table reports embedding and forecast changes under rewrites of earnings headlines, with the ridge coefficients fixed, 2004--2025. Rewrites multiply or divide reported numbers by ten, reverse an unambiguous directional expression, swap the firm name for that of the firm closest in capitalization in the same six-digit GICS industry, or remove the name and ticker. Forecast changes are in standard deviations of the test year's headline-level forecasts. Signed changes are positive when the forecast moves in the direction the rewrite implies, and Correct sign is the percentage moving that way. Both use only magnitude rewrites and direction flips of headlines whose wording has an unambiguous direction; Absolute forecast changes use all rewrites of each variant. Rank changes are the mean percentile-point change of the firm-day forecast (Absolute) and the percentage switching deciles (Decile). Cosine compares original and rewritten embeddings. In subspace is the squared projection of the normalized mean embedding change onto the leading 32 directions of the training-window category centroids or firm effects, the two magnitude rewrites sharing one pooled direction; random-direction benchmarks are 3.1\% for RoBERTa-large and 0.6\% for Llama-2 (13B).
\end{minipage}
\medskip

{\footnotesize
\textit{Panel A: Forecast and embedding changes}\\[2pt]
\begin{adjustbox}{max width=\linewidth, max totalheight=0.34\textheight}
\begin{tabular}{lccccccccc}
\toprule
\multicolumn{2}{c}{} & \multicolumn{3}{c}{Forecast change (sd)} & \multicolumn{2}{c}{Rank change} &  & \multicolumn{2}{c}{In subspace (\%)} \\
\cmidrule(lr){3-5} \cmidrule(lr){6-7} \cmidrule(lr){9-10}
Rewrite & Rewrites & Signed & Absolute & Correct sign (\%) & Absolute & Decile (\%) & Cosine & Category & Firm \\
\midrule
\multicolumn{10}{l}{\textit{RoBERTa}} \\
magnitude x10 & 539,436 & 0.13 & 0.30 & 66.9 & \phantom{0}1.7 & 16.9 & 0.999 & 19.7 & 13.9 \\
magnitude /10 & 539,436 & 0.23 & 0.36 & 73.0 & \phantom{0}2.2 & 20.5 & 0.999 & 19.7 & 13.9 \\
direction flipped & 466,798 & 1.79 & 1.77 & 99.8 & 10.5 & 71.5 & 0.999 & 74.4 & 16.9 \\
name swapped & 540,649 &  & 0.53 &  & \phantom{0}3.1 & 28.1 & 0.998 & 25.2 & 18.4 \\
name removed & 541,008 &  & 0.64 &  & \phantom{0}3.7 & 32.5 & 0.998 & 33.3 & 17.2 \\
\midrule
\multicolumn{10}{l}{\textit{Llama-2}} \\
magnitude x10 & 539,436 & 0.00 & 0.14 & 51.2 & \phantom{0}0.7 & \phantom{0}6.6 & 0.993 & 11.1 & \phantom{0}8.0 \\
magnitude /10 & 539,436 & 0.09 & 0.16 & 71.4 & \phantom{0}0.8 & \phantom{0}7.3 & 0.992 & 11.1 & \phantom{0}8.0 \\
direction flipped & 466,798 & 2.03 & 2.03 & 99.7 & 10.4 & 64.8 & 0.959 & 74.5 & 17.6 \\
name swapped & 540,649 &  & 0.74 &  & \phantom{0}3.5 & 29.3 & 0.916 & 38.0 & 39.6 \\
name removed & 541,008 &  & 0.77 &  & \phantom{0}3.4 & 28.7 & 0.874 & 38.4 & 46.9 \\
\bottomrule
\end{tabular}
\end{adjustbox}
\par\medskip
\textit{Panel B: Examples}\\[2pt]
\begin{adjustbox}{max width=\linewidth, max totalheight=0.21\textheight}
\begin{tabular}{lcc}
\toprule
Rewrite & Profit headline & Loss headline \\
\midrule
original & Microsoft 2Q EPS 14c \textgreater{}MSFT & Tesla 4Q Loss/Shr 69c \textgreater{}TSLA \\
magnitude x10 & Microsoft 2Q EPS 140c \textgreater{}MSFT & Tesla 4Q Loss/Shr 690c \textgreater{}TSLA \\
magnitude /10 & Microsoft 2Q EPS 1.4c \textgreater{}MSFT & Tesla 4Q Loss/Shr 6.9c \textgreater{}TSLA \\
direction flipped & Microsoft 2Q Loss/Shr 14c \textgreater{}MSFT & Tesla 4Q EPS 69c \textgreater{}TSLA \\
name swapped & Oracle 2Q EPS 14c & Ford 4Q Loss/Shr 69c \\
name removed & 2Q EPS 14c & 4Q Loss/Shr 69c \\
\bottomrule
\end{tabular}
\end{adjustbox}
}
\end{table}

\begin{table}[htbp]
\centering
\caption{Centroid Decomposition of the Embedding Forecast}
\label{tab:tabA_centroid}
\begin{minipage}{\linewidth}\footnotesize
This table decomposes the forecast of next open-to-open returns as $\hat y-\bar y=\beta'\bar E_c+\beta'(E-\bar E_c)$ with the ridge coefficients fixed, 2004--2025. $E$ is the centered headline embedding and $\bar E_c$ its mean over the training-window headlines of the category, zero for categories without any. Covariance shares divide each component's covariance with returns by the forecast's, and variance shares divide its variance by the forecast's; moments pool firm-days after removing day means. Covariance shares sum to 100; variance shares need not. Standalone columns use each component as a signal. Rank IC is the mean daily Spearman correlation with next open-to-open returns, in percent. SR is the annualized Sharpe ratio of equal-weighted (EW) and value-weighted (VW) decile long-short portfolios.
\end{minipage}
\medskip

{\footnotesize
\begin{adjustbox}{max width=\linewidth, max totalheight=0.75\textheight}
\begin{tabular}{llccccc}
\toprule
\multicolumn{2}{c}{} & \multicolumn{2}{c}{Share (\%)} & \multicolumn{3}{c}{Standalone performance} \\
\cmidrule(lr){3-4} \cmidrule(lr){5-7}
Model & Decomposition & $\mathrm{Cov}(\hat y^j,r)/\mathrm{Cov}(\hat y,r)$ & $\mathrm{Var}(\hat y^j)/\mathrm{Var}(\hat y)$ & Rank IC (\%) & EW SR & VW SR \\
\midrule
RoBERTa & Category centroid $\beta'\bar E_c$ & \phantom{0}82.0 & \phantom{0}40.4 & 2.79 & 5.64 & 2.07 \\
 & Within category $\beta'(E-\bar E_c)$ & \phantom{0}18.0 & \phantom{0}67.4 & 0.72 & 1.26 & 0.10 \\
 & Total forecast $\hat y$ & 100.0 & 100.0 & 2.41 & 4.92 & 1.21 \\
\midrule
Llama-2 & Category centroid $\beta'\bar E_c$ & \phantom{0}68.2 & \phantom{0}37.6 & 2.84 & 5.80 & 1.95 \\
 & Within category $\beta'(E-\bar E_c)$ & \phantom{0}31.8 & \phantom{0}64.9 & 1.12 & 1.56 & 0.66 \\
 & Total forecast $\hat y$ & 100.0 & 100.0 & 2.67 & 4.97 & 1.90 \\
\bottomrule
\end{tabular}
\end{adjustbox}
}
\end{table}

\begin{table}[htbp]
\centering
\caption{Forecasts Retrained on Templates and Centroids}
\label{tab:tabA_probe}
\begin{minipage}{\linewidth}\footnotesize
This table reports ridge forecasts of next open-to-open returns fitted separately to headline embeddings $E$, category templates $E_c^{\mathrm{tpl}}$, embeddings minus templates $E-E_c^{\mathrm{tpl}}$, category centroids $\bar E_c$, and within-category embeddings $E-\bar E_c$, 2004--2025. Templates are the sentences ``Company'' followed by the RavenPack category name, such as ``Company analyst ratings change positive,'' embedded like the headlines. A category's centroid is the mean embedding of its headlines in training years $T-8$ through $T-3$; unseen categories use the pooled mean. Each input uses the daily ridge's training and validation years and its own validation-selected penalty. DM $t$ and JK $z$ test rank-IC and Sharpe-ratio differences from the headline forecast. Rank IC is the mean daily Spearman correlation with next open-to-open returns, in percent. SR is the annualized Sharpe ratio of equal-weighted (EW) and value-weighted (VW) decile long-short portfolios.
\end{minipage}
\medskip

{\footnotesize
\begin{adjustbox}{max width=\linewidth, max totalheight=0.75\textheight}
\begin{tabular}{llcccccc}
\toprule
\multicolumn{2}{c}{} & \multicolumn{2}{c}{Rank IC (\%)} & \multicolumn{2}{c}{EW SR} & \multicolumn{2}{c}{VW SR} \\
\cmidrule(lr){3-4} \cmidrule(lr){5-6} \cmidrule(lr){7-8}
Model & Input & Mean & DM $t$ & SR & JK $z$ & SR & JK $z$ \\
\midrule
RoBERTa & Headline embedding $E$ & 2.41 &  & 4.92 &  & 1.21 &  \\
 & Category template $E^{\mathrm{tpl}}_c$ & 3.19 & \phantom{$-$0}9.24 & 6.15 & \phantom{$-$0}5.17 & 2.37 & \phantom{$-$}4.81 \\
 & $E-E^{\mathrm{tpl}}_c$ & 2.59 & \phantom{$-$0}3.84 & 5.42 & \phantom{$-$0}2.94 & 1.34 & \phantom{$-$}0.66 \\
 & Category centroid $\bar E_c$ & 3.15 & \phantom{$-$0}8.81 & 6.06 & \phantom{$-$0}4.85 & 2.07 & \phantom{$-$}3.54 \\
 & $E-\bar E_c$ & 0.76 & $-$21.65 & 1.47 & $-$15.86 & 0.23 & $-$3.84 \\
\midrule
Llama-2 & Headline embedding $E$ & 2.67 &  & 4.97 &  & 1.90 &  \\
 & Category template $E^{\mathrm{tpl}}_c$ & 3.21 & \phantom{$-$0}5.97 & 6.14 & \phantom{$-$0}4.89 & 2.13 & \phantom{$-$}0.86 \\
 & $E-E^{\mathrm{tpl}}_c$ & 2.92 & \phantom{$-$0}7.36 & 5.53 & \phantom{$-$0}4.43 & 1.95 & \phantom{$-$}0.28 \\
 & Category centroid $\bar E_c$ & 3.14 & \phantom{$-$0}5.33 & 6.10 & \phantom{$-$0}4.62 & 2.12 & \phantom{$-$}0.84 \\
 & $E-\bar E_c$ & 1.35 & $-$18.13 & 2.10 & $-$14.95 & 0.64 & $-$5.31 \\
\bottomrule
\end{tabular}
\end{adjustbox}
}
\end{table}

\begin{table}[htbp]
\centering
\caption{Content and Timing Decomposition of the Embedding Forecast}
\label{tab:tabA_decomposition}
\begin{minipage}{\linewidth}\footnotesize
This table reports the content and timing decomposition of embedding forecasts of next open-to-open returns, 2004--2025. A two-way projection of the headline embeddings of the up to six years before each test year on category and firm indicators gives category ($C$), firm ($F$), and idiosyncratic ($I$) components, and the forecast's ridge coefficients give $\hat y-\bar y=\hat y^C+\hat y^F+\hat y^I$, where $\bar y$ is the ridge intercept. $F$ is the firm effect for firms with at least ten headlines in the projection window, spread over at least two of its calendar years, and zero for other firms. The first block splits the category and idiosyncratic components into historical levels, which average their monthly means over the firm's news months in the preceding 12 months, and deviations; firm identity lies entirely in the level. The second block instead splits the idiosyncratic component into the firm's mean over its headlines of the preceding 20 calendar days, $\hat y^I_H$, and the remainder. Shares divide each component's covariance with returns, or its variance, by the forecast's. Moments pool firm-days after removing day means. Standalone columns use each component as a signal. Rank IC is the mean daily Spearman correlation with next open-to-open returns, in percent. SR is the annualized Sharpe ratio of equal-weighted (EW) and value-weighted (VW) decile long-short portfolios.
\end{minipage}
\medskip

{\footnotesize
\begin{adjustbox}{max width=\linewidth, max totalheight=0.75\textheight}
\begin{tabular}{lcccccccccc}
\toprule
 & \multicolumn{5}{c}{RoBERTa} & \multicolumn{5}{c}{Llama-2} \\
\cmidrule(lr){2-6} \cmidrule(lr){7-11}
 & \multicolumn{2}{c}{Share (\%)} & \multicolumn{3}{c}{Standalone} & \multicolumn{2}{c}{Share (\%)} & \multicolumn{3}{c}{Standalone} \\
\cmidrule(lr){2-3} \cmidrule(lr){4-6} \cmidrule(lr){7-8} \cmidrule(lr){9-11}
Decomposition & Cov & Var & Rank IC (\%) & EW SR & VW SR & Cov & Var & Rank IC (\%) & EW SR & VW SR \\
\midrule
\multicolumn{11}{l}{\textit{Full decomposition}} \\
Event category, level $\hat y^C_L$ & \phantom{00}2.2 & \phantom{00}3.4 & 0.54 & 0.38 & \phantom{$-$}0.10 & \phantom{00}1.9 & \phantom{00}4.2 & 0.47 & 0.26 & $-$0.06 \\
Event category, deviation $\hat y^C_D$ & \phantom{0}77.9 & \phantom{0}37.2 & 2.68 & 6.19 & \phantom{$-$}1.89 & \phantom{0}63.6 & \phantom{0}38.7 & 2.54 & 5.95 & \phantom{$-$}1.95 \\
Firm identity $\hat y^F$ (all level) & \phantom{00}6.3 & \phantom{0}21.4 & 0.47 & 0.51 & $-$0.33 & \phantom{0}15.7 & \phantom{0}32.7 & 1.09 & 0.93 & \phantom{$-$}0.88 \\
Idiosyncratic, level $\hat y^I_L$ & \phantom{00}1.1 & \phantom{00}5.2 & 0.20 & 0.50 & \phantom{$-$}0.47 & \phantom{00}1.7 & \phantom{00}6.0 & 0.19 & 0.33 & \phantom{$-$}0.28 \\
Idiosyncratic, deviation $\hat y^I_D$ & \phantom{0}12.5 & \phantom{0}43.5 & 0.53 & 1.36 & \phantom{$-$}0.21 & \phantom{0}17.1 & \phantom{0}35.5 & 0.39 & 1.33 & \phantom{$-$}0.71 \\
Total forecast $\hat y$ & 100.0 & 100.0 & 2.41 & 4.92 & \phantom{$-$}1.21 & 100.0 & 100.0 & 2.67 & 4.97 & \phantom{$-$}1.90 \\
\midrule
\multicolumn{11}{l}{\textit{The idiosyncratic component and the firm's own 20-day mean}} \\
20-day mean $\hat y^I_H$ & \phantom{00}2.6 & \phantom{0}19.9 & 0.32 & 0.36 & \phantom{$-$}0.53 & \phantom{00}4.4 & \phantom{0}19.3 & 0.22 & 0.46 & \phantom{$-$}0.30 \\
Remainder $\hat y^I-\hat y^I_H$ & \phantom{0}11.0 & \phantom{0}55.7 & 0.34 & 1.03 & \phantom{$-$}0.02 & \phantom{0}14.4 & \phantom{0}47.4 & 0.30 & 1.09 & \phantom{$-$}0.48 \\
\bottomrule
\end{tabular}
\end{adjustbox}
}
\end{table}

\begin{table}[htbp]
\centering
\caption{Factor Alphas of Forecast-Component Portfolios}
\label{tab:tabA_factors}
\begin{minipage}{\linewidth}\footnotesize
This table regresses daily equal- and value-weighted decile long-short returns on the Fama--French five factors and the momentum, short-term reversal, and long-term reversal factors, 2004--2025. Portfolios are sorted on the daily ridge forecast, on its category, firm, or idiosyncratic component from a two-way category--firm projection, or on its historical level $\hat y_L$ or deviation $\hat y_D$. The historical level is the mean of the firm's monthly average forecast over its news months in the preceding 12 months, computed with the current coefficients. Without such months it is the forecast from the projection's firm effect, or zero when the firm has $F=0$. The deviation is the remainder. Following CKX, portfolios keep the open-to-open trading rule but are valued at the close to match factor timing. $\alpha$ is annualized in percent; RMW and HML are profitability and value loadings. Newey--West $t$-statistics use 5 lags.
\end{minipage}
\medskip

{\footnotesize
\begin{adjustbox}{max width=\linewidth, max totalheight=0.75\textheight}
\begin{tabular}{lllcccccccccc}
\toprule
\multicolumn{3}{c}{} & \multicolumn{5}{c}{Equal-weighted} & \multicolumn{5}{c}{Value-weighted} \\
\cmidrule(lr){4-8} \cmidrule(lr){9-13}
Model & \multicolumn{2}{l}{Decomposition} & $\alpha$ & $t$ & $R^2$ & RMW & HML & $\alpha$ & $t$ & $R^2$ & RMW & HML \\
\midrule
RoBERTa & Content & Event category $\hat y^C$ & 79.2 & 25.31 & 0.01 & \phantom{$-$}0.12 & \phantom{$-$}0.04 & \phantom{$-$}30.7 & \phantom{$-$}9.48 & 0.03 & \phantom{$-$}0.16 & $-$0.03 \\
 &  & Firm identity $\hat y^F$ & \phantom{0}5.0 & \phantom{0}1.73 & 0.24 & \phantom{$-$}0.64 & \phantom{$-$}0.44 & \phantom{0}$-$6.2 & $-$1.93 & 0.10 & \phantom{$-$}0.31 & \phantom{$-$}0.35 \\
 &  & Idiosyncratic $\hat y^I$ & 15.0 & \phantom{0}6.09 & 0.02 & \phantom{$-$}0.08 & \phantom{$-$}0.07 & \phantom{$-$0}5.8 & \phantom{$-$}1.80 & 0.00 & $-$0.05 & \phantom{$-$}0.01 \\
\cmidrule(lr){2-13}
 & Timing & Level $\hat y_L$ & \phantom{0}4.1 & \phantom{0}1.29 & 0.26 & \phantom{$-$}0.79 & \phantom{$-$}0.46 & \phantom{0}$-$2.7 & $-$0.68 & 0.13 & \phantom{$-$}0.55 & \phantom{$-$}0.11 \\
 &  & Deviation $\hat y_D$ & 71.0 & 24.06 & 0.01 & $-$0.07 & $-$0.01 & \phantom{$-$}20.8 & \phantom{$-$}6.28 & 0.00 & $-$0.02 & \phantom{$-$}0.00 \\
\cmidrule(lr){2-13}
 &  & Total forecast $\hat y$ & 67.3 & 22.29 & 0.10 & \phantom{$-$}0.39 & \phantom{$-$}0.27 & \phantom{$-$}20.2 & \phantom{$-$}5.87 & 0.04 & \phantom{$-$}0.24 & \phantom{$-$}0.13 \\
\midrule
Llama-2 & Content & Event category $\hat y^C$ & 79.1 & 24.70 & 0.01 & \phantom{$-$}0.05 & $-$0.03 & \phantom{$-$}28.6 & \phantom{$-$}8.81 & 0.02 & \phantom{$-$}0.17 & $-$0.08 \\
 &  & Firm identity $\hat y^F$ & 12.8 & \phantom{0}4.12 & 0.16 & \phantom{$-$}0.54 & \phantom{$-$}0.20 & \phantom{$-$}14.6 & \phantom{$-$}4.09 & 0.08 & \phantom{$-$}0.06 & $-$0.15 \\
 &  & Idiosyncratic $\hat y^I$ & 18.4 & \phantom{0}6.53 & 0.02 & \phantom{$-$}0.07 & \phantom{$-$}0.08 & \phantom{$-$}14.0 & \phantom{$-$}3.78 & 0.01 & $-$0.14 & \phantom{$-$}0.05 \\
\cmidrule(lr){2-13}
 & Timing & Level $\hat y_L$ & 16.4 & \phantom{0}4.86 & 0.21 & \phantom{$-$}0.74 & \phantom{$-$}0.26 & \phantom{$-$}11.2 & \phantom{$-$}2.76 & 0.12 & \phantom{$-$}0.38 & $-$0.29 \\
 &  & Deviation $\hat y_D$ & 72.3 & 23.67 & 0.01 & $-$0.09 & $-$0.03 & \phantom{$-$}28.7 & \phantom{$-$}7.87 & 0.01 & $-$0.06 & \phantom{$-$}0.05 \\
\cmidrule(lr){2-13}
 &  & Total forecast $\hat y$ & 76.3 & 22.76 & 0.09 & \phantom{$-$}0.43 & \phantom{$-$}0.17 & \phantom{$-$}34.4 & \phantom{$-$}9.29 & 0.04 & \phantom{$-$}0.17 & $-$0.06 \\
\bottomrule
\end{tabular}
\end{adjustbox}
}
\end{table}

\begin{landscape}

\begin{table}[htbp]
\centering
\caption{Variance and Predictive Covariance by Principal-Component Band}
\label{tab:tabA_spectrum}
\begin{minipage}{\linewidth}\footnotesize
This table decomposes embedding variance and the forecast's covariance with next open-to-open returns across principal-component bands, ordered by decreasing training eigenvalue, 2004--2025. The first columns give each band's percentage of the total. $\bar\phi$ averages the ridge shrinkage factor $\lambda_k/(\lambda_k+\lambda)$ within the band, where $\lambda_k$ is the $k$th training eigenvalue and $\lambda$ the ridge penalty. Band-share columns give each component's percentage of the band's variance or covariance; PC $k$ contributes its ridge coefficient $w_k$ times the covariance of the component's coordinate with returns. Panel A uses category ($C$), firm ($F$), and idiosyncratic ($I$) components from a two-way projection on the ridge training window, $T-8$ through $T-3$. Panel B separates cells of RavenPack event group by headline direction into group centroids (Group), deviations of cell centroids from them (Direction), and within-cell residuals (Within cell). Variance and shrinkage average test years; covariance pools firm-days after removing day means.
\end{minipage}
\medskip

{\footnotesize
\textit{Panel A: Content decomposition}\\[2pt]
\begin{tabular}{llccccccccc}
\toprule
\multicolumn{2}{c}{} & \multicolumn{2}{c}{Share of total (\%)} &  & \multicolumn{3}{c}{\shortstack{Share of band \\ variance (\%)}} & \multicolumn{3}{c}{\shortstack{Share of band \\ covariance (\%)}} \\
\cmidrule(lr){3-4} \cmidrule(lr){6-8} \cmidrule(lr){9-11}
Model & PCs & Variance & Covariance & $\bar\phi$ & $C$ & $F$ & $I$ & $C$ & $F$ & $I$ \\
\midrule
RoBERTa & 1--8 & \phantom{0}42.4 & \phantom{00}2.6 & 0.98 & 77.8 & \phantom{0}2.4 & 19.2 & 187.8 & \phantom{0}$-$1.6 & $-$86.2 \\
 & 9--32 & \phantom{0}22.1 & \phantom{0}35.3 & 0.90 & 34.3 & 17.5 & 47.6 & \phantom{0}88.6 & \phantom{$-$0}1.5 & \phantom{$-$0}9.9 \\
 & 33--128 & \phantom{0}21.4 & \phantom{0}40.5 & 0.70 & 13.3 & 26.8 & 59.9 & \phantom{0}72.7 & \phantom{$-$0}7.6 & \phantom{$-$}19.7 \\
 & 129--512 & \phantom{0}11.5 & \phantom{0}19.9 & 0.32 & \phantom{0}6.4 & 20.6 & 73.1 & \phantom{0}70.0 & \phantom{$-$0}1.7 & \phantom{$-$}28.2 \\
 & 513--1024 & \phantom{00}2.6 & \phantom{00}1.7 & 0.11 & \phantom{0}2.1 & 15.3 & 82.6 & \phantom{0}40.9 & \phantom{$-$}16.6 & \phantom{$-$}42.5 \\
 & Total & 100.0 & 100.0 & 0.27 & 44.3 & 13.4 & 42.0 & \phantom{0}80.2 & \phantom{$-$0}4.2 & \phantom{$-$}15.6 \\
\midrule
Llama-2 & 1--8 & \phantom{0}45.0 & \phantom{00}5.9 & 0.98 & 76.3 & \phantom{0}2.7 & 20.6 & 102.0 & \phantom{$-$0}2.1 & \phantom{0}$-$4.1 \\
 & 9--32 & \phantom{0}16.7 & \phantom{0}27.9 & 0.90 & 42.8 & 14.5 & 41.2 & \phantom{0}60.3 & \phantom{$-$}14.1 & \phantom{$-$}25.6 \\
 & 33--128 & \phantom{0}12.4 & \phantom{0}46.1 & 0.67 & 20.0 & 22.7 & 57.4 & \phantom{0}75.0 & \phantom{$-$}13.8 & \phantom{$-$}11.2 \\
 & 129--512 & \phantom{0}10.8 & \phantom{0}14.3 & 0.34 & \phantom{0}6.2 & 31.6 & 62.3 & \phantom{0}52.5 & \phantom{$-$}11.7 & \phantom{$-$}35.8 \\
 & 513--5120 & \phantom{0}15.0 & \phantom{00}5.7 & 0.06 & \phantom{0}1.5 & 22.9 & 75.6 & \phantom{0}31.9 & \phantom{$-$}17.7 & \phantom{$-$}50.4 \\
 & Total & 100.0 & 100.0 & 0.10 & 45.0 & 13.3 & 41.3 & \phantom{0}66.8 & \phantom{$-$}13.1 & \phantom{$-$}20.0 \\
\bottomrule
\end{tabular}
\par\medskip
\textit{Panel B: Group $\times$ direction cells}\\[2pt]
\begin{tabular}{llccccccccc}
\toprule
\multicolumn{2}{c}{} & \multicolumn{2}{c}{Share of total (\%)} &  & \multicolumn{3}{c}{\shortstack{Share of band \\ variance (\%)}} & \multicolumn{3}{c}{\shortstack{Share of band \\ covariance (\%)}} \\
\cmidrule(lr){3-4} \cmidrule(lr){6-8} \cmidrule(lr){9-11}
Model & PCs & Variance & Covariance & $\bar\phi$ & Group & Direction & Within cell & Group & Direction & Within cell \\
\midrule
RoBERTa & 1--8 & \phantom{0}42.4 & \phantom{00}2.6 & 0.98 & 63.8 & 3.4 & 32.8 & 60.7 & 88.5 & $-$49.2 \\
 & 9--32 & \phantom{0}22.1 & \phantom{0}35.3 & 0.90 & 15.1 & 5.7 & 79.2 & 14.6 & 62.2 & \phantom{$-$}23.1 \\
 & 33--128 & \phantom{0}21.4 & \phantom{0}40.5 & 0.70 & \phantom{0}3.9 & 2.4 & 93.7 & \phantom{0}5.5 & 53.0 & \phantom{$-$}41.6 \\
 & 129--512 & \phantom{0}11.5 & \phantom{0}19.9 & 0.32 & \phantom{0}1.1 & 1.1 & 97.8 & \phantom{0}3.6 & 68.1 & \phantom{$-$}28.3 \\
 & 513--1024 & \phantom{00}2.6 & \phantom{00}1.7 & 0.11 & \phantom{0}0.2 & 0.3 & 99.5 & \phantom{0}0.1 & 48.6 & \phantom{$-$}51.4 \\
 & Total & 100.0 & 100.0 & 0.27 & 31.4 & 3.3 & 65.2 & \phantom{0}9.7 & 60.1 & \phantom{$-$}30.2 \\
\midrule
Llama-2 & 1--8 & \phantom{0}45.0 & \phantom{00}5.9 & 0.98 & 60.8 & 3.6 & 35.6 & 52.8 & 17.8 & \phantom{$-$}29.4 \\
 & 9--32 & \phantom{0}16.7 & \phantom{0}27.9 & 0.90 & 21.9 & 6.1 & 71.9 & 12.6 & 38.6 & \phantom{$-$}48.8 \\
 & 33--128 & \phantom{0}12.4 & \phantom{0}46.1 & 0.67 & \phantom{0}5.2 & 4.4 & 90.4 & \phantom{0}4.5 & 55.0 & \phantom{$-$}40.5 \\
 & 129--512 & \phantom{0}10.8 & \phantom{0}14.3 & 0.34 & \phantom{0}0.7 & 1.1 & 98.2 & \phantom{0}0.7 & 61.9 & \phantom{$-$}37.5 \\
 & 513--5120 & \phantom{0}15.0 & \phantom{00}5.7 & 0.06 & \phantom{0}0.1 & 0.2 & 99.7 & \phantom{0}1.7 & 22.6 & \phantom{$-$}75.7 \\
 & Total & 100.0 & 100.0 & 0.10 & 31.9 & 3.3 & 64.7 & \phantom{0}8.9 & 47.4 & \phantom{$-$}43.7 \\
\bottomrule
\end{tabular}
}
\end{table}

\begin{table}[htbp]
\centering
\caption{Forecast Decomposition by Event Group}
\label{tab:tabA_groups}
\begin{minipage}{\linewidth}\footnotesize
This table reports forecasts of next open-to-open returns within event groups with at least 50,000 headlines, 2004--2025. Rank IC is the mean daily Spearman correlation across the group's firm-days, in percent. Content shares use the category, firm, and idiosyncratic components of a two-way category--firm projection; timing shares use the historical level and the deviation from it. The historical level is the mean of the firm's monthly average forecast over its news months in the preceding 12 months. The category mean is the eight-year mean. Shares divide component covariance by forecast covariance within the group and are omitted when the latter has $|t|<2$, with 5 Newey--West lags. Contribution to CKX L--S is the group's contribution to the equal-weighted decile long-short return, in basis points per day; contributions sum to the portfolio return across all groups, including those omitted.
\end{minipage}
\medskip

{\footnotesize
\begin{tabular}{llccccccccc}
\toprule
\multicolumn{3}{c}{} & \multicolumn{2}{c}{Rank IC (\%)} & \multicolumn{5}{c}{Share of the predictive covariance (\%)} & Contribution to \\
\cmidrule(lr){4-5} \cmidrule(lr){6-10}
Model & Event group & Headlines & Category mean & Forecast & Category & Firm & Idiosyncratic & Level & Deviation & CKX L--S (bp) \\
\midrule
RoBERTa & Insider trading & 1,395,779 & \phantom{$-$}3.42 & \phantom{$-$}2.67 & \phantom{0}80.6 & \phantom{$-$0}1.7 & \phantom{$-$}17.7 & \phantom{$-$0}3.9 & \phantom{0}96.1 & \phantom{$-$}9.7 \\
 & Earnings & 1,054,174 & \phantom{$-$}2.93 & \phantom{$-$}2.82 & \phantom{0}51.2 & \phantom{$-$}15.7 & \phantom{$-$}33.2 & \phantom{$-$}15.2 & \phantom{0}84.8 & \phantom{$-$}3.0 \\
 & Revenues & \phantom{,0}432,552 & \phantom{$-$}1.32 & \phantom{$-$}2.32 & \phantom{0}18.1 & \phantom{$-$}35.3 & \phantom{$-$}46.7 & \phantom{$-$}32.1 & \phantom{0}67.9 & \phantom{$-$}1.3 \\
 & Technical analysis & \phantom{,0}431,954 & $-$0.19 & \phantom{$-$}0.50 & \phantom{0}42.3 & \phantom{$-$}21.1 & \phantom{$-$}36.5 & \phantom{$-$}32.6 & \phantom{0}67.4 & $-$0.0 \\
 & Analyst ratings & \phantom{,0}390,015 & \phantom{$-$}5.03 & \phantom{$-$}3.88 & 113.3 & \phantom{$-$0}5.6 & $-$18.9 & \phantom{$-$0}4.8 & \phantom{0}95.2 & \phantom{$-$}3.7 \\
 & Investor relations & \phantom{,0}333,417 & $-$1.16 & \phantom{$-$}0.70 &  &  &  &  &  & \phantom{$-$}0.5 \\
 & Price targets & \phantom{,0}251,315 & \phantom{$-$}1.29 & \phantom{$-$}1.01 & 110.0 & \phantom{$-$}13.3 & $-$23.3 & \phantom{$-$0}0.3 & \phantom{0}99.7 & \phantom{$-$}0.4 \\
 & Labor issues & \phantom{,0}197,779 & \phantom{$-$}2.47 & \phantom{$-$}2.38 & \phantom{0}54.0 & \phantom{$-$0}6.9 & \phantom{$-$}39.1 & \phantom{$-$}10.1 & \phantom{0}89.9 & \phantom{$-$}1.6 \\
 & Stock prices & \phantom{,0}188,260 & $-$0.60 & \phantom{$-$}2.02 & \phantom{0}29.8 & \phantom{$-$}17.7 & \phantom{$-$}52.5 & \phantom{$-$}37.2 & \phantom{0}62.8 & \phantom{$-$}0.6 \\
 & Products services & \phantom{,0}149,780 & \phantom{$-$}1.75 & \phantom{$-$}0.85 & \phantom{0}82.0 & \phantom{$-$}35.6 & $-$17.6 & \phantom{$-$}61.3 & \phantom{0}38.7 & \phantom{$-$}0.7 \\
 & Order imbalances & \phantom{,0}130,025 & \phantom{$-$}1.59 & \phantom{$-$}0.80 &  &  &  &  &  & $-$0.2 \\
 & Equity actions & \phantom{,0}129,964 & \phantom{$-$}4.50 & \phantom{$-$}4.57 & \phantom{0}77.9 & \phantom{$-$}19.7 & \phantom{$-$0}2.4 & \phantom{$-$}20.3 & \phantom{0}79.7 & \phantom{$-$}2.5 \\
 & Dividends & \phantom{,0}111,626 & \phantom{$-$}3.32 & \phantom{$-$}1.82 & \phantom{0}68.7 & $-$12.3 & \phantom{$-$}43.7 & $-$12.7 & 112.7 & \phantom{$-$}0.9 \\
 & Credit ratings & \phantom{,0}103,535 & \phantom{$-$}3.21 & \phantom{$-$}2.28 & \phantom{0}86.5 & \phantom{$-$}20.4 & \phantom{0}$-$6.9 & \phantom{$-$}24.0 & \phantom{0}76.0 & \phantom{$-$}0.4 \\
 & Acquisitions mergers & \phantom{,00}57,905 & $-$0.59 & \phantom{$-$}0.29 &  &  &  &  &  & \phantom{$-$}1.2 \\
\midrule
Llama-2 & Insider trading & 1,395,779 & \phantom{$-$}3.42 & \phantom{$-$}2.86 & \phantom{0}90.6 & \phantom{$-$0}1.6 & \phantom{$-$0}7.8 & \phantom{$-$0}1.2 & \phantom{0}98.8 & \phantom{$-$}6.7 \\
 & Earnings & 1,054,174 & \phantom{$-$}2.93 & \phantom{$-$}3.37 & \phantom{0}36.2 & \phantom{$-$}24.8 & \phantom{$-$}39.1 & \phantom{$-$}26.8 & \phantom{0}73.2 & \phantom{$-$}5.6 \\
 & Revenues & \phantom{,0}432,552 & \phantom{$-$}1.32 & \phantom{$-$}3.58 & \phantom{00}7.9 & \phantom{$-$}43.7 & \phantom{$-$}48.3 & \phantom{$-$}44.3 & \phantom{0}55.7 & \phantom{$-$}2.4 \\
 & Technical analysis & \phantom{,0}431,954 & $-$0.19 & \phantom{$-$}0.58 &  &  &  &  &  & $-$0.2 \\
 & Analyst ratings & \phantom{,0}390,015 & \phantom{$-$}5.03 & \phantom{$-$}3.93 & 101.6 & \phantom{$-$0}9.5 & $-$11.1 & \phantom{$-$}10.9 & \phantom{0}89.1 & \phantom{$-$}3.8 \\
 & Investor relations & \phantom{,0}333,417 & $-$1.16 & \phantom{$-$}0.65 &  &  &  &  &  & \phantom{$-$}0.4 \\
 & Price targets & \phantom{,0}251,315 & \phantom{$-$}1.29 & \phantom{$-$}1.27 &  &  &  &  &  & \phantom{$-$}0.2 \\
 & Labor issues & \phantom{,0}197,779 & \phantom{$-$}2.47 & \phantom{$-$}2.66 & \phantom{0}45.3 & \phantom{$-$}24.7 & \phantom{$-$}30.0 & \phantom{$-$}28.0 & \phantom{0}72.0 & \phantom{$-$}1.4 \\
 & Stock prices & \phantom{,0}188,260 & $-$0.60 & \phantom{$-$}2.94 & \phantom{0}13.3 & \phantom{$-$}25.6 & \phantom{$-$}61.1 & \phantom{$-$}43.0 & \phantom{0}57.0 & \phantom{$-$}0.8 \\
 & Products services & \phantom{,0}149,780 & \phantom{$-$}1.75 & \phantom{$-$}2.19 & \phantom{0}31.0 & \phantom{$-$}34.0 & \phantom{$-$}35.0 & \phantom{$-$}45.5 & \phantom{0}54.5 & \phantom{$-$}1.6 \\
 & Order imbalances & \phantom{,0}130,025 & \phantom{$-$}1.59 & \phantom{$-$}1.89 & \phantom{0}14.5 & \phantom{$-$}75.2 & \phantom{$-$}10.3 & \phantom{$-$}96.0 & \phantom{00}4.0 & $-$0.1 \\
 & Equity actions & \phantom{,0}129,964 & \phantom{$-$}4.50 & \phantom{$-$}5.18 & \phantom{0}67.0 & \phantom{$-$}30.5 & \phantom{$-$0}2.5 & \phantom{$-$}31.6 & \phantom{0}68.4 & \phantom{$-$}3.3 \\
 & Dividends & \phantom{,0}111,626 & \phantom{$-$}3.32 & \phantom{$-$}3.27 & \phantom{0}39.9 & \phantom{$-$}16.5 & \phantom{$-$}43.6 & \phantom{$-$}16.4 & \phantom{0}83.6 & \phantom{$-$}1.1 \\
 & Credit ratings & \phantom{,0}103,535 & \phantom{$-$}3.21 & \phantom{$-$}1.59 & \phantom{0}73.0 & \phantom{$-$}68.5 & $-$41.5 & \phantom{$-$}82.6 & \phantom{0}17.4 & \phantom{$-$}0.4 \\
 & Acquisitions mergers & \phantom{,00}57,905 & $-$0.59 & $-$0.44 &  &  &  &  &  & \phantom{$-$}1.5 \\
\bottomrule
\end{tabular}
}
\end{table}

\begin{table}[htbp]
\centering
\caption{Return Covariances of the Full Forecast Decomposition}
\label{tab:tabA_horizons}
\begin{minipage}{\linewidth}\footnotesize
This table reports return covariances of the category level and deviation, firm identity, and the idiosyncratic level and deviation, for signal years 2004--2025. Panel A uses the daily ridge forecast, Panel B its headline-weighted monthly average, and Panel C a ridge forecast fitted to monthly average embeddings. Category, firm, and idiosyncratic components come from a two-way category--firm projection of headline embeddings. Historical levels average the category and idiosyncratic components' monthly means over the firm's news months in the preceding 12 months; deviations are the remainders, and firm identity lies entirely in the level. Daily $h=0$ is the nontradable return to the eligible open and $h\ge1$ the $h$-th subsequent open-to-open return; monthly $h$ counts months after the signal month. Cov is the cross-sectional covariance times $10^6$; Share is the percentage of the forecast's covariance, omitted when the latter has $|t|<2$. Newey--West $t$-statistics use 5 daily or 6 monthly lags.
\end{minipage}
\medskip

{\footnotesize
\textit{Panel A: Daily forecast}\\[2pt]
\begin{adjustbox}{max width=\linewidth, max totalheight=0.21\textheight}
\begin{tabular}{llccccccccccccccc}
\toprule
\multicolumn{2}{c}{} & \multicolumn{3}{c}{Event category, level $\hat y^C_L$} & \multicolumn{3}{c}{Event category, deviation $\hat y^C_D$} & \multicolumn{3}{c}{Firm identity $\hat y^F$} & \multicolumn{3}{c}{Idiosyncratic, level $\hat y^I_L$} & \multicolumn{3}{c}{Idiosyncratic, deviation $\hat y^I_D$} \\
\cmidrule(lr){3-5} \cmidrule(lr){6-8} \cmidrule(lr){9-11} \cmidrule(lr){12-14} \cmidrule(lr){15-17}
Model & $h$ (trading days) & Cov & $t$ & Share (\%) & Cov & $t$ & Share (\%) & Cov & $t$ & Share (\%) & Cov & $t$ & Share (\%) & Cov & $t$ & Share (\%) \\
\midrule
RoBERTa & 0 & $-$0.01 & $-$1.56 & $-$0.5 & 2.65 & 42.31 & 96.9 & $-$0.11 & $-$6.36 & \phantom{0}$-$4.2 & $-$0.01 & $-$1.32 & \phantom{0}$-$0.4 & 0.22 & \phantom{0}9.09 & \phantom{0}8.1 \\
 & 1 & \phantom{$-$}0.02 & \phantom{$-$}1.92 & \phantom{$-$}1.9 & 0.72 & 28.73 & 78.3 & \phantom{$-$}0.06 & \phantom{$-$}2.79 & \phantom{$-$0}6.2 & \phantom{$-$}0.01 & \phantom{$-$}1.38 & \phantom{$-$0}1.2 & 0.11 & \phantom{0}6.34 & 12.4 \\
 & 2 & \phantom{$-$}0.01 & \phantom{$-$}1.33 & \phantom{$-$}5.6 & 0.14 & \phantom{0}7.47 & 68.2 & \phantom{$-$}0.01 & \phantom{$-$}0.51 & \phantom{$-$0}4.3 & \phantom{$-$}0.01 & \phantom{$-$}0.80 & \phantom{$-$0}3.0 & 0.04 & \phantom{0}2.26 & 18.8 \\
 & 3 & \phantom{$-$}0.01 & \phantom{$-$}0.63 & \phantom{$-$}4.6 & 0.05 & \phantom{0}2.85 & 44.5 & \phantom{$-$}0.03 & \phantom{$-$}1.61 & \phantom{$-$}23.8 & $-$0.01 & $-$0.76 & \phantom{0}$-$4.8 & 0.04 & \phantom{0}2.28 & 31.9 \\
 & 5 & \phantom{$-$}0.00 & \phantom{$-$}0.13 &  & 0.03 & \phantom{0}1.63 &  & \phantom{$-$}0.02 & \phantom{$-$}1.36 &  & $-$0.01 & $-$1.01 &  & 0.01 & \phantom{0}0.43 &  \\
\midrule
Llama-2 & 0 & $-$0.03 & $-$2.41 & $-$0.9 & 2.63 & 42.05 & 96.5 & $-$0.25 & $-$8.42 & \phantom{0}$-$9.2 & $-$0.05 & $-$3.73 & \phantom{0}$-$1.7 & 0.42 & 15.46 & 15.4 \\
 & 1 & \phantom{$-$}0.02 & \phantom{$-$}1.78 & \phantom{$-$}1.7 & 0.77 & 27.31 & 64.1 & \phantom{$-$}0.19 & \phantom{$-$}5.83 & \phantom{$-$}15.7 & \phantom{$-$}0.02 & \phantom{$-$}1.82 & \phantom{$-$0}1.7 & 0.20 & \phantom{0}9.40 & 16.9 \\
 & 2 & \phantom{$-$}0.01 & \phantom{$-$}1.36 & \phantom{$-$}5.0 & 0.14 & \phantom{0}6.48 & 48.1 & \phantom{$-$}0.06 & \phantom{$-$}2.25 & \phantom{$-$}21.2 & \phantom{$-$}0.03 & \phantom{$-$}2.66 & \phantom{$-$}10.0 & 0.05 & \phantom{0}2.39 & 15.8 \\
 & 3 & \phantom{$-$}0.00 & \phantom{$-$}0.32 & \phantom{$-$}2.0 & 0.04 & \phantom{0}1.83 & 23.2 & \phantom{$-$}0.08 & \phantom{$-$}2.93 & \phantom{$-$}47.2 & $-$0.00 & $-$0.32 & \phantom{0}$-$1.9 & 0.05 & \phantom{0}2.68 & 29.6 \\
 & 5 & $-$0.00 & $-$0.26 & $-$2.1 & 0.01 & \phantom{0}0.68 & 10.1 & \phantom{$-$}0.08 & \phantom{$-$}3.11 & \phantom{$-$}63.4 & \phantom{$-$}0.01 & \phantom{$-$}0.76 & \phantom{$-$0}5.5 & 0.03 & \phantom{0}1.67 & 23.1 \\
\bottomrule
\end{tabular}
\end{adjustbox}
\par\medskip
\textit{Panel B: Monthly average of the daily forecast}\\[2pt]
\begin{adjustbox}{max width=\linewidth, max totalheight=0.27\textheight}
\begin{tabular}{llccccccccccccccc}
\toprule
\multicolumn{2}{c}{} & \multicolumn{3}{c}{Event category, level $\hat y^C_L$} & \multicolumn{3}{c}{Event category, deviation $\hat y^C_D$} & \multicolumn{3}{c}{Firm identity $\hat y^F$} & \multicolumn{3}{c}{Idiosyncratic, level $\hat y^I_L$} & \multicolumn{3}{c}{Idiosyncratic, deviation $\hat y^I_D$} \\
\cmidrule(lr){3-5} \cmidrule(lr){6-8} \cmidrule(lr){9-11} \cmidrule(lr){12-14} \cmidrule(lr){15-17}
Model & $h$ (months) & Cov & $t$ & Share (\%) & Cov & $t$ & Share (\%) & Cov & $t$ & Share (\%) & Cov & $t$ & Share (\%) & Cov & $t$ & Share (\%) \\
\midrule
RoBERTa & 1 & \phantom{$-$}0.12 & \phantom{$-$}0.64 & 8.9 & \phantom{$-$}0.58 & \phantom{$-$}4.08 & \phantom{$-$}44.2 & \phantom{$-$}0.48 & \phantom{$-$}1.33 & 36.5 & 0.09 & 0.95 & \phantom{0}7.2 & \phantom{$-$}0.04 & \phantom{$-$}0.32 & \phantom{$-$0}3.2 \\
 & 2 & \phantom{$-$}0.10 & \phantom{$-$}0.53 &  & \phantom{$-$}0.23 & \phantom{$-$}1.47 &  & \phantom{$-$}0.44 & \phantom{$-$}1.31 &  & 0.09 & 0.76 &  & \phantom{$-$}0.30 & \phantom{$-$}2.68 &  \\
 & 4 & \phantom{$-$}0.03 & \phantom{$-$}0.14 &  & $-$0.22 & $-$1.65 &  & \phantom{$-$}0.28 & \phantom{$-$}0.93 &  & 0.09 & 0.83 &  & $-$0.07 & $-$0.58 &  \\
 & 6 & $-$0.03 & $-$0.14 &  & \phantom{$-$}0.17 & \phantom{$-$}1.11 &  & \phantom{$-$}0.23 & \phantom{$-$}0.82 &  & 0.05 & 0.44 &  & \phantom{$-$}0.10 & \phantom{$-$}0.82 &  \\
 & 12 & $-$0.00 & $-$0.02 &  & \phantom{$-$}0.27 & \phantom{$-$}2.03 &  & \phantom{$-$}0.17 & \phantom{$-$}0.52 &  & 0.01 & 0.16 &  & $-$0.13 & $-$1.02 &  \\
 & 18 & $-$0.03 & $-$0.17 &  & $-$0.10 & $-$0.70 &  & $-$0.07 & $-$0.22 &  & 0.05 & 0.37 &  & $-$0.06 & $-$0.37 &  \\
 & 24 & $-$0.10 & $-$0.52 &  & \phantom{$-$}0.29 & \phantom{$-$}1.92 &  & $-$0.05 & $-$0.13 &  & 0.03 & 0.27 &  & $-$0.18 & $-$1.12 &  \\
\midrule
Llama-2 & 1 & \phantom{$-$}0.19 & \phantom{$-$}0.89 & 6.7 & \phantom{$-$}0.58 & \phantom{$-$}3.46 & \phantom{$-$}20.5 & \phantom{$-$}1.50 & \phantom{$-$}2.79 & 53.3 & 0.51 & 3.27 & 18.3 & \phantom{$-$}0.04 & \phantom{$-$}0.23 & \phantom{$-$0}1.2 \\
 & 2 & \phantom{$-$}0.17 & \phantom{$-$}0.78 & 6.2 & \phantom{$-$}0.41 & \phantom{$-$}2.20 & \phantom{$-$}15.0 & \phantom{$-$}1.56 & \phantom{$-$}3.04 & 57.4 & 0.46 & 2.71 & 17.0 & \phantom{$-$}0.12 & \phantom{$-$}0.77 & \phantom{$-$0}4.4 \\
 & 4 & \phantom{$-$}0.08 & \phantom{$-$}0.36 & 4.3 & $-$0.24 & $-$1.34 & $-$12.3 & \phantom{$-$}1.56 & \phantom{$-$}3.36 & 81.0 & 0.54 & 3.98 & 27.8 & $-$0.01 & $-$0.11 & \phantom{0}$-$0.8 \\
 & 6 & \phantom{$-$}0.01 & \phantom{$-$}0.04 & 0.4 & \phantom{$-$}0.10 & \phantom{$-$}0.58 & \phantom{$-$0}4.6 & \phantom{$-$}1.07 & \phantom{$-$}2.27 & 51.0 & 0.52 & 3.04 & 24.8 & \phantom{$-$}0.40 & \phantom{$-$}2.10 & \phantom{$-$}19.2 \\
 & 12 & \phantom{$-$}0.07 & \phantom{$-$}0.31 & 3.1 & \phantom{$-$}0.31 & \phantom{$-$}1.97 & \phantom{$-$}14.1 & \phantom{$-$}1.21 & \phantom{$-$}2.14 & 55.6 & 0.45 & 2.94 & 20.9 & \phantom{$-$}0.14 & \phantom{$-$}0.86 & \phantom{$-$0}6.4 \\
 & 18 & $-$0.07 & $-$0.30 &  & $-$0.06 & $-$0.35 &  & \phantom{$-$}0.56 & \phantom{$-$}1.02 &  & 0.53 & 3.26 &  & \phantom{$-$}0.09 & \phantom{$-$}0.53 &  \\
 & 24 & $-$0.19 & $-$0.81 &  & \phantom{$-$}0.27 & \phantom{$-$}1.64 &  & \phantom{$-$}0.54 & \phantom{$-$}0.95 &  & 0.53 & 2.81 &  & \phantom{$-$}0.38 & \phantom{$-$}2.57 &  \\
\bottomrule
\end{tabular}
\end{adjustbox}
\par\medskip
\textit{Panel C: Monthly-trained forecast}\\[2pt]
\begin{adjustbox}{max width=\linewidth, max totalheight=0.27\textheight}
\begin{tabular}{llccccccccccccccc}
\toprule
\multicolumn{2}{c}{} & \multicolumn{3}{c}{Event category, level $\hat y^C_L$} & \multicolumn{3}{c}{Event category, deviation $\hat y^C_D$} & \multicolumn{3}{c}{Firm identity $\hat y^F$} & \multicolumn{3}{c}{Idiosyncratic, level $\hat y^I_L$} & \multicolumn{3}{c}{Idiosyncratic, deviation $\hat y^I_D$} \\
\cmidrule(lr){3-5} \cmidrule(lr){6-8} \cmidrule(lr){9-11} \cmidrule(lr){12-14} \cmidrule(lr){15-17}
Model & $h$ (months) & Cov & $t$ & Share (\%) & Cov & $t$ & Share (\%) & Cov & $t$ & Share (\%) & Cov & $t$ & Share (\%) & Cov & $t$ & Share (\%) \\
\midrule
RoBERTa & 1 & \phantom{$-$}0.51 & \phantom{$-$}0.35 &  & \phantom{$-$}3.67 & \phantom{$-$}2.30 &  & \phantom{$-$0}0.03 & \phantom{$-$}0.01 &  & \phantom{0}1.16 & 1.57 &  & \phantom{$-$}2.22 & \phantom{$-$}1.34 &  \\
 & 2 & \phantom{$-$}0.44 & \phantom{$-$}0.31 &  & \phantom{$-$}3.45 & \phantom{$-$}2.27 &  & \phantom{0}$-$0.06 & $-$0.03 &  & \phantom{0}0.52 & 0.58 &  & \phantom{$-$}1.24 & \phantom{$-$}0.81 &  \\
 & 4 & \phantom{$-$}0.80 & \phantom{$-$}0.57 &  & \phantom{$-$}0.02 & \phantom{$-$}0.01 &  & \phantom{0}$-$0.21 & $-$0.10 &  & \phantom{0}0.81 & 0.91 &  & \phantom{$-$}1.06 & \phantom{$-$}0.67 &  \\
 & 6 & $-$0.38 & $-$0.29 &  & \phantom{$-$}1.36 & \phantom{$-$}1.11 &  & \phantom{0}$-$1.29 & $-$0.57 &  & \phantom{0}1.21 & 1.49 &  & \phantom{$-$}1.35 & \phantom{$-$}0.93 &  \\
 & 12 & \phantom{$-$}0.79 & \phantom{$-$}0.60 &  & \phantom{$-$}1.95 & \phantom{$-$}1.13 &  & \phantom{0}$-$3.09 & $-$1.34 &  & \phantom{0}2.21 & 2.20 &  & \phantom{$-$}1.15 & \phantom{$-$}0.49 &  \\
 & 18 & \phantom{$-$}1.06 & \phantom{$-$}0.67 &  & $-$2.12 & $-$0.89 &  & \phantom{0}$-$4.45 & $-$1.92 &  & \phantom{0}1.13 & 1.08 &  & \phantom{$-$}0.93 & \phantom{$-$}0.69 &  \\
 & 24 & \phantom{$-$}1.12 & \phantom{$-$}0.94 &  & $-$0.12 & $-$0.08 &  & \phantom{0}$-$3.12 & $-$1.33 &  & \phantom{0}1.19 & 0.98 &  & \phantom{$-$}0.74 & \phantom{$-$}0.45 &  \\
\midrule
Llama-2 & 1 & $-$0.30 & $-$0.14 & \phantom{0}$-$1.2 & \phantom{$-$}4.71 & \phantom{$-$}1.75 & 18.2 & \phantom{$-$0}8.42 & \phantom{$-$}1.41 & 32.5 & \phantom{0}3.39 & 1.22 & 13.1 & \phantom{$-$}9.69 & \phantom{$-$}2.69 & \phantom{$-$}37.4 \\
 & 2 & $-$1.13 & $-$0.49 & \phantom{0}$-$7.3 & \phantom{$-$}4.22 & \phantom{$-$}1.60 & 27.4 & \phantom{$-$}10.25 & \phantom{$-$}1.59 & 66.6 & \phantom{0}4.21 & 1.48 & 27.4 & $-$2.17 & $-$0.71 & $-$14.1 \\
 & 4 & $-$2.21 & $-$0.95 &  & \phantom{$-$}4.29 & \phantom{$-$}2.05 &  & \phantom{$-$0}7.46 & \phantom{$-$}1.24 &  & \phantom{0}3.35 & 1.20 &  & $-$0.08 & $-$0.03 &  \\
 & 6 & $-$3.02 & $-$1.29 &  & $-$1.32 & $-$0.62 &  & \phantom{$-$0}5.83 & \phantom{$-$}0.89 &  & \phantom{0}3.71 & 1.08 &  & \phantom{$-$}6.03 & \phantom{$-$}1.47 &  \\
 & 12 & $-$2.50 & $-$1.06 & $-$13.5 & \phantom{$-$}4.73 & \phantom{$-$}2.10 & 25.5 & \phantom{$-$0}2.20 & \phantom{$-$}0.39 & 11.9 & 11.00 & 3.25 & 59.3 & \phantom{$-$}3.11 & \phantom{$-$}1.03 & \phantom{$-$}16.8 \\
 & 18 & $-$2.81 & $-$1.03 &  & $-$0.82 & $-$0.28 &  & \phantom{0}$-$5.10 & $-$0.98 &  & \phantom{0}7.00 & 2.43 &  & \phantom{$-$}1.00 & \phantom{$-$}0.25 &  \\
 & 24 & $-$0.49 & $-$0.22 &  & $-$0.13 & $-$0.05 &  & \phantom{0}$-$2.96 & $-$0.36 &  & \phantom{0}7.82 & 1.70 &  & \phantom{$-$}4.33 & \phantom{$-$}1.08 &  \\
\bottomrule
\end{tabular}
\end{adjustbox}
}
\end{table}

\begin{table}[htbp]
\centering
\caption{MSRR Strategy Decomposition with Re-estimated Weights}
\label{tab:tabA_reoptimized}
\begin{minipage}{\linewidth}\footnotesize
This table decomposes monthly maximum Sharpe ratio regression (MSRR) strategy returns with coefficients re-estimated at each horizon, for signal years 2004--2025. Firm-month embeddings are recursively standardized, residualized each month on 132 JKP characteristics, and combined out of sample by MSRR. News is assigned by arrival month. Signal month $\tau$ fixes the firms traded, those with news in month $\tau$, and the month of the characteristics. The strategy earns their returns in month $\tau+h$, the $h$-th subsequent return month, whether or not they have news then. For each $h$, the coefficients of signal month $\tau$ are fitted in an expanding window from 2000 to the horizon-$h$ portfolio returns of earlier signal months, using only returns realized by month $\tau+h-1$. Panel A separates the historical level, the mean of the firm's monthly average embedding over its news months in the preceding 12 months or the firm effect when the firm has no such months, from the deviation; Panel B also separates the category, firm, and idiosyncratic components of the two-way category--firm projection. Each component is residualized like the full embedding and uses the same coefficients, so component returns sum to the total. Ret is the annualized mean in percent after scaling the strategy to 10\% annual volatility at each horizon. Share is the percentage of the total mean return, omitted when the total has $|t|<2$; SR is annualized. Newey--West $t$-statistics use 6 lags.
\end{minipage}
\medskip

{\footnotesize
\textit{Panel A: Timing decomposition}\\[2pt]
\begin{adjustbox}{max width=\linewidth, max totalheight=0.32\textheight}
\begin{tabular}{llccccccccc}
\toprule
\multicolumn{2}{c}{} & \multicolumn{3}{c}{Embedding $E$} & \multicolumn{3}{c}{Level $E_L$} & \multicolumn{3}{c}{Deviation $E_D$} \\
\cmidrule(lr){3-5} \cmidrule(lr){6-8} \cmidrule(lr){9-11}
Model & $h$ (months) & SR & Ret & $t$ & Ret & $t$ & Share (\%) & Ret & $t$ & Share (\%) \\
\midrule
RoBERTa & 1 & 1.79 & 17.9 & \phantom{0}7.96 & \phantom{0}9.1 & \phantom{0}5.09 & 51.0 & \phantom{$-$}8.8 & \phantom{$-$}4.02 & 49.0 \\
 & 2 & 1.13 & 11.3 & \phantom{0}5.67 & \phantom{0}8.2 & \phantom{0}5.58 & 72.6 & \phantom{$-$}3.1 & \phantom{$-$}2.23 & 27.4 \\
 & 4 & 0.60 & \phantom{0}6.0 & \phantom{0}2.38 & \phantom{0}5.8 & \phantom{0}3.16 & 96.2 & \phantom{$-$}0.2 & \phantom{$-$}0.14 & \phantom{0}3.8 \\
 & 6 & 0.64 & \phantom{0}6.4 & \phantom{0}2.52 & \phantom{0}5.0 & \phantom{0}2.32 & 78.5 & \phantom{$-$}1.4 & \phantom{$-$}1.07 & 21.5 \\
 & 12 & 0.46 & \phantom{0}4.6 & \phantom{0}2.08 & \phantom{0}1.3 & \phantom{0}1.04 & 28.7 & \phantom{$-$}3.3 & \phantom{$-$}2.04 & 71.3 \\
 & 18 & 0.17 & \phantom{0}1.7 & \phantom{0}0.71 & \phantom{0}2.7 & \phantom{0}1.70 &  & $-$1.0 & $-$0.51 &  \\
 & 24 & 0.85 & \phantom{0}8.5 & \phantom{0}3.97 & \phantom{0}3.8 & \phantom{0}1.77 & 44.7 & \phantom{$-$}4.7 & \phantom{$-$}2.16 & 55.3 \\
\midrule
Llama-2 & 1 & 3.62 & 36.2 & 18.56 & 27.6 & 11.36 & 76.3 & \phantom{$-$}8.6 & \phantom{$-$}4.77 & 23.7 \\
 & 2 & 2.84 & 28.4 & 12.82 & 23.2 & 11.58 & 81.6 & \phantom{$-$}5.2 & \phantom{$-$}4.48 & 18.4 \\
 & 4 & 2.21 & 22.1 & \phantom{0}9.72 & 20.5 & 10.04 & 93.0 & \phantom{$-$}1.5 & \phantom{$-$}1.10 & \phantom{0}7.0 \\
 & 6 & 2.22 & 22.2 & \phantom{0}7.73 & 22.0 & 11.45 & 99.0 & \phantom{$-$}0.2 & \phantom{$-$}0.13 & \phantom{0}1.0 \\
 & 12 & 1.79 & 17.9 & \phantom{0}7.51 & 13.6 & \phantom{0}7.09 & 75.9 & \phantom{$-$}4.3 & \phantom{$-$}2.95 & 24.1 \\
 & 18 & 1.08 & 10.8 & \phantom{0}3.83 & 10.7 & \phantom{0}4.83 & 99.2 & \phantom{$-$}0.1 & \phantom{$-$}0.06 & \phantom{0}0.8 \\
 & 24 & 1.57 & 15.7 & \phantom{0}5.95 & 10.3 & \phantom{0}4.31 & 65.5 & \phantom{$-$}5.4 & \phantom{$-$}4.37 & 34.5 \\
\bottomrule
\end{tabular}
\end{adjustbox}
\par\medskip
\textit{Panel B: Full decomposition}\\[2pt]
\begin{adjustbox}{max width=\linewidth, max totalheight=0.36\textheight}
\begin{tabular}{llccccccccccccccc}
\toprule
\multicolumn{2}{c}{} & \multicolumn{3}{c}{Event category, level $C_L$} & \multicolumn{3}{c}{Event category, deviation $C_D$} & \multicolumn{3}{c}{Firm identity $F$} & \multicolumn{3}{c}{Idiosyncratic, level $I_L$} & \multicolumn{3}{c}{Idiosyncratic, deviation $I_D$} \\
\cmidrule(lr){3-5} \cmidrule(lr){6-8} \cmidrule(lr){9-11} \cmidrule(lr){12-14} \cmidrule(lr){15-17}
Model & $h$ (months) & Ret & $t$ & Share (\%) & Ret & $t$ & Share (\%) & Ret & $t$ & Share (\%) & Ret & $t$ & Share (\%) & Ret & $t$ & Share (\%) \\
\midrule
RoBERTa & 1 & \phantom{$-$}2.7 & \phantom{$-$}4.05 & \phantom{$-$}15.1 & \phantom{$-$}5.8 & \phantom{$-$}3.54 & \phantom{$-$}32.7 & \phantom{0}5.1 & \phantom{0}4.42 & 28.6 & 1.3 & 1.71 & \phantom{0}7.3 & \phantom{$-$}2.9 & \phantom{$-$}2.19 & \phantom{$-$}16.4 \\
 & 2 & \phantom{$-$}0.6 & \phantom{$-$}1.41 & \phantom{$-$0}5.3 & \phantom{$-$}1.1 & \phantom{$-$}1.13 & \phantom{$-$0}9.7 & \phantom{0}4.9 & \phantom{0}4.50 & 43.4 & 2.7 & 3.84 & 24.0 & \phantom{$-$}2.0 & \phantom{$-$}1.70 & \phantom{$-$}17.7 \\
 & 4 & \phantom{$-$}0.7 & \phantom{$-$}0.98 & \phantom{$-$}11.7 & \phantom{$-$}1.4 & \phantom{$-$}0.96 & \phantom{$-$}23.9 & \phantom{0}4.3 & \phantom{0}3.21 & 70.7 & 0.8 & 1.26 & 13.8 & $-$1.2 & $-$1.40 & $-$20.1 \\
 & 6 & \phantom{$-$}1.4 & \phantom{$-$}2.13 & \phantom{$-$}22.2 & $-$0.6 & $-$0.48 & \phantom{0}$-$8.7 & \phantom{0}2.6 & \phantom{0}1.61 & 39.7 & 1.1 & 1.35 & 16.6 & \phantom{$-$}1.9 & \phantom{$-$}1.74 & \phantom{$-$}30.2 \\
 & 12 & \phantom{$-$}0.2 & \phantom{$-$}0.38 & \phantom{$-$0}4.5 & \phantom{$-$}3.2 & \phantom{$-$}2.29 & \phantom{$-$}69.9 & \phantom{0}0.2 & \phantom{0}0.20 & \phantom{0}4.5 & 0.9 & 1.91 & 19.8 & \phantom{$-$}0.1 & \phantom{$-$}0.08 & \phantom{$-$0}1.3 \\
 & 18 & \phantom{$-$}1.2 & \phantom{$-$}1.40 &  & $-$0.6 & $-$0.37 &  & \phantom{0}0.7 & \phantom{0}0.50 &  & 0.8 & 1.32 &  & $-$0.4 & $-$0.48 &  \\
 & 24 & \phantom{$-$}0.6 & \phantom{$-$}0.82 & \phantom{$-$0}7.1 & \phantom{$-$}3.9 & \phantom{$-$}2.60 & \phantom{$-$}45.5 & \phantom{0}2.4 & \phantom{0}1.60 & 28.3 & 0.8 & 1.00 & \phantom{0}9.3 & \phantom{$-$}0.8 & \phantom{$-$}0.60 & \phantom{$-$0}9.8 \\
\midrule
Llama-2 & 1 & \phantom{$-$}1.6 & \phantom{$-$}2.34 & \phantom{$-$0}4.3 & \phantom{$-$}5.8 & \phantom{$-$}4.03 & \phantom{$-$}16.1 & 18.8 & 10.10 & 52.1 & 7.2 & 5.26 & 19.9 & \phantom{$-$}2.8 & \phantom{$-$}2.11 & \phantom{$-$0}7.6 \\
 & 2 & $-$0.0 & $-$0.11 & \phantom{0}$-$0.2 & \phantom{$-$}1.9 & \phantom{$-$}2.77 & \phantom{$-$0}6.6 & 16.1 & \phantom{0}8.95 & 56.7 & 7.1 & 6.00 & 25.1 & \phantom{$-$}3.3 & \phantom{$-$}3.10 & \phantom{$-$}11.7 \\
 & 4 & \phantom{$-$}0.1 & \phantom{$-$}0.19 & \phantom{$-$0}0.6 & \phantom{$-$}1.7 & \phantom{$-$}1.47 & \phantom{$-$0}7.5 & 12.8 & \phantom{0}8.71 & 57.8 & 7.6 & 5.23 & 34.6 & $-$0.1 & $-$0.11 & \phantom{0}$-$0.5 \\
 & 6 & \phantom{$-$}1.1 & \phantom{$-$}2.51 & \phantom{$-$0}5.1 & $-$0.6 & $-$0.78 & \phantom{0}$-$2.9 & 13.5 & \phantom{0}7.76 & 60.8 & 7.3 & 6.78 & 33.1 & \phantom{$-$}0.9 & \phantom{$-$}0.66 & \phantom{$-$0}3.9 \\
 & 12 & $-$0.1 & $-$0.24 & \phantom{0}$-$0.8 & \phantom{$-$}3.2 & \phantom{$-$}2.43 & \phantom{$-$}18.0 & \phantom{0}9.3 & \phantom{0}5.15 & 51.9 & 4.4 & 4.10 & 24.7 & \phantom{$-$}1.1 & \phantom{$-$}1.03 & \phantom{$-$0}6.1 \\
 & 18 & \phantom{$-$}0.4 & \phantom{$-$}0.86 & \phantom{$-$0}3.9 & \phantom{$-$}0.1 & \phantom{$-$}0.09 & \phantom{$-$0}0.7 & \phantom{0}7.7 & \phantom{0}4.21 & 72.1 & 2.5 & 3.21 & 23.2 & \phantom{$-$}0.0 & \phantom{$-$}0.01 & \phantom{$-$0}0.1 \\
 & 24 & $-$0.5 & $-$0.84 & \phantom{0}$-$3.1 & \phantom{$-$}3.4 & \phantom{$-$}3.49 & \phantom{$-$}21.4 & \phantom{0}7.8 & \phantom{0}3.77 & 49.2 & 3.0 & 4.07 & 19.3 & \phantom{$-$}2.1 & \phantom{$-$}2.51 & \phantom{$-$}13.2 \\
\bottomrule
\end{tabular}
\end{adjustbox}
}
\end{table}

\end{landscape}

\begin{table}[htbp]
\centering
\caption{MSRR Strategy Sharpe Ratios under Alternative Inputs}
\label{tab:tabA_constructions}
\begin{minipage}{\linewidth}\footnotesize
This table reports annualized Sharpe ratios of monthly maximum Sharpe ratio regression (MSRR) strategies on alternative text inputs, for signal months $\tau$ from January 2005 through 2025. Signal month $\tau$ fixes the firms traded, those with news in month $\tau$, and the month of the characteristics. The strategy earns their returns in month $\tau+h$, the $h$-th subsequent return month. The historical level $E_{L,\tau}$ is the mean of the firm's monthly average embedding over its news months from $\tau-12$ to $\tau-1$, or, when the firm has no such months, its firm effect from a two-way category--firm projection of the headline embeddings of the up to six years before the year of $\tau$. The deviation $E_{D,\tau}$ is the remainder of the month-$\tau$ embedding. The lagged level $E_{L,\tau-36}$ is the same mean over news months from $\tau-48$ to $\tau-37$. Lagged firm identity $F_{\tau-36}$ is the firm effect of the same projection fitted three years earlier, on the headlines of years $Y-9$ to $Y-4$ (from 2000 at the earliest), where $Y$ is the year of $\tau$; these headlines end at least 37 months before $\tau$. Each strategy trades only firms for which its input exists. Each input, including $E_{L,\tau-36}$ and $F_{\tau-36}$, is standardized with the coordinate means and standard deviations of all firm-month embeddings through month $\tau-1$, as in DKPT, and residualized on the month-$\tau$ values of 132 JKP characteristics. Rows with an industry control also include GICS-4 industry indicators in the residualization. Managed portfolios are combined out of sample by MSRR. For each input and $h$, the coefficients for signal month $\tau$ are fitted in an expanding window to the horizon-$h$ portfolio returns of all earlier signal months, using only returns realized by month $\tau+h-1$. Signal months start in January 2004 for every input, and the first 12 initialize the fit. News is assigned by arrival month.
\end{minipage}
\medskip

{\footnotesize
\begin{adjustbox}{max width=\linewidth, max totalheight=0.75\textheight}
\begin{tabular}{llccccccc}
\toprule
Model & Input & $h=1$ & $h=2$ & $h=4$ & $h=6$ & $h=12$ & $h=18$ & $h=24$ \\
\midrule
RoBERTa & Embedding $E$ & 1.53 & 1.01 & \phantom{$-$}0.42 & 0.55 & \phantom{$-$}0.48 & \phantom{$-$}0.06 & 0.55 \\
 & Level $E_L$ & 1.57 & 1.26 & \phantom{$-$}0.91 & 0.64 & \phantom{$-$}0.26 & $-$0.04 & 0.20 \\
 & Deviation $E_D$ & 0.83 & 0.42 & $-$0.05 & 0.29 & \phantom{$-$}0.20 & $-$0.07 & 0.42 \\
 & Level 36 months earlier, $E_{L,\,\tau-36}$ & 0.43 & 0.49 & \phantom{$-$}0.44 & 0.11 & \phantom{$-$}0.22 & \phantom{$-$}0.13 & 0.35 \\
 & Firm identity 36 months earlier, $F_{\tau-36}$ & 0.75 & 0.90 & \phantom{$-$}0.58 & 0.19 & $-$0.01 & $-$0.05 & 0.37 \\
 & Embedding $E$, industry control & 1.52 & 1.10 & \phantom{$-$}0.39 & 0.57 & \phantom{$-$}0.79 & \phantom{$-$}0.20 & 0.68 \\
 & Level $E_L$, industry control & 1.67 & 1.33 & \phantom{$-$}0.92 & 0.71 & \phantom{$-$}0.23 & $-$0.14 & 0.38 \\
 & Firm identity 36 months earlier, $F_{\tau-36}$, industry control & 0.84 & 0.88 & \phantom{$-$}0.62 & 0.17 & $-$0.19 & $-$0.16 & 0.28 \\
\midrule
Llama-2 & Embedding $E$ & 3.18 & 2.51 & \phantom{$-$}2.16 & 2.01 & \phantom{$-$}1.76 & \phantom{$-$}0.67 & 1.51 \\
 & Level $E_L$ & 3.51 & 3.21 & \phantom{$-$}3.04 & 2.69 & \phantom{$-$}1.48 & \phantom{$-$}1.09 & 0.75 \\
 & Deviation $E_D$ & 0.82 & 0.35 & $-$0.23 & 0.28 & \phantom{$-$}0.39 & $-$0.17 & 0.41 \\
 & Level 36 months earlier, $E_{L,\,\tau-36}$ & 1.96 & 1.78 & \phantom{$-$}1.89 & 1.44 & \phantom{$-$}0.99 & \phantom{$-$}1.02 & 0.81 \\
 & Firm identity 36 months earlier, $F_{\tau-36}$ & 2.36 & 2.30 & \phantom{$-$}2.20 & 1.85 & \phantom{$-$}1.20 & \phantom{$-$}1.20 & 0.88 \\
 & Embedding $E$, industry control & 3.31 & 2.88 & \phantom{$-$}2.24 & 2.28 & \phantom{$-$}2.01 & \phantom{$-$}0.84 & 1.72 \\
 & Level $E_L$, industry control & 3.77 & 3.45 & \phantom{$-$}3.46 & 2.93 & \phantom{$-$}1.85 & \phantom{$-$}1.24 & 0.80 \\
 & Firm identity 36 months earlier, $F_{\tau-36}$, industry control & 2.65 & 2.50 & \phantom{$-$}2.43 & 2.08 & \phantom{$-$}1.48 & \phantom{$-$}1.36 & 1.07 \\
\bottomrule
\end{tabular}
\end{adjustbox}
}
\end{table}

\begin{figure}[htbp]
\centering
\caption{Annual Performance of News-Based Signals}
\includegraphics[width=1\textwidth]{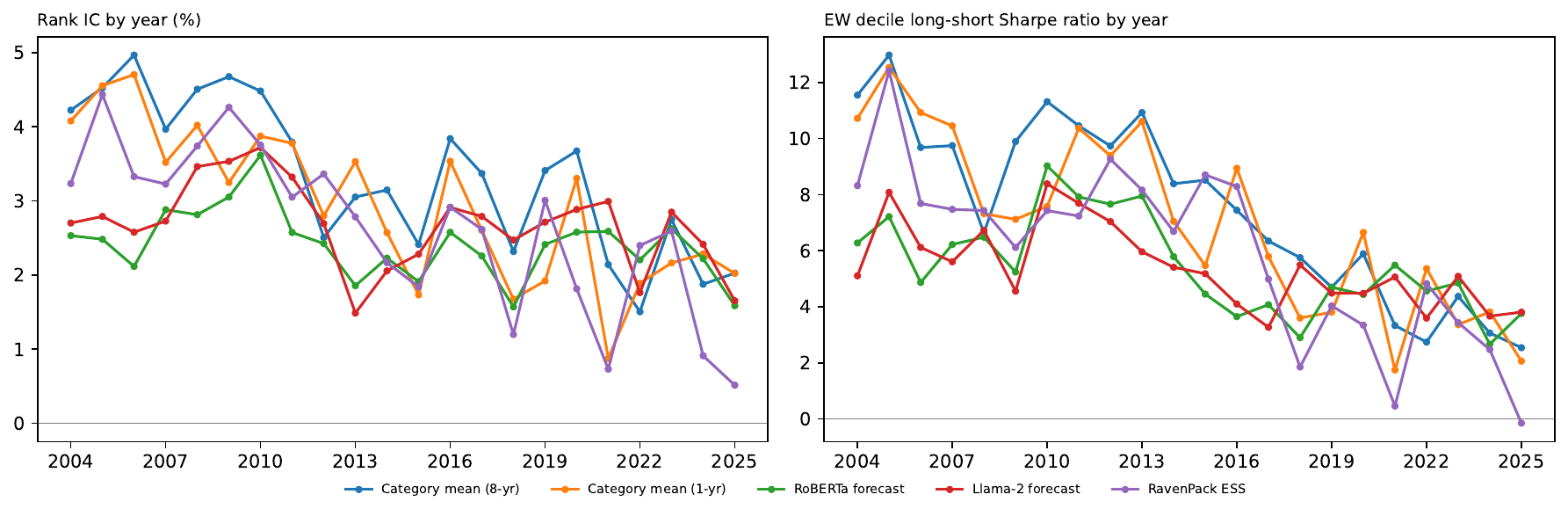}
\includegraphics[width=1\textwidth]{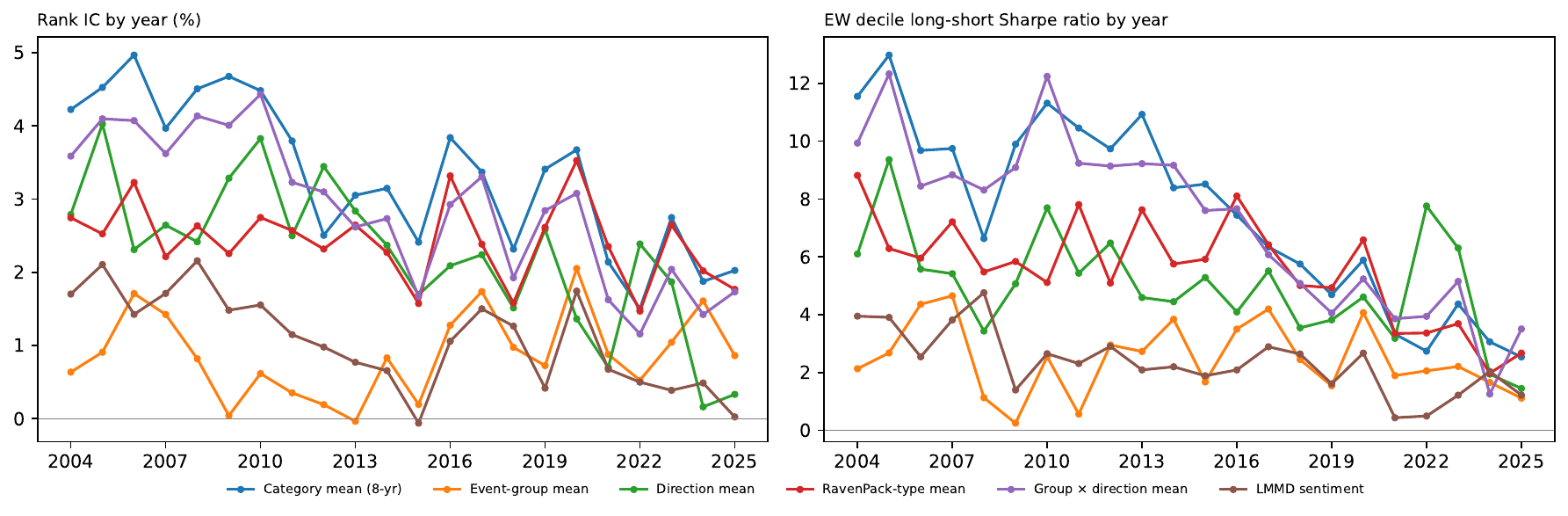}
\label{fig:annual_performance}
\begin{minipage}{\linewidth}\footnotesize
This figure reports annual performance of news-based signals, 2004--2025. The upper panels show one- and eight-year category forecasts, RoBERTa-large and Llama-2 (13B) ridge forecasts, and RavenPack event sentiment. The lower panels show forecasts formed from coarser RavenPack classifications and headline direction, and Loughran--McDonald dictionary sentiment, with the eight-year category forecast as the benchmark. Direction is positive, negative, or neutral under a fixed word rule. Sentiment is the positive-minus-negative dictionary word count divided by headline word count. Forecasts are averaged within firm-days. Left panels show mean daily Spearman correlations with next open-to-open returns, in percent, and right panels show annualized equal-weighted decile long-short Sharpe ratios.\end{minipage}
\end{figure}
\begin{figure}[htbp]
\centering
\caption{Cumulative Returns after News by Firm Size}
\includegraphics[width=0.85\textwidth]{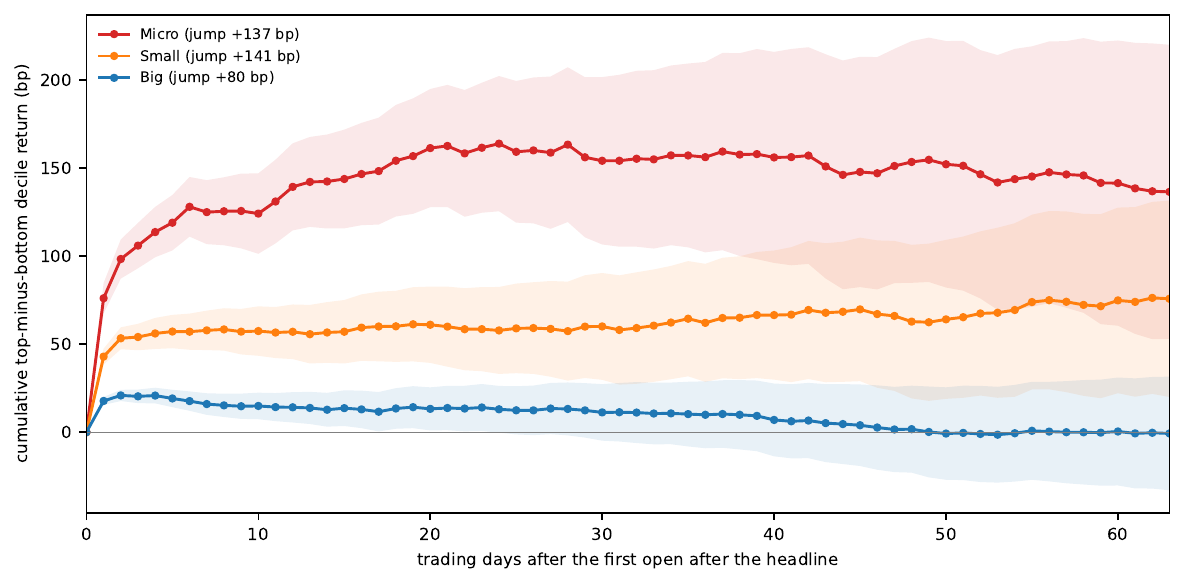}
\label{fig:figA_drift_size}
\begin{minipage}{\linewidth}\footnotesize
This figure plots cumulative returns of decile long-short portfolios sorted on the eight-year category mean within micro, small, and big stocks, 2004--2025. Micro, small, and big stocks are below the NYSE 20th capitalization percentile, between it and the NYSE median, and at or above the median, on the preceding trading day. Deciles are formed each day among the group's firm-days. The horizontal axis counts trading days after the first eligible open, and curves sum subsequent open-to-open returns, in basis points, averaged equally within deciles and across trading days. The sample includes headlines with an observed initial return and a complete 63-day return path. Legends report the jump, the initial return to the eligible open from the preceding close or, for news published during the trading session (09:30--16:00), from that session's open. Curves start at zero at the eligible open and exclude the jump. Shading shows 95\% Newey--West confidence intervals with $\max(5,h)$ lags at day $h$.
\end{minipage}
\end{figure}

\begin{figure}[htbp]
\centering
\caption{Cumulative Variance and Predictive Covariance by Principal Direction}
\label{fig:figA_spectrum}
\includegraphics[width=0.85\textwidth]{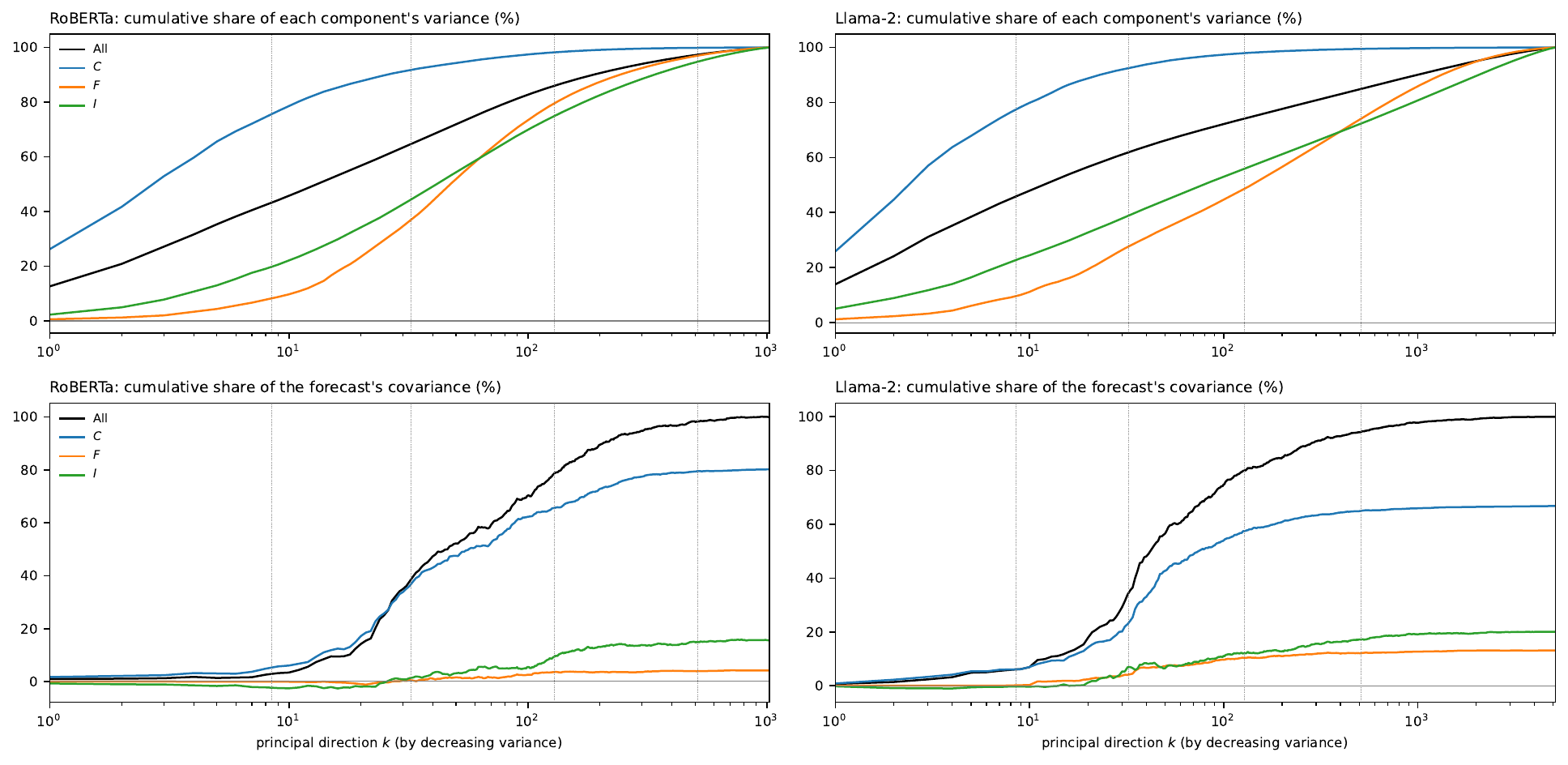}
\begin{minipage}{\linewidth}\footnotesize
This figure plots cumulative embedding variance and forecast-return covariance across principal directions, ordered by decreasing training variance on a logarithmic axis, 2004--2025. Components are category effects ($C$), firm effects ($F$), and idiosyncratic components ($I$) from a two-way projection on the ridge training window, and All is the full embedding (upper panels) or the forecast (lower panels). Upper panels normalize each component and the full embedding by its own total variance and average test years. Lower panels normalize all contributions by the forecast's covariance with next open-to-open returns, pooling firm-days after removing day means. Dotted lines separate principal-component bands.
\end{minipage}
\end{figure}

\begin{figure}[htbp]
\centering
\caption{Monthly Return Covariances of the Averaged Daily Forecast}
\includegraphics[width=1\textwidth]{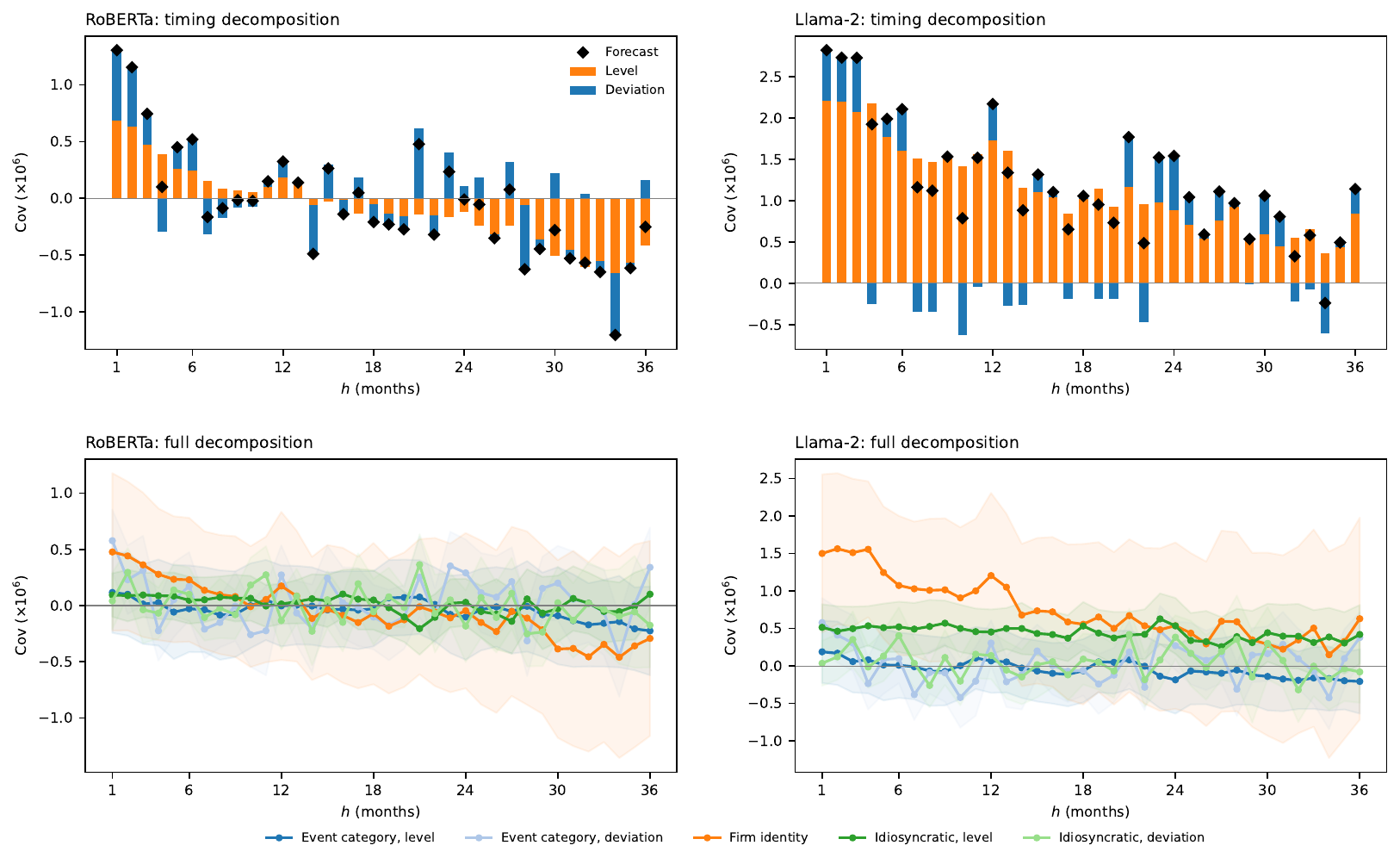}
\label{fig:fig2b}
\begin{minipage}{\linewidth}\footnotesize
This figure decomposes the covariance of the headline-weighted monthly average of daily forecasts with returns across monthly horizons, for signal years 2004--2025. News is assigned by trading month; horizons count months after the signal month, $h=1,\ldots,36$. The historical level is the mean of the firm's monthly average forecast over its news months in the preceding 12 months, computed with the current coefficients; without such months it is the forecast from the firm effect of a two-way category--firm projection of the headline embeddings of the up to six years before the signal year, or zero when the firm has $F=0$. The deviation is the remainder. Upper panels separate historical level and deviation; lower panels show the category, firm, and idiosyncratic components of the two-way projection, the first and last split into level and deviation. Diamonds show the total, to which the components sum. Covariances are cross-sectional, pooled across firm-months, and multiplied by $10^6$. Shading shows 95\% Newey--West confidence intervals with 6 lags.
\end{minipage}
\end{figure}

\begin{figure}[htbp]
\centering
\caption{Monthly Forecast Covariance with Returns across Horizons}
\includegraphics[width=1\textwidth]{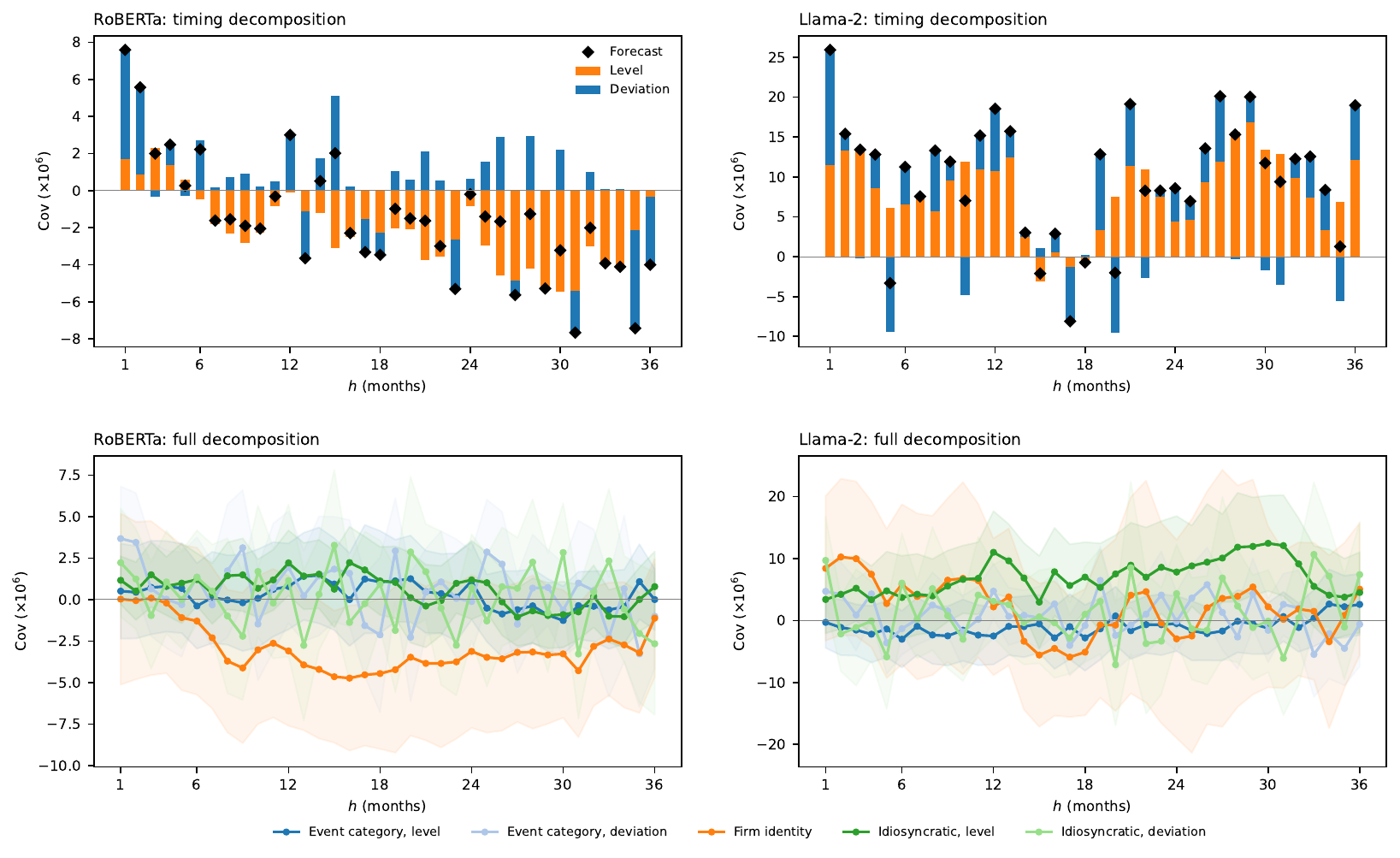}
\label{fig:fig2c}
\begin{minipage}{\linewidth}\footnotesize
This figure decomposes the covariance of the ridge forecast fitted to monthly average embeddings with returns across monthly horizons, for signal years 2004--2025. News is assigned by trading month; horizons count months after the signal month, $h=1,\ldots,36$. The historical level is the mean of the firm's monthly forecast over its news months in the preceding 12 months; the deviation is the remainder. Upper panels separate historical level and deviation; lower panels show the category, firm, and idiosyncratic components of the two-way projection, the first and last split into level and deviation. Diamonds show the total, to which the components sum. Covariances are cross-sectional, pooled across firm-months, and multiplied by $10^6$. Shading shows 95\% Newey--West confidence intervals with 6 lags.
\end{minipage}
\end{figure}
\begin{figure}[htbp]
\centering
\caption{Daily Forecast Level and Deviation across Horizons}
\includegraphics[width=1\textwidth]{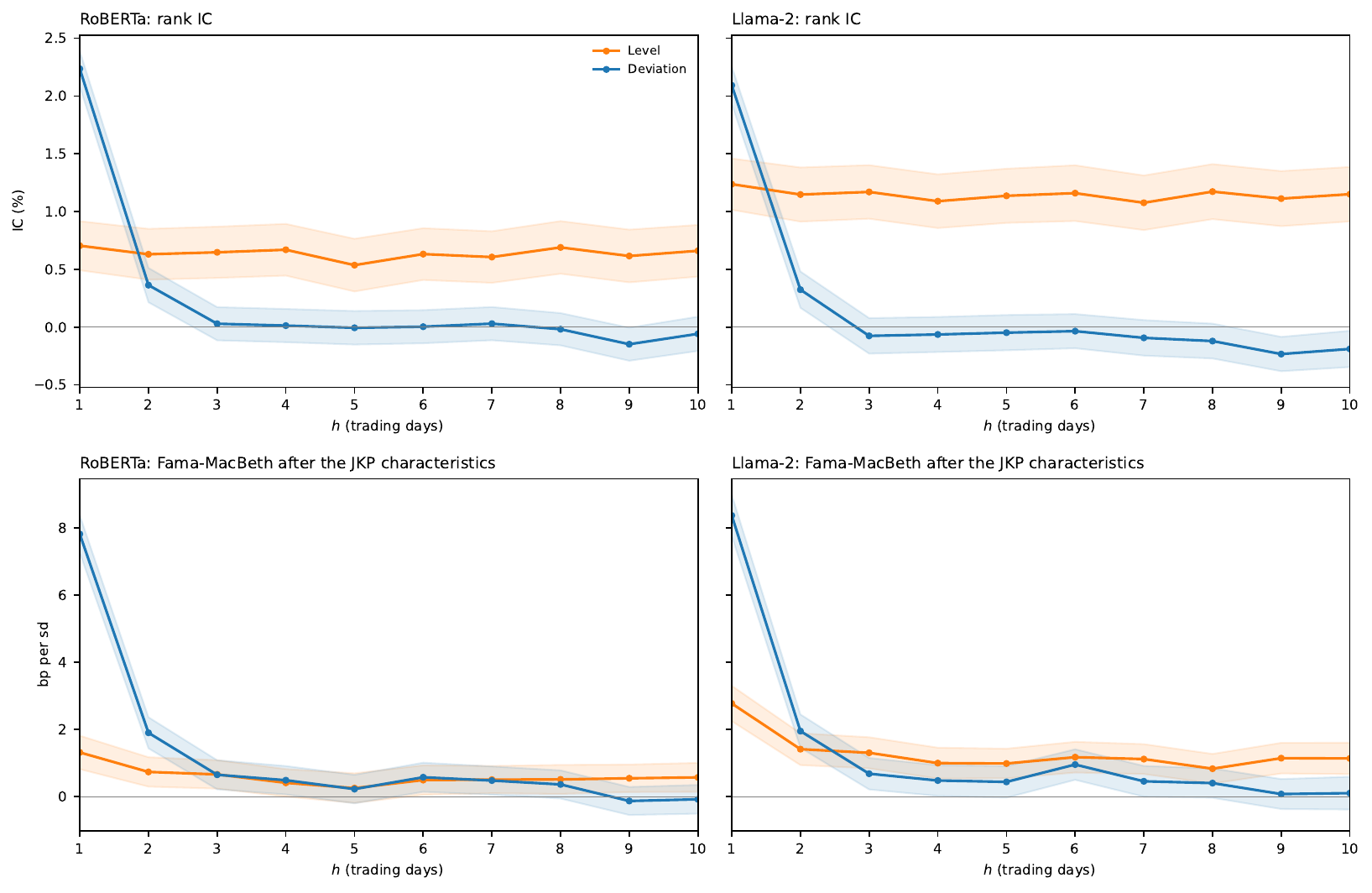}
\label{fig:figA_persistence_a}
\begin{minipage}{\linewidth}\footnotesize
This figure reports the historical level and the deviation of the daily ridge forecast as predictors of returns across trading-day horizons, for signal years 2004--2025. Horizon $h$ is the $h$-th open-to-open return after the news. The historical level is the mean of the firm's monthly average forecast over its news months in the preceding 12 months, computed with the current coefficients; without such months it is the forecast from the firm effect of a two-way category--firm projection of the headline embeddings of the up to six years before the signal year, or zero when the firm has $F=0$. The deviation is the remainder. Upper panels report each component's rank IC, the mean cross-sectional Spearman correlation with the horizon-$h$ return, in percent. Lower panels report the two components' coefficients in joint Fama--MacBeth regressions after both are residualized on JKP characteristics, in basis points per standard deviation; the residualization is estimated on the preceding month's firm-days with preceding month-end characteristics and applied to the current month. Shading shows 95\% Newey--West confidence intervals with 5 lags.
\end{minipage}
\end{figure}
\begin{figure}[htbp]
\centering
\caption{Averaged Daily Forecast Level and Deviation across Horizons}
\includegraphics[width=1\textwidth]{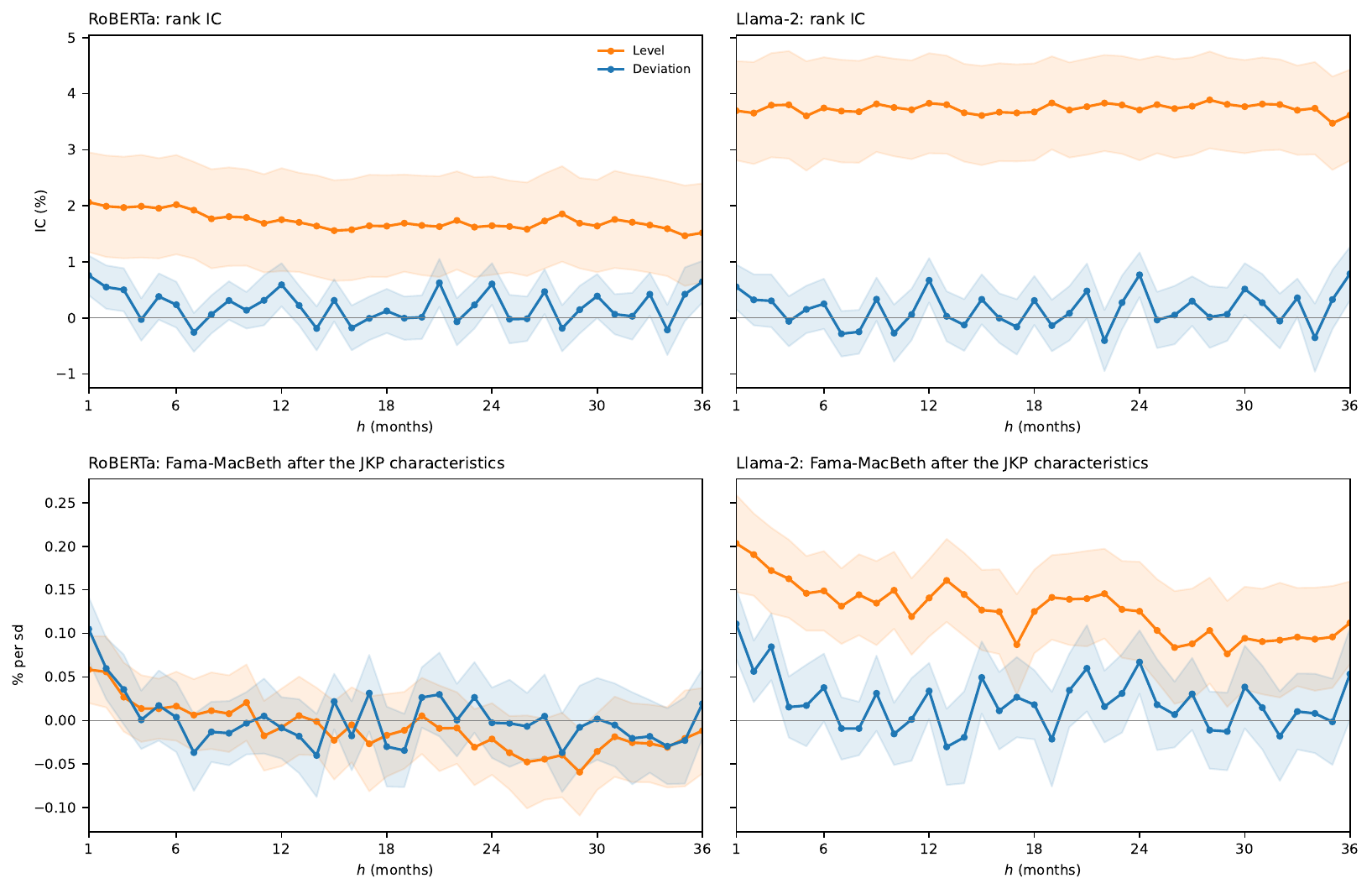}
\label{fig:figA_persistence_b}
\begin{minipage}{\linewidth}\footnotesize
This figure reports the historical level and the deviation of the headline-weighted monthly average of daily forecasts as predictors of returns across monthly horizons, for signal years 2004--2025. News is assigned by trading month; horizons count months after the signal month. The historical level is the mean of the firm's monthly average forecast over its news months in the preceding 12 months, computed with the current coefficients; without such months it is the forecast from the firm effect of a two-way category--firm projection of the headline embeddings of the up to six years before the signal year, or zero when the firm has $F=0$. The deviation is the remainder. Upper panels report each component's rank IC, the mean cross-sectional Spearman correlation with the horizon-$h$ return, in percent. Lower panels report the two components' coefficients in joint Fama--MacBeth regressions after both are residualized cross-sectionally on signal-month-end JKP characteristics, in percent per standard deviation. Shading shows 95\% Newey--West confidence intervals with 6 lags.
\end{minipage}
\end{figure}
\begin{figure}[htbp]
\centering
\caption{Monthly Forecast Level and Deviation across Horizons}
\includegraphics[width=1\textwidth]{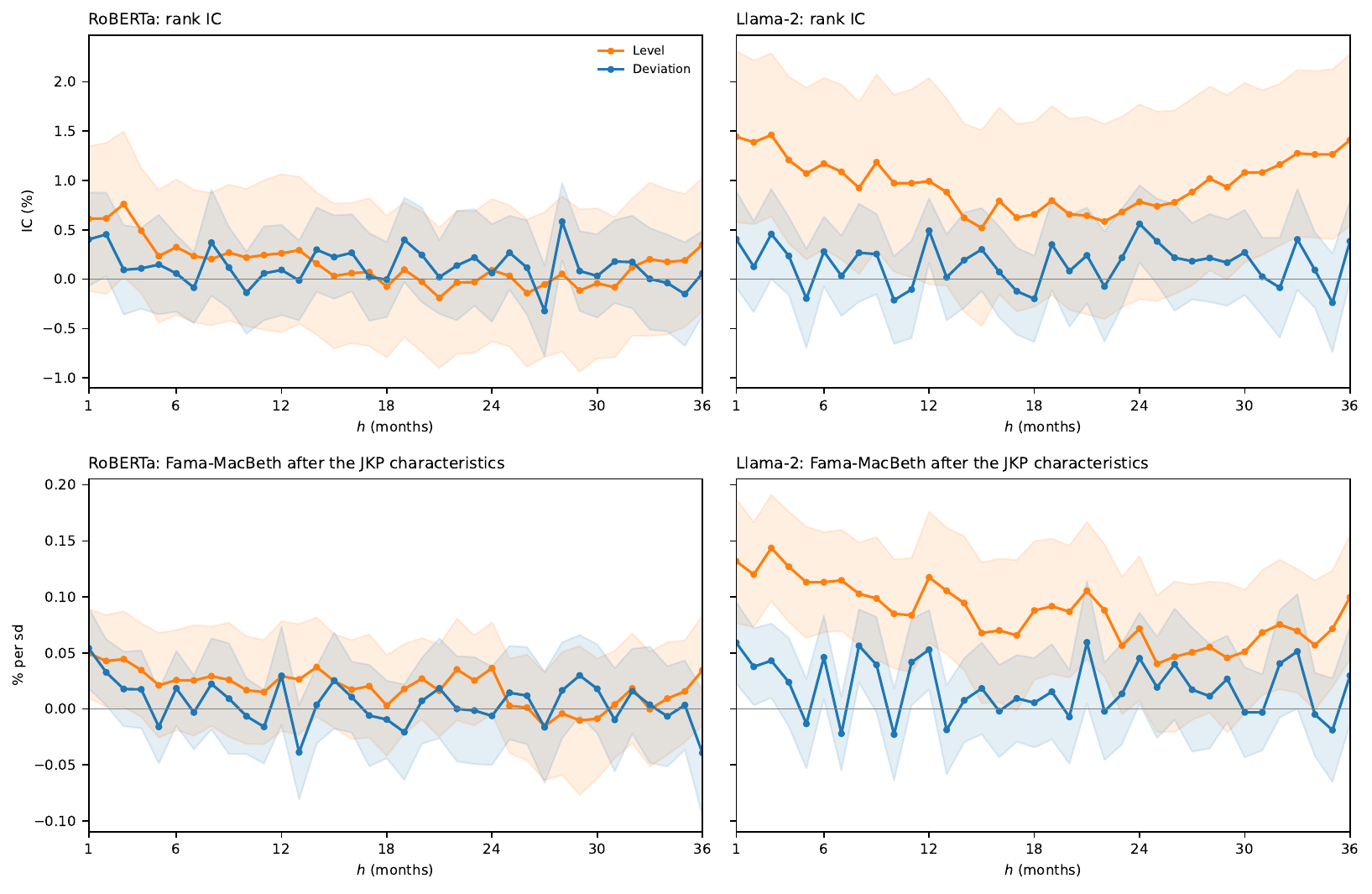}
\label{fig:figA_persistence_c}
\begin{minipage}{\linewidth}\footnotesize
This figure reports the historical level and the deviation of the ridge forecast fitted to monthly average embeddings as predictors of returns across monthly horizons, for signal years 2004--2025. News is assigned by trading month; horizons count months after the signal month. The historical level is the mean of the firm's monthly forecast over its news months in the preceding 12 months; the deviation is the remainder. Upper panels report each component's rank IC, the mean cross-sectional Spearman correlation with the horizon-$h$ return, in percent. Lower panels report the two components' coefficients in joint Fama--MacBeth regressions after both are residualized cross-sectionally on signal-month-end JKP characteristics, in percent per standard deviation. Shading shows 95\% Newey--West confidence intervals with 6 lags.
\end{minipage}
\end{figure}
\begin{figure}[htbp]
\centering
\caption{Averaged Daily Forecast Decomposition by Firm Size}
\includegraphics[width=\textwidth,height=0.75\textheight,keepaspectratio]{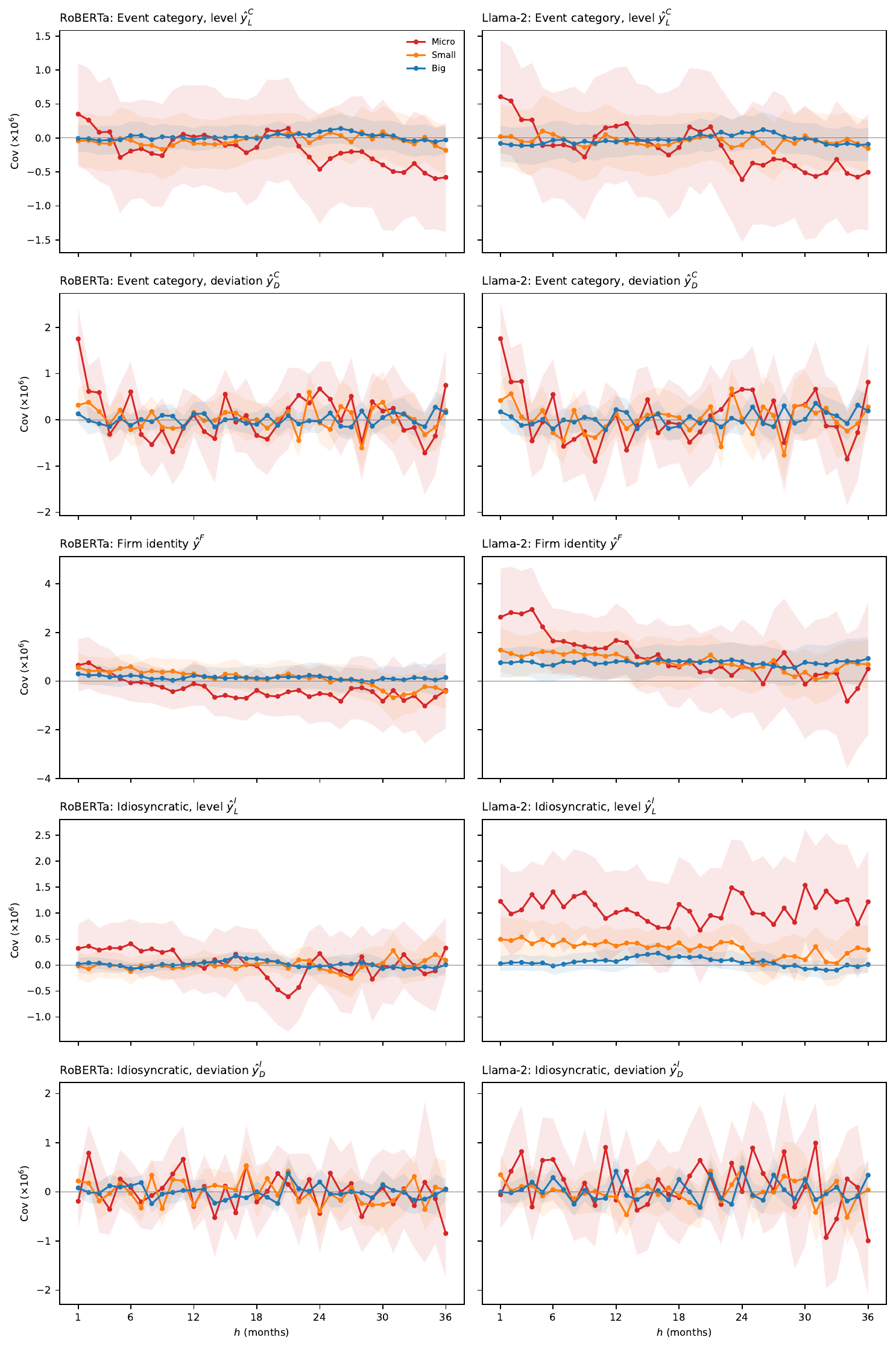}
\label{fig:figA_bridge_size}
\begin{minipage}{\linewidth}\footnotesize
This figure reports the covariance of each component of the headline-weighted monthly average of daily forecasts with returns $h$ months after the signal month, $h=1,\ldots,36$, within micro, small, and big stocks, for signal years 2004--2025. Components are the category level and deviation, firm identity, and the idiosyncratic level and deviation of the two-way category--firm projection. Historical levels average the category and idiosyncratic components' monthly means over the firm's news months in the preceding 12 months; firm identity lies entirely in the level, and deviations are the remainders. Size groups are JKP's at the signal month end: micro stocks are below the NYSE 20th capitalization percentile, small stocks between it and the NYSE median, and big stocks above the median. News is assigned by trading month. Covariances are cross-sectional within each month and group, pooled across firm-months, and multiplied by $10^6$. Shading shows 95\% Newey--West confidence intervals with 6 lags.
\end{minipage}
\end{figure}
\begin{figure}[htbp]
\centering
\caption{MSRR Strategy Return Decomposition by Firm Size}
\includegraphics[width=\textwidth,height=0.75\textheight,keepaspectratio]{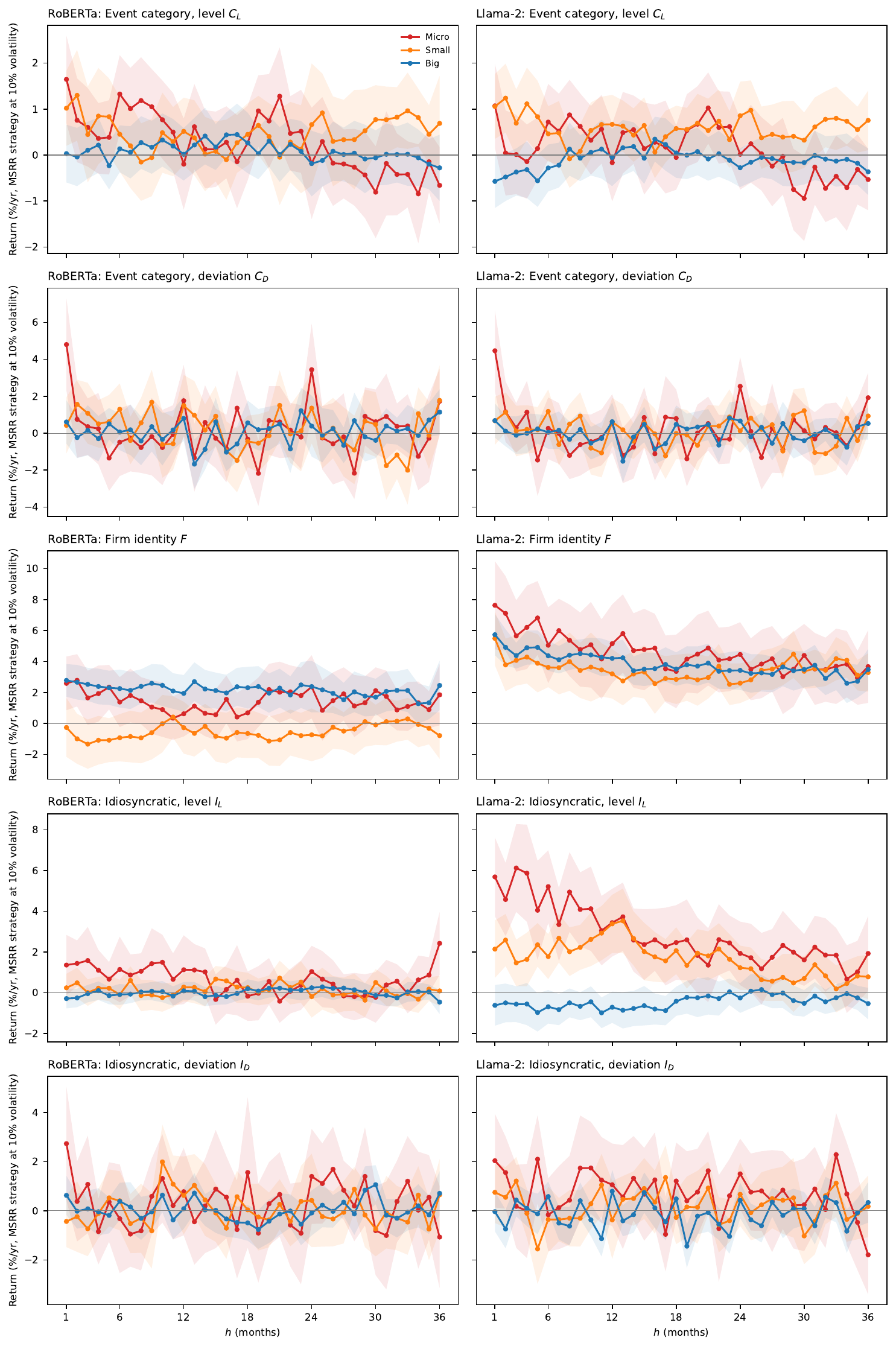}
\label{fig:figA_attribution_size}
\begin{minipage}{\linewidth}\footnotesize
This figure splits each component's contribution to monthly maximum Sharpe ratio regression (MSRR) strategy returns by the size group of each firm in the signal month, for signal years 2004--2025. Components are the category level and deviation, firm identity, and the idiosyncratic level and deviation of the two-way category--firm projection, each residualized each month on JKP characteristics. Historical levels average the category and idiosyncratic components' monthly means over the firm's news months in the preceding 12 months; firm identity lies entirely in the level, and deviations are the remainders. Size groups are JKP's at the signal month end: micro stocks are below the NYSE 20th capitalization percentile, small stocks between it and the NYSE median, and big stocks above the median. Each group's contribution sums its firms' returns under the same portfolio coefficients, so the three groups add up to the component's full-sample contribution. News is assigned by arrival month; horizon $h$ is the $h$-th subsequent return month, $h=1,\ldots,36$, and every horizon uses the signal month's $h=1$ MSRR coefficients. Returns are annualized percentages after scaling the strategy to 10\% annual volatility at each horizon. Shading shows 95\% Newey--West confidence intervals with 6 lags.
\end{minipage}
\end{figure}
\begin{figure}[htbp]
\centering
\caption{MSRR Signal Level and Deviation across Horizons}
\includegraphics[width=1\textwidth]{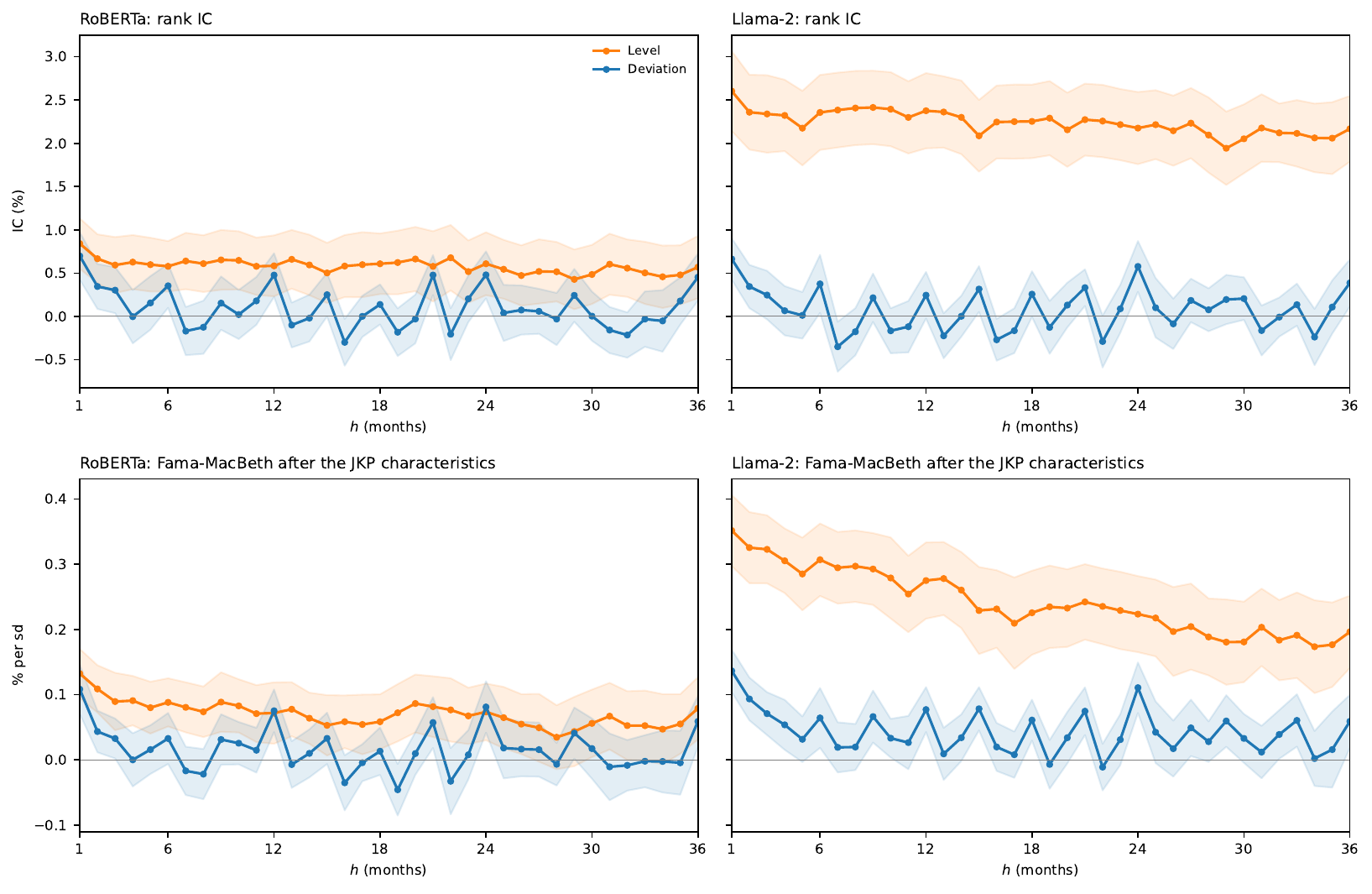}
\label{fig:figA_persistence_strategy}
\begin{minipage}{\linewidth}\footnotesize
This figure reports the historical level and the deviation of the monthly maximum Sharpe ratio regression (MSRR) signal as predictors of returns across monthly horizons, for signal years 2004--2025. Firm-month embeddings are residualized on JKP characteristics and combined out of sample using MSRR. The historical level is the mean of the firm's monthly average embedding over its news months in the preceding 12 months; the deviation is the remainder. Each is residualized like the embedding and projected onto its signal month's $h=1$ MSRR coefficients. News is assigned by arrival month; horizon $h$ is the $h$-th subsequent return month. Upper panels report each component's rank IC, the mean cross-sectional Spearman correlation with the horizon-$h$ return, in percent. Lower panels report the two components' coefficients in joint Fama--MacBeth regressions, in percent per standard deviation. Both panels use the JKP returns of firms with news in the signal month, with or without news in the return month. Shading shows 95\% Newey--West confidence intervals with 6 lags.
\end{minipage}
\end{figure}
\begin{figure}[htbp]
\centering
\caption{Return Regressions on Lagged Deviations across Horizons}
\includegraphics[width=0.8\textwidth]{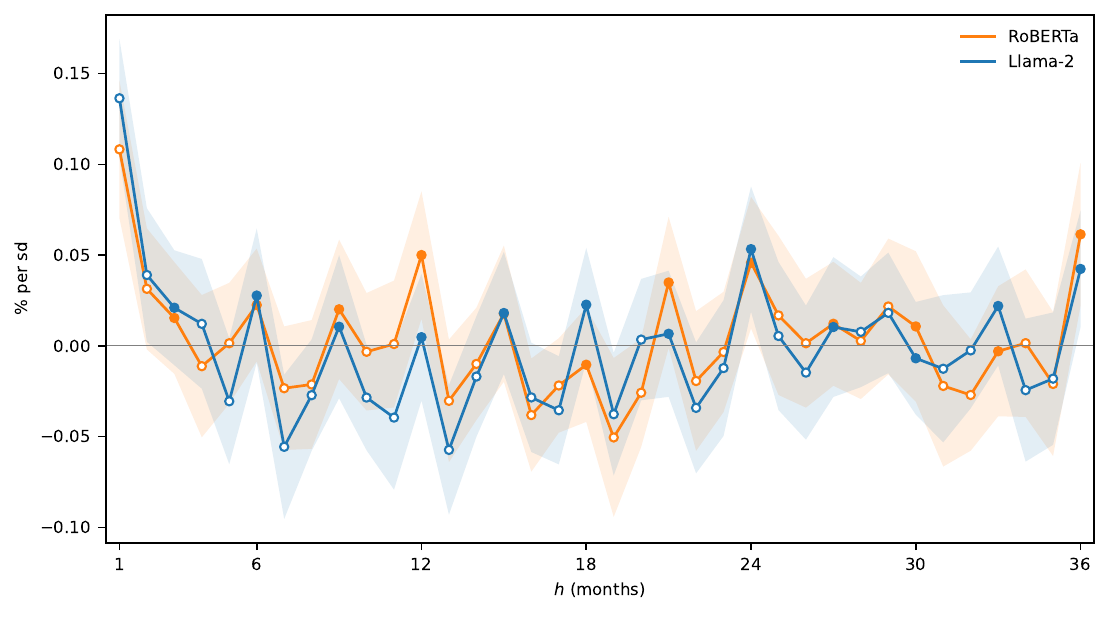}
\label{fig:figA_old_news}
\begin{minipage}{\linewidth}\footnotesize
This figure reports the coefficient on the deviation of month $\tau-k$ in monthly Fama--MacBeth regressions of next-month returns on this deviation and the historical level of month $\tau$, for firms with news in both months and signal years 2004--2025. The historical level is the mean of the firm's monthly average embedding over its news months in the preceding 12 months, or, without such months, the firm effect of a two-way category--firm projection of the headline embeddings of the up to six years before the signal year; the deviation is the remainder. News is assigned by arrival month. Each component is standardized and residualized on JKP characteristics, converted into a score by its inner product with the residualized-embedding strategy's maximum Sharpe ratio regression (MSRR) coefficients of its own signal month, and standardized cross-sectionally; coefficients are percentage returns per standard deviation of the score. The horizontal axis is the return month's distance from the deviation month, $h=k+1$. Filled markers are horizons that are multiples of three months. Shading shows 95\% Newey--West confidence intervals with 6 lags.
\end{minipage}
\end{figure}
\begin{figure}[htbp]
\centering
\caption{Rolling Rank IC of Monthly MSRR Levels and Deviations}
\includegraphics[width=0.85\textwidth]{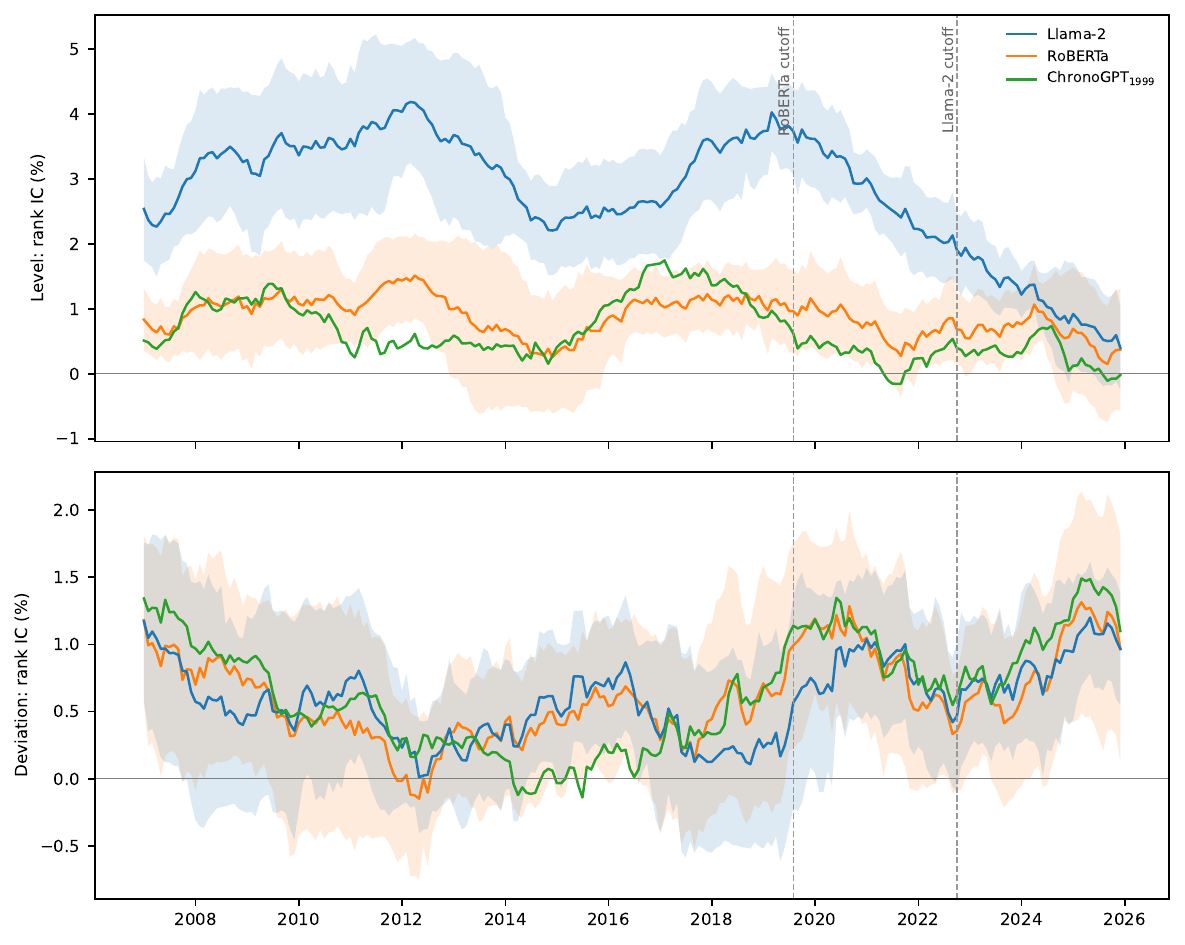}
\label{fig:figA_cutoff_monthly}
\begin{minipage}{\linewidth}\footnotesize
This figure reports the rank IC of the historical level (upper panel) and deviation (lower panel) of the monthly maximum Sharpe ratio regression (MSRR) signal for RoBERTa-large, Llama-2 (13B), and ChronoGPT$_{1999}$, for signal years 2004--2025. News is assigned by arrival month. The historical level is the mean of the firm's monthly average embedding over its news months in the preceding 12 months; the deviation is the remainder. Both are residualized on JKP characteristics and projected onto the signal month's $h=1$ MSRR coefficients, estimated out of sample. Rank IC is the cross-sectional Spearman correlation with the next month's JKP excess return, in percent, averaged over the 36 signal months ending at each date, among the signal months shared by the three encoders. Shading shows 95\% Newey--West confidence intervals for RoBERTa and Llama-2 with 6 lags within each window. Dashed lines mark the pretraining cutoffs of RoBERTa (July 2019) and Llama-2 (September 2022).
\end{minipage}
\end{figure}
\begin{figure}[htbp]
\centering
\caption{Cumulative Returns of Daily Forecast Portfolios by Component and Encoder}
\includegraphics[width=0.85\textwidth]{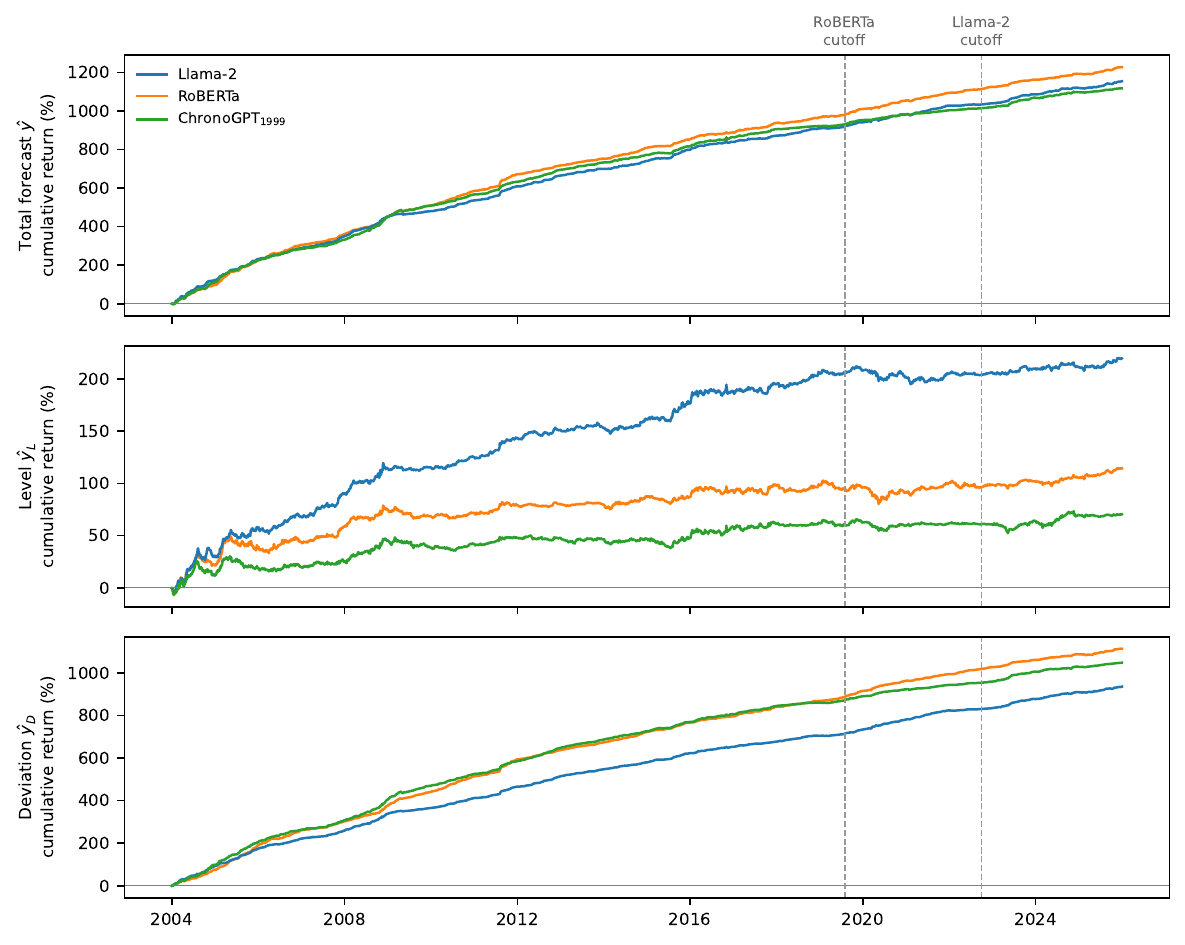}
\label{fig:figA_cutoff_returns_daily}
\begin{minipage}{\linewidth}\footnotesize
This figure plots cumulative returns of a daily long-short portfolio formed on the ridge forecast and of its historical level and deviation components for RoBERTa-large, Llama-2 (13B), and ChronoGPT$_{1999}$, for signal years 2004--2025. Portfolio weights are the forecast minus its cross-sectional mean on the day, divided by the sample's average number of firms per day, so the next open-to-open return splits into level and deviation components. The historical level is the mean of the firm's monthly average forecast over its news months in the preceding 12 months, computed with the current coefficients; without such months it is the forecast from the firm effect of a two-way category--firm projection of the headline embeddings of the up to six years before the signal year, or zero when the firm has $F=0$. The deviation is the remainder. Each encoder's portfolio is scaled to 10\% annual volatility over the plotted days, with the same scale for its components. Cumulative returns sum daily percentage returns over the trading days shared by the three encoders. Dashed lines mark the pretraining cutoffs of RoBERTa (July 2019) and Llama-2 (September 2022).
\end{minipage}
\end{figure}

\FloatBarrier
\clearpage
\section{Results with ChronoGPT$_{1999}$}
\label{sec:appendixB}

\begin{table}[htbp]
\centering
\caption{Category and Embedding Forecasts with ChronoGPT$_{1999}$}
\label{tab:tabB_residual}
\begin{minipage}{\linewidth}\footnotesize
This table reports the contribution of ChronoGPT$_{1999}$ embeddings beyond category forecasts of next open-to-open returns in signal years 2004--2025 and 2023--2025. Construction as in Table~\ref{tab:tab2}, except for the encoder. The 2023--2025 results evaluate the same forecasts. ChronoGPT$_{1999}$ is the December 1999 vintage of \citet{he2026chronologically}; embeddings average final-layer representations over non-padding tokens.
\end{minipage}
\medskip

{\footnotesize
\textit{Panel A: Fama--MacBeth regressions, standardized signals (bp per standard deviation)}\\[2pt]
\begin{adjustbox}{max width=\linewidth, max totalheight=0.31\textheight}
\begin{tabular}{lcccccc}
\toprule
 & \multicolumn{3}{c}{2004--2025} & \multicolumn{3}{c}{2023--2025} \\
\cmidrule(lr){2-4} \cmidrule(lr){5-7}
 & (1) & (2) & (3) & (4) & (5) & (6) \\
\midrule
Embedding forecast & \phantom{(0}6.38\phantom{)} & \phantom{(0}2.04\phantom{)} & \phantom{(0}2.80\phantom{)} & \phantom{(}6.16\phantom{)} & \phantom{(}3.10\phantom{)} & \phantom{(}4.03\phantom{)} \\
 & (22.40) & \phantom{0}(7.00) & \phantom{0}(9.58) & (6.71) & (3.40) & (4.34) \\
Category mean (8-yr) &  & \phantom{(}10.10\phantom{)} &  &  & \phantom{(}7.42\phantom{)} &  \\
 &  & (28.97) &  &  & (6.17) &  \\
Group $\times$ direction mean &  &  & \phantom{(0}8.55\phantom{)} &  &  & \phantom{(}5.20\phantom{)} \\
 &  &  & (27.56) &  &  & (6.29) \\
\bottomrule
\end{tabular}
\end{adjustbox}
\par\medskip
\textit{Panel B: Out of sample: the category mean plus a ridge forecast of its residual}\\[2pt]
\begin{adjustbox}{max width=\linewidth, max totalheight=0.24\textheight}
\begin{tabular}{llcccccccc}
\toprule
\multicolumn{2}{c}{} & \multicolumn{2}{c}{Rank IC (\%)} & \multicolumn{2}{c}{EW SR} & \multicolumn{2}{c}{VW SR} & \multicolumn{2}{c}{OOS $R^2$} \\
\cmidrule(lr){3-4} \cmidrule(lr){5-6} \cmidrule(lr){7-8} \cmidrule(lr){9-10}
Signal years & Forecast & Mean & DM $t$ & SR & JK $z$ & SR & JK $z$ & bp & $t$ \\
\midrule
2004--2025 & Category mean (8-yr) & 3.33 &  & 6.35 &  & 2.30 &  & 8.75 &  \\
 & + residual ridge, ChronoGPT$_{1999}$ & 3.30 & $-$0.89 & 6.99 & 5.17 & 2.69 & 2.85 & 8.18 & $-$1.88 \\
\midrule
2023--2025 & Category mean (8-yr) & 2.22 &  & 3.31 &  & 0.22 &  & 6.30 &  \\
 & + residual ridge, ChronoGPT$_{1999}$ & 2.38 & \phantom{$-$}1.16 & 5.00 & 3.49 & 0.40 & 0.32 & 6.56 & \phantom{$-$}0.28 \\
\bottomrule
\end{tabular}
\end{adjustbox}
}
\end{table}

\begin{table}[htbp]
\centering
\caption{Forecast Decomposition with ChronoGPT$_{1999}$}
\label{tab:tabB_split}
\begin{minipage}{\linewidth}\footnotesize
This table decomposes daily forecasts of next open-to-open returns from ChronoGPT$_{1999}$ in signal years 2004--2025 and 2023--2025. Construction as in Table~\ref{tab:tab6}, except for the encoder. The 2023--2025 results evaluate the same forecasts.
\end{minipage}
\medskip

{\footnotesize
\begin{adjustbox}{max width=\linewidth, max totalheight=0.75\textheight}
\begin{tabular}{llccccc}
\toprule
\multicolumn{2}{c}{} & \multicolumn{2}{c}{Share (\%)} & \multicolumn{3}{c}{Standalone performance} \\
\cmidrule(lr){3-4} \cmidrule(lr){5-7}
Signal years & Decomposition & $\mathrm{Cov}(\hat y^j,r)/\mathrm{Cov}(\hat y,r)$ & $\mathrm{Var}(\hat y^j)/\mathrm{Var}(\hat y)$ & Rank IC (\%) & EW SR & VW SR \\
\midrule
2004--2025 & Event category $\hat y^C$ & \phantom{$-$0}82.2 & \phantom{0}32.4 & 2.72 & \phantom{$-$}5.75 & \phantom{$-$}1.81 \\
 & Firm identity $\hat y^F$ & \phantom{$-$00}4.2 & \phantom{0}23.4 & 0.24 & \phantom{$-$}0.11 & \phantom{$-$}0.18 \\
 & Idiosyncratic $\hat y^I$ & \phantom{$-$0}13.6 & \phantom{0}58.2 & 0.25 & \phantom{$-$}1.03 & $-$0.01 \\
 & Total forecast $\hat y$ & \phantom{$-$}100.0 & 100.0 & 1.90 & \phantom{$-$}4.50 & \phantom{$-$}1.02 \\
\midrule
2023--2025 & Event category $\hat y^C$ & \phantom{$-$0}73.4 & \phantom{0}30.1 & 2.24 & \phantom{$-$}3.54 & $-$0.78 \\
 & Firm identity $\hat y^F$ & \phantom{00}$-$2.7 & \phantom{0}28.5 & 0.06 & $-$0.43 & $-$0.48 \\
 & Idiosyncratic $\hat y^I$ & \phantom{$-$0}29.3 & \phantom{0}48.0 & 0.55 & \phantom{$-$}1.74 & $-$0.05 \\
 & Total forecast $\hat y$ & \phantom{$-$}100.0 & 100.0 & 1.65 & \phantom{$-$}3.67 & \phantom{$-$}0.54 \\
\bottomrule
\end{tabular}
\end{adjustbox}
}
\end{table}

\begin{table}[htbp]
\centering
\caption{Forecast Return Covariances with ChronoGPT$_{1999}$}
\label{tab:tabB_horizons}
\begin{minipage}{\linewidth}\footnotesize
This table reports forecast covariances with returns across horizons for ChronoGPT$_{1999}$ in signal years 2004--2025 and 2023--2025. Construction as in Table~\ref{tab:tab7}, except for the encoder. The 2023--2025 results evaluate the same forecasts. Monthly horizons require at least 24 signal months.
\end{minipage}
\medskip

{\footnotesize
\textit{Panel A: Daily forecast}\\[2pt]
\begin{adjustbox}{max width=\linewidth, max totalheight=0.17\textheight}
\begin{tabular}{llcccccccc}
\toprule
\multicolumn{2}{c}{} & \multicolumn{2}{c}{Forecast} & \multicolumn{3}{c}{Level $\hat y_L$} & \multicolumn{3}{c}{Deviation $\hat y_D$} \\
\cmidrule(lr){3-4} \cmidrule(lr){5-7} \cmidrule(lr){8-10}
Signal years & $h$ (trading days) & Cov & $t$ & Cov & $t$ & Share (\%) & Cov & $t$ & Share (\%) \\
\midrule
2004--2025 & 0 & \phantom{$-$}2.28 & \phantom{$-$}33.16 & $-$0.10 & $-$4.38 & \phantom{0}$-$4.4 & \phantom{$-$}2.38 & \phantom{$-$}36.39 & 104.4 \\
 & 1 & \phantom{$-$}0.81 & \phantom{$-$}21.08 & \phantom{$-$}0.05 & \phantom{$-$}2.22 & \phantom{$-$0}6.3 & \phantom{$-$}0.76 & \phantom{$-$}23.76 & \phantom{0}93.7 \\
 & 2 & \phantom{$-$}0.19 & \phantom{$-$0}6.19 & \phantom{$-$}0.03 & \phantom{$-$}1.54 & \phantom{$-$}16.2 & \phantom{$-$}0.16 & \phantom{$-$0}6.41 & \phantom{0}83.8 \\
 & 3 & \phantom{$-$}0.07 & \phantom{$-$0}2.35 & \phantom{$-$}0.02 & \phantom{$-$}0.89 & \phantom{$-$}25.6 & \phantom{$-$}0.05 & \phantom{$-$0}2.14 & \phantom{0}74.4 \\
 & 5 & \phantom{$-$}0.07 & \phantom{$-$0}2.28 & \phantom{$-$}0.03 & \phantom{$-$}1.51 & \phantom{$-$}48.3 & \phantom{$-$}0.03 & \phantom{$-$0}1.52 & \phantom{0}51.7 \\
\midrule
2023--2025 & 0 & \phantom{$-$}1.15 & \phantom{$-$0}8.16 & $-$0.09 & $-$1.43 & \phantom{0}$-$7.8 & \phantom{$-$}1.24 & \phantom{$-$0}9.65 & 107.8 \\
 & 1 & \phantom{$-$}0.54 & \phantom{$-$0}5.84 & \phantom{$-$}0.05 & \phantom{$-$}0.95 & \phantom{$-$}10.0 & \phantom{$-$}0.49 & \phantom{$-$0}6.71 & \phantom{0}90.0 \\
 & 2 & \phantom{$-$}0.11 & \phantom{$-$0}1.35 & $-$0.04 & $-$0.70 &  & \phantom{$-$}0.14 & \phantom{$-$0}2.38 &  \\
 & 3 & \phantom{$-$}0.02 & \phantom{$-$0}0.24 & \phantom{$-$}0.02 & \phantom{$-$}0.39 &  & $-$0.00 & \phantom{0}$-$0.03 &  \\
 & 5 & $-$0.13 & \phantom{0}$-$1.76 & $-$0.02 & $-$0.41 &  & $-$0.10 & \phantom{0}$-$1.97 &  \\
\bottomrule
\end{tabular}
\end{adjustbox}
\par\medskip
\textit{Panel B: Monthly average of the daily forecast}\\[2pt]
\begin{adjustbox}{max width=\linewidth, max totalheight=0.19\textheight}
\begin{tabular}{llcccccccc}
\toprule
\multicolumn{2}{c}{} & \multicolumn{2}{c}{Forecast} & \multicolumn{3}{c}{Level $\hat y_L$} & \multicolumn{3}{c}{Deviation $\hat y_D$} \\
\cmidrule(lr){3-4} \cmidrule(lr){5-7} \cmidrule(lr){8-10}
Signal years & $h$ (months) & Cov & $t$ & Cov & $t$ & Share (\%) & Cov & $t$ & Share (\%) \\
\midrule
2004--2025 & 1 & \phantom{$-$}1.31 & \phantom{$-$}2.61 & \phantom{$-$}0.49 & \phantom{$-$}1.14 & 37.4 & \phantom{$-$}0.82 & \phantom{$-$}4.15 & 62.6 \\
 & 2 & \phantom{$-$}0.72 & \phantom{$-$}1.20 & \phantom{$-$}0.47 & \phantom{$-$}1.06 &  & \phantom{$-$}0.24 & \phantom{$-$}0.95 &  \\
 & 4 & \phantom{$-$}0.07 & \phantom{$-$}0.18 & \phantom{$-$}0.20 & \phantom{$-$}0.48 &  & $-$0.12 & $-$0.75 &  \\
 & 6 & \phantom{$-$}0.42 & \phantom{$-$}0.78 & \phantom{$-$}0.19 & \phantom{$-$}0.44 &  & \phantom{$-$}0.23 & \phantom{$-$}1.07 &  \\
 & 12 & \phantom{$-$}0.25 & \phantom{$-$}0.49 & \phantom{$-$}0.14 & \phantom{$-$}0.33 &  & \phantom{$-$}0.10 & \phantom{$-$}0.53 &  \\
 & 18 & $-$0.37 & $-$0.88 & $-$0.13 & $-$0.33 &  & $-$0.24 & $-$1.11 &  \\
 & 24 & \phantom{$-$}0.02 & \phantom{$-$}0.04 & $-$0.38 & $-$0.86 &  & \phantom{$-$}0.39 & \phantom{$-$}2.08 &  \\
\midrule
2023--2025 & 1 & \phantom{$-$}1.83 & \phantom{$-$}2.48 & \phantom{$-$}0.55 & \phantom{$-$}0.92 & 30.1 & \phantom{$-$}1.28 & \phantom{$-$}3.92 & 69.9 \\
 & 2 & \phantom{$-$}1.68 & \phantom{$-$}3.20 & \phantom{$-$}1.00 & \phantom{$-$}1.31 & 59.3 & \phantom{$-$}0.68 & \phantom{$-$}2.01 & 40.7 \\
 & 4 & \phantom{$-$}0.65 & \phantom{$-$}1.30 & \phantom{$-$}0.81 & \phantom{$-$}2.37 &  & $-$0.15 & $-$0.45 &  \\
 & 6 & \phantom{$-$}1.02 & \phantom{$-$}2.22 & \phantom{$-$}0.91 & \phantom{$-$}1.93 & 88.8 & \phantom{$-$}0.11 & \phantom{$-$}0.35 & 11.2 \\
 & 12 & \phantom{$-$}0.73 & \phantom{$-$}0.67 & \phantom{$-$}0.50 & \phantom{$-$}0.50 &  & \phantom{$-$}0.23 & \phantom{$-$}0.63 &  \\
\bottomrule
\end{tabular}
\end{adjustbox}
\par\medskip
\textit{Panel C: Monthly-trained forecast}\\[2pt]
\begin{adjustbox}{max width=\linewidth, max totalheight=0.19\textheight}
\begin{tabular}{llcccccccc}
\toprule
\multicolumn{2}{c}{} & \multicolumn{2}{c}{Forecast} & \multicolumn{3}{c}{Level $\hat y_L$} & \multicolumn{3}{c}{Deviation $\hat y_D$} \\
\cmidrule(lr){3-4} \cmidrule(lr){5-7} \cmidrule(lr){8-10}
Signal years & $h$ (months) & Cov & $t$ & Cov & $t$ & Share (\%) & Cov & $t$ & Share (\%) \\
\midrule
2004--2025 & 1 & \phantom{$-$}0.78 & \phantom{$-$}0.19 & $-$2.81 & $-$0.74 &  & \phantom{$-$}3.59 & \phantom{$-$}1.45 &  \\
 & 2 & $-$2.36 & $-$0.55 & $-$2.51 & $-$0.73 &  & \phantom{$-$}0.15 & \phantom{$-$}0.06 &  \\
 & 4 & $-$1.43 & $-$0.36 & $-$2.32 & $-$0.76 &  & \phantom{$-$}0.89 & \phantom{$-$}0.44 &  \\
 & 6 & $-$0.90 & $-$0.21 & $-$3.14 & $-$1.03 &  & \phantom{$-$}2.24 & \phantom{$-$}1.16 &  \\
 & 12 & \phantom{$-$}2.29 & \phantom{$-$}0.54 & $-$1.70 & $-$0.49 &  & \phantom{$-$}3.99 & \phantom{$-$}1.09 &  \\
 & 18 & $-$7.18 & $-$1.38 & $-$1.59 & $-$0.51 &  & $-$5.59 & $-$1.79 &  \\
 & 24 & \phantom{$-$}2.50 & \phantom{$-$}0.58 & \phantom{$-$}1.47 & \phantom{$-$}0.44 &  & \phantom{$-$}1.02 & \phantom{$-$}0.42 &  \\
\midrule
2023--2025 & 1 & $-$6.04 & $-$2.87 & $-$1.53 & $-$1.10 & 25.3 & $-$4.51 & $-$2.25 & 74.7 \\
 & 2 & \phantom{$-$}0.05 & \phantom{$-$}0.03 & $-$1.05 & $-$0.78 &  & \phantom{$-$}1.09 & \phantom{$-$}0.76 &  \\
 & 4 & $-$2.15 & $-$1.27 & $-$2.14 & $-$1.45 &  & $-$0.01 & $-$0.01 &  \\
 & 6 & $-$1.16 & $-$0.98 & $-$0.91 & $-$0.78 &  & $-$0.24 & $-$0.21 &  \\
 & 12 & $-$3.81 & $-$1.53 & $-$3.08 & $-$1.64 &  & $-$0.73 & $-$1.02 &  \\
\bottomrule
\end{tabular}
\end{adjustbox}
}
\end{table}

\begin{landscape}

\begin{table}[htbp]
\centering
\caption{MSRR Strategy Return Decomposition with ChronoGPT$_{1999}$}
\label{tab:tabB_attribution}
\begin{minipage}{\linewidth}\footnotesize
This table decomposes monthly MSRR strategy returns for ChronoGPT$_{1999}$ in signal years 2004--2025 and 2023--2025, with 263 and 35 signal months at $h=1$. Construction as in Table~\ref{tab:tab8}, except for the encoder. The 2023--2025 results evaluate the same strategy, scaled to 10\% annual volatility within that sample. Horizons with fewer than 24 signal months are omitted.
\end{minipage}
\medskip

{\footnotesize
\textit{Panel A: Timing decomposition}\\[2pt]
\begin{adjustbox}{max width=\linewidth, max totalheight=0.35\textheight}
\begin{tabular}{llccccccccc}
\toprule
\multicolumn{2}{c}{} & \multicolumn{3}{c}{Embedding $E$} & \multicolumn{3}{c}{Level $E_L$} & \multicolumn{3}{c}{Deviation $E_D$} \\
\cmidrule(lr){3-5} \cmidrule(lr){6-8} \cmidrule(lr){9-11}
Signal years & $h$ (months) & SR & Ret & $t$ & Ret & $t$ & Share (\%) & Ret & $t$ & Share (\%) \\
\midrule
2004--2025 & 1 & \phantom{$-$}1.38 & \phantom{$-$}13.8 & \phantom{$-$}5.85 & \phantom{$-$}6.0 & \phantom{$-$}3.56 & \phantom{$-$}43.5 & \phantom{$-$0}7.8 & \phantom{$-$}4.16 & \phantom{0}56.5 \\
 & 2 & \phantom{$-$}0.58 & \phantom{$-$0}5.8 & \phantom{$-$}2.94 & \phantom{$-$}4.5 & \phantom{$-$}2.65 & \phantom{$-$}77.7 & \phantom{$-$0}1.3 & \phantom{$-$}0.91 & \phantom{0}22.3 \\
 & 4 & \phantom{$-$}0.36 & \phantom{$-$0}3.6 & \phantom{$-$}1.51 & \phantom{$-$}3.8 & \phantom{$-$}2.45 &  & \phantom{0}$-$0.2 & $-$0.12 &  \\
 & 6 & \phantom{$-$}0.42 & \phantom{$-$0}4.2 & \phantom{$-$}1.82 & \phantom{$-$}3.6 & \phantom{$-$}2.40 &  & \phantom{$-$0}0.7 & \phantom{$-$}0.40 &  \\
 & 12 & \phantom{$-$}0.57 & \phantom{$-$0}5.7 & \phantom{$-$}2.82 & \phantom{$-$}2.4 & \phantom{$-$}1.24 & \phantom{$-$}41.2 & \phantom{$-$0}3.3 & \phantom{$-$}2.50 & \phantom{0}58.8 \\
 & 18 & \phantom{$-$}0.36 & \phantom{$-$0}3.6 & \phantom{$-$}1.81 & \phantom{$-$}1.4 & \phantom{$-$}0.82 &  & \phantom{$-$0}2.2 & \phantom{$-$}1.31 &  \\
 & 24 & \phantom{$-$}0.62 & \phantom{$-$0}6.2 & \phantom{$-$}2.52 & \phantom{$-$}3.0 & \phantom{$-$}1.42 & \phantom{$-$}47.6 & \phantom{$-$0}3.3 & \phantom{$-$}1.63 & \phantom{0}52.4 \\
\midrule
2023--2025 & 1 & \phantom{$-$}2.02 & \phantom{$-$}20.2 & \phantom{$-$}3.59 & \phantom{$-$}3.4 & \phantom{$-$}0.62 & \phantom{$-$}16.7 & \phantom{$-$}16.8 & \phantom{$-$}4.36 & \phantom{0}83.3 \\
 & 2 & \phantom{$-$}0.10 & \phantom{$-$0}1.0 & \phantom{$-$}0.16 & \phantom{$-$}2.8 & \phantom{$-$}0.73 &  & \phantom{0}$-$1.8 & $-$0.48 &  \\
 & 4 & $-$0.21 & \phantom{0}$-$2.1 & $-$0.30 & $-$0.2 & $-$0.05 &  & \phantom{0}$-$1.9 & $-$0.66 &  \\
 & 6 & $-$1.84 & $-$18.4 & $-$4.06 & \phantom{$-$}0.1 & \phantom{$-$}0.02 & \phantom{0}$-$0.5 & $-$18.5 & $-$3.46 & 100.5 \\
 & 12 & \phantom{$-$}0.59 & \phantom{$-$0}5.9 & \phantom{$-$}0.91 & \phantom{$-$}3.1 & \phantom{$-$}0.61 &  & \phantom{$-$0}2.8 & \phantom{$-$}0.64 &  \\
\bottomrule
\end{tabular}
\end{adjustbox}
\par\medskip
\textit{Panel B: Full decomposition}\\[2pt]
\begin{adjustbox}{max width=\linewidth, max totalheight=0.40\textheight}
\begin{tabular}{llccccccccccccccc}
\toprule
\multicolumn{2}{c}{} & \multicolumn{3}{c}{Event category, level $C_L$} & \multicolumn{3}{c}{Event category, deviation $C_D$} & \multicolumn{3}{c}{Firm identity $F$} & \multicolumn{3}{c}{Idiosyncratic, level $I_L$} & \multicolumn{3}{c}{Idiosyncratic, deviation $I_D$} \\
\cmidrule(lr){3-5} \cmidrule(lr){6-8} \cmidrule(lr){9-11} \cmidrule(lr){12-14} \cmidrule(lr){15-17}
Signal years & $h$ (months) & Ret & $t$ & Share (\%) & Ret & $t$ & Share (\%) & Ret & $t$ & Share (\%) & Ret & $t$ & Share (\%) & Ret & $t$ & Share (\%) \\
\midrule
2004--2025 & 1 & 2.5 & 3.91 & \phantom{$-$}18.0 & \phantom{$-$0}4.9 & \phantom{$-$}3.14 & 35.2 & \phantom{$-$}3.1 & \phantom{$-$}2.35 & \phantom{$-$}22.5 & \phantom{$-$}0.4 & \phantom{$-$}0.58 & \phantom{$-$0}3.0 & \phantom{$-$}2.9 & \phantom{$-$}3.05 & 21.2 \\
 & 2 & 2.2 & 3.72 & \phantom{$-$}38.5 & \phantom{$-$0}1.2 & \phantom{$-$}0.99 & 21.0 & \phantom{$-$}2.5 & \phantom{$-$}1.83 & \phantom{$-$}42.7 & $-$0.2 & $-$0.28 & \phantom{0}$-$3.6 & \phantom{$-$}0.1 & \phantom{$-$}0.06 & \phantom{0}1.3 \\
 & 4 & 1.2 & 2.18 &  & \phantom{$-$0}0.3 & \phantom{$-$}0.26 &  & \phantom{$-$}2.3 & \phantom{$-$}1.91 &  & \phantom{$-$}0.3 & \phantom{$-$}0.44 &  & $-$0.5 & $-$0.47 &  \\
 & 6 & 1.9 & 3.49 &  & \phantom{$-$0}0.7 & \phantom{$-$}0.58 &  & \phantom{$-$}2.0 & \phantom{$-$}1.51 &  & $-$0.4 & $-$0.57 &  & $-$0.0 & $-$0.00 &  \\
 & 12 & 0.8 & 1.19 & \phantom{$-$}13.8 & \phantom{$-$0}2.8 & \phantom{$-$}2.64 & 49.2 & \phantom{$-$}1.6 & \phantom{$-$}1.18 & \phantom{$-$}28.5 & $-$0.1 & $-$0.08 & \phantom{0}$-$1.1 & \phantom{$-$}0.5 & \phantom{$-$}0.42 & \phantom{0}9.6 \\
 & 18 & 0.6 & 1.16 &  & \phantom{$-$0}1.0 & \phantom{$-$}0.73 &  & \phantom{$-$}1.1 & \phantom{$-$}0.77 &  & $-$0.3 & $-$0.32 &  & \phantom{$-$}1.2 & \phantom{$-$}0.79 &  \\
 & 24 & 0.1 & 0.07 & \phantom{$-$0}1.0 & \phantom{$-$0}1.7 & \phantom{$-$}1.30 & 28.0 & \phantom{$-$}1.3 & \phantom{$-$}0.93 & \phantom{$-$}21.5 & \phantom{$-$}1.6 & \phantom{$-$}1.42 & \phantom{$-$}25.1 & \phantom{$-$}1.5 & \phantom{$-$}1.09 & 24.4 \\
\midrule
2023--2025 & 1 & 3.4 & 1.23 & \phantom{$-$}16.9 & \phantom{$-$}12.4 & \phantom{$-$}3.74 & 61.7 & $-$0.3 & $-$0.07 & \phantom{0}$-$1.3 & \phantom{$-$}0.2 & \phantom{$-$}0.14 & \phantom{$-$0}1.1 & \phantom{$-$}4.4 & \phantom{$-$}2.01 & 21.6 \\
 & 2 & 4.8 & 2.74 &  & \phantom{0}$-$2.1 & $-$0.57 &  & $-$1.1 & $-$0.31 &  & $-$0.9 & $-$0.88 &  & \phantom{$-$}0.3 & \phantom{$-$}0.11 &  \\
 & 4 & 1.3 & 0.79 &  & \phantom{$-$0}2.2 & \phantom{$-$}0.87 &  & $-$1.0 & $-$0.29 &  & $-$0.5 & $-$0.46 &  & $-$4.0 & $-$1.60 &  \\
 & 6 & 3.3 & 1.64 & $-$18.1 & $-$10.2 & $-$4.26 & 55.4 & $-$3.7 & $-$0.88 & \phantom{$-$}20.0 & \phantom{$-$}0.4 & \phantom{$-$}0.38 & \phantom{0}$-$2.4 & $-$8.3 & $-$1.61 & 45.0 \\
 & 12 & 2.2 & 0.71 &  & \phantom{$-$0}0.5 & \phantom{$-$}0.11 &  & \phantom{$-$}1.1 & \phantom{$-$}0.33 &  & $-$0.1 & $-$0.05 &  & \phantom{$-$}2.2 & \phantom{$-$}0.45 &  \\
\bottomrule
\end{tabular}
\end{adjustbox}
}
\end{table}
    
\end{landscape}

\begin{table}[htbp]
\centering
\caption{Alternative MSRR Strategy Inputs with ChronoGPT$_{1999}$}
\label{tab:tabB_constructions}
\begin{minipage}{\linewidth}\footnotesize
This table reports annualized Sharpe ratios of monthly MSRR strategies fitted to alternative text inputs from ChronoGPT$_{1999}$ in signal years 2004--2025 and 2023--2025. Construction as in the upper panels of Figure~\ref{fig:fig4_inputs}, except for the encoder. The 2023--2025 results evaluate the same strategies. News is assigned by arrival month, and $h$ denotes the $h$-th subsequent return month. Horizons with fewer than 24 signal months are blank.
\end{minipage}
\medskip

{\footnotesize
\begin{adjustbox}{max width=\linewidth, max totalheight=0.75\textheight}
\begin{tabular}{llccccccc}
\toprule
Signal years & Input & $h=1$ & $h=2$ & $h=4$ & $h=6$ & $h=12$ & $h=18$ & $h=24$ \\
\midrule
2004--2025 & Embedding $E$ & \phantom{$-$}1.35 & \phantom{$-$}0.14 & $-$0.10 & \phantom{$-$}0.20 & \phantom{$-$}0.28 & \phantom{$-$}0.15 & 0.55 \\
 & Level $E_L$ & \phantom{$-$}0.76 & \phantom{$-$}0.61 & \phantom{$-$}0.38 & \phantom{$-$}0.11 & \phantom{$-$}0.16 & \phantom{$-$}0.17 & 0.10 \\
 & Deviation $E_D$ & \phantom{$-$}0.84 & \phantom{$-$}0.41 & $-$0.28 & \phantom{$-$}0.18 & \phantom{$-$}0.40 & $-$0.03 & 0.49 \\
 & Level 36 months earlier, $E_{L,\,\tau-36}$ & \phantom{$-$}0.21 & \phantom{$-$}0.11 & \phantom{$-$}0.06 & $-$0.04 & $-$0.10 & $-$0.19 & 0.14 \\
 & Firm identity 36 months earlier, $F_{\tau-36}$ & \phantom{$-$}0.39 & \phantom{$-$}0.34 & \phantom{$-$}0.26 & $-$0.15 & \phantom{$-$}0.10 & $-$0.03 & 0.07 \\
\midrule
2023--2025 & Embedding $E$ & \phantom{$-$}2.26 & \phantom{$-$}0.43 & $-$0.21 & $-$1.64 & $-$0.11 &  &  \\
 & Level $E_L$ & \phantom{$-$}0.12 & \phantom{$-$}0.02 & $-$0.15 & $-$0.39 & $-$0.06 &  &  \\
 & Deviation $E_D$ & \phantom{$-$}1.37 & \phantom{$-$}0.27 & \phantom{$-$}0.45 & $-$0.42 & \phantom{$-$}0.01 &  &  \\
 & Level 36 months earlier, $E_{L,\,\tau-36}$ & $-$0.63 & $-$0.44 & $-$1.05 & $-$0.07 & \phantom{$-$}0.50 &  &  \\
 & Firm identity 36 months earlier, $F_{\tau-36}$ & \phantom{$-$}0.20 & \phantom{$-$}0.37 & \phantom{$-$}0.45 & $-$0.09 & \phantom{$-$}0.26 &  &  \\
\bottomrule
\end{tabular}
\end{adjustbox}
}
\end{table}

\begin{table}[htbp]
\centering
\caption{Valuations and Growth Forecasts with ChronoGPT$_{1999}$}
\label{tab:tabB_representation}
\begin{minipage}{\linewidth}\footnotesize
This table reports $R^2$ on held-out firms from cross-sectional regressions of year-end valuations and growth forecasts on firm-level news embeddings from ChronoGPT$_{1999}$ and firm characteristics, 2004--2025. Construction as in Table~\ref{tab:tab4}, except for the encoder and the inputs shown. News windows start no earlier than 2000.
\end{minipage}
\medskip

{\footnotesize
\textit{Panel A: Split sample}\\[2pt]
\begin{adjustbox}{max width=\linewidth, max totalheight=0.28\textheight}
\begin{tabular}{llccc}
\toprule
\multicolumn{2}{c}{} & \multicolumn{3}{c}{$R^2$ (\%)} \\
\cmidrule(lr){3-5}
Model & Inputs & $\log(P/B)$ & $\log(P/E)$ & LTG \\
\midrule
ChronoGPT$_{1999}$ & $F$ & 10.56 & \phantom{0}5.62 & \phantom{0}5.73 \\
 & Average embedding & 14.86 & \phantom{0}6.88 & \phantom{0}8.52 \\
 & Controls & 25.25 & 13.69 & 16.10 \\
 & $F$ + controls & 27.42 & 13.62 & 15.71 \\
 & Average embedding + controls & 34.11 & 13.82 & 15.89 \\
\bottomrule
\end{tabular}
\end{adjustbox}
\par\medskip
\textit{Panel B: Fitted on the previous window}\\[2pt]
\begin{adjustbox}{max width=\linewidth, max totalheight=0.28\textheight}
\begin{tabular}{llccc}
\toprule
\multicolumn{2}{c}{} & \multicolumn{3}{c}{$R^2$ (\%)} \\
\cmidrule(lr){3-5}
Model & Inputs & $\log(P/B)$ & $\log(P/E)$ & LTG \\
\midrule
ChronoGPT$_{1999}$ & $F$ & \phantom{0}9.13 & 3.82 & \phantom{0}4.00 \\
 & Average embedding & 12.87 & 4.77 & \phantom{0}5.44 \\
 & Controls & 20.12 & 7.74 & 10.61 \\
 & $F$ + controls & 21.95 & 7.72 & 10.25 \\
 & Average embedding + controls & 28.14 & 7.82 & 10.26 \\
\bottomrule
\end{tabular}
\end{adjustbox}
}
\end{table}

\begin{table}[htbp]
\centering
\caption{Firm Networks with ChronoGPT$_{1999}$}
\label{tab:tabB_networks}
\begin{minipage}{\linewidth}\footnotesize
This table reports adjusted $R^2$, in percent, from pooled regressions with year fixed effects of year-end firm characteristics on their mean over each firm's peers: the firms sharing its industry code, or the ten firms most similar in firm identity $F$, mean category component $C^{\mathrm{mix}}$, or average embedding $E^{\mathrm{avg}}$ from ChronoGPT$_{1999}$. Construction as in Table~\ref{tab:tab5}, except for the encoder. Panel A covers year ends in 2004--2025 with news windows starting no earlier than 2000, and Panel B covers year ends in 2023--2025 with news windows starting no earlier than 2023.
\end{minipage}
\medskip

{\footnotesize
\begin{adjustbox}{max width=\linewidth, max totalheight=0.75\textheight}
\begin{tabular}{lcccccccc}
\toprule
 & \multicolumn{5}{c}{Firm characteristics} & \multicolumn{3}{c}{Overlap} \\
\cmidrule(lr){2-6} \cmidrule(lr){7-9}
Signal & OI/Sales & OI/Assets & $\Delta$Sales & Market $\beta$ & Asset $\beta$ & SIC-3 & NAICS-4 & GICS-6 \\
\midrule
\multicolumn{9}{l}{\textit{Panel A: windows from 2000}} \\
SIC-3 & 10.63 & 33.94 & 6.71 & 26.39 & 32.18 &  & 78.3 & 62.3 \\
NAICS-4 & 10.50 & 33.83 & 6.91 & 25.71 & 32.15 & 76.3 &  & 59.2 \\
GICS-6 & 13.68 & 38.29 & 7.68 & 30.79 & 37.93 & 44.7 & 43.9 &  \\
$E^{\mathrm{avg}}$ & \phantom{0}7.24 & 23.62 & 2.69 & \phantom{0}9.62 & 14.80 & 12.8 & 13.2 & 12.1 \\
$F$ & \phantom{0}5.32 & 18.73 & 2.25 & \phantom{0}8.30 & 13.51 & 12.7 & 13.1 & 11.9 \\
$C^{\mathrm{mix}}$ & \phantom{0}8.99 & 35.49 & 3.04 & 11.52 & 15.28 & 10.8 & 11.2 & 10.5 \\
$E^{\mathrm{avg}}$, $E^{\mathrm{avg}}$ & \phantom{0}7.62 & 24.45 & 3.06 & 10.66 & 15.84 & 11.4 & 11.8 & 10.6 \\
$F$, $C^{\mathrm{mix}}$ & 10.44 & 38.68 & 3.35 & 15.39 & 20.98 & 11.7 & 12.1 & 11.1 \\
\midrule
\multicolumn{9}{l}{\textit{Panel B: windows from 2023}} \\
SIC-3 & \phantom{0}7.50 & 35.76 & 3.51 & 14.83 & 13.91 &  & 79.7 & 63.1 \\
NAICS-4 & \phantom{0}7.77 & 35.43 & 4.21 & 14.93 & 13.67 & 80.9 &  & 61.9 \\
GICS-6 & 12.27 & 40.34 & 3.62 & 20.83 & 18.12 & 46.2 & 44.2 &  \\
$E^{\mathrm{avg}}$ & \phantom{0}3.20 & 22.97 & 0.08 & \phantom{0}4.07 & \phantom{0}5.67 & 14.6 & 14.3 & 14.3 \\
$F$ & \phantom{0}2.35 & 19.85 & 0.30 & \phantom{0}2.91 & \phantom{0}4.52 & 14.1 & 13.8 & 13.8 \\
$C^{\mathrm{mix}}$ & \phantom{0}6.84 & 32.73 & 0.40 & \phantom{0}4.37 & \phantom{0}5.53 & 11.3 & 10.7 & 11.1 \\
$E^{\mathrm{avg}}$, $E^{\mathrm{avg}}$ & \phantom{0}3.64 & 23.62 & 0.31 & \phantom{0}4.81 & \phantom{0}6.62 & 13.4 & 13.1 & 13.0 \\
$F$, $C^{\mathrm{mix}}$ & \phantom{0}7.96 & 37.95 & 0.67 & \phantom{0}6.17 & \phantom{0}8.56 & 12.6 & 12.2 & 12.4 \\
\bottomrule
\end{tabular}
\end{adjustbox}
}
\end{table}

\FloatBarrier
\clearpage
\section{Results after the Pretraining Cutoffs}
\label{sec:appendixC}

\begin{table}[htbp]
\centering
\caption{Category and Embedding Forecasts after the Pretraining Cutoffs}
\label{tab:tabC_post_residual}
\begin{minipage}{\linewidth}\footnotesize
This table reports the contribution of embeddings beyond category forecasts of next open-to-open returns after pretraining. Construction as in Table~\ref{tab:tab2}, except that evaluation starts the January after each pretraining cutoff, covering 2020--2025 for RoBERTa-large and 2023--2025 for Llama-2 (13B). Panel B's category row uses RoBERTa's period, and each model's comparisons use its own period.
\end{minipage}
\medskip

{\footnotesize
\textit{Panel A: Fama--MacBeth regressions, standardized signals (bp per standard deviation)}\\[2pt]
\begin{adjustbox}{max width=\linewidth, max totalheight=0.33\textheight}
\begin{tabular}{lcccccc}
\toprule
 & \multicolumn{3}{c}{RoBERTa} & \multicolumn{3}{c}{Llama-2} \\
\cmidrule(lr){2-4} \cmidrule(lr){5-7}
 & (1) & (2) & (3) & (4) & (5) & (6) \\
\midrule
Embedding forecast & \phantom{(0}8.65\phantom{)} & \phantom{(}4.31\phantom{)} & \phantom{(}5.87\phantom{)} & \phantom{(}7.77\phantom{)} & \phantom{(}4.34\phantom{)} & \phantom{(}5.96\phantom{)} \\
 & (11.28) & (5.18) & (6.93) & (6.92) & (3.65) & (4.88) \\
Category mean (8-yr) &  & \phantom{(}8.49\phantom{)} &  &  & \phantom{(}6.47\phantom{)} &  \\
 &  & (9.34) &  &  & (5.38) &  \\
Group $\times$ direction mean &  &  & \phantom{(}5.74\phantom{)} &  &  & \phantom{(}4.37\phantom{)} \\
 &  &  & (7.89) &  &  & (5.06) \\
\bottomrule
\end{tabular}
\end{adjustbox}
\par\medskip
\textit{Panel B: Out of sample: the category mean plus a ridge forecast of its residual}\\[2pt]
\begin{adjustbox}{max width=\linewidth, max totalheight=0.22\textheight}
\begin{tabular}{lcccccccc}
\toprule
 & \multicolumn{2}{c}{Rank IC (\%)} & \multicolumn{2}{c}{EW SR} & \multicolumn{2}{c}{VW SR} & \multicolumn{2}{c}{OOS $R^2$} \\
\cmidrule(lr){2-3} \cmidrule(lr){4-5} \cmidrule(lr){6-7} \cmidrule(lr){8-9}
Forecast & Mean & DM $t$ & SR & JK $z$ & SR & JK $z$ & bp & $t$ \\
\midrule
Category mean (8-yr) & 2.33 &  & 3.73 &  & 0.79 &  & 6.52 &  \\
+ residual ridge, RoBERTa & 2.61 & 3.24 & 5.14 & 5.31 & 1.19 & 1.35 & 7.21 & 1.39 \\
+ residual ridge, Llama-2 & 2.71 & 2.38 & 4.94 & 3.05 & 0.70 & 0.70 & 7.23 & 0.77 \\
\bottomrule
\end{tabular}
\end{adjustbox}
}
\end{table}

\begin{table}[htbp]
\centering
\caption{Forecast Decomposition after the Pretraining Cutoffs}
\label{tab:tabC_post_split}
\begin{minipage}{\linewidth}\footnotesize
This table decomposes daily forecasts of next open-to-open returns after pretraining. Construction as in Table~\ref{tab:tab6}, except that evaluation starts the January after each pretraining cutoff, covering 2020--2025 for RoBERTa-large and 2023--2025 for Llama-2 (13B).
\end{minipage}
\medskip

{\footnotesize
\begin{adjustbox}{max width=\linewidth, max totalheight=0.75\textheight}
\begin{tabular}{llccccc}
\toprule
\multicolumn{2}{c}{} & \multicolumn{2}{c}{Share (\%)} & \multicolumn{3}{c}{Standalone performance} \\
\cmidrule(lr){3-4} \cmidrule(lr){5-7}
Model & Decomposition & $\mathrm{Cov}(\hat y^j,r)/\mathrm{Cov}(\hat y,r)$ & $\mathrm{Var}(\hat y^j)/\mathrm{Var}(\hat y)$ & Rank IC (\%) & EW SR & VW SR \\
\midrule
RoBERTa & Event category $\hat y^C$ & \phantom{0}77.4 & \phantom{0}50.2 & 2.36 & 4.53 & \phantom{$-$}1.18 \\
 & Firm identity $\hat y^F$ & \phantom{00}3.6 & \phantom{0}19.4 & 0.46 & 0.45 & $-$0.41 \\
 & Idiosyncratic $\hat y^I$ & \phantom{0}18.9 & \phantom{0}36.6 & 0.78 & 1.59 & \phantom{$-$}0.84 \\
 & Total forecast $\hat y$ & 100.0 & 100.0 & 2.30 & 4.25 & \phantom{$-$}1.39 \\
\midrule
Llama-2 & Event category $\hat y^C$ & \phantom{0}61.7 & \phantom{0}41.4 & 2.15 & 2.35 & $-$0.52 \\
 & Firm identity $\hat y^F$ & \phantom{00}5.5 & \phantom{0}31.1 & 0.51 & 0.54 & \phantom{$-$}0.71 \\
 & Idiosyncratic $\hat y^I$ & \phantom{0}32.8 & \phantom{0}35.3 & 1.12 & 1.89 & \phantom{$-$}1.22 \\
 & Total forecast $\hat y$ & 100.0 & 100.0 & 2.30 & 4.15 & \phantom{$-$}1.27 \\
\bottomrule
\end{tabular}
\end{adjustbox}
}
\end{table}

\begin{table}[htbp]
\centering
\caption{Forecast Return Covariances after the Pretraining Cutoffs}
\label{tab:tabC_post_horizons}
\begin{minipage}{\linewidth}\footnotesize
This table reports forecast covariances with returns after pretraining. Construction as in Table~\ref{tab:tab7}, except that evaluation starts the January after each pretraining cutoff, covering 2020--2025 for RoBERTa-large and 2023--2025 for Llama-2 (13B). Monthly horizons require at least 24 signal months.
\end{minipage}
\medskip

{\footnotesize
\textit{Panel A: Daily forecast}\\[2pt]
\begin{adjustbox}{max width=\linewidth, max totalheight=0.17\textheight}
\begin{tabular}{llcccccccc}
\toprule
\multicolumn{2}{c}{} & \multicolumn{2}{c}{Forecast} & \multicolumn{3}{c}{Level $\hat y_L$} & \multicolumn{3}{c}{Deviation $\hat y_D$} \\
\cmidrule(lr){3-4} \cmidrule(lr){5-7} \cmidrule(lr){8-10}
Model & $h$ (trading days) & Cov & $t$ & Cov & $t$ & Share (\%) & Cov & $t$ & Share (\%) \\
\midrule
RoBERTa & 0 & \phantom{$-$}1.64 & \phantom{$-$}19.44 & $-$0.06 & $-$1.63 & \phantom{0}$-$3.4 & \phantom{$-$}1.69 & \phantom{$-$}21.46 & 103.4 \\
 & 1 & \phantom{$-$}0.62 & \phantom{$-$}10.14 & \phantom{$-$}0.05 & \phantom{$-$}1.14 & \phantom{$-$0}8.4 & \phantom{$-$}0.57 & \phantom{$-$}12.15 & \phantom{0}91.6 \\
 & 2 & \phantom{$-$}0.13 & \phantom{$-$0}2.29 & \phantom{$-$}0.02 & \phantom{$-$}0.48 & \phantom{$-$}15.5 & \phantom{$-$}0.11 & \phantom{$-$0}2.78 & \phantom{0}84.5 \\
 & 3 & $-$0.01 & \phantom{0}$-$0.10 & \phantom{$-$}0.02 & \phantom{$-$}0.63 &  & $-$0.03 & \phantom{0}$-$0.75 &  \\
 & 5 & $-$0.05 & \phantom{0}$-$0.96 & $-$0.01 & $-$0.31 &  & $-$0.04 & \phantom{0}$-$1.08 &  \\
\midrule
Llama-2 & 0 & \phantom{$-$}1.50 & \phantom{$-$0}7.03 & $-$0.25 & $-$3.02 & $-$16.8 & \phantom{$-$}1.76 & \phantom{$-$0}8.70 & 116.8 \\
 & 1 & \phantom{$-$}0.91 & \phantom{$-$0}6.56 & \phantom{$-$}0.11 & \phantom{$-$}1.22 & \phantom{$-$}11.7 & \phantom{$-$}0.80 & \phantom{$-$0}7.94 & \phantom{0}88.3 \\
 & 2 & \phantom{$-$}0.34 & \phantom{$-$0}3.05 & \phantom{$-$}0.05 & \phantom{$-$}0.60 & \phantom{$-$}14.3 & \phantom{$-$}0.29 & \phantom{$-$0}3.36 & \phantom{0}85.7 \\
 & 3 & $-$0.02 & \phantom{0}$-$0.19 & \phantom{$-$}0.01 & \phantom{$-$}0.10 &  & $-$0.03 & \phantom{0}$-$0.35 &  \\
 & 5 & $-$0.13 & \phantom{0}$-$1.27 & $-$0.03 & $-$0.46 &  & $-$0.09 & \phantom{0}$-$1.20 &  \\
\bottomrule
\end{tabular}
\end{adjustbox}
\par\medskip
\textit{Panel B: Monthly average of the daily forecast}\\[2pt]
\begin{adjustbox}{max width=\linewidth, max totalheight=0.19\textheight}
\begin{tabular}{llcccccccc}
\toprule
\multicolumn{2}{c}{} & \multicolumn{2}{c}{Forecast} & \multicolumn{3}{c}{Level $\hat y_L$} & \multicolumn{3}{c}{Deviation $\hat y_D$} \\
\cmidrule(lr){3-4} \cmidrule(lr){5-7} \cmidrule(lr){8-10}
Model & $h$ (months) & Cov & $t$ & Cov & $t$ & Share (\%) & Cov & $t$ & Share (\%) \\
\midrule
RoBERTa & 1 & 1.00 & 1.05 & 0.46 & 0.62 &  & \phantom{$-$}0.55 & \phantom{$-$}1.64 &  \\
 & 2 & 1.51 & 2.17 & 0.89 & 1.32 & 59.1 & \phantom{$-$}0.62 & \phantom{$-$}3.32 & 40.9 \\
 & 4 & 1.20 & 1.76 & 1.09 & 1.97 &  & \phantom{$-$}0.11 & \phantom{$-$}0.37 &  \\
 & 6 & 1.00 & 1.54 & 1.08 & 2.09 &  & $-$0.09 & $-$0.33 &  \\
 & 12 & 1.32 & 1.87 & 1.10 & 1.74 &  & \phantom{$-$}0.22 & \phantom{$-$}1.08 &  \\
 & 18 & 0.47 & 0.49 & 0.89 & 1.12 &  & $-$0.43 & $-$1.16 &  \\
 & 24 & 0.59 & 0.90 & 0.45 & 0.84 &  & \phantom{$-$}0.14 & \phantom{$-$}0.41 &  \\
\midrule
Llama-2 & 1 & 3.10 & 4.00 & 2.03 & 2.44 & 65.5 & \phantom{$-$}1.07 & \phantom{$-$}2.25 & 34.5 \\
 & 2 & 3.10 & 3.15 & 2.80 & 3.51 & 90.1 & \phantom{$-$}0.31 & \phantom{$-$}0.70 & \phantom{0}9.9 \\
 & 4 & 0.92 & 1.60 & 1.80 & 2.72 &  & $-$0.88 & $-$1.47 &  \\
 & 6 & 1.65 & 1.15 & 0.87 & 1.06 &  & \phantom{$-$}0.78 & \phantom{$-$}0.96 &  \\
 & 12 & 1.03 & 0.87 & 1.04 & 1.15 &  & $-$0.01 & $-$0.01 &  \\
\bottomrule
\end{tabular}
\end{adjustbox}
\par\medskip
\textit{Panel C: Monthly-trained forecast}\\[2pt]
\begin{adjustbox}{max width=\linewidth, max totalheight=0.19\textheight}
\begin{tabular}{llcccccccc}
\toprule
\multicolumn{2}{c}{} & \multicolumn{2}{c}{Forecast} & \multicolumn{3}{c}{Level $\hat y_L$} & \multicolumn{3}{c}{Deviation $\hat y_D$} \\
\cmidrule(lr){3-4} \cmidrule(lr){5-7} \cmidrule(lr){8-10}
Model & $h$ (months) & Cov & $t$ & Cov & $t$ & Share (\%) & Cov & $t$ & Share (\%) \\
\midrule
RoBERTa & 1 & \phantom{0}$-$3.65 & $-$0.89 & $-$10.61 & $-$1.43 &  & \phantom{$-$0}6.95 & \phantom{$-$}1.13 &  \\
 & 2 & \phantom{0}$-$4.13 & $-$0.90 & $-$11.17 & $-$1.73 &  & \phantom{$-$0}7.05 & \phantom{$-$}2.06 &  \\
 & 4 & $-$12.16 & $-$1.67 & $-$10.35 & $-$1.58 &  & \phantom{0}$-$1.82 & $-$0.55 &  \\
 & 6 & $-$13.71 & $-$1.69 & $-$12.25 & $-$1.93 &  & \phantom{0}$-$1.46 & $-$0.47 &  \\
 & 12 & \phantom{0}$-$3.11 & $-$0.66 & \phantom{0}$-$6.05 & $-$1.13 &  & \phantom{$-$0}2.93 & \phantom{$-$}0.84 &  \\
 & 18 & \phantom{0}$-$3.05 & $-$0.49 & \phantom{0}$-$0.29 & $-$0.05 &  & \phantom{0}$-$2.76 & $-$1.04 &  \\
 & 24 & \phantom{$-$0}3.10 & \phantom{$-$}0.79 & \phantom{$-$0}3.84 & \phantom{$-$}0.66 &  & \phantom{0}$-$0.75 & $-$0.11 &  \\
\midrule
Llama-2 & 1 & \phantom{$-$}12.14 & \phantom{$-$}0.38 & $-$14.40 & $-$0.52 &  & \phantom{$-$}26.54 & \phantom{$-$}1.37 &  \\
 & 2 & \phantom{$-$0}6.34 & \phantom{$-$}0.19 & \phantom{$-$}11.34 & \phantom{$-$}0.31 &  & \phantom{0}$-$5.00 & $-$0.33 &  \\
 & 4 & \phantom{$-$0}2.61 & \phantom{$-$}0.07 & $-$10.34 & $-$0.32 &  & \phantom{$-$}12.95 & \phantom{$-$}0.71 &  \\
 & 6 & $-$37.87 & $-$1.04 & $-$17.64 & $-$0.39 &  & $-$20.23 & $-$0.94 &  \\
 & 12 & \phantom{$-$}23.33 & \phantom{$-$}0.65 & \phantom{$-$}16.96 & \phantom{$-$}0.40 &  & \phantom{$-$0}6.37 & \phantom{$-$}0.39 &  \\
\bottomrule
\end{tabular}
\end{adjustbox}
}
\end{table}

\begin{landscape}

\begin{table}[htbp]
\centering
\caption{MSRR Strategy Return Decomposition after the Pretraining Cutoffs}
\label{tab:tabC_post_attribution}
\begin{minipage}{\linewidth}\footnotesize
This table decomposes monthly MSRR strategy returns after pretraining, with 71 and 35 signal months at $h=1$. Construction as in Table~\ref{tab:tab8}, except that evaluation starts the January after each pretraining cutoff, covering 2020--2025 for RoBERTa-large and 2023--2025 for Llama-2 (13B), and the strategy is scaled to 10\% annual volatility in each model's evaluation sample. Horizons with fewer than 24 signal months are omitted.
\end{minipage}
\medskip

{\footnotesize
\textit{Panel A: Timing decomposition}\\[2pt]
\begin{adjustbox}{max width=\linewidth, max totalheight=0.35\textheight}
\begin{tabular}{llccccccccc}
\toprule
\multicolumn{2}{c}{} & \multicolumn{3}{c}{Embedding $E$} & \multicolumn{3}{c}{Level $E_L$} & \multicolumn{3}{c}{Deviation $E_D$} \\
\cmidrule(lr){3-5} \cmidrule(lr){6-8} \cmidrule(lr){9-11}
Model & $h$ (months) & SR & Ret & $t$ & Ret & $t$ & Share (\%) & Ret & $t$ & Share (\%) \\
\midrule
RoBERTa & 1 & \phantom{$-$}1.80 & \phantom{$-$}18.0 & \phantom{$-$}4.10 & \phantom{$-$}6.4 & \phantom{$-$}1.77 & \phantom{$-$}35.5 & \phantom{$-$}11.6 & \phantom{$-$}2.89 & \phantom{0}64.5 \\
 & 2 & \phantom{$-$}0.42 & \phantom{$-$0}4.2 & \phantom{$-$}1.11 & \phantom{$-$}7.2 & \phantom{$-$}2.55 &  & \phantom{0}$-$3.0 & $-$1.08 &  \\
 & 4 & \phantom{$-$}0.62 & \phantom{$-$0}6.2 & \phantom{$-$}1.12 & \phantom{$-$}4.5 & \phantom{$-$}1.65 &  & \phantom{$-$0}1.6 & \phantom{$-$}0.49 &  \\
 & 6 & \phantom{$-$}0.30 & \phantom{$-$0}3.0 & \phantom{$-$}0.56 & \phantom{$-$}4.9 & \phantom{$-$}1.71 &  & \phantom{0}$-$1.9 & $-$0.53 &  \\
 & 12 & \phantom{$-$}0.87 & \phantom{$-$0}8.7 & \phantom{$-$}2.15 & \phantom{$-$}3.8 & \phantom{$-$}1.08 & \phantom{$-$}43.2 & \phantom{$-$0}5.0 & \phantom{$-$}2.20 & \phantom{0}56.8 \\
 & 18 & \phantom{$-$}0.30 & \phantom{$-$0}3.0 & \phantom{$-$}0.87 & \phantom{$-$}2.3 & \phantom{$-$}0.64 &  & \phantom{$-$0}0.7 & \phantom{$-$}0.19 &  \\
 & 24 & \phantom{$-$}0.88 & \phantom{$-$0}8.8 & \phantom{$-$}2.17 & \phantom{$-$}8.1 & \phantom{$-$}1.92 & \phantom{$-$}92.0 & \phantom{$-$0}0.7 & \phantom{$-$}0.18 & \phantom{00}8.0 \\
\midrule
Llama-2 & 1 & \phantom{$-$}2.57 & \phantom{$-$}25.7 & \phantom{$-$}5.54 & \phantom{$-$}4.6 & \phantom{$-$}0.96 & \phantom{$-$}17.8 & \phantom{$-$}21.1 & \phantom{$-$}6.09 & \phantom{0}82.2 \\
 & 2 & \phantom{$-$}0.92 & \phantom{$-$0}9.2 & \phantom{$-$}1.24 & \phantom{$-$}7.5 & \phantom{$-$}1.65 &  & \phantom{$-$0}1.8 & \phantom{$-$}0.41 &  \\
 & 4 & \phantom{$-$}1.07 & \phantom{$-$}10.7 & \phantom{$-$}1.24 & \phantom{$-$}8.4 & \phantom{$-$}1.41 &  & \phantom{$-$0}2.4 & \phantom{$-$}0.36 &  \\
 & 6 & $-$1.16 & $-$11.6 & $-$2.50 & \phantom{$-$}4.0 & \phantom{$-$}1.11 & $-$34.0 & $-$15.6 & $-$5.00 & 134.0 \\
 & 12 & \phantom{$-$}0.10 & \phantom{$-$0}1.0 & \phantom{$-$}0.20 & $-$0.9 & $-$0.28 &  & \phantom{$-$0}1.9 & \phantom{$-$}0.47 &  \\
\bottomrule
\end{tabular}
\end{adjustbox}
\par\medskip
\textit{Panel B: Full decomposition}\\[2pt]
\begin{adjustbox}{max width=\linewidth, max totalheight=0.40\textheight}
\begin{tabular}{llccccccccccccccc}
\toprule
\multicolumn{2}{c}{} & \multicolumn{3}{c}{Event category, level $C_L$} & \multicolumn{3}{c}{Event category, deviation $C_D$} & \multicolumn{3}{c}{Firm identity $F$} & \multicolumn{3}{c}{Idiosyncratic, level $I_L$} & \multicolumn{3}{c}{Idiosyncratic, deviation $I_D$} \\
\cmidrule(lr){3-5} \cmidrule(lr){6-8} \cmidrule(lr){9-11} \cmidrule(lr){12-14} \cmidrule(lr){15-17}
Model & $h$ (months) & Ret & $t$ & Share (\%) & Ret & $t$ & Share (\%) & Ret & $t$ & Share (\%) & Ret & $t$ & Share (\%) & Ret & $t$ & Share (\%) \\
\midrule
RoBERTa & 1 & \phantom{$-$}2.0 & \phantom{$-$}1.61 & \phantom{$-$}11.3 & \phantom{$-$0}6.4 & \phantom{$-$}2.02 & \phantom{$-$}35.7 & \phantom{$-$}3.0 & \phantom{$-$}1.43 & 16.6 & 1.4 & 0.93 & \phantom{$-$0}7.6 & \phantom{$-$0}5.2 & \phantom{$-$}1.81 & 28.8 \\
 & 2 & \phantom{$-$}2.3 & \phantom{$-$}2.08 &  & \phantom{0}$-$2.6 & $-$1.13 &  & \phantom{$-$}2.3 & \phantom{$-$}0.98 &  & 2.5 & 2.17 &  & \phantom{0}$-$0.4 & $-$0.19 &  \\
 & 4 & \phantom{$-$}1.5 & \phantom{$-$}1.11 &  & \phantom{$-$0}1.5 & \phantom{$-$}0.57 &  & \phantom{$-$}1.7 & \phantom{$-$}0.86 &  & 1.4 & 1.18 &  & \phantom{$-$0}0.1 & \phantom{$-$}0.07 &  \\
 & 6 & \phantom{$-$}2.6 & \phantom{$-$}2.86 &  & \phantom{0}$-$3.6 & $-$1.44 &  & \phantom{$-$}0.6 & \phantom{$-$}0.34 &  & 1.7 & 0.95 &  & \phantom{$-$0}1.8 & \phantom{$-$}0.81 &  \\
 & 12 & \phantom{$-$}0.2 & \phantom{$-$}0.17 & \phantom{$-$0}2.9 & \phantom{$-$0}4.0 & \phantom{$-$}1.77 & \phantom{$-$}45.7 & \phantom{$-$}1.5 & \phantom{$-$}0.79 & 16.6 & 2.1 & 1.08 & \phantom{$-$}23.7 & \phantom{$-$0}1.0 & \phantom{$-$}0.37 & 11.1 \\
 & 18 & \phantom{$-$}1.3 & \phantom{$-$}0.92 &  & \phantom{0}$-$2.2 & $-$0.96 &  & \phantom{$-$}0.4 & \phantom{$-$}0.17 &  & 0.5 & 0.29 &  & \phantom{$-$0}2.8 & \phantom{$-$}0.84 &  \\
 & 24 & \phantom{$-$}2.3 & \phantom{$-$}1.03 & \phantom{$-$}26.2 & \phantom{0}$-$1.3 & $-$0.38 & $-$14.5 & \phantom{$-$}4.0 & \phantom{$-$}1.45 & 45.0 & 1.8 & 1.13 & \phantom{$-$}20.9 & \phantom{$-$0}2.0 & \phantom{$-$}0.80 & 22.5 \\
\midrule
Llama-2 & 1 & \phantom{$-$}2.2 & \phantom{$-$}0.77 & \phantom{$-$0}8.7 & \phantom{$-$}10.3 & \phantom{$-$}2.91 & \phantom{$-$}40.0 & \phantom{$-$}0.8 & \phantom{$-$}0.17 & \phantom{0}3.1 & 1.5 & 0.44 & \phantom{$-$0}6.0 & \phantom{$-$}10.8 & \phantom{$-$}2.59 & 42.2 \\
 & 2 & \phantom{$-$}2.3 & \phantom{$-$}1.53 &  & \phantom{0}$-$2.1 & $-$0.55 &  & $-$0.6 & $-$0.13 &  & 5.7 & 2.06 &  & \phantom{$-$0}3.9 & \phantom{$-$}1.79 &  \\
 & 4 & \phantom{$-$}1.6 & \phantom{$-$}0.71 &  & \phantom{$-$0}3.1 & \phantom{$-$}1.45 &  & \phantom{$-$}2.8 & \phantom{$-$}0.84 &  & 4.0 & 1.36 &  & \phantom{0}$-$0.8 & $-$0.12 &  \\
 & 6 & \phantom{$-$}1.2 & \phantom{$-$}0.67 & $-$10.5 & \phantom{0}$-$8.9 & $-$3.52 & \phantom{$-$}76.8 & $-$1.4 & $-$0.56 & 11.8 & 4.1 & 1.80 & $-$35.3 & \phantom{0}$-$6.6 & $-$5.03 & 57.2 \\
 & 12 & $-$0.2 & $-$0.18 &  & \phantom{0}$-$0.2 & $-$0.07 &  & $-$2.9 & $-$0.82 &  & 2.2 & 1.17 &  & \phantom{$-$0}2.2 & \phantom{$-$}0.51 &  \\
\bottomrule
\end{tabular}
\end{adjustbox}
}
\end{table}
    
\end{landscape}

\begin{table}[htbp]
\centering
\caption{Alternative MSRR Strategy Inputs after the Pretraining Cutoffs}
\label{tab:tabC_post_constructions}
\begin{minipage}{\linewidth}\footnotesize
This table reports annualized Sharpe ratios of monthly MSRR strategies fitted to alternative text inputs after pretraining. Construction as in the upper panels of Figure~\ref{fig:fig4_inputs}, except that evaluation starts the January after each pretraining cutoff, covering 2020--2025 for RoBERTa-large and 2023--2025 for Llama-2 (13B). News is assigned by arrival month, and $h$ denotes the $h$-th subsequent return month. Horizons with fewer than 24 signal months are blank.
\end{minipage}
\medskip

{\footnotesize
\begin{adjustbox}{max width=\linewidth, max totalheight=0.75\textheight}
\begin{tabular}{llccccccc}
\toprule
Model & Input & $h=1$ & $h=2$ & $h=4$ & $h=6$ & $h=12$ & $h=18$ & $h=24$ \\
\midrule
RoBERTa & Embedding $E$ & \phantom{$-$}1.59 & \phantom{$-$}0.65 & \phantom{$-$}0.36 & \phantom{$-$}0.26 & \phantom{$-$}0.18 & $-$0.48 & 0.21 \\
 & Level $E_L$ & \phantom{$-$}1.58 & \phantom{$-$}0.91 & \phantom{$-$}0.41 & \phantom{$-$}0.95 & \phantom{$-$}0.21 & $-$0.04 & 0.17 \\
 & Deviation $E_D$ & \phantom{$-$}0.93 & $-$0.08 & $-$0.19 & $-$0.16 & $-$0.21 & $-$0.89 & 0.17 \\
 & Level 36 months earlier, $E_{L,\,\tau-36}$ & $-$0.28 & $-$0.43 & $-$0.37 & $-$0.72 & $-$0.44 & \phantom{$-$}0.05 & 0.26 \\
 & Firm identity 36 months earlier, $F_{\tau-36}$ & \phantom{$-$}0.46 & \phantom{$-$}0.47 & \phantom{$-$}0.17 & $-$0.26 & $-$0.13 & \phantom{$-$}0.08 & 0.41 \\
\midrule
Llama-2 & Embedding $E$ & \phantom{$-$}2.22 & \phantom{$-$}1.40 & \phantom{$-$}1.17 & $-$0.77 & $-$0.39 &  &  \\
 & Level $E_L$ & \phantom{$-$}1.19 & \phantom{$-$}1.83 & \phantom{$-$}1.54 & \phantom{$-$}0.59 & $-$0.47 &  &  \\
 & Deviation $E_D$ & \phantom{$-$}1.48 & \phantom{$-$}0.54 & \phantom{$-$}0.09 & $-$0.77 & \phantom{$-$}0.08 &  &  \\
 & Level 36 months earlier, $E_{L,\,\tau-36}$ & \phantom{$-$}0.04 & \phantom{$-$}0.30 & \phantom{$-$}0.18 & $-$0.07 & \phantom{$-$}0.53 &  &  \\
 & Firm identity 36 months earlier, $F_{\tau-36}$ & \phantom{$-$}0.50 & \phantom{$-$}0.63 & $-$0.09 & $-$0.04 & \phantom{$-$}0.73 &  &  \\
\bottomrule
\end{tabular}
\end{adjustbox}
}
\end{table}

\begin{table}[htbp]
\centering
\caption{Valuations and Growth Forecasts after the Pretraining Cutoffs}
\label{tab:tabC_post_representation}
\begin{minipage}{\linewidth}\footnotesize
This table reports $R^2$ on held-out firms from cross-sectional regressions of year-end valuations and growth forecasts on firm-level embeddings of news published after each model's pretraining cutoff and firm characteristics. Construction as in Table~\ref{tab:tab4}, except for the inputs shown and that year ends are 2020--2025 for RoBERTa-large and 2023--2025 for Llama-2 (13B), and news windows start no earlier than the January after the cutoff. Panel B starts one year later.
\end{minipage}
\medskip

{\footnotesize
\textit{Panel A: Split sample}\\[2pt]
\begin{adjustbox}{max width=\linewidth, max totalheight=0.28\textheight}
\begin{tabular}{llccc}
\toprule
\multicolumn{2}{c}{} & \multicolumn{3}{c}{$R^2$ (\%)} \\
\cmidrule(lr){3-5}
Model & Inputs & $\log(P/B)$ & $\log(P/E)$ & LTG \\
\midrule
RoBERTa & $F$ & 17.74 & 12.85 & \phantom{0}3.01 \\
 & Average embedding & 24.47 & 15.36 & \phantom{0}5.47 \\
 & Controls & 22.86 & 17.96 & \phantom{0}4.70 \\
 & $F$ + controls & 32.99 & 18.77 & \phantom{0}5.04 \\
 & Average embedding + controls & 44.07 & 20.54 & \phantom{0}6.82 \\
\midrule
Llama-2 & $F$ & 24.41 & 22.65 & \phantom{0}7.26 \\
 & Average embedding & 31.45 & 23.52 & \phantom{0}9.64 \\
 & Controls & 22.24 & 16.36 & \phantom{0}4.73 \\
 & $F$ + controls & 42.25 & 22.91 & \phantom{0}9.29 \\
 & Average embedding + controls & 52.55 & 23.86 & 10.29 \\
\bottomrule
\end{tabular}
\end{adjustbox}
\par\medskip
\textit{Panel B: Fitted on the previous window}\\[2pt]
\begin{adjustbox}{max width=\linewidth, max totalheight=0.28\textheight}
\begin{tabular}{llccc}
\toprule
\multicolumn{2}{c}{} & \multicolumn{3}{c}{$R^2$ (\%)} \\
\cmidrule(lr){3-5}
Model & Inputs & $\log(P/B)$ & $\log(P/E)$ & LTG \\
\midrule
RoBERTa & $F$ & 13.12 & 10.14 & $-$0.83 \\
 & Average embedding & 19.13 & 12.26 & $-$2.24 \\
 & Controls & 16.23 & 12.03 & \phantom{$-$}0.64 \\
 & $F$ + controls & 25.56 & 13.19 & $-$0.66 \\
 & Average embedding + controls & 37.20 & 15.06 & $-$0.97 \\
\midrule
Llama-2 & $F$ & 20.73 & 15.81 & \phantom{$-$}2.61 \\
 & Average embedding & 27.90 & 17.34 & \phantom{$-$}4.90 \\
 & Controls & 20.11 & 10.09 & \phantom{$-$}3.41 \\
 & $F$ + controls & 37.74 & 15.02 & \phantom{$-$}2.28 \\
 & Average embedding + controls & 48.29 & 16.07 & \phantom{$-$}3.49 \\
\bottomrule
\end{tabular}
\end{adjustbox}
}
\end{table}

\begin{table}[htbp]
\centering
\caption{Firm Networks after the Pretraining Cutoffs}
\label{tab:tabC_post_networks}
\begin{minipage}{\linewidth}\footnotesize
This table reports adjusted $R^2$, in percent, from pooled regressions with year fixed effects of year-end firm characteristics on their mean over each firm's peers: the firms sharing its industry code, or the ten firms most similar in firm identity $F$, mean category component $C^{\mathrm{mix}}$, or average embedding $E^{\mathrm{avg}}$. Construction as in Table~\ref{tab:tab5}, except that news windows start no earlier than the January after each model's pretraining cutoff: Panel A uses RoBERTa-large for year ends in 2020--2025, and Panel B uses Llama-2 (13B) for year ends in 2023--2025.
\end{minipage}
\medskip

{\footnotesize
\begin{adjustbox}{max width=\linewidth, max totalheight=0.75\textheight}
\begin{tabular}{lcccccccc}
\toprule
 & \multicolumn{5}{c}{Firm characteristics} & \multicolumn{3}{c}{Overlap} \\
\cmidrule(lr){2-6} \cmidrule(lr){7-9}
Signal & OI/Sales & OI/Assets & $\Delta$Sales & Market $\beta$ & Asset $\beta$ & SIC-3 & NAICS-4 & GICS-6 \\
\midrule
\multicolumn{9}{l}{\textit{Panel A: RoBERTa, windows from 2020}} \\
SIC-3 & \phantom{0}8.35 & 37.89 & 5.95 & 17.31 & 16.37 &  & 79.5 & 63.4 \\
NAICS-4 & \phantom{0}8.19 & 37.61 & 6.28 & 17.15 & 16.04 & 80.6 &  & 61.1 \\
GICS-6 & 11.45 & 41.69 & 6.29 & 23.02 & 20.37 & 47.8 & 46.0 &  \\
$E^{\mathrm{avg}}$ & \phantom{0}8.68 & 46.75 & 3.00 & 16.30 & 17.13 & 29.1 & 29.6 & 29.8 \\
$F$ & \phantom{0}8.01 & 37.34 & 2.79 & 13.60 & 13.61 & 28.9 & 29.6 & 30.1 \\
$C^{\mathrm{mix}}$ & \phantom{0}9.40 & 48.04 & 2.65 & 10.25 & 11.84 & 17.7 & 17.2 & 16.4 \\
$E^{\mathrm{avg}}$, $E^{\mathrm{avg}}$ & 10.89 & 47.80 & 3.71 & 17.37 & 18.72 & 27.2 & 27.8 & 27.3 \\
$F$, $C^{\mathrm{mix}}$ & 12.16 & 52.59 & 3.96 & 18.67 & 18.35 & 23.1 & 23.2 & 23.0 \\
\midrule
\multicolumn{9}{l}{\textit{Panel B: Llama-2, windows from 2023}} \\
SIC-3 & \phantom{0}7.50 & 35.76 & 3.51 & 14.83 & 13.91 &  & 79.7 & 63.1 \\
NAICS-4 & \phantom{0}7.77 & 35.43 & 4.21 & 14.93 & 13.67 & 80.9 &  & 61.9 \\
GICS-6 & 12.27 & 40.34 & 3.62 & 20.83 & 18.12 & 46.2 & 44.2 &  \\
$E^{\mathrm{avg}}$ & 10.85 & 52.96 & 1.64 & 25.94 & 27.16 & 33.6 & 32.8 & 39.3 \\
$F$ & \phantom{0}9.63 & 50.70 & 2.00 & 29.29 & 28.43 & 37.7 & 36.6 & 44.9 \\
$C^{\mathrm{mix}}$ & 10.67 & 38.86 & 1.50 & \phantom{0}7.36 & \phantom{0}9.99 & 13.7 & 12.9 & 13.8 \\
$E^{\mathrm{avg}}$, $E^{\mathrm{avg}}$ & 13.31 & 53.69 & 2.80 & 27.18 & 28.42 & 30.5 & 29.7 & 34.7 \\
$F$, $C^{\mathrm{mix}}$ & 14.50 & 55.49 & 2.82 & 30.17 & 29.59 & 25.5 & 24.5 & 29.0 \\
\bottomrule
\end{tabular}
\end{adjustbox}
}
\end{table}

\FloatBarrier
\clearpage
\section{A Calibrated Framework with Look-Ahead Bias}
\label{sec:appendixD}

This section parameterizes the framework of Section~\ref{sec:framework} (Section~\ref{sec:appD1}). It then extends the framework to look-ahead bias, which we model as an encoder's memory of realized cash-flow news, the part of returns that is neither the response to published news nor the state premium (Section~\ref{sec:appD2}). Section~\ref{sec:appD3} calibrates the model. Notation follows Sections~\ref{sec:framework} and~\ref{sec:measure}.

\subsection{Parameterized Framework}
\label{sec:appD1}

\paragraph{Event drift.} We take the return response to a news signal in Section~\ref{sec:fw_drift} to be geometric and common across event categories, $\delta_h \propto \vartheta^{h-1}$, with $0 < \vartheta < 1$. With BBE's announcement response $\kappa$, the cumulative price response per unit of signal $h$ trading days after the first eligible open is
\begin{equation}\label{eq:D1}
\kappa + (1-\kappa)\left(1-\vartheta^h\right)
\end{equation}
so that $\delta_h = (1-\kappa)(1-\vartheta)\vartheta^{h-1}$ and the drift sums to $1-\kappa$.

\paragraph{Firm state.} Time $t$ counts trading days, and $\tau$ counts calendar months. Moments are taken across firms at a date. The firm state $z=\bar z+x$ of Section~\ref{sec:fw_long} has a permanent component $\bar z$ and a slowly evolving component $x$ with daily persistence $\rho$. It is stationary, with uncorrelated components and unit variance: $\Var(\bar z)+\Var(x)=1$. As in Section~\ref{sec:fw_long}, categories and details are independent across news items and of the state, and language noise is independent across news items and of all other terms.

\paragraph{Measurement.} We separate the mean embedding across news items, $\bar E$, from equation~\eqref{eq:measure}, so that every other term has mean zero. With the standardized detail $\tilde\nu = \nu/\sqrt{\Var(\nu)}$ and the standardized category signal $\tilde m_K = m_K/\sqrt{\Var(m_K)}$, the embedding is
\[
E = \bar E + a_K + q \tilde\nu e_\nu + g z e_z + \zeta, \qquad a_K = a \tilde m_K e_K + a_K^\perp
\]
Here the category vector $a_K\in\mathbb R^d$ loads on the unit vector $e_K$, and its remainder $a_K^\perp$ is unrelated to returns. The vectors $e_K$, $e_\nu$ and $e_z$ are orthogonal unit vectors, $a_K^\perp$ is orthogonal to them, and the language noise $\zeta$ is distributed $\mathcal N(0,\mathbf I_d)$. The loadings $a$, $q$ and $g$ measure the change in the embedding, in noise standard deviations, per standard deviation of each source.

Along $e_z$, the embedding is $gz$ plus unit-variance noise. The state accounts for the share
\[
\phi(g) = \frac{g^2}{1+g^2}
\]
of this coordinate's variance, its reliability as a measure of the state. The population forecast $\hat y=\beta'(E-\bar E)$, with $\beta = \Var(E)^{-1}\Cov(E,r_1)$, keeps the same share $\phi(g)$ of the state's contribution to returns. Likewise, $\phi(a)$ and $\phi(q)$ are the reliabilities of the category signal and the detail.\footnote{CKX estimate the forecast by ridge regression, which shrinks low-variance directions of the embedding more than high-variance ones. Because the calibration matches the covariances of the ridge forecast and its components, the calibrated loadings absorb this shrinkage, so $\phi(a)$, $\phi(q)$ and $\phi(g)$ combine each signal's reliability with the ridge's shrinkage along its direction.}

\paragraph{Decomposition components.} Let $\bar x^F$ be the mean of $x$ over the firm's news items in the projection window, and $\bar x^L=\mathcal L x$ the historical level of $x$. Suppose that
\[
\mathbb E[\tilde\nu\mid K]=\mathbb E[\tilde\nu\mid i]=0, \qquad \mathbb E[\zeta\mid K]=\mathbb E[\zeta\mid i]=0, \qquad \mathbb E[x-\bar x^F\mid K]=0
\]
where $i$ is the firm and expectations are taken over the news items of the projection window. The two-way projection of equation~\eqref{eq:content} then recovers
\[
C=a_K, \qquad F=g\left(\bar z+\bar x^F\right)e_z, \qquad I=g\left(x-\bar x^F\right)e_z+q\tilde\nu e_\nu+\zeta
\]
Applying $\mathcal L$ splits $I$ into $I_L$, whose state is $\bar x^L-\bar x^F$, and $I_D$, whose state is $x-\bar x^L$.

Suppose the firm's news is spread evenly over time. The projection window covers the six years before the test year. With $\tau=1,\ldots,12$ the month of the news in the test year and $\ell$ counting months before the news, the covariances of $\bar x^F$ and $\bar x^L$ with the state $h-1$ trading days later are $\bar\rho_F\rho^{h-1}\Var(x)$ and $\bar\rho_L\rho^{h-1}\Var(x)$, where
\[
\bar\rho_F = \frac{1}{12}\sum_{\tau=1}^{12}\frac{1}{72}\sum_{\ell=\tau}^{71+\tau}\rho^{21\ell},
\qquad
\bar\rho_L = \frac{1}{12}\sum_{\ell=1}^{12}\rho^{21\ell}.
\]
Table~\ref{tab:D1} collects the loadings for a firm with estimated $F$ and a historical level, where blank cells are zero.

\begin{table}[htbp]
\centering
\caption{Loadings of the Decomposition Components}
\label{tab:D1}
\begin{minipage}{\linewidth}\footnotesize
This table reports each decomposition component's loadings for a firm with estimated $F$ and a historical level.
\end{minipage}
\medskip

{\footnotesize
\begin{adjustbox}{max width=\linewidth}
\begin{tabular}{llcccc}
\toprule
 &  & $\tilde m_K$ & $\tilde\nu$ & $\bar z$ & $x$ \\
\midrule
Category level & $C_L$ &  &  &  &  \\
Category deviation & $C_D$ & $a$ &  &  &  \\
Firm identity & $F$ &  &  & $g$ & $g\bar\rho_F$ \\
Idiosyncratic level & $I_L$ &  &  &  & $g(\bar\rho_L-\bar\rho_F)$ \\
Idiosyncratic deviation & $I_D$ &  & $q$ &  & $g(1-\bar\rho_L)$ \\
\bottomrule
\end{tabular}
\end{adjustbox}
}
\end{table}

\paragraph{Predictive covariance by component.} Let $E^{(j)}$ be the embedding's component $j$, $\hat y^j=\beta'E^{(j)}$ the forecast's component $j$, and let $\Gamma_j(h)=\Cov(\hat y^j,r_h)$. Then
\begin{equation}\label{eq:D2}
\Gamma_{C_D}(h) = \phi(a)\delta_1\delta_h\Var(m_K)
\end{equation}
\begin{equation}\label{eq:D3}
\Gamma_F(h) = \phi(g)\pi^2\left(\Var(\bar z)+\bar\rho_F\rho^{h-1}\Var(x)\right)
\end{equation}
\begin{equation}\label{eq:D4}
\Gamma_{I_L}(h) = \phi(g)\pi^2(\bar\rho_L-\bar\rho_F)\rho^{h-1}\Var(x)
\end{equation}
\begin{equation}\label{eq:D5}
\Gamma_{I_D}(h) = \phi(q)\delta_1\delta_h\Var(\nu) + \phi(g)\pi^2(1-\bar\rho_L)\rho^{h-1}\Var(x)
\end{equation}
The category level averages the categories of the firm's news in earlier months. Because categories are independent across news items, it predicts returns only through the drift that this earlier news has left, which is negligible under~\eqref{eq:D1}. 
We therefore set $\Gamma_{C_L}(h)=0$ and assume that details of earlier news in $I_L$ carry no predictive covariance. In the data, the category level accounts for about 2\% of next-day predictive covariance. Its larger monthly predictability lies outside the model. Event drift enters only the deviations and fades at the rate $\vartheta$, within days, whereas the state decays at the rate $\rho$, over years.

\paragraph{Event categories and embeddings.} At $h=1$, the sum of \eqref{eq:D2} to \eqref{eq:D5} is the embedding forecast's predictive covariance. The category-mean forecast captures all of the category's drift, so its covariance is $\delta_1^2\Var(m_K)$. The category-mean forecast therefore has the higher next-day $R^2$ if and only if\footnote{At the calibrated parameters in Table~\ref{tab:D2}, the left side is 0.79 against 0.13 on the right for RoBERTa and 0.74 against 0.24 for Llama-2, in units of $10^{-6}$.}
\begin{equation}\label{eq:D6}
\left(1-\phi(a)\right)\delta_1^2\Var(m_K) > \phi(q)\delta_1^2\Var(\nu)+\phi(g)\pi^2
\end{equation}

\paragraph{Earnings surprises.} For earnings news, SUE proxies for the signal $s=m_K+\nu$. Given the category, the embedding measures $\nu$ through $q\tilde\nu$ plus unit-variance noise along $e_\nu$. Its gain over the category in explaining SUE is therefore
\begin{equation}\label{eq:D7}
R^2(s \mid E,K)-R^2(s \mid K) = \phi(q)\left(1-R^2(s \mid K)\right)
\end{equation}
Return covariances, as in \eqref{eq:D5}, involve only the product $\phi(q)\Var(\nu)$, so Section~\ref{sec:appD3} uses \eqref{eq:D7} to set $\phi(q)$ separately.

\paragraph{Finite samples.} CKX estimate $\beta$ by ridge regression with penalty $\lambda$ on $n_{\rm train}$ training news items, with embeddings $E_t$ and next-day returns $r_{1,t}$. Let $\Var(E)$ denote the embeddings' variance, the same in the training sample and the test period.

\begin{proposition}[Finite samples]\label{prop:D1}
Let $p_\lambda=\sum_k\left(\lambda_k/(\lambda_k+\lambda)\right)^2$ be the ridge's effective number of parameters, where $\lambda_k$ are the variances of the embedding's PCs, and let $\sigma_r^2=\Var(r_1)$. Suppose that (i) given the training embeddings, $\Var(r_{1,t})\approx\sigma_r^2$ and $\Cov(r_{1,t},r_{1,t'})=0$ for $t\neq t'$; (ii) the ridge's population coefficients $\left(\Var(E)+\lambda\mathbf I_d\right)^{-1}\Cov(E,r_1)$ are approximated by $\beta$; and (iii) the estimation error $\hat\beta-\beta$ has mean zero and is independent of test-period embeddings and returns. Then, to first order,
\begin{equation}\label{eq:D8}
\mathrm{IC}^2 \approx \frac{\mathrm{IC}^2_{\rm pop}}{1+p_\lambda\sigma_r^2/(n_{\rm train}\Gamma(1))}
\end{equation}
where $\Gamma(1)$, the sum of \eqref{eq:D2} to \eqref{eq:D5} at $h=1$, is the forecast's population covariance with next-day returns, and $\mathrm{IC}_{\rm pop}$ is its population correlation. The Mincer--Zarnowitz slope of returns on the estimated forecast is
\begin{equation}\label{eq:D9}
b_{\rm MZ} \approx \frac{1}{1+p_\lambda\sigma_r^2/(n_{\rm train}\Gamma(1))}
\end{equation}
For a target $s$ such as SUE, with $\Var(s \mid E)$ in place of $\sigma_r^2$, the expected out-of-sample $R^2$ is approximately
\[
R^2_{\rm pop}-\frac{p_\lambda}{n_{\rm train}}\frac{\Var(s \mid E)}{\Var(s)}
\]
\end{proposition}

\begin{proof}
The estimation error is
\[
\hat\beta-\beta=\left(\Var(E)+\lambda\mathbf I_d\right)^{-1}\left(\widehat{\Cov}(E,r_1)-\Cov(E,r_1)\right)
\]
Under (i), its covariance matrix is approximately
\[
\Var(\hat\beta-\beta)\approx\frac{\sigma_r^2}{n_{\rm train}}\left(\Var(E)+\lambda\mathbf I_d\right)^{-1}\Var(E)\left(\Var(E)+\lambda\mathbf I_d\right)^{-1}
\]
Because the error has mean zero and is independent of test-period embeddings and returns,
\[
\mathbb E\left[\Cov(\hat\beta'E,r_1)\right]=\beta'\Cov(E,r_1)=\Gamma(1)
\]
\[
\mathbb E\left[\Var(\hat\beta'E)\right]=\beta'\Var(E)\beta+\operatorname{tr}\left(\Var(E)\Var(\hat\beta-\beta)\right)\approx\Gamma(1)+\frac{p_\lambda\sigma_r^2}{n_{\rm train}}
\]
With $\mathrm{IC}^2_{\rm pop}=\Gamma(1)/\sigma_r^2$, ratios of these expected moments give \eqref{eq:D8} and \eqref{eq:D9}. For $s$, the out-of-sample $R^2$ is
\[
R^2_{\rm oos}=\frac{2\Cov(\hat s,s)-\Var(\hat s)}{\Var(s)}
\]
and the same steps give the last expression.
\end{proof}

The IC falls short of its population value, and the Mincer--Zarnowitz slope lies below one, when $p_\lambda\sigma_r^2/(n_{\rm train}\Gamma(1))$ is large, as it is for a weak signal estimated with many effective parameters from few news items. For the same reason, an embedding's out-of-sample gain in explaining SUE lies below its population gain \eqref{eq:D7}.

\paragraph{Monthly aggregation.} As in Section~\ref{sec:monthagg}, the firm's monthly embedding in month $\tau$ averages its $n$ news items in the month, which arrive on trading days $t_1,\ldots,t_n$:
\[
E^m_\tau=\frac{1}{n}\sum_{k=1}^{n}E_{t_k}
\]
We write $E^m$ when the month is clear.
Let $r^m_h$ be the firm's return in the $h$th month after the signal month. The state is constant within the month and contributes $\pi_m z$ to the next month's expected return, with $\pi_m=21\pi$. Monthly persistence is $\rho_m=\rho^{21}$, and $\bar\rho_F$ and $\bar\rho_L$ average powers of $\rho_m$.

Along $e_z$, the monthly embedding is $gz$ plus the average of the $n$ items' noise, whose variance is $1/n$. Along $e_K$ and $e_\nu$, averaging divides the variance of the signal and of the noise alike by $n$. The month's return contains the drift of all $n$ items and the premium of their common state, so averaging leaves each direction's covariance with it unchanged. Pooled across firm-months, the noise variance along $e_z$ is $\mathbb E[1/n]$. The monthly moments are therefore those of a firm-month with $\bar n$ items, where $1/\bar n=\mathbb E[1/n]$, so that $\bar n$ is the harmonic mean of the number of news items per firm-month in the data.

News arrives evenly over the month's 21 trading days. For news whose $\ell$th post-news return is the first in the next month, $\ell=1,\ldots,21$, the $h$th month after the signal month contains the drift of trading days $21(h-1)+\ell$ through $21h+\ell-1$. Averaged over $\ell$, the drift realized in that month per unit of signal is
\footnote{More generally, monthly aggregation sums the daily return responses to a news signal and cannot itself create persistence. A persistent monthly response to the same signal therefore requires a correspondingly persistent daily response, rather than arising from temporal aggregation alone.}
\[
\bar\delta_h = \frac{1}{21}\sum_{\ell=1}^{21}\sum_{h'=21(h-1)+\ell}^{21h+\ell-1}\delta_{h'} = \frac{(1-\kappa)\left(1-\vartheta^{21}\right)^2}{21(1-\vartheta)}\vartheta^{21(h-1)}
\]
At the calibrated $\vartheta$, the first month after the signal month receives about 6\% of the post-announcement drift, and later months receive almost none. For the projection of $r^m_1$ on $E^m$, with coefficients $\beta^m=\Var(E^m)^{-1}\Cov(E^m,r^m_1)$, the components' covariances with $r^m_h$ are
\begin{equation}\label{eq:D10}
\Gamma^m_{C_D}(h) = \bar n\phi(a)\bar\delta_1\bar\delta_h\Var(m_K)
\end{equation}
\begin{equation}\label{eq:D11}
\Gamma^m_F(h) = \phi(\sqrt{\bar n}g)\pi_m^2\left(\Var(\bar z)+\bar\rho_F\rho_m^{h-1}\Var(x)\right)
\end{equation}
\begin{equation}\label{eq:D12}
\Gamma^m_{I_L}(h) = \phi(\sqrt{\bar n}g)\pi_m^2(\bar\rho_L-\bar\rho_F)\rho_m^{h-1}\Var(x)
\end{equation}
\begin{equation}\label{eq:D13}
\Gamma^m_{I_D}(h) = \bar n\phi(q)\bar\delta_1\bar\delta_h\Var(\nu) + \phi(\sqrt{\bar n}g)\pi_m^2(1-\bar\rho_L)\rho_m^{h-1}\Var(x)
\end{equation}
and the model sets $\Gamma^m_{C_L}(h)$ to zero. Averaging raises the state's effective loading from $g$ to $\sqrt{\bar n}g$ and leaves $\phi(a)$ and $\phi(q)$ unchanged, while little drift remains after the signal month. Daily and monthly predictability can therefore be dominated by different sources: event categories at the next-day horizon and the historical level at monthly horizons.

Panel B of Table~\ref{tab:tab7} reports the monthly average of the daily forecast, which retains the daily regression coefficients rather than re-estimating them at the monthly frequency. The covariance of its $F+I_L$ component with $r_h^m$ is
\begin{equation}\label{eq:D14}
\phi(g)\pi\pi_m\left(\Var(\bar z)+\bar\rho_L\rho_m^{h-1}\Var(x)\right)
\end{equation}
The historical level's share of the state's predictive covariance,
\[
\frac{\Var(\bar z)+\bar\rho_L\rho_m^{h-1}\Var(x)}{\Var(\bar z)+\rho_m^{h-1}\Var(x)}
\]
rises toward one with the horizon when $\Var(\bar z)>0$. Beyond drift, the deviation predicts returns only through the state's change since the past year, which later historical levels absorb. Let $U_\tau=x_\tau-\rho_m x_{\tau-1}$ denote the innovation to the evolving state in the signal month, so that $\Var(U)=(1-\rho_m^2)\Var(x)$. It is the newly arriving information that DKPT's news shocks aim to capture. The historical level does not yet reflect it, so it enters only the deviation, with predictive covariance
\[
\phi(\sqrt{\bar n}g)\,\pi_m^2(1-\rho_m^2)\rho_m^{h-1}\Var(x).
\]
This replaces $1-\bar\rho_L$ in the state term of \eqref{eq:D13} by $1-\rho_m^2$, so $U_\tau$ accounts for a share $(1-\rho_m^2)/(1-\bar\rho_L)$ of the deviation's state term, about a third at the calibrated values. Its share of the predictive covariance of the whole state, summed over firm identity, the idiosyncratic level, and the deviation, is
\[
\frac{(1-\rho_m^2)\rho_m^{h-1}\Var(x)}{\Var(\bar z)+\rho_m^{h-1}\Var(x)},
\]
about 2\% at every horizon up to 36 months, including the first. In the data, the deviation's first-month predictability comes mainly from the category deviation, especially among micro stocks, whose event drift lasts longer.

\paragraph{Maximum Sharpe ratio regression.} In a signal month, let $N$ be the number of stocks with news. Let $E^\perp_i\in\mathbb R^d$ be stock $i$'s input to DKPT's residualized-embedding strategy: $E^m$ residualized on firm characteristics and a constant, or only centered across stocks. Let $r^m_{h,i}$ be the stock's return in the $h$th month, with $\mu_i=\mathbb E[r^m_{1,i} \mid E^\perp]$. The $k$th managed portfolio weights stocks by the $k$th coordinate of $E^\perp_i$, so the vector of managed-portfolio returns is
\[
f=\sum_{i=1}^{N}E^\perp_i r^m_{1,i}
\]
MSRR chooses the weights $b$ on these portfolios to maximize the Sharpe ratio of $b'f$, and the strategy holds stock $i$ with weight $b'E^\perp_i$.

\begin{proposition}[MSRR as a projection]\label{prop:D2}
Suppose that, conditional on $E^\perp$, return innovations are uncorrelated across stocks and have a common variance $\sigma^2$. Approximating cross-sectional sample moments by their population counterparts and abstracting from variation in conditional portfolio means, the MSRR weights satisfy
\begin{equation}\label{eq:D15}
b \propto \Var(E^\perp)^{-1}\Cov(E^\perp,r^m_1),
\end{equation}
and the strategy's squared Sharpe ratio is approximately
\begin{equation}\label{eq:D16}
\mathrm{SR}^2 \approx
\frac{N}{\sigma^2}
\Cov(E^\perp,r^m_1)'
\Var(E^\perp)^{-1}
\Cov(E^\perp,r^m_1).
\end{equation}
\end{proposition}

\begin{proof}
MSRR regresses a constant on the managed-portfolio returns $f$, so its population weights satisfy
\[
b\propto\mathbb E[ff']^{-1}\mathbb E[f]
\propto\Var(f)^{-1}\mathbb E[f],
\]
where the second proportionality follows from the Sherman--Morrison formula.

Since $E^\perp$ is centered across stocks,
\[
\mathbb E[f\mid E^\perp]
=\sum_i E^\perp_i\mu_i
\approx N\Cov(E^\perp,r^m_1).
\]
Under the assumed return covariance structure,
\[
\Var(f\mid E^\perp)
=\sigma^2\sum_i E^\perp_i E_i^{\perp\prime}
\approx N\sigma^2\Var(E^\perp).
\]
Ignoring the variation in $\mathbb E[f\mid E^\perp]$ across samples gives
$\Var(f)\approx N\sigma^2\Var(E^\perp)$.
Substitution yields \eqref{eq:D15}, while
$\mathrm{SR}^2=\mathbb E[f]'\Var(f)^{-1}\mathbb E[f]$
gives \eqref{eq:D16}.
\end{proof}

Proposition~\ref{prop:D2} links the MSRR strategy to the predictive covariances of the embedding components. Without shrinkage, its weights are proportional to the coefficients of a linear return projection. The portfolio weight on stock $i$ is therefore proportional to its forecast, and the decomposition of the embedding induces the same decomposition of portfolio weights.

Let $\hat y_i^{m,j}$ denote component $j$'s contribution to the forecast. Its contribution to the strategy's return at horizon $h$, $R_h^{(j)}$, the term of component $j$ in the attribution of Section~\ref{sec:attribution}, satisfies
\begin{equation}\label{eq:D17}
\mathbb E[R_h^{(j)}]
=\Upsilon\Cov(\hat y^{m,j},r_h^m)
=\Upsilon\,\Gamma_j^m(h),
\end{equation}
where the factor $\Upsilon$ is encoder-specific but common across components and horizons. It cancels in return ratios, allowing cross-component ratios to calibrate the allocation of predictive covariance and cross-horizon ratios to calibrate the state's persistence and permanent share. These covariances are given by \eqref{eq:D10}--\eqref{eq:D13} for centered embeddings; the extension to embeddings residualized on firm characteristics is derived below.

\paragraph{Residualization on characteristics.}

We examine how DKPT's residualization affects the predictive covariance of each embedding component and the strategy's Sharpe ratio. Let $X$ denote characteristics observed in the signal month, and let $z_X$ be the projection of the state $z$ on $X$ and a constant. In our decomposition, the firm state enters $F$, $I_L$, and $I_D$ through
\[
z^F=\bar z+\bar x^F,\qquad
z^{I_L}=\bar x^L-\bar x^F,\qquad
z^{I_D}=x-\bar x^L,
\]
respectively. These terms sum to $z$. Let $z_X^j$ denote the projection of each $z^j$ on $X$ and a constant, so that $\sum_j z_X^j=z_X$. So far, the embedding along $e_z$ contains only $gz$ and language noise. To study residualization, we allow characteristics to move the text along this coordinate without reflecting the state. Along $e_z$, the monthly embedding is then $gz+\psi'X$ plus language noise of variance $1/\bar n$, where $\psi'X$ is this characteristic-driven text. The rest of the appendix, including the calibration, sets $\psi=0$. Residualization removes $\psi'X$, so the residualized monthly targets do not depend on $\psi$.

\begin{proposition}[Residualization on characteristics]\label{prop:D3}
Let $R^2(z\mid X)$ denote the fraction of state variance explained by $X$. Suppose that $a_K$, $\tilde\nu$, and $\zeta$ are uncorrelated with $X$, and that $X$ contains no additional information about expected returns conditional on the news and the state:
\[
\mathbb E[r_h^m\mid K,\nu,\bar z,x,X]
=
\mathbb E[r_h^m\mid K,\nu,\bar z,x].
\]
Residualizing each embedding component on $X$ and a constant leaves its event-drift contribution unchanged. For $j\in\{F,I_L,I_D\}$, the state contribution to its predictive covariance becomes
\begin{equation}\label{eq:D18}
\Gamma_j^{m\perp}(h)
=
\phi\left(\sqrt{\bar n(1-R^2(z\mid X))}\,g\right)
\pi_m^2
\Cov\left(z^j-z_X^j,\bar z+\rho_m^{h-1}x\right).
\end{equation}
\end{proposition}

\begin{proof}
Residualization removes $\psi'X$ and leaves the state coordinate $g(z-z_X)$ plus language noise. Since $\Var(z)=1$ and $\Var(z-z_X)=1-R^2(z\mid X)$, the residualized coordinate has variance
\[
\Var(E_z^\perp)
=
g^2(1-R^2(z\mid X))+\frac{1}{\bar n},
\]
where $E_z^\perp=e_z'E^\perp$. By the conditional-mean assumption, its covariance with next-month returns is
\[
\Cov(E_z^\perp,r_1^m)
=
g\pi_m\Cov(z-z_X,z)
=
g\pi_m(1-R^2(z\mid X)).
\]
The linear projection coefficient is the ratio of these moments. Multiplying it by the coordinate's state part gives
\[
\phi\left(\sqrt{\bar n(1-R^2(z\mid X))}\,g\right)
\pi_m(z-z_X).
\]
By linearity, this splits across components as $z^j-z_X^j$. The state $h$ months later is $\bar z+\rho_m^{h-1}x$ plus innovations uncorrelated with information in the signal month, yielding \eqref{eq:D18}.
\end{proof}

Equation \eqref{eq:D18} shows two effects of residualization on state predictability. First, characteristics remove the part of the state they explain, reducing the remaining state's predictive covariance with returns. Second, because language noise is unchanged, the remaining state is measured less reliably, further reducing its contribution to the forecast. At the same time, residualization can improve the Sharpe ratio by removing embedding variation unrelated to returns.

\paragraph{Sharpe ratios.} DKPT report that residualizing on JKP characteristics raises the Sharpe ratio of their strategy from 1.7 for the demeaned embedding to 3.1. A strategy on the fitted value of the embedding on characteristics, which they call ``predictable news,'' earns a lower Sharpe ratio than one on the residual (DKPT, Figure~4). Yet characteristics themselves predict returns. Under the conditional-mean assumption of Proposition~\ref{prop:D3}, characteristics predict returns only through $z$. If $\psi'X$ is also uncorrelated with $z$, as Proposition~\ref{prop:D4} assumes, it does not predict returns.

\begin{proposition}[Sharpe ratios after projection on characteristics]\label{prop:D4}
Let $\mathrm{SR}_z$ denote the Sharpe ratio of a strategy that trades on the state itself, with $\mathrm{SR}_z^2\approx(N/\sigma^2)\pi_m^2$, and let $\mathrm{SR}_X$ denote the Sharpe ratio of MSRR on the characteristics. Along $e_z$, let $\mathrm{SR}_{E\mid X}$, $\mathrm{SR}_\perp$, and $\mathrm{SR}$ denote the Sharpe ratios of the single-coordinate strategies on the fitted part, the residual, and the embedding. Suppose that the assumptions of Proposition~\ref{prop:D3} hold and that $\Cov(\psi'X,z)=0$. Then
\[
\begin{aligned}
\frac{\mathrm{SR}_X^2}{\mathrm{SR}_z^2}
&\approx
R^2(z\mid X),\\
\frac{\mathrm{SR}_{E\mid X}^2}{\mathrm{SR}_z^2}
&\approx
R^2(z\mid X)\,
\frac{g^2R^2(z\mid X)}{g^2R^2(z\mid X)+\Var(\psi'X)},\\
\frac{\mathrm{SR}_\perp^2}{\mathrm{SR}_z^2}
&\approx
\left(1-R^2(z\mid X)\right)
\frac{g^2\left(1-R^2(z\mid X)\right)}{g^2\left(1-R^2(z\mid X)\right)+1/\bar n},\\
\frac{\mathrm{SR}^2}{\mathrm{SR}_z^2}
&\approx
\frac{g^2}{g^2+\Var(\psi'X)+1/\bar n}.
\end{aligned}
\]
\end{proposition}

\begin{proof}
By Proposition~\ref{prop:D2}, applied with each signal in place of $E^\perp$, a single-coordinate strategy has a squared Sharpe ratio of approximately $N/\sigma^2$ times the squared covariance of its signal with $r_1^m$, divided by the signal's variance. Under the assumptions of Proposition~\ref{prop:D3}, this covariance is $\pi_m$ times the signal's covariance with $z$. We use $\Var(z)=1$ and $\Var(z_X)=\Cov(z_X,z)=R^2(z\mid X)$. Because $z-z_X$ is uncorrelated with $X$, $\Cov(\psi'X,z_X)=\Cov(\psi'X,z)=0$.

By \eqref{eq:D15}, MSRR on the characteristics has a signal proportional to $z_X$, whose covariance with $z$ and variance both equal $R^2(z\mid X)$. The fitted part is $gz_X+\psi'X$, with covariance $gR^2(z\mid X)$ and variance $g^2R^2(z\mid X)+\Var(\psi'X)$. The residual is $g(z-z_X)$ plus language noise, with covariance $g(1-R^2(z\mid X))$ and variance $g^2(1-R^2(z\mid X))+1/\bar n$, as in the proof of Proposition~\ref{prop:D3}. The embedding is $gz+\psi'X$ plus language noise, with covariance $g$ and variance $g^2+\Var(\psi'X)+1/\bar n$. The ratio of each signal's squared covariance with $z$ to its variance gives the four expressions.
\end{proof}

Each ratio is the share of the state that the signal contains, times the reliability with which the signal measures that share. The first two ratios explain why predictable news earns little. MSRR on the characteristics and the fitted part contain the same share of the state, $R^2(z\mid X)$. MSRR weights the characteristics by how well they predict returns, whereas the fitted part weights them by their loadings in the text, which adds $\psi'X$, a term unrelated to returns. When $\Var(\psi'X)$ is large relative to $g^2R^2(z\mid X)$, the fitted part earns little although the characteristics predict returns. Language noise does not enter the fitted part and plays no role in this comparison.

The last two ratios show how residualization changes the Sharpe ratio. Removing $\psi'X$ raises it, and removing $gz_X$ lowers it. With $R^2(z\mid X)=0$, only the first effect operates, and removing three-quarters of the coordinate's variance doubles the Sharpe ratio. With $\psi=0$, only the second operates, and $\mathrm{SR}_\perp^2/\mathrm{SR}^2\le1-R^2(z\mid X)$. A large $\Var(\psi'X)$ therefore produces both of DKPT's findings: the fitted part earns little, and removing it raises the Sharpe ratio. MSRR on the embedding cannot obtain this gain by itself, because $\psi'X$ lies on the same coordinate as $gz$. The gain does not show that the residual isolates new information. By \eqref{eq:D18}, the state that characteristics do not span stays in the residual, in firm identity, the idiosyncratic level and the idiosyncratic deviation.

\paragraph{Firm identity and valuations.} Let log price-to-book at the end of year $T$ be linear in the state, with an error uncorrelated with the state and the embedding. Let firm identity be estimated on a window of $\bar n^F$ news items ending then, where $\bar n^F$ is the harmonic mean across firms, with $z^F_T=\bar z+\bar x^F_T$. In a finite window, firm identity's state coordinate is $gz^F_T$ plus the mean of its items' noise, of variance $1/\bar n^F$. The population $R^2$ of firm identity's state coordinate is
\begin{equation}\label{eq:D20}
R^2_T = R^2\left(\log(P/B)_T \mid z_T\right)\Corr\left(z^F_T,z_T\right)^2\phi\left(\sqrt{\bar n^F\Var(z^F_T)}g\right)
\end{equation}
Only the third term depends on the encoder, so the ratio of two encoders' $R^2$ at the same date equals the ratio of their third terms. Because firm identity averages many news items, both third terms are close to one at the calibrated loadings, so the ratio is a weak check on the state loadings rather than an independent estimate of them.

\subsection{Look-Ahead Bias}
\label{sec:appD2}

\paragraph{Sources of look-ahead bias.}
Let $\mathcal I_t$ denote investors' information through $t$, and let $\mathbb E_t[\cdot]=\mathbb E[\cdot\mid\mathcal I_t]$. At the monthly frequency, $\mathbb E_\tau$ denotes expectations given information at the end of month $\tau$. For a return forecast $\hat y_t$ constructed from the embedding $E_t$, with coefficients fixed, define its look-ahead component as
\[
\hat y_t^{\mathrm{LA}}=\hat y_t-\mathbb E_t[\hat y_t].
\]
Look-ahead bias arises when this component covaries with returns that are unpredictable from investors' information (see also \citealp{sarkar2024lookahead}). Its covariance with future returns can be decomposed as
\begin{equation}\label{eq:D21}
\Cov(\hat y_t^{\mathrm{LA}},r_h)
=
\pi\Cov(\hat y_t^{\mathrm{LA}},z_{t+h-1})
+\Cov(\hat y_t^{\mathrm{LA}},r_h-\pi z_{t+h-1}).
\end{equation}

The first term arises when an encoder pretrained on later text captures information about the priced firm state that was unavailable to investors at $t$. The predictive value of this information need not disappear when returns move beyond the encoder's cutoff. Knowledge of the permanent state remains useful, while knowledge of the evolving state decays gradually with its persistence. We capture this channel through the state loading $g$ rather than model it separately, since persistent state information alone does not explain the relative decline in historical-level predictability across the cutoff documented in Section~\ref{sec:cutoff}.

The second term captures information about returns beyond the state premium. We focus on this channel because memory of realized returns can generate a sharp decline in predictability when return dates pass the encoder's cutoff. Consistent with this mechanism, \citet{lopez2025memorization} and \citet{gao2025detecting} document sharp declines in LLM recall of realized financial outcomes after training cutoffs. We model this channel through the encoder's memory of cash-flow news.

\paragraph{Return dynamics with cash-flow news.}
To specify which realized returns an encoder can remember, we start from the log-linear decomposition of \citet{campbell1991variance}, built on \citet{campbell1988dividend}, which expresses unexpected monthly returns as cash-flow news minus discount-rate news. With $r^m_{i,\tau}$ firm $i$'s log return in month $\tau$,
\[
r^m_{i,\tau+1}-\mathbb E_\tau r^m_{i,\tau+1}
=N_{CF,i,\tau+1}-N_{DR,i,\tau+1}.
\]
Cash-flow news is the revision in expected future dividend growth, while discount-rate news is the revision in expected future returns:
\[
N_{DR,i,\tau+1}
=(\mathbb E_{\tau+1}-\mathbb E_\tau)\sum_{k\ge1}\chi^k r^m_{i,\tau+1+k},
\]
where $\chi$ is the discount coefficient of the log-linearization. These expectation revisions may be induced by published news. In our framework, the response to a published news item is captured by its category $K$ and details $\nu$.

In our model, firm $i$'s state in month $\tau$ is $z_{i,\tau}=\bar z_i+x_{i,\tau}$, with evolving component $x_{i,\tau}$. It contributes $\pi_m z_{i,\tau}$ to the next month's expected return, while published news generates the drift specified above. To derive state-related discount-rate news, we assume that investors' beliefs about the permanent state remain fixed, while innovations in the evolving state are reflected in their expected returns. The monthly state dynamics then imply
\[
(\mathbb E_{\tau+1}-\mathbb E_\tau)z_{i,\tau+k}
=\rho_m^{k-1}(x_{i,\tau+1}-\rho_m x_{i,\tau}).
\]
The state-related component of discount-rate news is thus
\[
N_{DR,i,\tau+1}^{(z)}
=\frac{\pi_m\chi}{1-\chi\rho_m}(x_{i,\tau+1}-\rho_m x_{i,\tau}).
\]

For tractability, we approximate additional unexpected returns with daily cash-flow-news shocks $\theta_{i,t}$. We assume that $\mathbb E_{t-1}[\theta_{i,t}]=0$ and $\Var(\theta_{i,t})=\sigma_\theta^2/21$, and that these shocks are independent across trading days and of the state, categories, news details, and language noise. Their monthly sum has variance $\sigma_\theta^2$. This simplified specification does not separately model the return impact of state-related discount-rate news or its covariance with innovations in the firm state.

The return $h$ trading days after a news item published at $t$, conditional on the realized cash-flow-news shock, satisfies
\begin{equation}\label{eq:D22}
\mathbb E[r_h\mid K,\nu,\bar z,x,\theta_{i,t+h}]
=\delta_h(m_K+\nu)+\pi(\bar z+\rho^{h-1}x)+\theta_{i,t+h}.
\end{equation}
The three terms capture the drift following published news, the state premium, and the cash-flow-news shock realized on that trading day. Unlike the first two terms, the future shock is unpredictable from investors' information at $t$, but may be remembered by an encoder pretrained on later text.

We calibrate $\sigma_\theta^2$ using the variance of cumulative monthly returns over 36 months, net of the variance of the modeled state premia. We use 36 months, the longest horizon of the monthly targets. This gives the approximation
\begin{equation}\label{eq:D23}
36\sigma_\theta^2\approx\Var\left(\sum_{\ell=1}^{36}r^m_{i,\tau+\ell}\right)-\pi_m^2\left[36^2\Var(\bar z)+\Var(x)\sum_{\ell=1}^{36}\sum_{\ell'=1}^{36}\rho_m^{|\ell-\ell'|}\right].
\end{equation}
Returns are monthly log returns, demeaned and winsorized within month, and the state is measured at the start of each month. The residual variance also includes published-news drift and discount-rate news, as well as covariance terms omitted from the simplified model.

To gauge the importance of the omitted discount-rate component, note that the state dynamics imply
\[
\Var(N_{DR}^{(z)})
=\pi_m^2\Var(x)(1-\rho_m^2)\left(\frac{\chi}{1-\chi\rho_m}\right)^2.
\]
Using the parameter values calibrated in Section~\ref{sec:appD3} and $\chi=0.967^{1/12}$ from \citet{vuolteenaho2002drives}, this variance is approximately 18\% of $\sigma_\theta^2$.

\paragraph{Memory of cash-flow news.}
Let $t^\ast$ denote the encoder's pretraining cutoff, and let $N_\theta$ denote the length of its memory window in months, assuming 21 trading days per month. We represent the encoder's memory of firm $i$'s realized cash-flow news by
\[
\tilde\theta_i^\ast
=\frac{1}{\sigma_\theta\sqrt{N_\theta}}
\sum_{t=t^\ast-21N_\theta+1}^{t^\ast}\theta_{i,t},
\]
which has unit variance. Let $\eta\ge0$ govern the strength of this memory in the embedding. The measurement equation becomes
\[
E_t=\bar E+a_K+q\tilde\nu e_\nu+gz e_z+\eta\tilde\theta_i^\ast e_\theta+\zeta_t,
\]
where $e_\theta$ is a unit direction orthogonal to the other signal directions and to $a_K^\perp$. The loading $\eta$ is in the same units as $a$, $q$, and $g$.

We set $N_\theta=360$ months, or $7{,}560$ trading days, for both encoders. The cutoff month is September 2022 for Llama-2 and July 2019 for RoBERTa, with the latter following CKX. We take $t^\ast$ to be the last trading day of the cutoff month. Both memory windows begin before January 2000 and include all of the sample's cash-flow news through their respective cutoffs. ChronoGPT, with its 1999 cutoff, remembers none of the sample's cash-flow news.

For a daily return $r_h$ following a news item published at $t$, \eqref{eq:D22} implies
\begin{equation}\label{eq:D24}
\Cov(\eta\tilde\theta_i^\ast,r_h)
=\frac{\eta\sigma_\theta}{21\sqrt{N_\theta}}
\mathbbm{1}\{t^\ast-21N_\theta<t+h\le t^\ast\}.
\end{equation}
This covariance contributes to the second term of \eqref{eq:D21} and vanishes when the return date exceeds the encoder's cutoff.

Our specification captures firm-level memory shared across news items. Memory of returns following individual news items, by contrast, would vary across headlines and could contribute to the deviation component. The absence of a cutoff-related decline in deviation predictability (Section~\ref{sec:cutoff}) suggests that such news-specific memory is not the main source of look-ahead bias.

\paragraph{Memory in the forecast.}
Because training returns contain information captured by the encoder's memory, the forecast assigns weight to $e_\theta$. If all training months fall within the memory window, the daily regression coefficient is $\eta\sigma_\theta/[21\sqrt{N_\theta}(1+\eta^2)]$, yielding a daily memory covariance of
\begin{equation}\label{eq:D25}
M=\phi(\eta)\frac{\sigma_\theta^2}{21^2N_\theta}.
\end{equation}
With partial coverage, the coefficient scales with the fraction of training months within the window. Direct monthly projection instead yields
\begin{equation}\label{eq:D26}
M^m=\phi(\sqrt{\bar n}\eta)\frac{\sigma_\theta^2}{N_\theta},
\end{equation}
compared with $21M$ from averaging daily forecasts. Since $\phi(\eta)\le1$, daily memory covariance is bounded by $\sigma_\theta^2/(21^2N_\theta)$, decreasing with the memory window's length. Residualization against signal-date characteristics leaves the memory coordinate's covariance with future returns unchanged, because future cash-flow shocks are independent of those characteristics. We also take $\tilde\theta_i^\ast$ to be uncorrelated with the characteristics, so that \eqref{eq:D26} is unchanged as well; past shocks in $\tilde\theta_i^\ast$ that characteristics reflect would lower the memory coordinate's variance and raise the memory covariance.

\paragraph{Memory in the decomposition components.}
Because memory is constant across a firm's news items, it enters the same components as the permanent state. When a firm effect is estimated, firm identity absorbs the memory covariance $M_t(h)$ in addition to the baseline covariance $\Gamma_F(h)$ in \eqref{eq:D3}, where $M_t(h)$ equals \eqref{eq:D25} scaled by the share of training months within the memory window if the return date $t+h$ is remembered, and zero otherwise. The historical level $E_L$ inherits the same memory covariance, while, for such a firm, the other components have none. 

\paragraph{The cutoff comparison.}
Predictability can change across an encoder's pretraining cutoff because of both memory and changes in baseline predictability. To distinguish these effects, we compare changes in each encoder's IC with those of a control encoder evaluated on the same firm-days.

For each comparison, let $p\in\{1,2\}$ denote the periods before and after the chosen split. For encoder $o$ and component $j$, let $\bar\Gamma_{j,p}^o$ and $\bar M_{j,p}^o$ denote the mean baseline and memory covariances in period $p$. Define
\[
\omega_j^o
=\frac{\bar\Gamma_{j,1}^o}
{\bar\Gamma_{j,1}^o+\bar M_{j,1}^o},
\qquad
\xi_j^o
=\frac{\bar\Gamma_{j,2}^o}{\bar\Gamma_{j,1}^o}.
\]
The first, the memory-free share, is the fraction of period-1 predictive covariance not attributable to memory, and $1-\omega_j^o$ is the memory share. The second measures the change in baseline predictability.

We allow the state-premium parameter $\pi$ and cross-sectional state variances to differ across periods, while holding the encoder's signal loadings fixed. Applying the population projection in \eqref{eq:D3} separately to each period gives, for example,\footnote{The squared dependence on $\pi$ assumes period-specific population projections. With fixed forecast coefficients, the covariance depends linearly on the period-specific state premium instead.}
\[
\xi_F^o
=\frac{\pi_2^2\left[\Var_2(\bar z)+\bar\rho_F\rho^{h-1}\Var_2(x)\right]}
{\pi_1^2\left[\Var_1(\bar z)+\bar\rho_F\rho^{h-1}\Var_1(x)\right]}.
\]
When an encoder's memory ceases to predict returns in period 2, $\bar M_{j,2}^o=0$, and its covariance ratio becomes
\begin{equation}\label{eq:D27}
\frac{\bar\Gamma_{j,2}^o+\bar M_{j,2}^o}
{\bar\Gamma_{j,1}^o+\bar M_{j,1}^o}
=\xi_j^o\omega_j^o.
\end{equation}
Thus, the change reflects both baseline predictability and the loss of memory.

Our empirical measure of predictive content is rank IC rather than covariance.\footnote{Unlike covariance, rank IC is invariant to monotonic transformations of forecasts and is less sensitive to extreme observations, making it more suitable for comparing predictive performance across encoders and periods.} Under joint normality and small correlations, rank IC is approximately proportional to Pearson correlation, with a factor common to all encoders and periods. Approximating period-average ICs using period-level predictive covariances and dispersions gives
\[
\mathrm{IC}_{j,p}^o
\propto
\frac{\bar\Gamma_{j,p}^o+\bar M_{j,p}^o}
{\sigma_{j,p}^o\sigma_{r,p}},
\]
where $\sigma_{j,p}^o$ and $\sigma_{r,p}$ denote the period-level cross-sectional dispersions of the forecast component and returns.

We assume that, absent memory, the two encoders would exhibit the same proportional change in rank IC for each component:\footnote{Memory-related predictability disappears at the pretraining cutoff date, creating a sharp break that may dominate other changes in baseline predictability.}
\[
\xi_j^{\mathrm{Llama2}}\frac{\sigma_{j,1}^{\mathrm{Llama2}}}{\sigma_{j,2}^{\mathrm{Llama2}}}
=
\xi_j^{\mathrm{RoBERTa}}\frac{\sigma_{j,1}^{\mathrm{RoBERTa}}}{\sigma_{j,2}^{\mathrm{RoBERTa}}}.
\]
After an encoder's pretraining cutoff, its memory covariance vanishes, so its memory-free share computed for that period equals one.

In the model, this assumption restricts changes in the composition of each forecast component. For $F$ and $E_L$, encoder $o$'s forecast component in period $p$ is a state term $g\,e_z'\beta^o_p\,z^j$ plus terms uncorrelated with the state: language noise, the memory coordinate and, for $E_L$, the category level. Here $\beta^o_p$ is the forecast's coefficient vector, $z^j$ is the part of the state in component $j$, with $z^{E_L}=\bar z+\bar x^L$, and the subscript $p$ marks moments and the premium in period $p$. Let $S^o_{j,p}$ denote the state term's share of the component's cross-sectional variance:
\[
S^o_{j,p}=\frac{\left(g\,e_z'\beta^o_p\right)^2\Var_p(z^j)}{\left(\sigma^o_{j,p}\right)^2}.
\]
Because $\bar\Gamma^o_{j,p}=g\,e_z'\beta^o_p\,\pi_p\Cov_p(z^j,z)$ for next-day returns, the coefficient and the loading cancel:
\[
\xi^o_j\frac{\sigma^o_{j,1}}{\sigma^o_{j,2}}
=\frac{\pi_2\Cov_2\left(z^j,z\right)}{\pi_1\Cov_1\left(z^j,z\right)}
\sqrt{\frac{\Var_1(z^j)\,S^o_{j,2}}{\Var_2(z^j)\,S^o_{j,1}}}.
\]
Both encoders predict returns for the same firms and periods, so all factors other than $S$ are common to both. The assumption therefore holds if and only if $S^o_{j,2}/S^o_{j,1}$ is the same for both. If loadings, coefficients up to scale, and the variance of terms unrelated to the state stay fixed, this ratio depends only on $S^o_{j,1}$ and $\Var_2(z^j)/\Var_1(z^j)$, and equals one when $\Var(z^j)$ is unchanged. At the calibrated parameters, the state term accounts for 24\% of Llama-2's firm-identity forecast variance before January 2023 and 99\% of RoBERTa's. Memory accounts for most of Llama-2's remaining variance. With equal memory-free shares, Llama-2's firm-identity ratio of 0.32 would require $\Var(z^F)$ to fall by about 93\%.

We first split at January 2023, the first January after Llama-2's cutoff, comparing Llama-2 with RoBERTa. Both encoders are past their pretraining cutoffs in period 2, giving
\begin{equation}\label{eq:D28}
\mathrm{RIC}_j
\equiv
\frac{\mathrm{IC}_{j,2}^{\mathrm{Llama2}}/\mathrm{IC}_{j,1}^{\mathrm{Llama2}}}
{\mathrm{IC}_{j,2}^{\mathrm{RoBERTa}}/\mathrm{IC}_{j,1}^{\mathrm{RoBERTa}}}
=
\frac{\omega_j^{\mathrm{Llama2}}}{\omega_j^{\mathrm{RoBERTa}}}.
\end{equation}

We then split at January 2020, the first January after RoBERTa's cutoff, using Llama-2 as the control and restricting the sample to dates through September 2022. RoBERTa crosses its pretraining cutoff, while both periods precede Llama-2's cutoff. Assuming Llama-2's memory-free share is the same in both periods, we obtain
\begin{equation}\label{eq:D29}
\mathrm{RIC}_j^*
\equiv
\frac{\mathrm{IC}_{j,2}^{\mathrm{RoBERTa}}/\mathrm{IC}_{j,1}^{\mathrm{RoBERTa}}}
{\mathrm{IC}_{j,2}^{\mathrm{Llama2}}/\mathrm{IC}_{j,1}^{\mathrm{Llama2}}}
=\omega_j^{\mathrm{RoBERTa},*}.
\end{equation}
The asterisk distinguishes the periods used in the second comparison.

\begin{figure}[htbp]
\centering
\caption{Relative IC Ratios across January 2023 and at Placebo Splits}
\includegraphics[width=\textwidth]{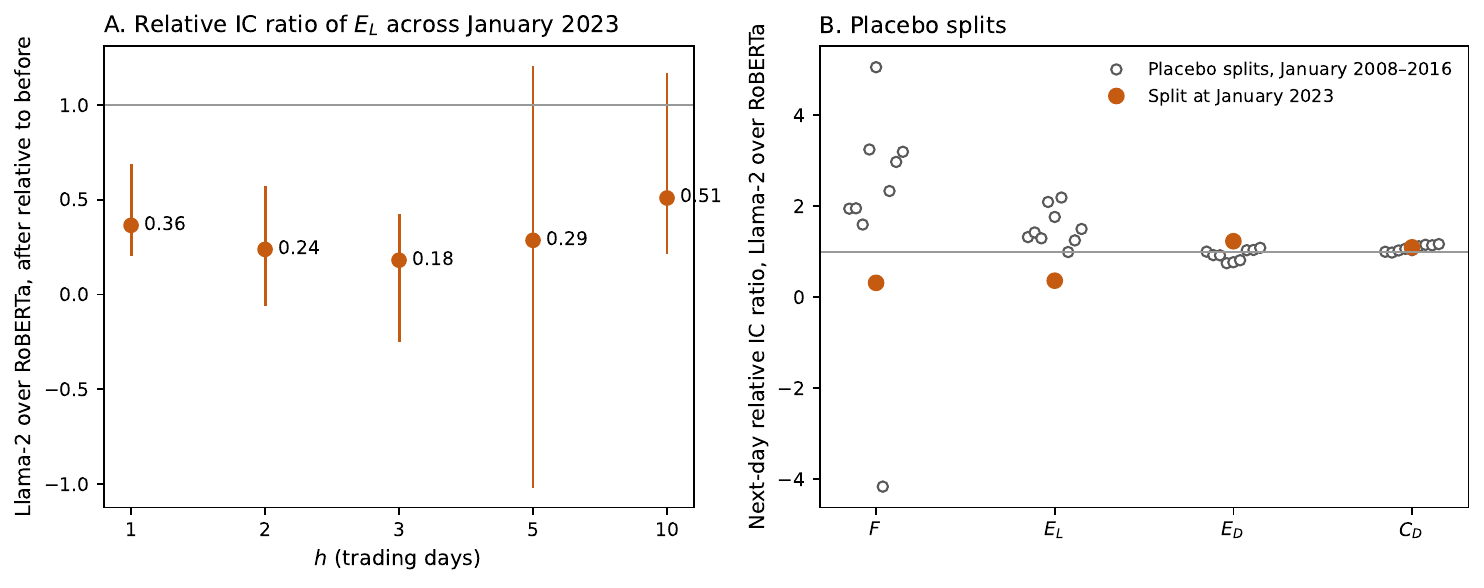}
\label{fig:D1}
\begin{minipage}{\linewidth}\footnotesize
This figure reports Llama-2's relative IC ratios against RoBERTa on common trading days. The ratio divides Llama-2's ratio of its later to its earlier mean daily rank IC by the same ratio for RoBERTa. Panel A gives the historical level's ratio across January 2023 at horizon $h$ trading days, with 90\% bootstrap intervals from 40-day moving blocks drawn within periods. Panel B gives next-day ratios for firm identity $F$, the historical level $E_L$, the deviation $E_D$, and the category deviation $C_D$ at a split each January from 2008 to 2016, on days from 2004 through three years after the split (hollow), and across January 2023 (filled).
\end{minipage}
\end{figure}

Because memory is concentrated in firm identity and the historical level, their relative IC ratios should be lower than those of the deviation components. Category deviation contains no memory, and the overall deviation contains little, so their ratios also provide a check on the common-change assumption. Figure~\ref{fig:D1} repeats the Llama-2 comparison at placebo splits each January from 2008 to 2016, when both encoders remember both periods. Category-deviation ratios range from 0.98 to 1.17 and overall-deviation ratios from 0.75 to 1.09, while firm-identity and historical-level ratios mostly exceed one. Their ratios across January 2023 lie below all placebo ratios of the historical level and all but one of firm identity's.

\paragraph{Full-forecast comparisons.}
CKX compare portfolio performance across an encoder's pretraining cutoff, while DKPT compare strategy performance across ChronoGPT vintages.\footnote{DKPT and \citet{kelly2026scaling} find similar strategy Sharpe ratios for point-in-time and full-sample versions of their models. In our framework, the effect of later pretraining depends on an encoder's ability to extract persistent firm information and its loading on memory. These effects can differ across models, particularly as representation capacity changes.} Such comparisons may obscure look-ahead bias concentrated in the historical level.

In our model, memory enters the historical level. Let $\Gamma$ denote total predictive covariance and $\Delta\Gamma$ its change attributable to memory, holding other sources of predictability fixed. Since $\Delta\Gamma=\Delta\Gamma_{E_L}$,
\begin{equation}\label{eq:D30}
\frac{\Delta\Gamma}{\Gamma}
=\frac{\Gamma_{E_L}}{\Gamma}
\frac{\Delta\Gamma_{E_L}}{\Gamma_{E_L}}.
\end{equation}
The effect on full-forecast predictability is scaled by the historical level's share of total predictive covariance. This share is small for CKX's next-day forecasts, where category deviations dominate. For DKPT's MSRR strategy, changes in Sharpe ratios also depend on portfolio optimization and return volatility, so they need not track changes in historical-level predictability. Component-level comparisons thus provide a more direct assessment of look-ahead bias.

\paragraph{Lagged historical level.}
We next consider how the predictive performance of the historical level changes when measured 36 months earlier. Under stationarity and comparable news coverage, the lagged and current levels have the same variance, while the lagged level's covariance with future state premia is reduced by the factor
\begin{equation}\label{eq:D31}
\frac{\operatorname{Var}(\bar z)+\bar\rho_L\rho_m^{36}\operatorname{Var}(x)}
{\operatorname{Var}(\bar z)+\bar\rho_L\operatorname{Var}(x)}.
\end{equation}
The permanent state remains informative, while the evolving state's contribution decays with $\rho_m^{36}$. In the data, the lagged level is available for 83\% as many firm-months as the current-level signal.\footnote{The ratio is computed among firm-months with news and next-month returns. Lagged coverage requires news during months $\tau-48$ through $\tau-37$, while current-level coverage requires either an estimated firm effect or news during months $\tau-12$ through $\tau-1$.} By \eqref{eq:D16}, the Sharpe ratio without memory is approximately $\sqrt{0.83}$ times \eqref{eq:D31} times that of the current-level strategy, assuming comparable signal quality across the two samples.

Memory, unlike the evolving state, is constant across a firm's news items and does not decay when the historical level is lagged. Lagging therefore reduces the historical level's state-related predictability without similarly reducing its memory-related predictability. Under the reoptimized MSRR weights in \eqref{eq:D15}, each independent predictive direction contributes to expected strategy returns in proportion to its squared covariance with returns divided by its variance. Holding signal variances fixed, the state contribution relative to memory declines by the square of the factor in \eqref{eq:D31}.

\subsection{Calibration}
\label{sec:appD3}

We calibrate the framework to assess the magnitudes of the mechanisms underlying the empirical results. Returns around news announcements inform event drift, while the distribution of predictability across components and horizons informs the persistence and measurement of firm information. Changes in predictability around pretraining cutoffs help distinguish persistent firm information from memory. The calibration proceeds in two steps: preliminary parameter calibration followed by a joint search for the remaining parameters.

\paragraph{Preliminary calibration.}
Panel A of Table~\ref{tab:D2} reports the parameters fixed before the joint search. We determine $\kappa$ and $\vartheta$ from the category long--short portfolio's announcement-day return and its cumulative returns from the first eligible open through trading days 1 and 10. Up to the portfolio's return per unit of signal, these correspond to the two terms of \eqref{eq:D1}.

We set $\operatorname{Var}(m_K)$ using the category-mean forecast's next-day predictive covariance, $\delta_1^2\operatorname{Var}(m_K)$. A forecast's long--short return is proportional to its covariance with returns divided by its standard deviation, while its raw Fama--MacBeth slope is the covariance divided by its variance. Thus, the squared long--short return divided by the slope is proportional to predictive covariance. We treat the proportionality constant as common across forecasts evaluated on the same firm-days. We use this relation to infer the category-mean covariance from each embedding forecast's covariance and average the estimates from RoBERTa and Llama-2.

For Llama-2, we set $\phi(q)$ so that \eqref{eq:D7} equals its out-of-sample SUE $R^2$ gain of 3.2 percentage points, with event categories explaining 1.9\% of SUE variation. The out-of-sample gain lies below its population counterpart by Proposition~\ref{prop:D1}. Daily moments identify only $\phi(g)\pi^2$, so we fix Llama-2's state loading at $g=1$. The memory window follows \citet{crane2025total}, while the news-count parameters are measured directly from the data. In \eqref{eq:D8}, we set $\sigma_r$ so that RoBERTa's model IC equals its data IC.

\begin{table}[htbp]
\centering
\caption{Calibrated Parameters}
\label{tab:D2}
\begin{minipage}{\linewidth}\footnotesize
This table reports the calibrated parameters for ChronoGPT$_{1999}$, RoBERTa-large and Llama-2 (13B). Panel A lists values set before the joint search, Panel B the values from the search, and Panel C quantities implied by them. Source gives the moments or equations that mainly determine each value. The category signal and the detail are in percent of value, and the state premium is in basis points per trading day. $\Upsilon$ converts monthly covariances times $10^6$ into percent per year. The levels of $\phi(g)$ and $\pi$ are conditional on fixing Llama-2's $g$ at one. The last two rows are averages over the strategies' signal months, including those after each encoder's pretraining cutoff.
\end{minipage}
\medskip

{\footnotesize
\begin{adjustbox}{max width=\linewidth}
\begin{tabular}{llccccl}
\toprule
Parameter & Description & Common & ChronoGPT$_{1999}$ & RoBERTa & Llama-2 & Source \\
\midrule
\multicolumn{7}{@{}l}{\textit{Panel A: Set before the search}} \\
$\kappa$ & Announcement response & 0.68 &  &  &  & Category long--short return, \eqref{eq:D1} \\
$\vartheta$ & Daily drift decay & 0.26 &  &  &  & Category long--short return, \eqref{eq:D1} \\
$\mathrm{sd}(m_K)$ & Category signal & 0.52 &  &  &  & Category-mean long--short return and slope \\
$q$ & Detail loading &  &  &  & 0.18 & SUE $R^2$ gain, \eqref{eq:D7} \\
$g$ & State loading &  &  &  & 1 & Normalization \\
$N_\theta$ & Memory window, months &  &  & 360 & 360 & \citet{crane2025total} \\
$\bar n$ & News items per firm-month & 3.86 &  &  &  & Harmonic mean in the data \\
$\bar n^F$ & News items per firm, valuation windows & 189 &  &  &  & Data, \eqref{eq:D20} \\
\midrule
\multicolumn{7}{@{}l}{\textit{Panel B: Joint search}} \\
$\mathrm{sd}(\nu)$ & Detail & 1.06 &  &  &  & Daily $I_D$, \eqref{eq:D34} \\
$\pi$ & State premium & 2.7 &  &  &  & Daily $F$ net of memory, \eqref{eq:D32} \\
$\Var(\bar z)$ & Permanent share & 0.29 &  &  &  & $F$, $I_L$, \eqref{eq:D32}--\eqref{eq:D36} \\
$\rho_m$ & Monthly state persistence & 0.983 &  &  &  & Monthly components, \eqref{eq:D35}--\eqref{eq:D37} \\
$a$ & Category-signal loading &  & 0.88 & 0.95 & 1.03 & Daily $C_D$, \eqref{eq:D2} \\
$q$ & Detail loading &  & 0.14 & 0.14 &  & Daily $I_D$, \eqref{eq:D34} \\
$g$ & State loading &  & 0.39 & 0.43 &  & Daily $F$, $I_L$, monthly components \\
$\eta$ & Memory loading &  &  & 0.12 & 2.20 & IC ratios (Llama-2), monthly $F$, $I_L$ (RoBERTa) \\
$\Upsilon$ & Monthly strategy scale &  & 1.47 & 6.39 & 6.12 & Monthly components, \eqref{eq:D17} \\
$\sigma_\theta^2$ & Cash-flow-news variance, monthly & 0.011 &  &  &  & \eqref{eq:D23}, jointly with $\pi$ \\
\midrule
\multicolumn{7}{@{}l}{\textit{Panel C: Implied}} \\
 & Half-life of $x$, months & 39 &  &  &  & $\ln 2/(-\ln\rho_m)$ \\
$\phi(g)$ & Reliability of the state measure &  & 0.13 & 0.16 & 0.50 & $g^2/(1+g^2)$ \\
$\phi(\eta)$ & Reliability of the memory measure &  &  & 0.01 & 0.83 & $\eta^2/(1+\eta^2)$ \\
 & Memory's share of the first-month mean return, MSRR on $E_L$ (\%) &  & 0 & 24 & 47 & \eqref{eq:D16}, \eqref{eq:D26} \\
 & Memory's share of the first-month mean return, MSRR on $E_{L,\tau-36}$ (\%) &  & 0 & 40 & 66 & \eqref{eq:D16}, \eqref{eq:D31} \\
\bottomrule
\end{tabular}
\end{adjustbox}
}
\end{table}

\paragraph{Coverage shares.}
The model in Section~\ref{sec:appD1} describes a representative firm with news throughout the projection window, whereas the empirical decomposition retains all firms in the CKX and DKPT samples, including those with short news histories. For a firm covered only in the final year of the projection window, $\bar x^F$ and $\bar x^L$ would reflect similar news periods, causing $F$ to capture recent information rather than long-run firm information. To maintain a comparable interpretation of $F$ across firms, Section~\ref{sec:decomp} sets $F$ to zero for firms with fewer than ten news items in the projection window or news in only one calendar year, without excluding these firms from the sample.

For firms with $F=0$, persistent firm information that would otherwise enter $F$, including the permanent state, $\bar x^F$, and memory, instead enters $I_L$ if the firm has news in the preceding 12 months and $I_D$ otherwise. The remaining historical variation in the evolving state, $\bar x^L-\bar x^F$, likewise enters $I_L$ when trailing news is available and $I_D$ otherwise. Table~\ref{tab:D3} reports the shares $\varpi_F$, $\varpi_{I_L}$, and $\varpi_{I_D}$ associated with these assignments, together with the share $H$ of observations with news in the preceding 12 months.\footnote{Firm effects are determined from each test year's two-way projection. Shares are computed over firm-days for daily forecasts and over firm-months with news and next-month returns for monthly strategies, averaging news items within each firm-day or firm-month.} Assuming comparable average predictive covariance per observation across the three groups, these shares allocate persistent firm information's pooled covariance across $F$, $I_L$, and $I_D$.

\begin{table}[htbp]
\centering
\caption{Coverage Shares and Component Assignment}
\label{tab:D3}
\begin{minipage}{\linewidth}\footnotesize
This table reports the shares of firm-days and firm-months by whether the firm effect is estimated and whether the observation has a historical level. Component is where persistent firm information enters, and $H$ is the share of observations with news in the preceding 12 months.
\end{minipage}
\medskip

{\footnotesize
\begin{adjustbox}{max width=\linewidth}
\begin{tabular}{lcccc}
\toprule
 & Share & Component & Firm-days & Firm-months \\
\midrule
Firm effect estimated & $\varpi_F$ & $F$ & 0.936 & 0.912 \\
$F=0$, historical level available & $\varpi_{I_L}$ & $I_L$ & 0.061 & 0.081 \\
$F=0$, no historical level & $\varpi_{I_D}$ & $I_D$ & 0.003 & 0.007 \\
News available in the preceding 12 months & $H$ & & 0.996 & 0.991 \\
\bottomrule
\end{tabular}
\end{adjustbox}
}
\end{table}

Using the firm-day coverage shares in Table~\ref{tab:D3}, the daily predictive covariances of $F$, $I_L$, and $I_D$, including memory, which are the model values for the daily targets of the joint search, are
\begin{equation}\label{eq:D32}
\Gamma_{F,t}(h) = \varpi_F\left[\phi(g)\pi^2\left(\Var(\bar z)+\bar\rho_F\rho^{h-1}\Var(x)\right)+M_t(h)\right]
\end{equation}
\begin{equation}\label{eq:D33}
\begin{split}
\Gamma_{I_L,t}(h) &= \varpi_{I_L}\left[\phi(g)\pi^2\left(\Var(\bar z)+\bar\rho_F\rho^{h-1}\Var(x)\right)+M_t(h)\right]\\
&\quad + H\phi(g)\pi^2(\bar\rho_L-\bar\rho_F)\rho^{h-1}\Var(x)
\end{split}
\end{equation}
\begin{equation}\label{eq:D34}
\begin{split}
\Gamma_{I_D,t}(h) &= \phi(q)\delta_1\delta_h\Var(\nu) + \varpi_{I_D}\left[\phi(g)\pi^2\left(\Var(\bar z)+\bar\rho_F\rho^{h-1}\Var(x)\right)+M_t(h)\right]\\
&\quad + \phi(g)\pi^2\left((1-H)(\bar\rho_L-\bar\rho_F)+1-\bar\rho_L\right)\rho^{h-1}\Var(x)
\end{split}
\end{equation}
Let $M^m_\tau(h)$ equal \eqref{eq:D26} for a remembered return month and zero otherwise. Using the firm-month coverage shares, the monthly predictive covariances, which give the model values for the monthly targets up to the scale $\Upsilon$, are
\begin{equation}\label{eq:D35}
\Gamma^m_{F,\tau}(h) = \varpi_F\left[\phi(\sqrt{\bar n}g)\pi_m^2\left(\Var(\bar z)+\bar\rho_F\rho_m^{h-1}\Var(x)\right)+M^m_\tau(h)\right]
\end{equation}
\begin{equation}\label{eq:D36}
\begin{split}
\Gamma^m_{I_L,\tau}(h) &= \varpi_{I_L}\left[\phi(\sqrt{\bar n}g)\pi_m^2\left(\Var(\bar z)+\bar\rho_F\rho_m^{h-1}\Var(x)\right)+M^m_\tau(h)\right]\\
&\quad + H\phi(\sqrt{\bar n}g)\pi_m^2(\bar\rho_L-\bar\rho_F)\rho_m^{h-1}\Var(x)
\end{split}
\end{equation}
\begin{equation}\label{eq:D37}
\begin{split}
\Gamma^m_{I_D,\tau}(h) &= \bar n\phi(q)\bar\delta_1\bar\delta_h\Var(\nu) + \varpi_{I_D}\left[\phi(\sqrt{\bar n}g)\pi_m^2\left(\Var(\bar z)+\bar\rho_F\rho_m^{h-1}\Var(x)\right)+M^m_\tau(h)\right]\\
&\quad + \phi(\sqrt{\bar n}g)\pi_m^2\left((1-H)(\bar\rho_L-\bar\rho_F)+1-\bar\rho_L\right)\rho_m^{h-1}\Var(x)
\end{split}
\end{equation}
The memory covariances $\bar M_{j,p}^o$ in \eqref{eq:D27}--\eqref{eq:D29} correspond to the period averages of the memory terms in \eqref{eq:D32}--\eqref{eq:D34}. 

Although news from firms assigned $F=0$ accounts for only a small share of observations, their persistent firm information enters $I_L$ when historical news is available. Because, in the data, $I_L$'s predictive covariance is less than half that of $F$, this information can account for a sizeable share of $I_L$'s predictability. The permanent state of these firms can thus help explain the persistence of $I_L$'s returns across months.

\paragraph{Joint search.}
The second step minimizes
\[
\sum_{\mathcal B}\frac{1}{|\mathcal B|}\sum_{j\in\mathcal B}
\left(\frac{\hat m_j-m_j}{\mathrm{se}(\hat m_j)}\right)^2,
\]
where $\hat m_j$ is an empirical moment, $m_j$ its model counterpart, and $\mathrm{se}(\hat m_j)$ its standard error. The search chooses 16 parameters, with $\sigma_\theta^2$ solved jointly, to match the 49 moments of Table~\ref{tab:D4} below its first row. Each block $\mathcal B$ receives equal weight, regardless of the number of moments it contains. We use diagonal standard-error weights rather than estimating the full covariance matrix of the moments. Band standard errors are Newey--West, with five daily or six monthly lags plus the band's width minus one. 
For relative IC ratios, we infer standard errors from their 90\% confidence intervals under a normal approximation.

The cash-flow-news variance is determined by \eqref{eq:D23} within the joint search. The variance of 36-month returns is 0.415, estimated from seven non-overlapping windows starting in January 2004. At the calibrated parameters, state premia account for about 9\% of this variance, yielding $\sigma_\theta^2=0.011$. All 64 scrambled Sobol starts over wide parameter ranges reach the reported minimum.

\begin{table}[htbp]
\centering
\caption{Targeted Moments and Fit}
\label{tab:D4}
\begin{minipage}{\linewidth}\footnotesize
This table reports the moments targeted in the joint search, for ChronoGPT$_{1999}$, RoBERTa-large and Llama-2 (13B). Strategy returns are the residualized-embedding strategy's fixed-weight component returns before any horizon-specific volatility scaling, which by Proposition~\ref{prop:D2} equal $\Upsilon$ times the model's monthly covariances. Horizon bands are averages. Equation gives the model value. RMS $t$ is the root mean squared ratio of the data-minus-model difference to the standard error over the block's moments. The first row is matched exactly before the joint search.
\end{minipage}
\medskip

{\footnotesize
\begin{adjustbox}{max width=\linewidth}
\begin{tabular}{lcccc}
\toprule
Moment & Equation & Horizons & Count & RMS $t$ \\
\midrule
Cumulative category long--short return & \eqref{eq:D1} & Days 0, 1, 10 & 3 & 0.00 \\
\multicolumn{5}{@{}l}{\textit{Daily covariance with returns}} \\
$C_D$ & \eqref{eq:D2} & Day 1; days 2--10 & 6 & 0.71 \\
$I_D$ & \eqref{eq:D34} & Day 1; days 2--10 & 6 & 0.98 \\
$F$ & \eqref{eq:D32} & Days 2--10 & 3 & 0.73 \\
$I_L$ & \eqref{eq:D33} & Days 2--10 & 3 & 0.68 \\
\multicolumn{5}{@{}l}{\textit{MSRR strategy return}} \\
$F$ & \eqref{eq:D35} & Months 2--12, 13--24, 25--36 & 9 & 0.71 \\
$I_L$ & \eqref{eq:D36} & Months 2--12, 13--24, 25--36 & 9 & 1.14 \\
$I_D$ & \eqref{eq:D37} & Months 2--12, 13--24, 25--36 & 9 & 0.58 \\
\multicolumn{5}{@{}l}{\textit{Relative IC ratios of $F$ and $E_L$}} \\
Llama-2, January 2023 & \eqref{eq:D28} & Next day & 2 & 0.23 \\
RoBERTa, January 2020 & \eqref{eq:D29} & Next day & 2 & 0.26 \\
\midrule
Sobol starts reaching the minimum & & & 64 of 64 & \\
\bottomrule
\end{tabular}
\end{adjustbox}
}
\end{table}

\paragraph{Calibration targets and identification.}
We target the first horizon only where the model attributes predictability to event drift, namely daily $C_D$ and $I_D$. State-related daily targets and all monthly targets begin at horizon 2, except the relative IC ratios, which use next-day rank ICs and, unlike covariances, show no day-one excess. Little event drift remains after the signal month, so the monthly targets primarily discipline state persistence, the permanent share, and the allocation of predictability across components.

Day-one $C_D$ determines the category-signal loading $a$. Given Llama-2's $\phi(q)$, its day-one $I_D$ determines $\Var(\nu)$, while the other encoders' day-one $I_D$ determines their $q$. The decay of monthly returns across horizon bands and the split between $F$ and $I_L$ determine $\rho_m$ and the permanent share. Llama-2's memory loading is primarily determined by its relative IC ratios. RoBERTa's memory loading balances its monthly $F$, which favors more memory, against the relative IC ratios from both cutoff comparisons and its monthly $I_L$, which constrain it.

The state premium connects the cutoff comparisons to the other calibration targets. With Llama-2's $g=1$, $\pi$ is primarily determined by its daily $F$ covariance net of memory, while the relative IC ratios determine the memory share of that covariance. The state premium also enters the return-variance restriction in \eqref{eq:D23}. Llama-2's $\phi(\eta)=0.83$ places its memory covariance at 83\% of the bound in Section~\ref{sec:appD2}, so extending the memory window would leave less room for memory to match the cutoff ratios.

The monthly calibration involves two approximations. First, each encoder's strategy has a single scale $\Upsilon$, defined in \eqref{eq:D17}, which absorbs the overall scale of MSRR shrinkage. Because ridge shrinkage varies across predictive directions, using one $\Upsilon$ for all components and horizons is an approximation. The monthly targets therefore discipline the other parameters through relative component magnitudes and term structures. Second, monthly model values assume that the characteristics are uncorrelated with the permanent state and with current and past values of the evolving state, so that $z_X^j=0$ for each component and \eqref{eq:D18} reduces to \eqref{eq:D11}--\eqref{eq:D13}. Residualization leaves $\rho_m$ unchanged, so the decay of monthly predictability still identifies state persistence as long as the characteristics do not span the evolving state. The calibrated permanent share and allocation across components, however, depend on this assumption.

The excluded first-horizon moments provide checks on the model. In the data, firm identity's day-one covariance is about two to three times its average over days 2--10 for all three encoders, including ChronoGPT, whereas its rank IC is nearly unchanged: 0.23\% versus 0.26\% for ChronoGPT, 0.47\% versus 0.47\% for RoBERTa, and 1.09\% versus 1.02\% for Llama-2. This suggests that the excess day-one covariance reflects differences in return magnitudes or tails rather than stronger rank predictability. The model's state component, whose covariance changes little across days, does not capture this pattern.

Monthly $I_D$ also exhibits a pronounced first-horizon difference. Its month-one return exceeds its month-two return for all three encoders. For RoBERTa, the return standard deviation is similar across months 1, 2, and 3, indicating that the difference is primarily in mean returns. Refitting MSRR weights to month-two returns raises $I_D$'s annualized month-two return before volatility scaling from 2.3\% to 32.6\% for RoBERTa and from 20.9\% to 57.3\% for Llama-2, while RoBERTa's firm identity changes little, from 83.3\% to 79.9\%. Thus, weights fitted to month-one returns capture substantially less of the deviation's predictive content in month two than weights fitted to month-two returns. The calibrated model, with little event drift beyond the signal month and a single direction for the persistent state, cannot account for the magnitude of this horizon dependence.

\paragraph{Alternative specifications.}
Table~\ref{tab:D5} examines the sensitivity of the calibrated memory shares and state persistence to alternative specifications. We reduce the variance available to memory by the variance of state-related discount-rate news,\footnote{The variance of state-related discount-rate news is about 18\% of $\sigma_\theta^2$ (Section~\ref{sec:appD2}). Since memory covariance depends on $\sigma_\theta^2/N_\theta$, this reduction is equivalent to extending the memory window from 30 to about 37 years.} include month-one $F$ and $I_L$ returns among the targets, and set $\varpi_F=1$ instead of using the empirical coverage shares. We also allow observed characteristics to explain part of the state,\footnote{We set $R^2(z\mid X)=0.29$, the valuation $R^2$ implied by \eqref{eq:D20} for RoBERTa. We do not use Llama-2's 0.47, which may reflect remembered valuations. At the baseline loadings, this assumption implies that characteristics explain about 12\% of RoBERTa's and 23\% of Llama-2's variance along the monthly state coordinate.} impose $\Var(\bar z)=0$, or allow separate monthly scales $\Upsilon$ for $F$ and $I_L$.

Llama-2's memory shares remain between 67\% and 76\% for $F$ and between 61\% and 66\% for $E_L$, constrained by the same relative IC ratios targeted in every specification. RoBERTa's corresponding shares range from zero to 31\% and from zero to 22\%, respectively, remaining below half of Llama-2's in every specification. Reducing the variance available to memory pushes Llama-2's memory reliability $\phi(\eta)$ to its upper bound, as less variance available for memory requires a more reliable memory measure to match the cutoff ratios.

Long-horizon state persistence is also relatively stable. The firm state's 36-month autocorrelation ranges from 0.58 to 0.71, compared with 0.67 in the baseline, even as the evolving state's half-life varies from 11 to 67 months. In particular, setting $\Var(\bar z)=0$ still produces a 36-month autocorrelation of 0.69, with a half-life of 67 months. The separate-scale specification yields a similar autocorrelation but has multiple local minima, with only 38 of 64 starts reaching the reported minimum. Thus, the calibration consistently implies persistent firm information, although its division between permanent and evolving states is sensitive to the specification.

\begin{table}[htbp]
\centering
\caption{Calibration under Alternative Specifications}
\label{tab:D5}
\begin{minipage}{\linewidth}\footnotesize
This table reports the joint search under alternative specifications. Memory share is the percentage of the component's next-day predictive covariance attributable to memory, $1-\omega_j^o$, before January 2023 for Llama-2 and January 2020 for RoBERTa. $\rho_z(36)=\Var(\bar z)+\rho_m^{36}\Var(x)$ is the autocorrelation of the firm state at 36 months, the longest horizon of the monthly targets. Loss is the minimized criterion of the joint search. Each search runs from 64 starting values.
\end{minipage}
\medskip

{\footnotesize
\begin{adjustbox}{max width=\linewidth}
\begin{tabular}{lccccccccc}
\toprule
 & \multicolumn{2}{c}{Llama-2 memory (\%)} & \multicolumn{2}{c}{RoBERTa memory (\%)} & Llama-2 & Half-life of $x$ & & & Convergence to \\
\cmidrule(lr){2-3}\cmidrule(lr){4-5}
Specification & $F$ & $E_L$ & $F$ & $E_L$ & $\phi(\eta)$ & (months) & $\rho_z(36)$ & Loss & same minimum \\
\midrule
Baseline & 70 & 62 & 11 & 8 & 0.83 & 39 & 0.67 & 4.72 & 64/64 \\
$0.82\,\sigma_\theta^2$ in memory & 69 & 62 & 9 & 7 & 1.00 & 39 & 0.68 & 4.54 & 64/64 \\
$h=1$ targeted for $F$ and $I_L$ & 69 & 61 & 9 & 7 & 0.83 & 36 & 0.67 & 5.17 & 64/64 \\
$\varpi_F=1$ & 72 & 61 & 21 & 14 & 0.88 & 45 & 0.58 & 5.56 & 64/64 \\
$R^2(z\mid X)=0.29$ & 76 & 66 & 31 & 22 & 0.89 & 47 & 0.60 & 5.32 & 64/64 \\
$\Var(\bar z)=0$ & 72 & 64 & 18 & 13 & 0.86 & 67 & 0.69 & 4.80 & 64/64 \\
Separate $\Upsilon$ for $F$ and $I_L$ & 67 & 62 & 0 & 0 & 0.81 & 11 & 0.71 & 3.05 & 38/64 \\
\bottomrule
\end{tabular}
\end{adjustbox}
}
\end{table}

\end{appendices}

\end{document}